\documentclass[a4paper,11pt,twoside]{article}
\usepackage[utf8]{inputenc}
\usepackage[T1]{fontenc}
\usepackage[english]{babel}
\usepackage[hmargin={32mm,32mm},vmargin={32mm,35mm}]{geometry}
\usepackage{amsmath}
\usepackage{amsfonts}
\usepackage{amssymb}
\usepackage{enumerate}
\usepackage{graphicx}
\usepackage{tocloft}
\usepackage{bbm}
\usepackage{feynmf}
\usepackage{epsfig}
\usepackage{color}
\usepackage{slashed}
\usepackage{xspace}
\usepackage{url}
\usepackage{undertilde}
\usepackage{accents}
\usepackage[makeroom]{cancel}
\usepackage[hidelinks]{hyperref}
\usepackage{afterpage}
\usepackage{cite}
\usepackage{listings}
\usepackage{caption}
\usepackage{epstopdf}
\usepackage{bookmark}

\newcommand{\bra}[1]{\langle #1 |}
\newcommand{\ket}[1]{|#1\rangle}
\newcommand{\braket}[2]{\langle #1 | #2 \rangle}
\newcommand{\bbraket}[2]{\bigl\langle #1 \big| #2 \bigr\rangle}

\newcommand{\bbra}[1]{\bigl\langle #1 \bigr|}
\newcommand{\bket}[1]{\bigl|#1\bigr\rangle}

\newcommand{\be}{\begin{equation}}
\newcommand{\ee}{\end{equation}}
\newcommand{\bea}{\begin{eqnarray}}
\newcommand{\eea}{\end{eqnarray}}
\newcommand{\bal}{\begin{align}}
\newcommand{\eal}{\end{align}}

\newcommand{\eg}{e.g.\@\xspace}
\newcommand{\ie}{i.e.\@\xspace}
\newcommand{\Eq}[1]{Eq.\@\xspace\eqref{#1}}
\newcommand{\Eqs}[1]{Eqs.\@\xspace\eqref{#1}}
\newcommand{\Fig}[1]{Fig.\@\xspace\ref{#1}}
\newcommand{\Figs}[1]{Figs.\@\xspace\ref{#1}}

\newcommand{\updown}[2]{^{#1}_{\phantom{#1}#2}}
\newcommand{\downup}[2]{_{#1}^{\phantom{#1}#2}}

\DeclareMathOperator{\Tr}{{\rm Tr}}
\newcommand{\half}{\tfrac{1}{2}}

\newcommand{\Id}{\mathbbm{1}}
\newcommand{\R}{\mathbbm{R}}
\newcommand{\C}{\mathbbm{C}}

\newcommand{\makeSymbol}[1]{\mathord{\vcenter{\hbox{#1}}}}
\newcommand{\RealSymb}[2]{\makeSymbol{\includegraphics[scale=#2]{#1}}}

\numberwithin{equation}{section}

\newsavebox{\mybox}

\newcommand{\su}{\mathfrak{su}(2)}
\newcommand{\const}{\mathrm{const.}}
\newcommand{\Cyl}{\mathrm{Cyl}}
\newcommand{\D}[4]{{D^{(#1)#2}}_{#3}(#4)}
\newcommand{\Db}[4]{{D^{(#1)#2}}_{#3}\bigl(#4\bigr)}
\newcommand{\CG}[4]{C^{(#1#2)}{}\downup{#3}{#4}}
\newcommand{\CGi}[4]{C^{(#1#2)}{}\updown{#3}{#4}}
\newcommand{\dde}{\frac{d}{d\epsilon}\bigg|_{\epsilon=0}}
\newcommand{\Sigm}[4]{({\sigma^{(#1)}_{#2}}){}\updown{#3}{#4}}
\newcommand{\Tau}[4]{({\tau^{(#1)}_{#2}}){}\updown{#3}{#4}}
\newcommand{\threej}[6]{\begin{pmatrix} #1&#2&#3 \\ #4&#5&#6 \end{pmatrix}}
\newcommand{\sixj}[6]{\begin{Bmatrix} #1&#2&#3 \\ #4&#5&#6 \end{Bmatrix}}
\newcommand{\ninej}[9]{\begin{Bmatrix} #1&#2&#3 \\ #4&#5&#6 \\ #7&#8&#9 \end{Bmatrix}}

\newcommand{\sis}[1]{\phantomsection \addcontentsline{toc}{section}{#1}}
\addto\captionsenglish{\renewcommand{\contentsname}{{\LARGE Contents}}}

\newcounter{nopage}
\newenvironment{nopage}
 {\clearpage\stepcounter{nopage}
  
  \thispagestyle{empty}}
 {\clearpage\addtocounter{page}{-1}}

\begin{document}

\begin{nopage}

$\phantom{x}$

\vspace{48pt}

\begin{center}

{\Huge \textbf{Dynamics in canonical models of}}

\vspace{4pt}

{\Huge \textbf{loop quantum gravity}}

\vspace{30pt}

{\huge Ilkka M\"akinen}

\vspace{30pt}

\begin{figure}[h]
	\centering
		\includegraphics[width=0.4\textwidth]{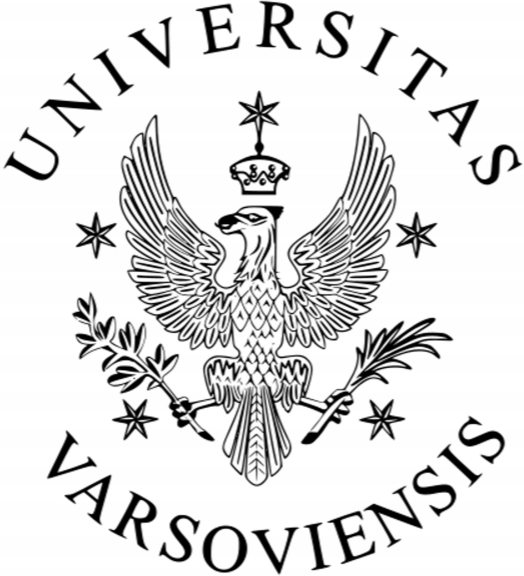}
\end{figure}

\vspace{24pt}

{\LARGE Dissertation for the degree of} 

\vspace{4pt}

{\LARGE Doctor of Philosophy}

\vspace{18pt}

{\Large Faculty of Physics} 

\vspace{4pt}

{\Large University of Warsaw}

\vspace{18pt}

{\Large May 2019}

\end{center}

\vspace{32pt}

{\large

\noindent Supervisor:

\vspace{2pt}

\noindent prof.\@\xspace dr hab.\@\xspace Jerzy Lewandowski (University of Warsaw)

\vspace{2pt}

\noindent Referees:

\vspace{2pt}

\noindent prof.\@\xspace dr hab.\@\xspace Jerzy Kowalski--Glikman (University of Wroc\l aw)

\vspace{2pt}

\noindent dr.\@\xspace Simone Speziale (Aix-Marseille University)

}

\end{nopage}

\begin{nopage}

\noindent{\LARGE \textbf{Acknowledgements}} \\

\noindent As a journey this long is nearing its completion, even I can be sufficiently moved to give thanks to those who have helped me at various points along the way:

My supervisor, prof.\@\xspace Jerzy Lewandowski, who offered me a place in his group at a time when it was not obvious that I would be able to find a position as a PhD student in loop quantum gravity. I am also grateful for the greatest possible freedom he has allowed me in carrying out my academic work, while always being available to provide guidance whenever I have been in need of it.

My referees, prof.\@\xspace Jerzy Kowalski--Glikman and dr.\@\xspace Simone Speziale, for the time they have devoted to reviewing this thesis, whose page count has repeatedly exceeded all of my expectations.

The friends, colleagues and collaborators, whose worldlines have overlapped with mine throughout my years at the Faculty of Physics in Warsaw. There are far too many of you to acknowledge all of you by name, but I am particularly thankful to Emanuele for introducing me to the art of little diagrams and for reminding me of the importance of imagination and fantasy in physics, to Jędrek and Marcin for patiently guiding me through various practical problems of everyday life when I arrived in Poland as a clueless newcomer, and most of all to Mehdi, an invaluable collaborator and a trusty friend, whose endless optimism -- both in physics and in life in general -- continues to provide me with an often-needed source of encouragement.

The Polish disc golf community and the all friends I have made there. You have offered me a welcome distraction from the rigors of my scientific work, and in this way you, more than anything else, have made Poland feel like a second home to me.

Mike Gottlieb, Robert van Leeuwen, Kimmo Tuominen, and others who encouraged me to keep pursuing what interests me the most in physics, even when an eyebrow or two was being raised at the adventurous young student who had set his mind on entering the fascinating world of loop quantum gravity.

The Jenny and Antti Wihuri Foundation and the Narodowe Centrum Nauki, whose financial support has made this work possible.

Finally, and most importantly, my family -- the unyielding bedrock on which I have stood securely while making my way to where I am today -- and above all my parents, who have always supported me fully, unconditionally and in every possible way in everything I have ever taken upon myself to do.

\end{nopage}

\begin{nopage}

\enlargethispage{\baselineskip}

\noindent{\LARGE \textbf{Abstract}} \\

\noindent Loop quantum gravity is a tentative but relatively well-developed attempt at a manifestly background independent theory of quantum gravity. The kinematical structure of loop quantum gravity is well understood, with the elementary kinematical states of the theory having a physically appealing interpretation as states describing quantized, discrete spatial geometries. However, formulating the dynamics of the theory in a fully satisfactory way has proven to be surprisingly challenging, and to a significant extent, the issue of dynamics still remains an open problem in loop quantum gravity.

In this thesis we consider the problem of dynamics in canonical loop quantum gravity, primarily in the context of deparametrized models, in which a scalar field is taken as a physical time variable for the dynamics of the gravitational field. The dynamics of the quantum states of the gravitational field is then generated directly by a physical Hamiltonian operator, instead of being implicitly defined through the kernel of a Hamiltonian constraint. The deparametrized approach therefore bypasses many of the considerable technical difficulties one encounters when working with the constraint operator.

We introduce a new construction of a Hamiltonian operator for loop quantum gravity, which has both mathematical and practical advantages in comparison to earlier proposals. Most importantly, our Hamiltonian can be constructed as a symmetric operator, and is therefore a mathematically consistent candidate for a generator of unitary physical time evolution in deparametrized models. Furthermore, by employing the recently introduced curvature operator in our construction, we obtain a significant practical simplification in the structure of the Hamiltonian operator.

After having completed the definition of several variations of the new Hamiltonian, we develop methods for evaluating the dynamics generated by a given physical Hamiltonian in an approximate form, even if an exact solution to the eigenvalue problem of the Hamiltonian cannot be achieved. In particular, we show that for large values of the Barbero--Immirzi parameter, the Euclidean part of the Hamiltonian can be considered as a perturbation over the comparatively simple curvature operator, and hence standard perturbation theory can be used to approximate the spectrum of the complete Hamiltonian in terms of the known eigenvalues and eigenstates of the curvature operator. We illustrate the use of our approximation methods with simple numerical calculations.

We also introduce a new representation for intertwiners in loop quantum gravity. By considering intertwiners projected onto coherent states of angular momentum, we obtain a description in which intertwiners are represented as polynomials of certain complex variables, and operators in loop quantum gravity are expressed as differential operators acting on these variables. This makes it possible to describe the action of the Hamiltonian geometrically, in terms of the unit vectors parametrizing the angular momentum coherent states, as opposed to complicated, non-intuitive expressions involving combinations of the Wigner $nj$-symbols of $SU(2)$ recoupling theory. The new representation could therefore turn out to be a useful tool for analyzing the dynamics in loop quantum gravity.

In addition to reviewing the results of the author's scientific work, this thesis also gives a thorough introduction to the basic framework of canonical loop quantum gravity, as well as a self-contained presentation of the graphical formalism for $SU(2)$ recoupling theory, which is the invaluable tool for performing practical calculations in loop quantum gravity in the spin network basis. The author therefore hopes that parts of this thesis could serve as a comprehensible source of information for anyone interested in learning the elements of loop quantum gravity.

\end{nopage}

\begin{nopage}

\noindent{\LARGE \textbf{Streszczenie}} \\

\noindent Pętlowa grawitacja kwantowa jest pierwszą i stosunkowo zaawansowaną próbą konstrukcji jawnie niezależnej od tła teorii grawitacji kwantowej. Kinematyczna struktura pętlowej grawitacji kwantowej jest dobrze poznana, przy czym elementarne stany kinematyczne teorii mają fizycznie dobrze zrozumiałą interpretację jako stany opisujące skwantowane, dyskretne geometrie przestrzenne. Sformułowanie dynamiki teorii w sposób w pełni zadowalający okazało się jednak zaskakująco trudne i w znacznym stopniu kwestia dynamiki nadal pozostaje otwartym problemem w kwantowej grawitacji pętli.

W rozprawie tej rozważamy problem dynamiki w kanonicznej kwantowej grawitacji pętlowej głównie w kontekście modeli deparametryzowanych, w których pole skalarne pełni funkcję czasu fizycznego dla dynamiki pola grawitacyjnego. Dynamika stanów kwantowych pola grawitacyjnego jest następnie generowana bezpośrednio przez fizyczny operator Hamiltona, zamiast być domyślnie definiowana przez jądro hamiltonowskiego operatora więzów. W związku z tym podejście zdeparametryzowane pozwala ominąć wiele znacznych trudności technicznych napotykanych podczas pracy z więzami hamil-tonowskimi.

Wprowadzamy nową konstrukcję operatora hamiltonowskiego dla pętlowej grawitacji kwantowej, która ma zarówno zalety matematyczne, jak i praktyczne w porównaniu z wcześniejszymi propozycjami. Co najważniejsze, nasz Hamiltonian może być skonstruowany jako operator symetryczny, a zatem jest matematycznie spójnym kandydatem na generator unitarnej ewolucji czasowej w zdeparametryzowanych modelach. Ponadto, dzięki zastosowaniu niedawno wprowadzonego w naszej konstrukcji operatora krzywizny, jesteśmy w stanie dokonać znacznego i praktycznego uproszczenia w strukturze operatora hamiltonowskiego.

Po zakończeniu analizy różnych możliwych definicji nowego hamiltonianu, opracowujemy przybliżone metody badań dynamiki generowanej przez dany fizyczny hamiltonian przydatne, gdy nie można uzyskać dokładnego rozwiązania problemu wartości własnych hamiltonianu. W szczególności pokazujemy, że dla dużych wartości parametru Barbero--Immirzi, część euklidesową hamiltonianu można uznać za zaburzenie względnie prostego operatora krzywizny, a zatem można zastosować standardową teorię zaburzeń w celu przybliżenia widma pełnego hamiltonianu pod względem znanych wartości własnych i stanów własnych operatora krzywizny. Zilustrowaliśmy wykorzystanie naszych metod aproksymacji za pomocą prostych obliczeń numerycznych.

Wprowadzamy również nową reprezentację dla operatorów splatających w pętlowej grawitacji kwantowej. Rozpatrując operatory splatające rzutowane na stany koherentne momentu pędu, otrzymujemy opis, w którym operatory splatające są reprezentowane jako wielomiany pewnych zmiennych zespolonych, a operatory w kwantowej grawitacji pętlowej są wyrażane jako operatory różnicowe działające na te zmienne. Umożliwia to opisanie działania hamiltonianu geometrycznie, pod względem znormalizowanych wektorów parametryzujących stany koherentne momentu pędu, w przeciwieństwie do skomplikowanych, nieintuicyjnych wyrażeń obejmujących kombinacje symboli Wignera $nj$ teorii sprzężenia $SU(2)$. Nowa reprezentacja może zatem okazać się użytecznym narzędziem do analizy dynamiki w kwantowej grawitacji pętli.

Oprócz przeglądu wyników pracy naukowej autora, rozprawa zawiera również do-kładne wprowadzenie do formalizmu kanonicznej kwantowej grawitacji pętlowej, a także samodzielną prezentację metod graficznych dla teorii sprzężenia $SU(2)$, która jest nieocenionym narzędziem do wykonywania praktycznych obliczeń w kwantowej grawitacji pętlowej w bazie sieci spinowej. Autor ma zatem nadzieję, że część rozprawy może służyć jako zrozumiałe źródło wiedzy dla wszystkich zainteresowanych nauką elementów kwantowej grawitacji pętlowej.

\end{nopage}

\clearpage\addtocounter{page}{-1}

\renewcommand{\contentsname}{{\LARGE Contents} \vspace{12pt}}

\pagestyle{empty}
{
  \renewcommand{\thispagestyle}[1]{}
  \tableofcontents
}
\clearpage
\pagestyle{plain}

\clearpage\addtocounter{page}{-1}

\begin{nopage}

$\phantom{x}$

\vspace{96pt}

\vspace{24pt}

\noindent\textbf{\huge Introduction} 

\vspace{72pt}

\begin{quote}

The history of science is rich in the example of the fruitfulness of bringing two sets of techniques, two sets of ideas, developed in separate contexts for the pursuit of new truth, into touch with one another.

\begin{flushright}
-- J.\@\xspace Robert Oppenheimer
\end{flushright}
\end{quote}

\end{nopage}

\pagenumbering{arabic}
\setcounter{page}{1}

\sis{Introduction}

\subsection*{The problem of quantum gravity}

\enlargethispage{\baselineskip}

\noindent Quantum mechanics -- including quantum field theory as the generalization to systems with an infinite number of degrees of freedom -- and the general theory of relativity are the two cornerstones on which our current understanding of the physical world is founded. Besides providing us with a beautiful geometrical picture of gravitation, general relativity has found physical applications \eg in cosmology and astrophysics. The numerous accomplishments of quantum mechanics arguably culminate in the extraordinary empirical success of the standard model of particle physics.

However, despite being supported by vast amounts of experimental evidence, the fact remains that general relativity and quantum mechanics are fundamentally inconsistent with each other. On one hand, conventional quantum field theories describe excitations of quantum fields propagating over a fixed, non-dynamical background spacetime. On the other hand, one of the central elements of general relativity is the idea that spacetime itself is a dynamical entity. The geometry of spacetime is encoded in the metric tensor, which is a dynamical field, its dynamics being governed by the Einstein equation. In turn, quantum theory has taught us that any dynamical field is inherently a quantum object, and the excitations of the field appear as discrete quanta.

Taken together, the ideas of general relativity and quantum mechanics therefore suggest that spacetime itself should be quantized, with the quantum nature of spacetime presumably becoming apparent at sufficiently small distance scales. General relativity's picture of spacetime as a smooth classical manifold would then be only an approximate notion, which could not be expected to continue being valid down to arbitrarily small scales. The singularities which are present in general relativity, for example in black holes and in the beginning of the universe, can be taken as a hint that this is indeed the case. That is, the singularities should not be seen as legitimate predictions of general relativity, but rather as indications that in these extreme physical circumstances the classical description of spacetime provided by general relativity simply breaks down, and the theory is not able to give a reliable account of the situation.

This state of affairs therefore opens up a fascinating challenge: To understand general relativity and quantum mechanics in terms of a single, consistent underlying framework. This problem, the problem of quantum gravity, is arguably the most significant unsolved problem in fundamental theoretical physics today.

Historically, the methods of conventional quantum field theory were brought to bear on the problem of quantum gravity soon after they had found great success in the case of quantum electrodynamics. A line of research led by DeWitt attempted to formulate a quantum theory of gravity by employing a perturbative expansion of the metric around flat Minkowski spacetime (or possibly some other fixed background metric). However, it was eventually determined that the divergences from which the resulting theory suffers are non-renormalizable, and therefore the theory itself is not a satisfactory candidate for a fundamental theory of physics.

In hindsight, the failure of conventional quantum field theory to produce a sensible theory of quantum gravity is perhaps not very surprising. It simply seems to confirm that the conflict is irreparably large between the fundamental ideas of general relativity and the assumptions inherent in the framework of perturbative quantum field theory. A theory in which the quantized gravitational field is represented as something like a graviton field propagating over a flat background spacetime seems rather questionable in light of the fact that spacetime is dynamical in general relativity -- in a sense, the gravitational field is nothing but the geometry of spacetime itself.

The above considerations present a rather strong reason to expect that a successful theory of quantum gravity, which conforms to the basic lessons learned from general relativity, should be formulated as a background independent quantum field theory. That is, the theory should be expressed in a way in which no reference is made to any fixed, non-dynamical background spacetime.

\subsection*{Loop quantum gravity}

Loop quantum gravity is a promising and relatively well-developed proposal for a manifestly background independent theory of quantum gravity. Essentially, loop quantum gravity is an attempt to answer the question of whether we can find a common framework which is able to accommodate the fundamental ideas of both general relativity and quantum mechanics. In this sense, loop quantum gravity is less ambitious than certain other approaches to the problem of quantum gravity, most notably string theory, which seek to formulate a theory of quantum gravity by developing a unified theory of all the fundamental interactions.

Nevertheless, the strategy of taking two well-established theories and asking what they imply when considered together has repeatedly proven to be tremendously successful in the history of physics, often leading to insights far beyond what could have been inferred on the basis of either theory alone. Special relativity was born out of combining Galilean relativity with Maxwell's electrodynamics, quantum field theory resulted from merging non-relativistic quantum mechanics with special relativity, while combining quantum electrodynamics and the theory of weak interactions into a unified description of the electroweak interaction was a crucial step in the development of the standard model of particle physics. In the same way, loop quantum gravity can be seen as the result of combining quantum mechanics and general relativity.

The kinematical framework of loop quantum gravity is essentially the outcome of a straightforward canonical quantization of the Ashtekar formulation of general relativity, following the ideas laid out by Dirac on the quantization of generally covariant theories. Thus, the only basic inputs which go into the construction of the theory are quantum mechanics and general relativity, including in particular the idea that background independence is of fundamental importance. Loop quantum gravity does not require the introduction of any radically new physical assumptions -- such as additional spacetime dimensions which would have to be compactified in an ad hoc manner, or the existence of supersymmetric particles which continue to evade the best efforts of the experimentalists to detect them -- for the internal consistency of the theory\footnote{On the other hand, the existence of extra dimensions or supersymmetry can be incorporated into loop quantum gravity, as shown in \cite{dim1, dim2, dim3, dim4, susy1, susy2, susy3}.}.

At the kinematical level, the structure of loop quantum gravity is well understood. The kinematical Hilbert space of the theory is spanned by the so-called spin network states, which have a compelling physical interpretation as states describing discrete, quantized spatial geometries\footnote{This interpretation is derived by studying the eigenvalue problem of operators corresponding to geometric quantities such as volumes of regions and areas of surfaces. The discrete nature of spatial geometry is therefore a result of the theory; it is not a hypothesis inserted by hand during the construction of the formalism.}. In this way, loop quantum gravity is able to provide a concrete realization of the idea of the quantized gravitational field as a dynamical object, whose excitations are the elementary quanta out of which spacetime itself is built.

\subsection*{The issue of dynamics}

While the kinematics of loop quantum gravity is by now well established, the question of formulating the dynamics of the theory in a satisfactory manner has proven to be a source of considerable difficulties. According to Dirac's quantization algorithm, the dynamics should be specified by promoting the Hamiltonian constraint of classical general relativity into an operator in the quantum theory and looking for the states which are annihilated by this operator. The states belonging to the kernel of the constraint operator form the physical Hilbert space of the theory, and the dynamics is determined by the scalar product on the physical Hilbert space, which defines transition amplitudes between physical states.

The first step of the program, namely the construction a well-defined Hamiltonian constraint operator, was accomplished by Thiemann already more than twenty years ago. However, to this day very little concrete knowledge has been established concerning the structure of the physical Hilbert space defined by Thiemann's constraint operator, or any of its variants constructed later by other authors. From a practical perspective, the task of deriving any non-trivial solutions of the constraint in explicit form (let alone ''interesting'' solutions, to which one could associate a clear physical interpretation) has turned out to be extremely challenging.

The difficulties encountered in working with the Hamiltonian constraint have motivated many practitioners of loop quantum gravity to look for alternative ways of formulating the dynamics of the theory. Perhaps the most popular among these is the spin foam formalism, also often referred to as covariant loop quantum gravity, which completely abandons the canonical formulation of the dynamics, introducing instead a particular implementation of the path integral for general relativity, which enables one to define the dynamics by associating transition amplitudes to spin network states.

In this work, the problem of dynamics is considered within the framework of the canonical theory, and the main tool used for trying to find an adequate solution is the so-called method of deparametrization. In practice, this consists of considering loop quantum gravity coupled to a suitable scalar field, and using the scalar field as a physical, relational time variable, with respect to which the dynamics of the gravitational field is described.

On a technical level, a deparametrized model of loop quantum gravity trades the Hamiltonian constraint for a physical Hamiltonian operator, which generates time evolution of spin network states with respect to the time defined by the reference scalar field. Thus, most of the problems associated with extracting solutions of the constraint and understanding the structure of the physical Hilbert space are bypassed in the deparametrized context. Conceptually, deparametrization can be seen as a particular way of circumventing the infamous ''problem of time'' in general relativity. While no physically meaningful information is contained in the evolution of the geometric degrees of freedom with respect to an essentially arbitrary time coordinate, the way in which spatial geometry changes in relation to the value of the scalar field nevertheless has an intrinsic, physically significant meaning.

The problem of dynamics is certainly not the only open issue in loop quantum gravity today. Among the other challenges, we may mention the problem of the continuum limit of the theory, \ie understanding how a seemingly smooth classical geometry can be recovered from the fundamentally discrete quantum states of loop quantum gravity. Extracting falsifiable physical predictions from the formalism is another problem which is poorly understood at the moment, but which must necessarily be dealt with in order to ascertain whether loop quantum gravity is not only a mathematically consistent theory of quantum gravity, but also provides a physically correct description of the observable world. However, interesting as these questions are, they are outside the scope of the present work. As far as the open problems of loop quantum gravity go, this thesis is confined to the issue of looking for a satisfactory formulation of the dynamics in the context of the canonical theory.

Moreover, even though this work is entirely concerned with loop quantum gravity, our discussion would be distinctly incomplete if we did not at least mention the other principal lines of research within which an answer to the question of quantum gravity is currently being sought. Asymptotic safety, causal dynamical triangulations, causal sets, group field theory, noncommutative geometry, and others are all motivated by their respective philosophies, and come with their own sets of strengths and weaknesses. Each of them has the potential to contribute valuable insights to the problem at hand, even if they ultimately turn out to not be the correct answer to the question of the quantum theory of gravity.

\subsection*{Outline of this thesis}

The central theme of this thesis is the use of deparametrized models of general relativity, in which a scalar field plays the role of a physical time variable for the dynamics of the gravitational field, as a tool for formulating the dynamics of loop quantum gravity in a satisfactory manner. In the author's opinion, there are two main criteria which a model of loop quantum gravity has to satisfy in order to be considered ''satisfactory''. On one hand, it is certainly necessary to pay a certain level of attention to mathematical details while constructing the model, so as to ensure that the resulting model will not suffer from any serious mathematical inconsistencies. On the other hand, it is equally important for the structure of the model to be simple enough that there can be at least a reasonable hope that the model could one day be used to make concrete calculations in order to investigate the physical content of the model and assess its viability as a theory of physics. For a physicist, having a model with which one cannot make calculations about physics is hardly better than having no model at all.

A presentation of the author's scientific work on the dynamics of deparametrized models of loop quantum gravity could certainly have been accomplished in far fewer pages than this thesis actually contains. The discrepancy is explained by the secondary purpose of this work: To provide a reasonably self-contained overview of loop quantum gravity in its canonical formulation. In particular, it is my hope that parts of this thesis could serve as an accessible introduction to the more practical aspects of canonical loop quantum gravity\footnote{As far as the mathematical and conceptual aspects of the theory are concerned, several excellent expositions, such as the books of Thiemann \cite{Thiemann} and Rovelli \cite{Rovelli}, are already available.} for students who are entering the field, or are otherwise interested in learning the elements of the subject in a deeper than superficial level. (Sometimes it feels like it was only yesterday that I was in that situation myself.)

With the above considerations in mind, the material in this work (after the present introductory chapter) is divided into four parts. The first part comprises the first ten chapters of the thesis, and primarily serves as an introduction to the kinematical framework of loop quantum gravity. Chapters \ref{ch:canonicalgr} and \ref{ch:holonomyflux} set the classical foundations for the quantum theory by presenting the relevant canonical formulations of general relativity, and the proper choice of classical variables to be quantized. Chapters \ref{ch:Hkin} and \ref{ch:operators} introduce the kinematical Hilbert space of loop quantum gravity and the elementary operators thereon. Chapter \ref{ch:gauss} deals with gauge invariance, and Chapter \ref{ch:spinnetwork} describes the basic structure of the spin network states, which are the elementary gauge invariant quantum states of the theory. The implementation of diffeomorphism invariance is discussed in Chapter \ref{ch:diff}. In Chapter \ref{ch:geometric} we introduce operators corresponding to basic geometric quantities: areas, volumes, angles and lengths.

In Chapter \ref{ch:CS} we present the complexifier coherent states introduced to loop quantum gravity by Thiemann. This sets the stage for Chapter \ref{ch:CSoperators}, which concludes the first part of the thesis, and in which we use the ideas of ''coherent state quantization'' to propose an alternative prescription to define the basic operators of loop quantum gravity, and study the properties of the operators constructed in this way. The results presented in Chapter \ref{ch:CSoperators} were obtained by the author together with collaborators, and they have been published in the article \cite{paper3}. They are not directly relevant to the main topic of this thesis, but they have been included anyway, since it is reasonable to expect that the techniques developed in Chapter \ref{ch:CSoperators} could be used to investigate the dynamics of loop quantum gravity once the technology associated with coherent states in the theory becomes more fully developed.

The second part of this work consists of Chapters \ref{ch:dynamics}--\ref{ch:coherent3j}, and deals with the issue of dynamics in canonical loop quantum gravity, particularly in the context of deparametrized models. Chapter \ref{ch:dynamics} contains a general introduction to the problem of dynamics. Chapter \ref{ch:deparametrized} gives a concise presentation of the classical theory of the deparametrized models which are considered later in the work; in particular, it establishes the form of the physical Hamiltonian which must be quantized in order to define the dynamics of the quantum theory. Chapter \ref{ch:Thiemann} reviews Thiemann's construction of the Hamiltonian constraint operator. On one hand, this provides as a prototype for how Hamiltonians for loop quantum gravity can be constructed. On the other hand, by studying Thiemann's Hamiltonian we uncover the reason why it cannot serve as a mathematically consistent physical Hamiltonian in deparametrized models, and hence motivate the need to search for an alternative way of constructing the physical Hamiltonian.

The main results of this thesis, which are presented in Chapters \ref{ch:Hphys}--\ref{ch:coherent3j}, are as follows:
\begin{itemize}
\item Construction of a physical Hamiltonian operator for loop quantum gravity deparametrized with respect to a free scalar field. The essential new feature of the construction is a new technique for regularizing the Euclidean part of the Hamiltonian. The new regularization ensures that the adjoint of the Euclidean part is available as a densely defined operator, making it possible to construct a symmetric Hamiltonian, which can be a mathematically consistent candidate for the generator of physical time evolution. Another aspect of the construction is the use of the curvature operator introduced earlier by members of the Warsaw group in place of the Lorentzian part of Thiemann's Hamiltonian. This step leads to a considerable simplification in the structure of the Hamiltonian. (Chapter \ref{ch:Hphys})
\item Extension of the ideas introduced in Chapter \ref{ch:Hphys} to the case of the Hamiltonian constraint. The constraint operator can be consistently defined on the so-called vertex Hilbert space proposed by Lewandowski and Sahlmann. The resulting operator not only serves as the constraint operator in the vacuum theory, but also provides a physical Hamiltonian operator for loop quantum gravity deparametrized with respect to non-rotational dust. (Chapter \ref{ch:constraint})
\item An explicit computation of the matrix elements of the Hamiltonian (or, more precisely, the individual operators out of which the Hamiltonian is constructed) in the spin network basis. The recoupling theory of $SU(2)$, and in particular the associated graphical formalism, is the central tool used in the calculations. (Chapter \ref{ch:elements})
\item Development of approximation methods which can be used to compute time evolution in deparametrized models, even if an exact knowledge of the spectrum of the physical Hamiltonian is not available. The main approximation scheme is based on the observation that for sufficiently large values of the Barbero--Immirzi parameter, the Euclidean part of the Hamiltonian can be considered as a perturbation over the comparatively much simpler curvature operator. Standard time-independent perturbation theory can then be used to approximate the spectrum of the entire Hamiltonian in terms of the numerically accessible eigenvalues and eigenstates of the curvature operator. (Chapter \ref{ch:approximation})
\item A novel description of intertwiners in loop quantum gravity, based on projecting intertwiners onto the basis of angular momentum coherent states instead of the conventional basis of magnetic indices. Within this formalism, intertwiners can be expressed as polynomials of certain complex variables, while operators in loop quantum gravity, in particular the Hamiltonian, can be formulated as differential operators acting on these variables. The main motivation for this work is the hope that the action of the Hamiltonian could become more transparent when it is expressed geometrically, in the language of the unit vectors associate to the spin coherent states, as opposed to non-intuitive expressions involving combinations of the Wigner $nj$-symbols of $SU(2)$ recoupling theory. (Chapter \ref{ch:coherent3j})
\end{itemize}
In the third part of this thesis, we summarize the work presented in it, and assess the significance of the results obtained. We also point out the central questions which have been left unanswered in this work, and mention other possible directions for future research. The fourth part of the thesis consists of an appendix in which we give a self-contained presentation of those aspects of $SU(2)$ representation and recoupling theory which are relevant to practical calculations in loop quantum gravity. In particular, we present a detailed introduction to the graphical methods which constitute an invaluable tool for computing matrix elements of loop quantum gravity operators in the spin network basis.  

The material presented in Chapters \ref{ch:Hphys}, \ref{ch:constraint}, \ref{ch:approximation} and \ref{ch:coherent3j} is based on the work published respectively in the articles \cite{paper1}, \cite{paper2}, \cite{paper5} and \cite{paper4}. The details of computing matrix elements of the Hamiltonian have not previously been published anywhere, though the matrix elements given in Chapter \ref{ch:elements} formed the basis for the numerical calculations presented in \cite{paper5}. In addition to the articles mentioned above, various aspects of the author's work on the dynamics of canonical loop quantum gravity have also been discussed in the conference proceedings \cite{proc2}, \cite{proc3} and \cite{proc1}.

\begin{nopage}

\sis{Part 1. The kinematical framework of loop quantum gravity}

$\phantom{x}$

\vspace{96pt}

\noindent\textbf{\LARGE Part 1 \\}

\vspace{24pt}

\noindent\textbf{\huge The kinematical framework of \vspace{10pt} \\
loop quantum gravity} 

\vspace{72pt}

\begin{quote}

How did I get here? Trickle becomes stream; tributaries run together, gathering force. The march of ideas carves channels into the landscape -- ideas borne by individuals who are in turn swept away by its current. This river is our history. We walk in paths worn down by those who came before us. Each of us arrives midstream, joining a procession so entrenched as to appear as that's just how it is. From deep within these grooves, it's hard to imagine people just like us set it all in motion.

\begin{flushright}
-- Nick Sousanis
\end{flushright}
\end{quote}

\end{nopage}

\section{Canonical formulations of general relativity}\label{ch:canonicalgr}

The classical theory of general relativity is encoded in the Einstein-Hilbert action
\be\label{S_EH}
S[g] = \frac{1}{16\pi G}\int d^4x\,\sqrt{-g}R,
\ee
which leads to the Einstein equations upon variation with respect to the metric $g_{\mu\nu}$. The fundamental symmetry of general relativity is diffeomorphism invariance, which is reflected as the invariance of the action \eqref{S_EH} under general coordinate transformations $x^\mu \to y^\mu(x)$. In particular, a natural notion of time does not exist in the theory. ''Time'' is simply one of the spacetime coordinates $x^\mu$; as such, it is subject to arbitrary reparametrizations and lacks any physically distinguished meaning.

\subsection{Foliation of spacetime}

The absence of ''time'' in general relativity means that a necessary prelude for a canonical quantization of the theory is to artificially introduce a time variable into the formalism. To achieve the splitting of the spacetime manifold $M$ into ''space'' and ''time'', one introduces a set of spacelike hypersurfaces $\{\Sigma_t\}$, which foliate the entire manifold $M$. On each spatial surface $\Sigma$, we have the induced metric
\be\label{q_mn}
q_{\mu\nu} = g_{\mu\nu} + n_\mu n_\nu,
\ee
where $n_\mu$ is the unit normal of $\Sigma$ (\ie $n^\mu n_\mu = -1$). The metric $q_{\mu\nu}$ is determined uniquely by the conditions $q_{\mu\nu} n^\nu = 0$, and $g_{\mu\nu}m^\nu = q_{\mu\nu}m^\nu$ for any vector $m^\mu$ tangent to $\Sigma$. Taken in the form $q^\mu_\nu = \delta^\mu_\nu + n^\mu n_\nu$, the induced metric acts as a projection operator, whose action on any vector removes the component orthogonal to $\Sigma$.

By itself, the decomposition of spacetime into the surfaces $\{\Sigma_t\}$ does not enable one to view fields on spacetime as fields on space evolving with respect to the artificial time parameter labeling the surfaces, even if the surfaces are defined as the constant surfaces of some time function $t$ on spacetime. What is lacking from the picture is a well-defined way of associating points on different surfaces $\Sigma_t$ to each other. Such an association can be provided by choosing a ''time-evolution vector field'' $t^\mu$, which must satisfy $t^\mu\nabla_\mu t = 1$, and whose integral curves identify points on different surfaces $\Sigma_t$ as the same spatial point at different moments of time. Once the vector field $t^\mu$ is chosen, time derivatives of spatial tensor fields are naturally defined as
\be
\dot T\updown{\mu_1\cdots \mu_m}{\nu_1\cdots \nu_n} = q^{\mu_1}_{\alpha_1}\cdots q^{\mu_m}_{\alpha_m}q^{\nu_1}_{\beta_1}\cdots q^{\nu_n}_{\beta_n} {\cal L}_t T\updown{\alpha_1\cdots \alpha_m}{\beta_1\cdots \beta_n}
\ee
(where the projections are typically necessary to ensure that the resulting tensor is again spatial).

In general, the time-evolution vector field can be decomposed into components orthogonal and tangential to the spatial surface as
\be
t^\mu = Nn^\mu + N^\mu.
\ee
The freedom in choosing $t^\mu$ is now parametrized by the lapse function $N = -t^\mu n_\mu$ and the shift vector $N^\mu = q^{\mu\nu}t_\nu$. The freedom to choose arbitrary coordinates in the spacetime formulation of general relativity is reflected as the ability to choose the lapse and the shift arbitrarily.

\begin{figure}[t]
	\centering
		\includegraphics[width=0.5\textwidth]{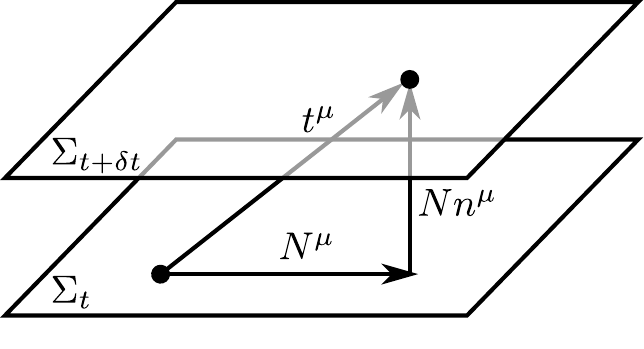}
		\caption{Foliation of spacetime.}
		\label{fig1-1}
\end{figure}

To express the spacetime metric in this formalism, we start from
\be
g^{\mu\nu} = q^{\mu\nu} - n^\mu n^\nu = q^{\mu\nu} - \frac{1}{N^2}(t^\mu - N^\mu)(t^\nu-N^\nu).
\ee
Writing this as a matrix, in coordinates where $t^\mu = (1,0)$ and $N^\mu = (0,N^a)$, we find
\be
g^{\mu\nu} = \begin{pmatrix} g^{00} & g^{a0} \\ g^{0b} & g^{ab} \end{pmatrix} = \begin{pmatrix} -1/N^2 & N^a/N^2 \\ N^b/N^2 & q^{ab} - N^aN^b/N^2\end{pmatrix}.
\ee
The inverse of this matrix is given by
\be
g_{\mu\nu} = \begin{pmatrix} -N^2 + N^aN_a & N_a \\ N_b & q_{ab} \end{pmatrix},
\ee
and hence the spacetime metric takes the form
\be\label{3+1metric}
ds^2 = g_{\mu\nu} dx^\mu dx^\nu = -N^2\,dt^2 + q_{ab}\bigl(dx^a + N^a dt\bigr)\bigl(dx^b + N^b dt\bigr).
\ee
Here we see explicitly that spacetime geometry, described by the ten components of $g_{\mu\nu}$, has been traded for the geometry of the spatial surfaces, described by the six components of $q_{ab}$, together with the four functions $(N,N^a)$ encoding the way in which the spatial surfaces are embedded in spacetime.

\subsection{Curvature of the spatial surfaces}

The spatial metric $q_{ab}$ has its associated covariant derivative, which is determined by the compatibility condition $D_aq_{bc} = 0$. Equivalently, the spatial covariant derivative is given by the projection of the spacetime covariant derivative down to the spatial surface:
\be\label{D_a}
D_aT\updown{b_1\cdots b_m}{c_1\cdots c_n} = q^{b_1}_{\mu_1}\cdots q^{b_m}_{\mu_m} q^{\nu_1}_{c_1}\cdots q^{\nu_n}_{c_n} q^\lambda_a \nabla_\lambda T\updown{\mu_1\cdots\mu_m}{\nu_1\cdots\nu_n}.
\ee
Recalling that $q^{\mu\nu}n_\nu=0$, we can verify that
\be
D_aq_{bc} = q^\mu_b q^\nu_c q^\lambda_a \nabla_\lambda\bigl(g_{\mu\nu} + n_\mu n_\nu\bigr) = 0.
\ee
Furthermore, for any scalar function $\phi$ on $\Sigma$, we have
\be
(D_aD_b - D_bD_a)\phi = q^\mu_a q^\nu_b\Bigl((\partial_\mu\partial_\nu - \partial_\nu\partial_\mu)\phi + (\Gamma^\lambda_{\mu\nu} - \Gamma^\lambda_{\nu\mu})\partial_\lambda\phi\Bigr) = 0,
\ee
where $\Gamma^\lambda_{\mu\nu}$ are the Christoffel symbols of the spacetime metric $g_{\mu\nu}$. Therefore $D_a$ is indeed the unique torsionless covariant derivative compatible with $q_{ab}$.

The spatial covariant derivative defines a spatial curvature tensor through the relation
\be
(D_aD_b - D_bD_a)v^c = {}^{(3)}\!R\updown{c}{dab}v^d.
\ee
The tensor ${}^{(3)}\!R\updown{a}{bcd}$ measures the intrinsic curvature of the spatial surface. Information about the curvature associated with the way in which the spatial surface is embedded in spacetime is contained in the extrinsic curvature tensor
\be
K_{ab} = D_an_b = q^\mu_a q^\nu_b \nabla_\mu n_\nu.
\ee
Equivalently, the extrinsic curvature can be defined as the Lie derivative of the spatial metric along the normal vector $n^\mu$:
\be\label{K = Lq}
K_{ab} = \frac{1}{2}{\cal L}_{\vec n} q_{ab}.
\ee
This shows that $K_{ab}$ is a symmetric tensor. Furthermore, the time derivative of the spatial metric is related to the extrinsic curvature by\footnote{For reasons of space, we refrain from displaying the proofs of several statements made in this chapter. They can be found \eg in \cite{Bojowald} or \cite{Thiemann}, from which most of the material in this chapter has been gathered.}
\be\label{dot q}
\dot q_{ab} = 2NK_{ab} + D_aN_b + D_bN_a.
\ee

There exist various curvature theorems relating the intrinsic and extrinsic curvature of the spatial surface to the four-dimensional spacetime curvature tensor. We have the Gauss equation
\be\label{Gausseq}
q^\alpha_a q^\beta_b q^\gamma_c q^\delta_d R_{\alpha\beta\gamma\delta} = {}^{(3)}\!R_{abcd} + K_{ac}K_{bd} - K_{ad}K_{bc},
\ee
the Codazzi equation
\be
q^\alpha_a q^\beta_b q^\gamma_c n^\mu R_{\mu\alpha\beta\gamma} = D_cK_{ab} - D_bK_{ac},
\ee
and the Ricci equation
\be
n^\mu n^\nu R_{a\mu b \nu} = -{\cal L}_{\vec n}K_{ab} + K_{ac}K^c_b + D_aa_b + a_aa_b,
\ee
where $a_a = n^\mu\nabla_\mu n_a$. Furthermore, contraction of the Ricci equation yields
\be\label{contrRicci}
n^\mu n^\nu R_{\mu\nu} = K^2 - K_{ab}K^{ab} + \nabla_\mu v^\mu,
\ee
where $K = K^a_a$ and $v^\mu = n^\nu\nabla_\nu n^\mu - n^\mu\nabla_\nu n^\nu$.

\subsection{The ADM formulation}

With the help of the curvature theorems of the previous section, the Einstein--Hilbert action can be rewritten by eliminating the spacetime metric in favour of the variables associated with the 3+1 decomposition of spacetime. Using \Eqs{Gausseq} and \eqref{contrRicci}, and exploiting the symmetries of the Riemann tensor, we find
\begin{align}
R &= g^{\mu\nu}g^{\lambda\sigma}R_{\mu\lambda\nu\sigma} \notag \\
&=\bigl(q^{\mu\nu}q^{\lambda\sigma} - q^{\mu\nu}n^\lambda n^\sigma - n^\mu n^\nu q^{\lambda\sigma} + n^\mu n^\nu n^\lambda n^\sigma\bigr)R_{\mu\lambda\nu\sigma} \notag \\
&= q^{ab}q^{cd}q^\mu_a q^\nu_b q^\lambda_c q^\sigma_d R_{\mu\lambda\nu\sigma} - \Bigl[\bigl(g^{\mu\nu} + n^\mu n^\nu\bigr) n^\lambda n^\sigma + n^\mu n^\nu\bigl(g^{\lambda\sigma} + n^\lambda n^\sigma\bigr)\Bigr]R_{\mu\lambda\nu\sigma} \notag \\
&= q^{ab}q^{cd}\Bigl({}^{(3)}\!R_{acbd} + K_{ab}K_{cd} - K_{ac}K_{bd}\Bigr) - 2n^\mu n^\nu R_{\mu\nu} \notag \\
&= {}^{(3)}\!R + K_{ab}K^{ab} - K^2 - 2\nabla_\mu v^\mu. \label{R->R3}
\end{align}
Noting also that $\det g = -N^2\det q$, and neglecting the boundary term arising from the last term in \Eq{R->R3}, we conclude that the Lagrangian of the Einstein--Hilbert action becomes
\be\label{L_ADM}
L = \frac{1}{16\pi G} \int d^3x\,N\sqrt q\Bigl(K_{ab}K^{ab} - K^2 + {}^{(3)}\!R\Bigr),
\ee
with a ''kinetic term'' involving the extrinsic curvature, and a ''potential term'' depending only on the spatial metric. Considering the spatial metric as the configuration variable, the associated canonical momentum can be extracted as
\be
p^{ab} = \frac{\delta L}{\delta\dot h_{ab}} = \frac{1}{2N}\frac{\delta L}{\delta K_{ab}} = \frac{1}{16\pi G}\sqrt q\bigl(K^{ab} - Kq^{ab}\bigr).
\ee
The canonical Hamiltonian
\be
H = \int d^3x\,p^{ab}\dot q_{ab} - L,
\ee
then comes out in the form
\be\label{H_ADM}
H = \int d^3x\,\biggl[16\pi G\frac{N}{\sqrt q}\Bigl(p_{ab}p^{ab} - \frac{1}{2}p^2\Bigr) - \frac{1}{16\pi G}N\sqrt q{}^{(3)}\!R + 2p^{ab}D_aN_b\biggr].
\ee
Since the Lagrangian \eqref{L_ADM} does not contain time derivatives of $N$ or $N^a$, the lapse and the shift appear as Lagrange multipliers in the Hamiltonian \eqref{H_ADM}. The equations of motion of $N$ and $N^a$ are equivalent to the constraint equations $C=0$ and $C_a=0$, where
\be\label{C_a-ADM}
C_a = -2D_bp^b_a
\ee
is called the diffeomorphism constraint (or vector constraint), while
\be\label{C-ADM}
C = 16\pi G\frac{1}{\sqrt q}\Bigl(p_{ab}p^{ab} - \frac{1}{2}p^2\Bigr) - \frac{1}{16\pi G}\sqrt q{}^{(3)}\!R
\ee
is the Hamiltonian constraint (or scalar constraint). Only the components of the spatial metric $q_{ab}$ have a dynamical equation of motion, given by $\dot q_{ab} = \{q_{ab},H\}$. In fact, by considering contractions of the Gauss and Codazzi equations, one finds
\be
C_a = -\frac{\sqrt q}{8\pi G}n^\mu q^\nu_a G_{\mu\nu}  \qquad \text{and} \qquad  \qquad C = -\frac{\sqrt q}{8\pi G}n^\mu n^\nu G_{\mu\nu},
\ee
where $G_{\mu\nu}$ is the Einstein tensor. This shows that the constraint equations $C=0$ and $C_a=0$ are equivalent to the four Einstein equations $G_{00}=0$ and $G_{0a}=0$, while the equations $G_{ab}=0$ are encoded in the equations of motion of the spatial metric.

The total Hamiltonian 
\be\label{H_ADM2}
H = \int d^3x\,\bigl(NC + N^aC_a\bigr)
\ee
is only a linear combination of the constraints, and therefore vanishes when the equations of motion are satisfied. This is a general feature in the mechanics of reparametrization-invariant systems, and is a reflection of the fact that evolution with respect to the time parameter $t$ is not a true, physical time evolution, but merely a kind of gauge transformation, since the time parameter can be arbitrarily redefined through coordinate transformations.

\subsection{Analysis of the constraints}

To uncover the geometric significance of the diffeomorphism and Hamiltonian constraint, it is convenient to study them in the smeared form
\begin{align}
C(\vec N) &= \int d^3x\,N^aC_a, \\
C(N) &= \int d^3x\,NC.
\end{align}
Straightforward calculations show that the Poisson brackets between the canonical variables and the diffeomorphism constraint are given by
\be
\{q_{ab},C(\vec N)\} = {\cal L}_{\vec N}q_{ab}, \qquad \{p^{ab},C(\vec N)\} = {\cal L}_{\vec N}p^{ab},
\ee
indicating that the diffeomorphism constraint indeed generates diffeomorphisms along the spatial surface. For the Hamiltonian constraint, it is still relatively simple to find
\be
\{q_{ab},C(N)\} = {\cal L}_{N\vec n}q_{ab}.
\ee
However, a much more involved calculation, a detailed outline of which is given in \cite{Thiemann}, is required to establish the result
\be
\{p^{ab},C(N)\} = {\cal L}_{N\vec n}p^{ab} + \frac{1}{2}NCq^{ab} - \frac{1}{16\pi G}N\sqrt q\bigl(q^{a\mu}q^{b\nu} - q^{ab}q^{\mu\nu}\bigr)R_{\mu\nu}.
\ee
Therefore the Hamiltonian constraint generates diffeomorphisms in the direction orthogonal to the spatial surface, though this interpretation is valid only when the equations of motion are satisfied.

The Poisson brackets of the constraints among themselves are given by
\begin{subequations}\label{ADMalgebra}
\begin{align}
\{C(\vec M),C(\vec N)\} &= C({\cal L}_{\vec M}\vec N)\\
\{C(\vec M),C(N)\} &= C({\cal L}_{\vec M}N) \\
\{C(M),C(N)\} &= C(\vec X[q]) \label{C,C ADM}
\end{align}
\end{subequations}
where in the last equation we have the phase space dependent vector field
\be
X^a[q] = q^{ab}(M\partial_bN - N\partial_bM).
\ee
We see that the algebra of the constraints closes, in the sense that the Poisson brackets between the constraints involve only the constraints themselves. This means that the ''constraint surface'' (\ie the region of phase space in which the constraints are satisfied) is preserved under the action of the constraints. In other words, the evolution generated by the Hamiltonian \eqref{H_ADM2} preserves the constraint surface, and no further constraints arise from having to require that the constraints imposed at some initial time will continue to hold at later times. In the language of Dirac's theory of constrained systems \cite{Dirac}, the set of constraints $(C_a,C)$ is first class.

The canonical formulation of general relativity described in this and the previous section was developed in 1962 by Arnowitt, Deser and Misner \cite{ADM1, ADM2}, whose work was largely motivated by the desire to find a satisfactory classical starting point for a canonical quantization of general relativity. A substantial amount of work was subsequently done on the quantization of the ADM formulation, particularly by DeWitt \cite{DeWitt1, DeWitt2, DeWitt3}. In the quantum theory, the momenta $p^{ab}$ act as functional derivatives on wave functionals of the spatial metric, and therefore major problems arise both from the highly non-polynomial way in which the Hamiltonian constraint \eqref{C-ADM} depends on the canonical variables, and from insufficient mathematical understanding of the ''space of metrics'' on which wave functionals of the form $\Psi[q_{ab}]$ are defined. In fact, no mathematically rigorous quantization of the ADM formulation has ever emerged, prompting the search for alternative, more easily quantizable formulations of canonical general relativity.

\subsection{An intermediate formulation}

A new set of variables for canonical general relativity was introduced by Ashtekar in 1986 \cite{Ashtekar1986, Ashtekar1987}. The formulation of general relativity in Ashtekar's variables is the classical starting point underlying loop quantum gravity. Following \cite{Thiemann}, we will make our way from the ADM variables $(q_{ab},p^{ab})$ to the Ashtekar variables by means of an appropriate canonical transformation.\footnote{The Ashtekar variables can also be derived by performing a 3+1 decomposition of the Holst \mbox{action\cite{Holst}}
\[
S[e,\omega] = \frac{1}{16\pi G} \int d^4x\,(\det e)e^\mu_I e^\nu_J\biggl(F_{\mu\nu}^{IJ}[\omega] - \frac{1}{2\beta}\epsilon\updown{IJ}{KL}F_{\mu\nu}^{KL}[\omega]\biggr).
\]
Here the first term is the Palatini action in the tetrad formalism, and the second term (with an arbitrary coupling constant $1/2\beta$) vanishes identically when $F_{\mu\nu}^{IJ}$ is the Riemann curvature, and therefore has no effect on the classical equations of motion. For details of the derivation, see \eg \cite{Bojowald}.
} The transformation is carried out in two steps. The first step starts with the introduction of an orthonormal triad $e^a_i(x)$ for the spatial metric, 
\be
q_{ab}e^a_i e^b_j = \delta_{ij},
\ee
the goal being to use (a suitably modified version of) the triad as one half of the new pair of canonical variables. The co-triad $e_a^i = q_{ab}e^{bi}$ is the inverse of $e^a_i$ both with respect to the spatial index and the internal index:
\be
e_a^i e^a_j = \delta^i_j, \qquad e_a^i e_i^b = \delta_a^b.
\ee
The spatial metric and its inverse can then be expressed in terms of the triad as $q_{ab} = e_a^i e_{bi}$ and $q^{ab} = e^a_i e^{bi}$.

We then define the densitized triad
\be
E^a_i = \sqrt q e^a_i
\ee
as the prospective new canonical variable. The inverse spatial metric is related to the densitized triad by
\be
qq^{ab} = E^a_i E^{bi}.
\ee
By taking the time derivative of this equation, one finds
\be
\dot q^{ab} = \frac{1}{q}\bigl(\dot E^a_i E^{bi} + E^a_i\dot E^{bi} - q^{ab}\dot E^c_i E^{ci}\bigr),
\ee
which can be used to show that the canonical term in the ADM action becomes
\be
\int d^4x\,p^{ab}\dot q_{ab} = \frac{1}{8\pi G}\int d^4x\,E^a_i\dot K_a^i,
\ee
where $K_a^i$ is the extrinsic curvature in the form
\be
K_a^i = K_{ab}e^{bi}.
\ee
This suggests that $E^a_i/8\pi G$ can indeed be taken as a canonical momentum, with the corresponding configuration variable being $K_a^i$. 

For a more precise demonstration of this claim, one must show that the Poisson brackets
\begin{subequations}\label{K,E}
\begin{align}
\{K_a^i(x),K_b^j(y)\} &= 0 \\
\{E^a_i(x),E^b_j(y)\} &= 0 \\ 
\{K_a^i(x),E^b_j(y)\} &= 8\pi G\delta_a^b\delta^i_j\delta(x,y)
\end{align}
\end{subequations}
are equivalent to the canonical Poisson brackets of the ADM variables, namely
\begin{subequations}
\begin{align}
\{q_{ab}(x),q_{cd}(y)\} &= 0 \label{q,q}\\
\{p^{ab}(x),p^{cd}(y)\} &= 0 \label{p,p}\\ 
\{q_{ab}(x),p^{cd}(y)\} &= \delta_{(a}^c\delta_{b)}^d \delta(x,y). \label{q,p}
\end{align}
\end{subequations}
It turns out that \eqref{q,q} and \eqref{q,p} are automatically satisfied, but the remaining bracket \eqref{p,p} is reproduced only if the constraint
\be
E^a_iK^{bi} = E^b_iK^{ai}
\ee
(which fixes $K_{ab}$ to be a symmetric tensor) holds. The new constraint, known as the Gauss constraint, can be written in the equivalent form
\be\label{G-KE}
G_i = \frac{1}{8\pi G}\epsilon\downup{ij}{k}K_a^jE^a_k = 0,
\ee
which suggests that it is the generator of rotations associated with the invariance of the spatial metric under internal $SU(2)$ rotations of the triad. Further calculations show that the diffeomorphism and Hamiltonian constraints are expressed in terms of the new variables as
\be\label{D-KE}
C_a = \frac{1}{8\pi G}\Bigl(D_b\bigl(K_a^i E^b_i\bigr) - D_a\bigl(K_b^iE^b_i\bigr)\Bigr)
\ee
and
\be\label{C-KE}
C = -\frac{1}{16\pi G}\biggl(\frac{E^a_iE^b_j}{\sqrt{\det E}}\bigl(K_a^iK_b^j - K_a^jK_b^i\bigr) + \sqrt{\det E}\,{}^{(3)}\!R\biggr).
\ee

\subsection{Ashtekar variables}

In order to pass from the intermediate variables $(K_a^i,E^a_i)$ to the proper Ashtekar variables, we start by defining a covariant derivative in the internal space as
\be
D_av^i = \partial_a v^i + {\omega_a}\updown{i}{j}v^j.
\ee
The explicit form of the connection ${\omega_a}\updown{i}{j}$ is determined by the requirement that the covariant derivative is compatible with the triad,
\be\label{D_ae_b}
D_ae_b^i = \partial_a e_b^i - \Gamma^c_{ab}e_c^i + {\omega_a}\updown{i}{j}e_b^j = 0.
\ee
From ${\omega_a}\updown{i}{j}$ we construct the so-called spin connection
\be
\Gamma_a^i = -\frac{1}{2}\epsilon^{ijk}\omega_{ajk}.
\ee
Its explicit expression in terms of the densitized triad reads
\begin{align}
\Gamma_a^i &= \frac{1}{2}\epsilon^{ijk}e^b_k\Bigl(\partial_b e_{aj} - \partial_a e_{bj} + e^c_je_a^l\partial_b e_{cl}\Bigr)\notag \\
&= \frac{1}{2}\epsilon^{ijk}E^b_k\biggl(\partial_b E_{aj} - \partial_a E_{bj} + E^c_jE_a^l\partial_b E_{cl} + E_{aj}\frac{\partial_b\det E}{\det E}\biggr) \label{Gamma_a^i}
\end{align}
(where $E_a^i = e_a^i/\sqrt q$ is the inverse of $E^a_i$). The Ashtekar connection is then defined as
\be\label{A_a^i}
A_a^i = \Gamma_a^i + \beta K_a^i.
\ee
In Ashtekar's original formulation, the value of the parameter $\beta$ was fixed to $\beta = \pm i$, since this choice turns out to significantly simplify the form of the Hamiltonian constraint. However, this comes at the price of having to impose reality conditions, which turn out to be difficult to deal with, especially in the context of quantum theory. The connection \eqref{A_a^i} with arbitrary values of $\beta$ was considered later by Immirzi \cite{Immirzi} (for complex $\beta$) and Barbero \cite{Barbero} (for real $\beta$). For this reason the parameter $\beta$, which is now taken to be real-valued but otherwise arbitrary, is known as the Barbero--Immirzi parameter.

The Ashtekar connection turns out to be canonically conjugate to the densitized triad, even though this certainly does not seem obvious at a first sight, when looking at the explicit form of the spin connection \eqref{Gamma_a^i}. A hint that $A_a^i$ and $E^a_i$ might be a pair of canonical variables can be derived by considering the antisymmetric part of \Eq{D_ae_b},
\be
\partial_{[a}e_{b]}^i + \epsilon\updown{i}{jk}\Gamma^j_{[a}e_{b]}^k = 0.
\ee
Taking the time derivative of this equation and contracting suitably with the triad, one can show that
\be\label{EdGamma}
E^a_i\dot\Gamma_a^i = -\frac{1}{2}\epsilon\downup{i}{jk}(\det e)e^a_je^b_kD_a\dot e_b^i.
\ee
Hence $E^a_i\dot\Gamma_a^i$ is a total derivative, so up to a boundary term we have
\be
\frac{1}{8\pi G}\int d^3x\,E^a_i\dot K_a^i = \frac{1}{8\pi\beta G}\int d^3x\,E^a_i\dot A_a^i,
\ee
suggesting that $A_a^i$ and $E^a_i/8\pi\beta G$ are canonically conjugate. To verify this, one must check that the canonical Poisson brackets
\begin{subequations}
\begin{align}
\{A_a^i(x),A_b^j(y)\} &= 0 \\
\{E^a_i(x),E^b_j(y)\} &= 0 \\ 
\{A_a^i(x),E^b_j(y)\} &= 8\pi\beta G\delta_a^b\delta^i_j\delta(x,y)
\end{align}
\end{subequations}
follow from the brackets \eqref{K,E} of the intermediate variables. The only bracket which is not trivially satisfied is the first one,
\be
\{A_a^i(x),A_b^j(y)\} = 8\pi\beta G\biggl(\frac{\delta\Gamma_a^i(x)}{\delta E^b_j(y)} - \frac{\delta\Gamma_b^j(x)}{\delta E^b_j(y)}\biggr).
\ee
We see that a sufficient condition for the right-hand side to vanish is that the spin connection $\Gamma_a^i$ can be derived from a generating function $\Phi$ as $\Gamma_a^i(x) = {\delta\Phi}/{\delta E^a_i(x)}$. A generating function which does the trick is given by $\Phi = \int d^3x\,E^a_i\Gamma_a^i$, since \Eq{EdGamma} indicates that the variation $E^a_i\delta\Gamma_a^i$ is a total derivative. This concludes the demonstration that the variables $A_a^i$ and $E^a_i/8\pi\beta G$ are canonical.

\subsection{Constraints in the Ashtekar formulation}

We will now translate the constraints \eqref{G-KE}--\eqref{C-KE} into the Ashtekar variables. Defining the covariant derivative of the Ashtekar connection as
\be
{\cal D}_av^i = \partial_a v^i + \epsilon\updown{i}{jk} A_a^jv^k,
\ee
we can write the Gauss constraint in the form
\be
G_i = \frac{1}{8\pi\beta G}{\cal D}_aE^a_i,
\ee
since $E^a_i$ is constant with respect to the covariant derivative of the spin connection.

The key to expressing the diffeomorphism and Hamiltonian constraint in terms of the Ashtekar variables is the identity
\be\label{FR-identity}
F_{ab}^i = R_{ab}^i + \beta\bigl(D_aK_b^i - D_bK_a^i\bigr) + \beta^2\epsilon\updown{i}{jk}K_a^jK_b^k,
\ee
which relates to each other the curvature tensors of the spin connection and the Ashtekar connection,
\begin{align}
R_{ab}^i &= \partial_a\Gamma_b^i - \partial_b\Gamma_a^i + \epsilon\updown{i}{jk}\Gamma_a^j\Gamma_b^k, \\
F_{ab}^i &= \partial_a A_b^i - \partial_b A_a^i + \epsilon\updown{i}{jk}A_a^jA_b^k.
\end{align}
For the diffeomorphism constraint \eqref{D-KE}, we contract \Eq{FR-identity} with $E^b_i$, obtaining
\be\label{FR-contr}
\beta\Bigl(D_a\bigl(E^b_iK_b^i\bigr) - D_b\bigl(K_a^iE^b_i\bigr)\Bigr) = F_{ab}^iE^b_i - R_{ab}^iE^b_i - \beta^2\epsilon\updown{i}{jk}K_a^jK_b^kE^b_i.
\ee
The last term on the right-hand side is proportional to the Gauss constraint \eqref{G-KE}, and we will therefore ignore it. For the second term on the right, the structure equations $de^i + \Gamma\updown{i}{j}\wedge e^j = 0$ and $R\updown{i}{j} = d\Gamma\updown{i}{j} + \Gamma\updown{i}{k}\wedge\Gamma\updown{k}{j}$ imply $R\updown{i}{j}\wedge e^j = 0$, from which it follows that $R_{ab}^ie^b_i = 0$. Hence \Eq{FR-contr} indicates that the diffeomorphism constraint \eqref{D-KE} is expressed in terms of the Ashtekar variables as
\be\label{C_a-AE}
C_a = \frac{1}{8\pi\beta G}F_{ab}^iE^b_i.
\ee
For the Hamiltonian constraint, contraction of \Eq{FR-identity} with $\epsilon\downup{i}{jk}E^a_jE^b_k$ gives
\be\label{FR-contr2}
\epsilon\updown{ij}{k}E^a_i E^b_jF_{ab}^k = \beta^2\bigl(K_a^iK_b^j - K_a^jK_b^i\bigr)E^a_iE^b_j - (\det E){}^{(3)}\! R,
\ee
where a term involving the Gauss constraint has again been dropped. Using \Eq{FR-contr2}, we can eliminate either one of the terms in \Eq{C-KE} in favour of the other. In this way we obtain the two equivalent expressions
\be\label{C-AE-1}
C = \frac{1}{16\pi G}\frac{E^a_iE^b_j}{\sqrt{\det E}}\Bigl(\epsilon\updown{ij}{k}F_{ab}^k - (1+\beta^2)\bigl(K_a^iK_b^j - K_a^jK_b^i\bigr)\Bigr)
\ee
and
\be\label{C-AE-2}
C = \frac{1}{16\pi G}\frac{1}{\beta^2}\biggl(\frac{\epsilon\updown{ij}{k}E^a_iE^b_jF_{ab}^k}{\sqrt{\det E}} + (1+\beta^2)\sqrt{\det E}\,{}^{(3)}\! R\biggr)
\ee
for the Hamiltonian constraint in the Ashtekar variables. (In \Eq{C-AE-1}, $K_a^i$ should of course be understood as a function of the variables $A_a^i$ and $E^a_i$.) \Eq{C-AE-1} is the more well-known form of the constraint, at least in the loop quantum gravity literature, but \Eq{C-AE-2} will play a significant role later in this work, providing an important ingredient for a new formulation of the dynamics in loop quantum gravity.

The smeared Gauss constraint
\be\label{G-ash}
G(\lambda) = \int d^3x\,\lambda^iG_i
\ee
generates local $SU(2)$ gauge transformations on the canonical variables. This is indicated by the Poisson brackets
\be
\{A_a^i,G(\lambda)\} = -{\cal D}_a\lambda^i, \qquad \{E^a_i,G(\lambda)\} = \epsilon\downup{ij}{k}\lambda^j E^a_k,
\ee
where the expressions on the right-hand side are the variations of the $\su$-valued objects $A_a = A_a^i\tau_i$ and $E^a = E^a_i\tau^i$ under an infinitesimal version of the gauge transformation $A_a \to gA_a g^{-1} + g\partial_a g^{-1}$, $E^a \to gE^ag^{-1}$.

Diffeomorphisms along the spatial surface are generated by the modified diffeomorphism constraint
\be\label{D-ash}
D(\vec N) = \int d^3x\,N^a\bigl(C_a - A_a^iG_i\bigr),
\ee
in which a multiple of the Gauss constraint has been subtracted from \eqref{C_a-AE}. The Poisson brackets of the canonical variables with the modified constraint \eqref{D-ash} read
\be
\{A_a^i,D(\vec N)\} = {\cal L}_{\vec N}A_a^i, \qquad \{E^a_i,D(\vec N)\} = {\cal L}_{\vec N}E^a_i.
\ee

The Poisson brackets of the constraints \eqref{G-ash} and \eqref{D-ash} and the Hamiltonian constraint $C(N) = \int d^3x\,NC$ between themselves are given by
\begin{subequations}
\begin{align}
\{G(\lambda),G(\mu)\} &= G([\lambda,\mu]) \\
\{D(\vec N),G(\lambda)\} &= G({\cal L}_{\vec N}\lambda) \\
\{D(\vec M),D(\vec N)\} &= D({\cal L}_{\vec M}\vec N) \\
\{D(\vec M),C(N)\} &= C({\cal L}_{\vec M}N) \\
\{G(\lambda),C(N)\} &= 0 \\
\{C(M),C(N)\} &= D(\vec Y[E]) + G(\rho[A,E]).
\end{align}
\end{subequations}
In the first equation $[\lambda,\mu]^i = \epsilon\updown{i}{jk}\lambda^j\mu^k$ is the $SU(2)$ commutator, and in the last equation
\be
Y^a[E] = \beta^2\frac{E^a_iE^{bi}}{\det E}(M\partial_a N - N\partial_a M),
\ee
and an explicit expression for the function $\rho[A,E]$ can be found \eg in \cite{status}. We see that the algebra of the constraints is again first class in Dirac's terminology.

\newpage

\section{The elementary classical variables}\label{ch:holonomyflux}

The quantization of any classical theory must start with the choice of an appropriate set of elementary variables on the classical phase space, which form the starting point for the construction of the quantum theory. In the case of general relativity, these variables should be defined in a background-independent manner, without reference to a metric or any other fixed background structures on the spatial manifold $\Sigma$. Further desirable properties of the chosen variables would be simple behavior under spatial diffeomorphisms and internal gauge transformations, as well as sufficient smearing so that the Poisson bracket between the basic variables (which is to be promoted into a commutator in the quantum theory) is non-singular.

The Ashtekar variables $(A_a^i,E^a_i)$ do not provide a suitable starting point for quantization by themselves. However, natural geometric objects related to the Ashtekar connection and the densitized triad are holonomies of the connection along curves, and fluxes of the triad through surfaces in the spatial manifold. Loop quantum gravity is based on the choice of these holonomies and fluxes as the elementary classical variables. This choice turns out to fulfill all the requirements outlined above. Moreover, it has the additional advantage that similar variables are well-known in the context of lattice gauge theories, so any insights that have been gained there may potentially be useful also in the case of loop quantum gravity.

\subsection{The holonomy}

The holonomy $h_e[A]$ is the $SU(2)$-valued parallel propagator of the Ashtekar connection $A_a^i$ along a curve $e$ in the spatial manifold $\Sigma$. In other words, if $e$ is parametrized by a parameter $s\in[0,1]$, and if $u$ is any (constant) vector in the $SU(2)$ representation space, the covariant derivative of the vector $u(s) = h_{e(s)}u$ along $e$ must vanish. From this it follows that the holonomy satisfies the differential equation
\be\label{holonomy-de}
\frac{d}{ds}h_{e(s)}[A] + \dot e^a(s)A_a\bigl(e(s)\bigr)h_{e(s)}[A] = 0
\ee
with the initial condition $h_{e(0)} = 1$. Here $A_a = A_a^i\tau_i$, with $\tau_i$ the anti-Hermitian generators of $SU(2)$ (normalized according to $\Tr(\tau_i\tau_j) = -\half\delta_{ij}$). Integrating the above equation from 0 to $s$ gives
\be
h_{e(s)}[A] = 1-\int_0^s ds'\,\dot e^a(s')A_a\bigl(e(s')\bigr)h_{e(s')}[A].
\ee
By repeatedly iterating this equation and evaluating the solution at $s=1$, we find the explicit expression
\be\label{holonomy-exp}
h_e[A] = \sum_n (-1)^n\,\int_0^1 ds_1\int_0^{s_1} ds_2\dots \int_0^{s_{n-1}}ds_n\,\dot e^{a_1}(s_1)A_{a_1}\bigl(e(s_1)\bigr)\cdots \dot e^{a_n}(s_n)A_{a_n}\bigl(e(s_n)\bigr),
\ee
or
\be\label{holonomy}
h_e[A] = {\cal P}\exp\biggl(-\int_e A\biggr),
\ee
${\cal P}\,\exp$ denoting the path-ordered exponential, with smaller values of the path parameter ordered to the right.

From \Eq{holonomy} it is immediate to see that the holonomy satisfies the property
\be\label{h^-1}
h_{e^{-1}}[A] = h_e^{-1}[A],
\ee
where $e^{-1}$ denotes $e$ taken with the opposite orientation. Moreover, if $e_1$ and $e_2$ are two curves such that the endpoint of $e_1$ coincides with the beginning point of $e_2$, we have
\be\label{h2h1}
h_{e_2}[A]h_{e_1}[A] = h_{e_2\circ e_1}[A],
\ee
where $e_2\circ e_1$ is the curve composed of $e_1$ followed by $e_2$. 

A further property of the holonomy, which we will need later on, is the relation between the holonomy around a small closed loop and the curvature of the Ashtekar connection. This relation can be extracted from the basic formula \eqref{holonomy-exp} as follows. Suppose that $\alpha$ is a closed loop of infinitesimal coordinate area $\epsilon^2$. By \Eq{holonomy-exp}, the holonomy around the loop is given by
\begin{align}
h_\alpha[A] = 1 &- \int_0^1 ds\,\dot\alpha^a(s)A_a\bigl(\alpha(s)\bigr) \notag \\
&+ \int_0^1 ds\int_0^s dt\,\dot\alpha^a(s)A_a\bigl(\alpha(s)\bigr)\dot\alpha^b(t)A_b\bigl(\alpha(t)\bigr) + {\cal O}(\epsilon^3). \label{hol-loop1}
\end{align}
For concreteness, let us choose coordinates such that $\alpha$ lies in the $(x^1,x^2)$-plane and passes through the origin of the coordinate system. Then we can expand the connection around the point $x=0$ as $A_a\bigl(\alpha(s)\bigr) = A_a(0) + \alpha^b(s)\partial_b A_a(0) + {\cal O}(\epsilon^2)$ (throughout the calculation we are dropping terms which will not contribute to the final result at lowest nontrivial order). Inserting this into \Eq{hol-loop1} and evaluating the integrals $\int_0^1 ds\,\dot\alpha^a(s) = 0$ and $\int_0^s dt\,\dot\alpha^a(t) = \alpha^a(s)-\alpha^a(0)$, we find
\be
h_\alpha[A] = 1 - \Bigl(\partial_bA_a(0) - A_a(0)A_b(0)\Bigr)\int_0^1 ds\,\dot\alpha^a(s)\alpha^b(s) + {\cal O}(\epsilon^3).
\ee
The integral $\int_0^1 ds\,\dot\alpha^a(s)\alpha^b(s)$ is evidently antisymmetric in $a$ and $b$, so we can write
\be
h_\alpha[A] = 1 - F_{12}(0)\int_0^1 ds\,\dot\alpha^1(s)\alpha^2(s) + {\cal O}(\epsilon^3).
\ee
Now the remaining integral $\int_0^1 ds\,\dot\alpha^1(s)\alpha^2(s) = \int_\alpha \alpha^2\,d\alpha^1$ simply gives the area enclosed by the loop $\alpha$, \ie $\epsilon^2$. Therefore we conclude
\be\label{h-loop}
h_\alpha[A] = 1 - \epsilon^2F_{12}(0) + {\cal O}(\epsilon^3),
\ee
which is the relation we were looking for (and of course at lowest order in $\epsilon$, the curvature may be evaluated at any point on or inside of $\alpha$). The infinitesimal relation \eqref{h-loop} can be seen as a special case of the non-Abelian Stokes' theorem (see \eg \cite{Stokes}), which relates the path-ordered exponential of the connection around an arbitrary loop to the so-called surface-ordered exponential of the curvature over a surface bounded by the loop (the surface ordering becoming irrelevant in the limit where the size of the loop shrinks to zero).

\subsection{The flux}

The conjugate variable to the holonomy is given by the flux of the densitized triad $E^a_i$ across a (two-dimensional) surface $S$ in $\Sigma$. Introducing coordinates $(\sigma^1,\sigma^2)$ on the surface, so that the location of the surface in $\Sigma$ is given by $x^a(\sigma^1,\sigma^2)$, the flux is defined as
\be\label{flux}
E_i(S) = \int_S d^2\sigma\,n_a(\sigma)E^a_i\bigl(x(\sigma)\bigr)
\ee
with $n_a(\sigma) = \epsilon_{abc}(\partial x^b/\partial\sigma^1)(\partial x^c/\partial\sigma^2)$ the normal one-form on $S$. (Note that no metric on $\Sigma$ is needed in order to define $n_a$.) More generally, making use of an $\su$-valued function $f = f^i\tau_i$, we may consider the smeared flux variable
\be\label{sm-flux}
E_f(S) = \int_S d^2\sigma\,n_a(\sigma)f^i\bigl(x(\sigma)\bigr) E^a_i\bigl(x(\sigma)\bigr).
\ee
Notice that in the definitions \Eq{holonomy} and \Eq{flux} the connection and the densitized triad are integrated respectively against one-dimensional curves and two-dimensional surfaces. Hence the total degree of smearing is just enough to absorb the delta function present in the Poisson bracket $\{A_a^i(x),E^b_j(y)\} = 8\pi\beta G\delta_a^b\delta^i_j\delta(x,y)$, making the Poisson bracket between the holonomy and the flux non-singular.

\subsection{Poisson bracket of the elementary variables}

In order to compute the Poisson bracket between the holonomy and the flux, we must first find the functional derivative of the holonomy with respect to the connection. This can be done by considering the functional derivative of the defining equation \eqref{holonomy-de} \cite{variations}:
\be\label{holonomy-de-fd}
\frac{d}{ds}\frac{\delta h_{e(s)}}{\delta A_a^i(x)} + \dot e^b(s)A_b\bigl(e(s)\bigr)\frac{\delta h_{e(s)}}{\delta A_a^i(x)} = -\dot e^a(s)\delta^{(3)}\bigl(x,e(s)\bigr)\tau_ih_{e(s)}.
\ee
This can be solved with the ansatz
\be
\frac{\delta h_{e(s)}}{\delta A_a^i(x)} = h_{e(s)}\Lambda(s),
\ee
which gives, after inserting into \Eq{holonomy-de-fd} and using \Eq{holonomy-de}
\be
h_{e(s)}\frac{d\Lambda(s)}{ds}= -\dot e^a(s)\delta^{(3)}\bigl(x,e(s)\bigr)\tau_ih_{e(s)}.
\ee
By inspection, the solution of this equation is
\be
\Lambda(s) = -\int_0^s dt\,\dot e^a(t)\delta^{(3)}\bigl(x,e(t)\bigr)h_{e(t)}^{-1}\tau_ih_{e(t)} + \const
\ee
and the requirement $\delta h_{e(0)}/\delta A_a^i(x) = 0$ shows that the constant is equal to zero. Setting then $s=1$ and introducing the notation $e(s,s_0)$ for the segment of $e$ extending from parameter value $s_0$ to parameter value $s$, we obtain the result
\be\label{holonomy-fd}
\frac{\delta h_e[A]}{\delta A_a^i(x)} = -\int_0^1 ds\,\dot e^a(s)\delta^{(3)}\bigl(x,e(s)\bigr)h_{e(1,s)}\tau_i h_{e(s,0)}.
\ee
With the help of \Eq{holonomy-fd}, and following \cite{Rovelli}, we may now compute the Poisson bracket between $h_e[A]$ and $E_i(S)$. We have
\be\label{bracket-integral}
\bigl\{h_e[A],E_i(S)\bigr\} = -8\pi\beta G\int_0^1 ds\,\int_S d^2\sigma\,\dot e^a(s)n_a(\sigma)\delta^{(3)}\bigl(x(\sigma),e(s)\bigr)h_{e(1,s)}\tau_i h_{e(s,0)}.
\ee
Here the integral clearly vanishes if the curve $e$ does not intersect the surface $S$. The integral also vanishes if $e$ intersects $S$ tangentially, in which case the tangent vector $\dot e^a(s)$ is orthogonal to the normal $n_a(\sigma)$ at the intersection point. In the case of a single transversal intersection, the factor
\be
\epsilon_{abc}\frac{\partial e^a(s)}{\partial s}\frac{\partial x^b(\sigma)}{\partial\sigma^1}\frac{\partial x^c(\sigma)}{\partial\sigma^2}
\ee
is the Jacobian of the coordinate transformation $(\sigma,s) \to x^a(\sigma) + e^a(s)$ around the intersection point. After performing the change of variables, we find that the integral is equal to $\pm 1$, the sign depending on the relative orientation of $S$ and $e$. We have therefore shown that\footnote{We have assumed that the point where $e$ intersects $S$ is not the beginning or ending point of $e$. In case it is, the result \eqref{h-E-bracket} is multiplied by $\half$, because the delta function in \eqref{bracket-integral} now gets integrated only over half of its domain.}
\be\label{h-E-bracket}
\bigl\{h_e[A],E_i(S)\bigr\} = -8\pi\beta G\nu(S,e)h_{e(1,s_0)}\tau_i h_{e(s_0,0)},
\ee
where $s_0$ is the value of the curve parameter at the intersection point, and the factor $\nu(S,e)$ is defined as
\be
\nu(S,e) = \begin{cases} +1 & \text{if the orientation of $S$ agrees with the orientation of $e$} \\
-1 & \text{if the orientations of $S$ and $e$ are opposite} \\
0 & \text{if $e$ intersects $S$ tangentially or not at all} \end{cases}
\ee
If $S$ and $e$ intersect at multiple points, then each intersection contributes a term of the form \eqref{h-E-bracket}.

For the purpose of choosing appropriate classical variables for the construction of the quantum theory, the most important feature of the result \eqref{h-E-bracket} is that the algebra of the holonomy and the flux closes, in the sense that their Poisson bracket depends on a finite number of the basic variables. As can be seen from \Eq{bracket-integral}, this would not be the case if one had chosen a three-dimensional smearing of the densitized triad.

\subsection{Gauge transformations and diffeomorphisms}

Let us then consider the behavior of the holonomy and the flux under $SU(2)$ gauge transformations and spatial diffeomorphisms. Under a local gauge transformation given by the group element $g(x)\in SU(2)$, the corresponding non-smeared variables transform as
\be\label{A^g}
A_a(x) \to g(x) A_a(x)g^{-1}(x) + g(x)\partial_a g^{-1}(x) 
\ee
and
\be\label{E^g}
E^a(x) \to g(x)E^a(x)g^{-1}(x).
\ee
To find the transformation law of the holonomy, suppose that the vector $u(s)$ satisfies the equation
\be
\frac{d}{ds}u(s) + \dot e^a(s)A_a\bigl(e(s)\bigr)u(s) = 0,
\ee
from which it follows that $u(s) = h_{e(s)}[A]u(0)$. If we introduce the gauge transformed vector $u_g(s) = g\bigl(e(s)\bigr)u(s)$, then a direct calculation shows that $u_g(s)$ satisfies
\be
\frac{d}{ds}u_g(s) + \dot e^a(s)A_a^{(g)}\bigl(e(s)\bigr)u_g(s) = 0
\ee
with the gauge transformed connection $A_a^{(g)} = gA_ag^{-1} + g\partial_ag^{-1}$. It follows that $u_g(s) = h_{e(s)}[A^{(g)}]u_g(0)$, from which we obtain, after setting $s=1$, inserting $u_g=gu$ and rearranging,
\be\label{h^g}
h_e[A^{(g)}] = g\bigl(t(e)\bigr)h_e[A]g^{-1}\bigl(s(e)\bigr),
\ee
where $s(e)$ and $t(e)$ denote respectively the beginning point (source) and endpoint (target) of the curve $e$.

The transformation law of the flux is the most simply stated in terms of the smeared flux variable \eqref{sm-flux}. Observing that $f^iE^a_i = -2\Tr(fE^a)$, we immediately find
\be
E_f^{(g)}(S) = E_{g^{-1}fg}(S).
\ee
For a diffeomorphism $\phi$ on the spatial manifold $\Sigma$, it is straightforward to show that
\begin{align}
h_e[A^{(\phi)}] &= h_{\phi(e)}[A], \label{h^phi} \\
E^{(\phi)}_f &= E_{\phi^{-1}(f)}\bigl(\phi(S)\bigr).
\end{align}
We therefore see that the holonomy and the flux have simple and natural transformation properties under gauge transformations and diffeomorphisms. The non-local transformation law of the flux under gauge transformations turns out to not present a problem for constructing gauge-invariant operators in the quantum theory, ultimately because all ''interesting'' operators can be expressed in terms of fluxes associated to infinitesimally small surfaces.

\subsection{Parallel transported flux variable}

We close this chapter on the choice of classical variables by mentioning a possible alternative definition of the flux variable. This definition requires us to choose a set of curves $\{p_{x_0\leftarrow x}\}$ which connect each point $x$ on $\Sigma$ (or at least each point on the surface $S$) to a fixed point $x_0$, which may lie on $S$ or outside of it. The ''parallel transported'' flux variable is then defined as
\be\label{tr-flux}
E^{(p)}(S) = \int_S d^2\sigma\,n_a(\sigma) h_{p_{x_0\leftarrow x(\sigma)}} E^a\bigl(x(\sigma)\bigr) h_{p_{x_0\leftarrow x(\sigma)}}^{-1}.
\ee
The main advantage of this definition is the local behavior of the parallel transported flux under gauge transformations: under the transformation \eqref{A^g}--\eqref{E^g} we simply have
\be
E^{(p)}(S) \to g(x_0)E^{(p)}(S)g^{-1}(x_0).
\ee
Even so, the variable \eqref{tr-flux} is probably not an ideal choice for the variable on which the construction of the quantum theory is based, since the presence of holonomies makes it technically more complicated, and the choice of the paths $\{p_{x_0\leftarrow x}\}$ introduces a considerable amount of arbitrariness into the construction. The operator corresponding to the variable \eqref{tr-flux} will nevertheless be a useful quantity in the quantum theory obtained by choosing the holonomy and the simple flux \eqref{flux} as the fundamental classical variables.

\section{The kinematical Hilbert space}\label{ch:Hkin}

A general framework for quantizing classical theories with constraints has been formulated by Dirac in \cite{Dirac}. The first step of Dirac's algorithm is to define a kinematical Hilbert space, on which the Poisson brackets between the elementary classical variables -- given by \Eq{h-E-bracket} in the case of loop quantum gravity -- are represented by the commutation relations between the corresponding elementary operators. The following step of the program consists of imposing the constraints in the quantum theory by constructing operators corresponding to the constraints and looking for the sector of the kinematical Hilbert space which is annihilated by all of the constraint operators. Once the physical Hilbert space has been constructed in this way, information about the physical content of the theory will be encoded in the set of physical observables (\ie operators commuting with all the constraints), and in the scalar product of the physical Hilbert space, which defines transition amplitudes between the physical states.

\subsection{Cylindrical functions}

The kinematical Hilbert space of loop quantum gravity is defined in terms of so-called cylindrical functions of a (generalized, distributional) connection $A$ (see \cite{status} and references therein). Suppose $\Gamma$ is a graph consisting of $N$ oriented edges $e_1,\dots,e_N$, embedded in the spatial manifold $\Sigma$. Then a cylindrical function is a functional of the form
\be\label{cyl-f}
\Psi_\Gamma[A] = \psi\bigl(h_{e_1}[A],\dots,h_{e_N}[A]\bigr),
\ee
where $\psi$ is any (complex-valued) function on $SU(2)^N$. More precisely, the function $\Psi_\Gamma[A]$ is called cylindrical with respect to the graph $\Gamma$. We denote the set of all functions cylindrical with respect to $\Gamma$ by $\Cyl_\Gamma$, and the set of all cylindrical functions (with respect to any graph) by $\Cyl$.

There is considerable freedom in choosing the graph on which a given cylindrical function is defined. In particular, any function that is cylindrical with respect to a graph $\Gamma$ is also cylindrical with respect to any larger graph $\Gamma'$ which contains all the edges of $\Gamma$. Explicitly, the function \eqref{cyl-f} can be written as a cylindrical function on $\Gamma'$ as
\be
\Psi_{\Gamma'}[A] = \psi'\bigl(h_{e_1}[A],\dots,h_{e_N}[A],h_{e_{N+1}}[A],\dots,h_{e_{N'}}[A]\bigr),
\ee
where $e_{N+1},\dots,e_{N'}$ are the edges of $\Gamma'$ that are not contained in $\Gamma$, and the function $\psi'(g_1,\dots,g_N,g_{N+1},\dots,g_{N'})$ is equal to $\psi(g_1,\dots,g_N)$ regardless of the values of the arguments $g_{N+1},\dots,g_{N'}$. We may also consider reversing the orientation of an edge in $\Gamma$, as well as artificially splitting a single edge into two by introducing a ''trivial'' bivalent node. The properties \eqref{h^-1} and \eqref{h2h1} indicate that these operations correspond respectively to replacing a $g$ with $g^{-1}$, and replacing a pair of arguments $g_1$, $g_2$ with the single argument $g_1g_2$, in the function $\psi$ of \Eq{cyl-f}.

\subsection{Scalar product on $\Cyl$}

A scalar product on $\Cyl$ can be naturally defined using the (normalized) Haar measure on $SU(2)$ \cite{ALmeasure1, ALmeasure2}. To begin with, for two functions of the form \eqref{cyl-f}, both cylindrical with respect to the same graph $\Gamma$, we may simply define
\be\label{cyl-sp}
\braket{\Psi_\Gamma}{\Phi_\Gamma} = \int dg_1\,\dots\,dg_N\,\overline{\psi(g_1,\dots,g_N)}\phi(g_1,\dots,g_N).
\ee
In order to define the scalar product between two functions $\Psi_{\Gamma_1}$ and $\Phi_{\Gamma_2}$, cylindrical with respect to two different graphs $\Gamma_1$ and $\Gamma_2$, we note that the freedom described above allows us to take any graph $\Gamma_{12}$ that contains both $\Gamma_1$ and $\Gamma_2$ as subgraphs, and view $\Psi_{\Gamma_1}$ and $\Phi_{\Gamma_2}$ as cylindrical functions on $\Gamma_{12}$. The scalar product between the two functions can then be defined as
\be\label{cyl-sp-diff}
\braket{\Psi_{\Gamma_1}}{\Phi_{\Gamma_2}} \equiv \braket{\Psi_{\Gamma_{12}}}{\Phi_{\Gamma_{12}}}
\ee
with the right-hand side given by \Eq{cyl-sp}. The normalization of the Haar measure guarantees that the value of $\braket{\Psi_{\Gamma_1}}{\Phi_{\Gamma_2}}$ does not depend on how the graph $\Gamma_{12}$ is chosen. Moreover, the transformation laws \eqref{h^g} and \eqref{h^phi} of the holonomy, together with the left and right invariance of the Haar measure, imply that the scalar product defined here is gauge and diffeomorphism invariant. The kinematical Hilbert space of loop quantum gravity is then defined as the Cauchy completion of $\Cyl$ with respect to the norm arising from the scalar product \eqref{cyl-sp}:
\be\label{H_kin-decomp}
{\cal H}_{\rm kin} = \overline{\Cyl}.
\ee

\subsection{Decomposition into orthogonal subspaces}

The space of all cylindrical functions can clearly be decomposed into subspaces of functions cylindrical with respect to a given graph, \ie $\Cyl = \cup_\Gamma\,\Cyl_\Gamma$. However, because any cylindrical function can be considered as cylindrical with respect to many different graphs, the subspaces $\Cyl_\Gamma$ are not orthogonal to each other. A decomposition into orthogonal subspaces can nevertheless be obtained by suitably restricting the set of graphs included in the decomposition, as well as the set of functions belonging to the subspace associated to each graph \cite{ALMMT}. 

First of all we note that reversing the orientation of any number of edges of a graph does not produce independent cylindrical functions, since the matrix elements of an inverse $SU(2)$ matrix $g^{-1}$ are not independent from those of $g$, but are related to them by $(g^{-1})\updown{A}{B} = \epsilon^{AC}\epsilon_{BD}g\updown{D}{C}$. Therefore graphs which differ from each other only by orientation of some of their edges should not be counted as two different graphs in the orthogonal decomposition of $\Cyl$.

Secondly we must remove the redundancy which arises from being able to arbitrarily enlarge the graph on which a cylindrical function is defined. Due to the freedom to introduce trivial edges into the graph, the subspace of a graph $\Gamma$ should only contain functions that have a non-trivial dependence on each holonomy associated with the edges of $\Gamma$. Furthermore, the freedom to split an edge into two by introducing a trivial bivalent node dictates that if the graph does contain a bivalent node at which two edges $e_1$ and $e_2$ meet, then the corresponding cylindrical function must depend on the holonomies of the two edges in some other way and not only through the matrix product $h_{e_2}h_{e_1}$. (If the dependence on the two holonomies were only through their product, then the function would be cylindrical also with respect to the smaller graph from which the bivalent node is absent.)

Hence we have argued that a decomposition of $\Cyl$ into orthogonal subspaces is given by
\be
\Cyl = \bigcup_{[\Gamma]}\,{\Cyl}'_{\Gamma},
\ee
where the union is taken over equivalence classes $[\Gamma]$ of graphs up to orientation of the edges, and the space ${\Cyl}'_{\Gamma}$ contains only those functions \eqref{cyl-f} for which $\psi(g_1,\dots,g_N)$ is not constant with respect to any of its arguments, and which do not contain any trivial bivalent nodes of the type described above.

\subsection{Basis states}

The form of the scalar product \eqref{cyl-sp} implies that $\Cyl_e$, the space of cylindrical functions based on a single edge $e$, is essentially the space $L_2(SU(2),dg_{\rm Haar})$. Therefore we can find a basis on $\Cyl_e$ by using the Peter-Weyl theorem, which says that an orthonormal basis on $L_2(G,dg_{\rm Haar})$, with $G$ any compact Lie group, is given by (suitably normalized) matrix elements of the irreducible representation matrices of the group. In the case of $SU(2)$, the functions
\be\label{basis_e}
\sqrt{d_j}\D{j}{m}{n}{h_e}
\ee
provide an orthonormal basis on $\Cyl_e$. Since the space $\Cyl_\Gamma$ is essentially a tensor product of the spaces $\Cyl_e$ over the edges of $\Gamma$, it follows that an orthonormal basis on $\Cyl_\Gamma$ is formed by products of the functions \eqref{basis_e},
\be\label{basis_G}
f^\Gamma_{\{j_e\},\{m_e\},\{n_e\}}(h_{e_1},\dots,h_{e_N}) = \prod_{e\in\Gamma} \sqrt{d_{j_e}}\D{j_e}{m_e}{n_e}{h_e}.
\ee

Let us then introduce a useful generalization of the basis \eqref{basis_G}, which can be used, among other things, to find a convenient orthonormal basis on the reduced space $\Cyl'_\Gamma$, and consequently on the entire space $\Cyl$. Suppose the spin labels belonging to the edges of $\Gamma$ are fixed, and consider a node of $\Gamma$ with $N$ incoming edges $(e_1,\dots,e_N)$ carrying spins $(j_1,\dots,j_N)$, and $N'$ outgoing edges $(e'_1,\dots,e'_{N'})$ carrying spins $(j'_1,\dots,j'_{N'})$. To this node we associate a set of tensors of the form $t^{n_1\cdots n_{N'}}_{m_1\cdots m_N}$ (carrying a lower index for each edge coming in to the node and an upper index for each edge going out of the node), which are required to form an orthonormal basis of the space ${\cal H}_{j_1}\otimes\dots\otimes{\cal H}_{j_N}\otimes{\cal H}_{j'_1}^*\otimes\dots\otimes{\cal H}_{j_{N'}}^*$ but can be otherwise chosen arbitrarily. After such an assignment of tensors is carried out at each node of the graph, we may define the functions
\be\label{basis_G2}
f^\Gamma_{\{j_e\},\{t_v\}}(h_{e_1},\dots,h_{e_N}) = \biggl(\prod_{v\in\Gamma} \bigl(t_v\bigr)^{n^v_1\cdots n^v_{N'_v}}_{m^v_1\cdots m^v_{N_v}}\biggr)\biggl(\prod_{e\in\Gamma} \sqrt{d_{j_e}}\D{j_e}{m_e}{n_e}{h_e}\biggr),
\ee
where the contraction of $SU(2)$ indices is carried out according to the structure of the graph $\Gamma$. (That is, for each $\D{j_e}{m_e}{n_e}{h_e}$, the index $m_e$ corresponds to one index of the tensor $t_v$ at the ending node of $e$, with which it is contracted. The index $n_e$ is similarly contracted with the tensor at the beginning node of $e$.) The functions \eqref{basis_G2} also form an orthonormal basis of $\Cyl_\Gamma$, and in fact the previous basis \eqref{basis_G} can be obtained from \eqref{basis_G2} by making a suitable (and rather simple) choice of the tensors $t_v$.

The conditions by which the space $\Cyl_\Gamma$ is reduced to $\Cyl'_\Gamma$ are particularly simple to state in terms of the functions \eqref{basis_G2}. The space $\Cyl'_\Gamma$ is spanned by those states of the form \eqref{basis_G2} for which none of the spins associated to the edges of the graph are equal to zero. Furthermore, for each bivalent node of the graph, the tensor associated to the node must not be proportional to the unit tensor. Equivalently, using the language of $SU(2)$ recoupling theory, the spins of the two edges which meet at the bivalent node must not be coupled to total spin zero by the tensor at the node.

\newpage

\section{Operators on ${\cal H}_{\rm kin}$}\label{ch:operators}

In this chapter we will introduce a number of basic operators on the kinematical Hilbert space. In particular, we will define operators corresponding to holonomies and fluxes, and show that their commutator correctly reproduces the fundamental classical Poisson bracket \eqref{h-E-bracket}.

\subsection{Holonomy operator}

The holonomy operator acts on cylindrical functions by multiplication:
\be\label{D*Psi}
\D{j}{m}{n}{h_e}\Psi_\Gamma[A] = \D{j}{m}{n}{h_e}\psi(h_{e_1},\dots,h_{e_N}).
\ee
Note that the character of the result depends on whether the edge $e$ is among the edges of the graph $\Gamma$. If $e$ is not contained in $\Gamma$, then the right-hand side of \eqref{D*Psi} is a cylindrical function on the graph $\Gamma\cup e$; in effect, the action of the holonomy operator has added a new edge to the graph on which the cylindrical function is based. If $e$ coincides with one of the edges of $\Gamma$, the function \eqref{D*Psi} is still an element of $\Cyl_\Gamma$. In this case, the basic tool for computing the explicit action of the holonomy operator in the basis \eqref{basis_G} or \eqref{basis_G2} is the Clebsch--Gordan series of $SU(2)$:
\be\label{CGser}
\D{j_1}{m_1}{n_1}{g}\D{j_2}{m_2}{n_2}{g} = \sum_{jmn} \CGi{j_1j_2}{j}{m_1m_2}{m}\CG{j_1j_2}{j}{n_1n_2}{n}\D{j}{m}{n}{g},
\ee
where $\CG{j_1j_2}{j}{m_1m_2}{m}$ and $\CGi{j_1j_2}{j}{m_1m_2}{m}$ denote the $SU(2)$ Clebsch--Gordan coefficients\footnote{This somewhat unusual notation is chosen in order to display the correct index structure of equations involving $SU(2)$ magnetic indices. For a full discussion of this point, we refer the reader to the Appendix. Briefly, letting $\ket{j_1m_1;j_2m_2}$ and $\ket{j_1j_2;jm}$ denote the ''uncoupled'' and ''coupled'' states in ${\cal H}_{j_1}\otimes{\cal H}_{j_2}$, the Clebsch--Gordan coefficient is given by $\CG{j_1j_2}{j}{m_1m_2}{m} = \braket{j_1j_2;jm}{j_1m_1;j_2m_2}$; therefore the index $m$ is a dual (or upper) index. In usual physics literature the inverse coefficient $\CGi{j_1j_2}{j}{m_1m_2}{m} = \braket{j_1m_1;j_2m_2}{j_1j_2;jm}$ is normally not distinguished from $\CG{j_1j_2}{j}{m_1m_2}{m}$, because the two are numerically equal to each other when standard phase conventions are used.}. The remaining case, where $e$ partially overlaps with an edge (or multiple edges) of $\Gamma$, can be reduced to the two cases discussed above by using the multiplicative property \eqref{h2h1} of the holonomy. Naturally any cylindrical function of $A$ also gives rise to a well-defined multiplicative operator on ${\cal H}_{\rm kin}$.

On the other hand, the connection $A_a^i$ itself does not exist as a well-defined operator on the kinematical Hilbert space. Since the holonomy of the connection along a path $e(\epsilon)$ of infinitesimal coordinate length $\epsilon$ is given by $h_{e(\epsilon)}[A] = 1 - \epsilon\dot e^aA_a^i\tau_i + {\cal O}(\epsilon^2)$, one could try to define a connection operator in the quantum theory through a suitable limit of the form
\be\label{A-op-limit}
\lim_{\epsilon\to 0} \frac{h_{e(\epsilon)}-1}{\epsilon}\Psi_\Gamma[A].
\ee
However, the action of the operator $h_{e(\epsilon)}$ on the function $\Psi_\Gamma[A]$ will always modify either the graph $\Gamma$ over which the function is cylindrical, or the spin quantum number associated to an edge of $\Gamma$ (depending on whether $e(\epsilon)$ coincides with an edge of $\Gamma$). For this reason, the function $h_{e(\epsilon)}\Psi_\Gamma[A]$ will typically be orthogonal to $\Psi_\Gamma[A]$ under the scalar product \eqref{cyl-sp}--\eqref{cyl-sp-diff} even for arbitrarily small values of $\epsilon$, and therefore the limit \eqref{A-op-limit} fails to exist in general.

\subsection{The operator $J_i^{(x,e)}$}

At this point it is convenient to define several auxiliary operators, which will facilitate the discussion of the flux operator -- \ie the operator corresponding to the classical variable \eqref{flux} -- and later turn out to be useful in other ways as well. To begin with, we define the so-called left- and right-invariant vector fields,
\be\label{L_iandR_i}
L_i\psi(g) = i\dde\psi\bigl(ge^{\epsilon\tau_i}\bigr), \qquad R_i\psi(g) = i\dde\psi\bigl(e^{-\epsilon\tau_i}g\bigr),
\ee
which are operators on $L_2\bigl(SU(2)\bigr)$ (or $\Cyl_e$). In the literature one sometimes also encounters the operators
\be
L_\xi\psi(g) = i\dde\psi\bigl(ge^{\epsilon\xi}\bigr), \qquad R_\xi\psi(g) = i\dde\psi\bigl(e^{-\epsilon\xi}g\bigr),
\ee
where $\xi = \xi^i\tau_i$ is an arbitrary element of $\su$.

Using the left- and right-invariant vector fields, we further define a set of very useful operators $J_i^{(x,e)}$ on ${\cal H}_{\rm kin}$. Each of these operators carries an $SU(2)$ vector index $i$ and is labeled by a point $x$ and an edge $e$ such that $x$ is either the beginning or the ending point of $e$. The action of these operators on cylindrical functions is defined in the following way. If $\Gamma$ is a graph which contains $e$ as one of its edges, then the operator $J_i^{(x,e)}$ is declared to act on elements of $\Cyl_\Gamma$ as
\be
J_i^{(x,e)}\Psi_\Gamma[A] = \begin{cases} L_i^{(e_k)}\psi(h_{e_1},\dots,h_{e_k},\dots,h_{e_N}) & \text{if $e=e_k$ and $e$ begins at $x$} \\
R_i^{(e_k)}\psi(h_{e_1},\dots,h_{e_k},\dots,h_{e_N}) & \text{if $e=e_k$ and $e$ ends at $x$} \end{cases}
\ee
where $L_i^{(e_k)}$ and $R_i^{(e_k)}$ act on the holonomy $h_{e_k}$. In the case where $e$ is not an edge of $\Gamma$, or $x$ is not a node of $\Gamma$, we set $J_i^{(x,e)}\Psi_\Gamma[A] = 0.$

It is immediate to see that the action of the operators $J_i^{(x,e)}$ on holonomies is given by
\be\label{J*h_e}
J_i^{(x,e)}\D{j}{m}{n}{h_e} = \begin{cases} i\D{j}{m}{\mu}{h_e}\Tau{j}{i}{\mu}{n} & \text{if $x = s(e)$} \\
-i\Tau{j}{i}{m}{\mu}\D{j}{\mu}{n}{h_e} & \text{if $x = t(e)$} \end{cases}
\ee
Furthermore, they satisfy the $SU(2)$ algebra
\be\label{[J,J]}
\bigl[J_i^{(x,e)},J_j^{(x',e')}\bigr] = \delta_{x,x'}\delta_{e,e'}i\epsilon\downup{ij}{k}J_k^{(x,e)}.
\ee
Of course, using $L_\xi$ and $R_\xi$ in place of $L_i$ and $R_i$, we may also define the operators $J_\xi^{(x,e)}$.

\subsection{Flux operator}

We are now ready to discuss the flux operator, \ie the operator resulting from quantizing the classical function \eqref{flux}. Let us try to compute the action of the operator on a basis state of the form \eqref{basis_G}. Applying the quantization rule $A_a^i \to 8\pi\beta G(-i\,\delta/\delta A_a^i)$, we obtain
\be
E_i(S)f^{\Gamma}(h_{e_1},\dots,h_{e_N}) = 8\pi\beta G\int_S d^2\sigma\,n_a(\sigma)(-i)\frac{\delta}{\delta A_a^i\bigl(x(\sigma)\bigr)}f^{\Gamma}(h_{e_1},\dots,h_{e_N}).
\ee
\newpage
\noindent Recalling \Eq{holonomy-fd}, it is evident that each edge which intersects the surface $S$ will contribute a term of the form
\be
8\pi\beta Gi\int d^2\sigma\,n_a(\sigma)\int_0^1 ds\,\dot e^a(s)\delta^{(3)}\bigl(x,e(s)\bigr)h_{e(1,s)}\tau_ih_{e(s,0)}.
\ee
The integral appearing here has already been encountered in \Eq{bracket-integral}, and is equal to $\half\nu(S,e)h_e\tau_i$ or $-\half\nu(S,e)\tau_i h_e$, depending on whether the edge $e$ intersects $S$ at its beginning point or endpoint. As before, edges intersecting $S$ tangentially, or not intersecting it at all, do not give a contribution. Comparing the result with the definition of $J_i^{(x,e)}$, we see that the action of the flux operator can be expressed in the form
\be\label{E*Psi}
E_i(S)\Psi_\Gamma[A] = 8\pi\beta G\sum_{x\in S}\sum_{\text{$e$ at $x$}} \frac{1}{2}\kappa(S,e) J_i^{(x,e)}\Psi_\Gamma[A],
\ee
where the geometric factor $\kappa(S,e)$ is defined as
\be
\kappa(S,e) = \begin{cases} +1 & \text{if $e$ lies above $S$} \\
-1 & \text{if $e$ lies below $S$} \\
0 & \text{if $e$ intersects $S$ tangentially or not at all}
\end{cases}
\ee
Here ''above'' and ''below'' are understood with respect to the direction defined by the normal vector of the surface. The uncountable sum over all the points of $S$ may look unsettling at a first sight, but the expression \eqref{E*Psi} is nevertheless well defined, since the sum receives non-vanishing contributions only from the finite number of points at which the edges of the graph $\Gamma$ intersect the surface $S$.

On the other hand, the densitized triad $E^a_i(x)$ is not a well-defined operator on ${\cal H}_{\rm kin}$ (precisely as was the case with the holonomy and the connection). This is because the functional derivative $\delta h_e[A]/\delta A_a^i(x)$, given by \Eq{holonomy-fd}, contains a delta function and is therefore not a normalizable element of the kinematical Hilbert space.

Using \Eqs{D*Psi} and \eqref{E*Psi}, it is a simple calculation to verify that the commutator between the holonomy and flux operators is correctly related to the Poisson bracket between the corresponding classical variables. For example, in the case where the edge $e$ has a single intersection with the surface $S$, one finds
\be
\bigl[D^{(j)}(h_e),E_i(S)\bigr] = -i(8\pi\beta G)\nu(S,e)D^{(j)}\bigl(h_{e(1,s_0)}\bigr)\tau^{(j)}_iD^{(j)}\bigl(h_{e(s_0,0)}\bigr)
\ee
(where, as in \Eq{h-E-bracket}, $s_0$ is the value of the curve parameter at the intersection point).

\subsection{Generalizations of the flux operator}

By following the steps that led to \Eq{E*Psi}, one may also construct the operators corresponding to the smeared and parallel transported flux variables \eqref{sm-flux} and \eqref{tr-flux}. For the smeared flux operator, an entirely similar calculation that led to \eqref{E*Psi} produces the result
\be
E_f(S)\Psi_\Gamma[A] = 8\pi\beta G\sum_{x\in S}\sum_{\text{$e$ at $x$}} \frac{1}{2}\kappa(S,e) J_{f(x)}^{(x,e)}\Psi_\Gamma[A].
\ee
In the case of the parallel transported flux, we take
\be\label{E_i^p class}
E_i^{(p)}(S) = -2\Tr\Bigl(\tau_i E^{(p)}(S)\Bigr)
\ee
as the classical variable to be quantized. In order to derive the form of the resulting operator, it is convenient to start by using the relation
\be\label{tau-transf}
g\tau_ig^{-1} = R\updown{k}{i}(g)\tau_k,
\ee
where $R(g)$ is the $\R^3$ rotation matrix corresponding to the $SU(2)$ element $g$, to rewrite the variable \eqref{E_i^p class} as
\be
E_i^{(p)}(S) = \int_S d^2\sigma\,n_a(\sigma)R\updown{k}{i}\bigl(h^{-1}_{p_{x_0\leftarrow x(\sigma)}}\bigr)E^a_k\bigl(x(\sigma)\bigr).
\ee
To find the action of the operator on a holonomy, it now suffices to essentially repeat the calculation performed in the previous section (assuming a factor ordering where the functional derivative is ordered to the right). For instance, if the beginning point of $e$ lies on $S$, we find
\be
E_i^{(p)}(S)h_e = 8\pi\beta G\frac{i}{2}\nu(S,e)h_eh^{-1}_{p_{x_0\leftarrow s(e)}}\tau_ih_{p_{x_0\leftarrow s(e)}},
\ee
where \Eq{tau-transf} has been used in reverse to eliminate the rotation matrix. From this (and the corresponding calculation for the case where the edge $e$ ends on $S$) we may conclude that the parallel transported flux operator takes the form
\be\label{E_i^p}
E_i^{(p)}(S)\Psi_\Gamma[A] = 8\pi\beta G\sum_{x\in S}\sum_{\text{$e$ at $x$}} \frac{1}{2}\kappa(S,e) J_{h_{p_{x_0\leftarrow x}}^{-1}\tau_i h_{p_{x_0\leftarrow x}}}^{(x,e)}\Psi_\Gamma[A].
\ee
Note that this operator is given in terms of operators $J_{\xi(x)}^{(x,e)}$ where $\xi(x) = h_{p_{x_0\leftarrow x}}^{-1}\tau_i h_{p_{x_0\leftarrow x}}$ is an $A$-dependent function. Therefore, while the operator $E_i(S)$ maps each space $\Cyl_\Gamma$ into itself (in other words, preserves the graph of the cylindrical function on which it acts), the same might not always be the case for the operator $E_i^{(p)}(S)$.

\subsection{Uniqueness of the kinematical representation}

The kinematical representation of holonomies and fluxes, described in this and the preceding chapter and commonly referred to as the Ashtekar--Lewandowski representation, has several peculiar properties. For example, an operator representing the Ashtekar connection does not exist, and as we will see later, geometric operators corresponding to quantities such as areas of surfaces and volumes of regions in the spatial manifold $\Sigma$, turn out to have purely discrete spectra. On the other hand, the construction of the representation may seem somewhat arbitrary, so it is natural to ask how much freedom there would be in possibly constructing different, inequivalent kinematical representations for loop quantum gravity.

An answer to this question is provided by a uniqueness theorem -- the so-called LOST theorem \cite{LOST, Fleischhack} -- which states that under certain rather general assumptions, the Ashtekar--Lewandowski representation is (up to unitary equivalence) the only representation of the holonomy-flux algebra. Diffeomorphism invariance, which enters through the assumption that the kinematical Hilbert space contains at least one diffeomorphism invariant state, plays a crucial role in the proof of the theorem. This result is all the more remarkable if one recalls that no corresponding uniqueness theorem holds in conventional quantum field theories, in which many different, inequivalent representations may generally exist.

Alternative kinematical representations for loop quantum gravity have nevertheless been introduced by circumventing one or more of the assumptions of the LOST theorem. The Koslowski--Sahlmann representation \cite{Koslowski, Sahlmann} violates the LOST theorem's assumptions about diffeomorphisms by introducing a fixed, classical background geometry. The flux representation of Dittrich and Geiller \cite{fluxrep, fluxrep2, fluxrep3} is based on an altogether different choice of elementary classical variables.

\section{Gauge invariance}\label{ch:gauss}

The kinematical Hilbert space of loop quantum gravity and the elementary operators on it having been set, the next stage of the quantization program consists of imposing the Gauss, diffeomorphism and Hamiltonian constraints. In other words, one is looking for a precise way of implementing the steps indicated by the schematic diagram
\[
{\cal H}_{\rm kin} \quad \xrightarrow{\quad G_i=0 \quad} \quad {\cal H}_G \quad \xrightarrow{\quad C_a=0 \quad} \quad {\cal H}_{\rm diff} \quad \xrightarrow{\quad C=0 \quad} \quad  {\cal H}_{\rm phys}.
\]
The first two steps of the diagram, corresponding to the Gauss and diffeomorphism constraints, are well understood. On the other hand, finding a satisfactory and practically manageable way to deal with the Hamiltonian constraint has proven rather elusive, and forms the main subject of the second part of this work. In this chapter we take the relatively simple example of the Gauss constraint as an opportunity to illustrate the various possible approaches to the problem of solving a given constraint. The diffeomorphism constraint will be discussed in Chapter \ref{ch:diff}.

\subsection{Solution by inspection}

The ideal situation would be that the gauge transformations generated by the constraint are understood in sufficient detail that the set of states invariant under these transformations can simply be written down by inspection. In the case of the Gauss constraint, the associated gauge transformations are local $SU(2)$ gauge transformations. We recall that under a transformation specified by the gauge function $g(x)\in SU(2)$, the holonomy transforms as
\be\label{h[Ag]}
h_e[A^{(g)}] = g\bigl(t(e)\bigr)h_e[A]g^{-1}\bigl(s(e)\bigr).
\ee
Using this, we may consider how a gauge transformation affects a generic state such as the state $f^{\Gamma}_{\{j_e\},\{t_v\}}$ of \Eq{basis_G2}. Let us focus on a specific node $v$ of $\Gamma$. \Eq{h[Ag]} shows that each edge ending at $v$ contributes a $D^{(j_e)}(g_v)$, while each edge starting from $v$ contributes a $D^{(j'_e)}(g_v^{-1})$, where $g_v$ is the value of the gauge function at the node. Hence the gauge transformation effectively replaces the tensor $t_v$ with the gauge transformed tensor
\begin{align}
\bigl(t_v^{(g)}\bigr)^{n_1\cdots n_{N'}}_{m_1\cdots m_N} &= \D{j_1}{\mu_1}{m_1}{g_v}\dots\D{j_N}{\mu_N}{m_N}{g_v} \notag \\
&\quad\times\D{j'_1}{n_1}{\nu_1}{g^{-1}_v}\dots\D{j'_N}{n_{N'}}{\nu_{N'}}{g^{-1}_v}\bigl(t_v\bigr)^{\nu_1\cdots \nu_{N'}}_{\mu_1\cdots \mu_N} \label{invtensor}
\end{align}
at each node of the graph, but otherwise preserves the form of the state $f^{\Gamma}_{\{j_e\},\{t_v\}}$.

From these considerations it is clear that a set of gauge invariant states can be obtained by restricting the set of tensors $\{t_v\}$ to those that satisfy $t_v^{(g)} = t_v$ for every $g\in SU(2)$. In particular, an orthonormal basis on the gauge invariant subspace of $\Cyl_\Gamma$, which we denote by ${\cal H}_G^{\Gamma}$, is given by the states
\be\label{invstate}
\psi^\Gamma_{\{j_e\},\{\iota_v\}}(h_{e_1},\dots,h_{e_N}) = \biggl(\prod_{v\in\Gamma} \iota_v\biggr)^{m_1\cdots m_N}_{n_1\cdots n_N}\biggl(\prod_{e\in\Gamma} \sqrt{d_{j_e}} \D{j_e}{m_e}{n_e}{h_e}\biggr),
\ee
provided that the tensors $\iota_v$ at each node $v$ are taken from any orthonormal basis of the space ${\rm Inv}\,\bigl({\cal H}_{j_1}\otimes\dots\otimes{\cal H}_{j_N}\otimes{\cal H}_{j'_1}^*\otimes\dots\otimes{\cal H}_{j_{N'}}^*\bigr)$, \ie the subspace of ${\cal H}_{j_1}\otimes\dots\otimes{\cal H}_{j_N}\otimes{\cal H}_{j'_1}^*\otimes\dots\otimes{\cal H}_{j_{N'}}^*$ whose elements are invariant under $SU(2)$ in the sense of \Eq{invtensor}. The space of all gauge invariant cylindrical functions, denoted by $\Cyl_G$, is then spanned by the states \eqref{invstate} on all possible graphs $\Gamma$, and finally the entire gauge invariant Hilbert space ${\cal H}_G$ is defined as the completion of $\Cyl_G$ (with respect to the norm arising from the scalar product \eqref{cyl-sp}).

\subsection{The Gauss constraint operator}

The approach most in line with Dirac's ideas on quantizing a constrained theory is to systematically construct a quantum operator corresponding to the constraint and then try to find all the states that are annihilated by the operator. For the Gauss constraint, following the treatment of \cite{Thiemann}, we start by performing an integration by parts in the classical expression $G(\lambda) = \int d^3x\,\lambda^i{\cal D}_aE^a_i$, which leads us to consider the operator
\be
G(\lambda) = -i\int d^3x\,{\cal D}_a\lambda^i(x)\frac{\delta}{\delta A_a^i(x)}.
\ee
In order to derive the action of the operator on a generic cylindrical function, let us first calculate how $G(\lambda)$ acts on a single holonomy. Recalling \Eq{holonomy-fd}, we find
\be
G(\lambda)h_e = -i\int_0^1 ds\,\dot e^a(s)\,\Bigl[\partial_a\lambda^i\bigl(e(s)) + \epsilon\updown{i}{jk}A_a^j\bigl(e(s)\bigr)\lambda^k\bigl(e(s)\bigr)\Bigr]h_{e(1,s)}\tau_ih_{e(s,0)},
\ee
the delta function arising from the functional derivative having been absorbed by the integration over $\Sigma$. Introducing the notation $\lambda = \lambda^i\tau_i$ and $A = \dot e^a A_a^i\tau_i$, we can write this as
\be\label{Gh_e}
G(\lambda)h_e = -i\int_0^1 ds\,h_{e(1,s)}\biggl(\frac{d}{ds}\lambda\bigl(e(s)\bigr) + \bigl[A\bigl(e(s)\bigr),\lambda\bigl(e(s)\bigr)\bigr]\biggr)h_{e(s,0)}.
\ee
If we now use the holonomy equation \eqref{holonomy-de} in the form
\be
\frac{d}{ds}h_{e(s,0)} = -A\bigl(e(s)\bigr)h_{e(s,0)}
\ee
and
\be
\frac{d}{ds}h_{e(1,s)} = \frac{d}{ds}h^{-1}_{e(s,1)} = h_{e(1,s)}A\bigl(e(s)\bigr)
\ee
we see that \Eq{Gh_e} reduces to
\begin{align}
G(\lambda)h_e &= -i\int_0^1 ds\,\frac{d}{ds}\Bigl(h_{e(1,s)}\lambda\bigl(e(s)\bigr)h_{e(s,0)}\Bigr) = -i\Bigl[\lambda\bigl(t(e)\bigr)h_e - h_e\lambda\bigl(s(e)\bigr)\Bigr] \notag \\
&= \Bigl(\lambda^i\bigl(t(e)\bigr)J_i^{(t(e),e)} + \lambda^i\bigl(s(e)\bigr)J_i^{(s(e),e)} \Bigr)h_e.
\end{align}
From this we may deduce that the action of the Gauss constraint operator on a general cylindrical function is given by
\be\label{G-op}
G(\lambda)\Psi_\Gamma[A] = \sum_{v\in\Gamma}\lambda^i(v)\sum_{\text{$e$ at $v$}} J_i^{(v,e)}\Psi_\Gamma[A],
\ee
or
\be
G(\lambda)\Psi_\Gamma[A] = \sum_{v\in\Gamma} \lambda^i(v)G_i^{(v)}\Psi_\Gamma[A],
\ee
where the Gauss operator associated to a single node is
\be\label{G_v}
G_i^{(v)} = \sum_{\text{$e$ at $v$}} J_i^{(v,e)}.
\ee
Having constructed the constraint operator $G(\lambda)$, the following step of the program is to try to find all the states which are annihilated by the operator. Using \Eq{G-op} to act on the generic basis state \eqref{basis_G2}, we find that $G(\lambda)$ annihilates the state provided that the tensor $t_v$ at each node satisfies the condition
\begin{align}
\Tau{j_1}{i}{\mu}{m_1}\bigl(t_v\bigr)^{n_1\cdots n_{N'}}_{\mu m_2\cdots m_N} &+ \ldots + \Tau{j_N}{i}{\mu}{m_N}\bigl(t_v\bigr)^{n_1\cdots n_{N'}}_{m_1\cdots m_{N-1}\mu} \notag \\
&-\Tau{j'_1}{i}{n_1}{\nu}\bigl(t_v\bigr)^{\nu n_2\dots n_{N'}}_{m_1\cdots m_N} - \ldots - \Tau{j'_{N'}}{i}{n_{N'}}{\nu}\bigl(t_v\bigr)^{n_1\dots n_{N'-1}\nu}_{m_1\cdots m_N} = 0. \label{invtensor-inf}
\end{align}
But this condition simply requires $t_v$ to be invariant under $SU(2)$, since \eqref{invtensor-inf} is equivalent to \eqref{invtensor} for an infinitesimal gauge transformation, arising from a gauge function of the form $g = \Id + \epsilon^i\tau_i + {\cal O}(\epsilon^2)$. Hence the more methodical approach has reproduced the states \eqref{invstate} as the solutions of the Gauss constraint.

\subsection{Group averaging}

Yet another way to approach the problem of finding the solutions of a constraint is the so-called procedure of group averaging \cite{ALMMT}. A particular virtue of this approach is that group averaging may possibly be used even if a well-defined constraint operator cannot be constructed for some reason. The idea is to construct solutions of the constraint by suitably averaging arbitrary cylindrical functions with respect to the action of the gauge transformations associated with the constraint. In the example of the Gauss constraint, let us consider the (at this point formal) projection operator
\be\label{P_G}
P_G = \int [dg]\,U(g),
\ee
where $U(g)$ is the unitary operator which implements gauge transformations on cylindrical functions, and $[dg]$ is some appropriate integration measure constructed using the $SU(2)$ Haar measure. Due to the group property of the operators $U(g)$ and the invariance of the Haar measure, the operator $P_G$ should satisfy $U(g)P_G = P_G$. Therefore, given any cylindrical function $\Psi$, we would expect the function $P_G\Psi$ to be gauge invariant.

The gauge transformation of a cylindrical function on a graph $\Gamma$ is completely determined by the values of the gauge function $g(x)$ at the nodes of $\Gamma$. Therefore an appropriate definition of the operator \eqref{P_G} on $\Cyl_\Gamma$ would be
\be
P_G\big|_\Gamma = \int dg_{v_1}\dots dg_{v_M}\,U(g).
\ee
Recalling the transformation law \eqref{h[Ag]}, we see that the function
\be\label{P_Gpsi}
P_G\psi(h_{e_1},\dots,h_{e_N}) = \int dg_{v_1}\dots dg_{v_M}\,\psi\bigl(g_{t(e_1)}h_{e_1}g^{-1}_{s(e_1)},\dots,g_{t(e_N)}h_{e_N}g^{-1}_{s(e_N)}\bigr)
\ee
is indeed gauge invariant.

In order to derive the form of the gauge invariant functions \eqref{invstate} from the group averaging approach, let us expand the function $\psi$ in \eqref{P_Gpsi} in the basis \eqref{basis_G} as
\be\label{psi-exp}
\psi(h_{e_1},\dots,h_{e_N}) = \sum_{\{j_e\},\{m_e\},\{n_e\}} c_{\{j_e\},\{m_e\},\{n_e\}}f^\Gamma_{\{j_e\},\{m_e\},\{n_e\}}(h_{e_1},\dots,h_{e_N}).
\ee
Inserting this into \Eq{P_Gpsi} and using the identity
\begin{align}
\int dg\,&\D{j_1}{m_1}{n_1}{g}\cdots\D{j_N}{m_N}{n_N}{g} \notag \\
&\times\D{j_1'}{m_1'}{n_1'}{g^{-1}}\cdots\D{j_N'}{m_N'}{n_N'}{g^{-1}} = \sum_\iota \iota^{m_1\dots m_N}_{m_1'\dots m_N'} \iota^{n_1'\dots n_N'}_{n_1\cdots n_N},
\end{align}
where the sum over $\iota$ runs through any (real-valued) orthonormal basis of the intertwiner space ${\rm Inv}\,\bigl({\cal H}_{j_1}\otimes\dots\otimes{\cal H}_{j_N}\otimes{\cal H}_{j'_1}^*\otimes\dots\otimes{\cal H}_{j_{N'}}^*\bigr)$, we find that group averaging of the function \eqref{psi-exp} indeed produces a function of the form
\be
P_G\psi(h_{e_1},\dots,h_{e_N}) = \sum_{\{j_e\},\{\iota_v\}} c_{\{j_i\},\{\iota_v\}}\biggl(\prod_{v\in\Gamma} \iota_v\biggr)\cdot\biggl(\prod_{e\in\Gamma} \sqrt{d_{j_e}} \D{j_e}{m_e}{n_e}{h_e}\biggr). 
\ee
Thus we once again obtain the result that gauge invariant functions on $\Cyl$ are given by linear combinations of the states \eqref{invstate}.

\section{Spin network states}\label{ch:spinnetwork}

\begin{figure}[t]
	\centering
		\includegraphics[width=0.4\textwidth]{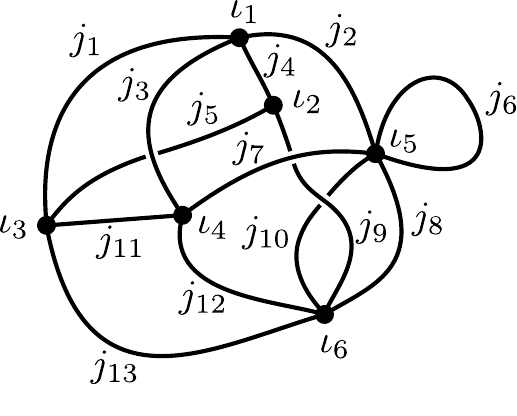}
		\caption{A spin network state.}
	\label{fig6-0}
\end{figure}

The states
\be\label{spinnetwork}
\psi^\Gamma_{\{j_e\},\{\iota_v\}}(h_{e_1},\dots,h_{e_N}) = \biggl(\prod_{v\in\Gamma} \iota_v\biggr)\cdot\biggl(\prod_{e\in\Gamma} \sqrt{d_{j_e}} \D{j_e}{m_e}{n_e}{h_e}\biggr),
\ee
which span the gauge-invariant Hilbert space ${\cal H}_G$, are called spin network states \cite{sn1, sn2, sn3}. The ''quantum numbers'' labeling a spin network state are the graph $\Gamma$, a spin $j_e$ on each edge of the graph, and an intertwiner $\iota_v$ at each node of the graph, as shown in \Fig{fig6-0}. We will also use the notation $\ket{\Gamma,j_e,\iota_v}$ for the state \eqref{spinnetwork} as an abstract state vector in ${\cal H}_G$. A physical interpretation of spin network states as states of quantized spatial geometry will be brought out in Chapter \ref{ch:geometric}, where we study operators corresponding to various geometric quantities on the spatial manifold $\Sigma$.

\subsection{The spin network decomposition}

Spin network states form an orthonormal\footnote{With respect to the scalar product defined by \Eq{cyl-sp}. Since the gauge invariant Hilbert space is a proper subspace of the kinematical Hilbert space, the scalar product on ${\cal H}_G$ is immediately given by that on ${\cal H}_{\rm kin}$, and does not require any further attention.} basis on the gauge invariant Hilbert space, provided that suitable restrictions, similar to those discussed below \Eq{H_kin-decomp}, are put into place. If we require that
\begin{itemize}
\item Two graphs differing from each other only by different orientation of some of their edges are identified with each other,
\item A graph containing a number of trivial bivalent nodes is identified with the graph obtained by removing the trivial nodes, and
\item A spin network state, some of whose edges carry zero spin, is identified with the corresponding state on the graph from which the $j=0$ edges are removed,
\end{itemize}
we have the orthogonal decomposition
\be
{\cal H}_G = \bigcup_{[\Gamma]}\,{{\cal H}_G^\Gamma}',
\ee
where $[\Gamma]$ denotes equivalence classes of graphs up to orientation of edges and removal of trivial bivalent nodes, and ${{\cal H}_G^\Gamma}'$ is the space of spin network states which are based on the graph $\Gamma$ and carry a non-trivial spin on each of their edges.

\subsection{Intertwiners}

In order to perform technical calculations in loop quantum gravity, it is essential to have a detailed understanding of spin network states, and in particular of intertwiners, or the invariant tensors residing at the nodes of a spin network graph. The main tool for calculating with spin networks is the recoupling theory of $SU(2)$ (see \eg \cite{BrinkSatchler, Varshalovich}), which is well known to physicists in the context of angular momentum in quantum mechanics. A comprehensive review of the subject is given in the Appendix; here we will summarize some of its most important elements.

The space of intertwiners associated to a three-valent node, ${\rm Inv}\,\bigl({\cal H}_{j_1}\otimes{\cal H}_{j_2}\otimes{\cal H}_{j_3}\bigr)$, is one-dimensional, provided that the spins $j_1$, $j_2$ and $j_3$ satisfy the Clebsch--Gordan conditions: $|j_1-j_2|\leq j_3\leq j_1+j_2$ and $j_1+j_2+j_3$ is an integer. The single intertwiner which spans the space ${\rm Inv}\,\bigl({\cal H}_{j_1}\otimes{\cal H}_{j_2}\otimes{\cal H}_{j_3}\bigr)$ is given by the Wigner 3$j$-symbol
\be\label{int3v}
\iota_{m_1m_2m_3} = \threej{j_1}{j_2}{j_3}{m_1}{m_2}{m_3}, 
\ee
which is known from the theory of angular momentum as the object which couples three angular momenta into vanishing total angular momentum.

Indices of intertwiners can be raised and lowered using the epsilon tensors $\epsilon^{(j)}_{mn}$ and $\epsilon_{\phantom{m}}^{(j)mn}$, which map between the spaces ${\cal H}_j$ and ${\cal H}_j^*$, and both of whose components with respect to the preferred basis of ${\cal H}_j$ are equal to $(-1)^{j-m}\delta_{m,-n}$. For example, the space ${\rm Inv}\,\bigl({\cal H}_{j_1}^*\otimes{\cal H}_{j_2}\otimes{\cal H}_{j_3}\bigr)$ is spanned by the intertwiner
\be
\iota\updown{m_1}{m_2m_3} = \epsilon_{\phantom{m}}^{(j_1)m_1m}\iota_{mm_2m_3},
\ee
which is just the Clebsch--Gordan coefficient $\CG{j_2j_3}{j_1}{m_2m_3}{m_1}$, up to a multiplicative factor.

Intertwiners of higher valence can be built using the basic intertwiner \eqref{int3v} and the epsilon tensor. For example, by contracting two three-valent intertwiners with epsilon, we obtain the four-valent intertwiner
\be\label{int4-12}
\bigl(\iota_{12}^{(k)}\bigr)_{m_1m_2m_3m_4} = \threej{j_1}{j_2}{k}{m_1}{m_2}{m}\epsilon_{\phantom{m}}^{(k)mn}\threej{k}{j_3}{j_4}{n}{m_3}{m_4}.
\ee
As the ''internal'' spin $k$ ranges over all the values allowed by the Clebsch--Gordan conditions, the intertwiners \eqref{int4-12} form a basis in the space of four-valent intertwiners ${\rm Inv}\,\bigl({\cal H}_{j_1}\otimes{\cal H}_{j_2}\otimes{\cal H}_{j_3}\otimes{\cal H}_{j_4}\bigr)$. This basis is orthogonal but not normalized, since the norm of the intertwiner \eqref{int4-12} is equal to $1/\sqrt{d_k}$. Another basis of the same intertwiner space can be obtained by choosing to couple the spins $j_1$ and $j_3$ to the internal spin:
\be\label{int4-13}
\bigl(\iota_{13}^{(l)}\bigr)_{m_1m_2m_3m_4} = \threej{j_1}{j_3}{l}{m_1}{m_3}{m}\epsilon_{\phantom{m}}^{(l)mn}\threej{l}{j_2}{j_4}{n}{m_2}{m_4}.
\ee
The bases \eqref{int4-12} and \eqref{int4-13} are related to each other by
\be\label{chbasis}
\bket{\iota_{12}^{(k)}} = \sum_l d_l(-1)^{j_2+j_3+k+l}\sixj{j_1}{j_2}{k}{j_4}{j_3}{l}\bket{\iota_{13}^{(l)}},
\ee
where the object with curly brackets is the Wigner 6$j$-symbol.

By continuing to attach three-valent intertwiners to each other by contraction with epsilon, one can obtain intertwiners of arbitrarily high valence. In general, the basis of the $N$-valent intertwiner space ${\rm Inv}\,\bigl({\cal H}_{j_1}\otimes\cdots\otimes{\cal H}_{j_N}\bigr)$ constructed according to this scheme is labeled by $N-3$ internal spins. In the language of angular momentum, the values of the internal spins determine the eigenvalues of the operators $\bigl(J^{(1)} + J^{(2)}\bigr)^2$, $\bigl(J^{(1)} + J^{(2)} + J^{(3)})^2$, $\dots$, $\bigl(J^{(1)} + \dots + J^{(N-2)}\bigr)^2$.

\subsection{$SU(2)$ graphical calculus}\label{sec:SU2calc}

A graphical notation, which has been developed for $SU(2)$ recoupling theory, is tremendously powerful for calculations involving intertwiners. The basic tensors $\delta_m^n$ and $\epsilon^{(j)}_{mn}$ are represented graphically by\footnote{The graphical diagrams used for $SU(2)$ calculations should not be confused with generic drawings of spin network states, such as \Fig{fig6-0}. In particular, the arrow representing the epsilon tensor has nothing to do with the orientation of a spin network edge.}
\vspace{-12pt}
\begin{align}
\delta_m^n &\quad = \quad \RealSymb{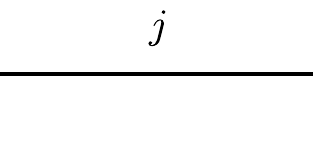}{0.6} \\
\epsilon^{(j)}_{mn} &\quad = \quad \RealSymb{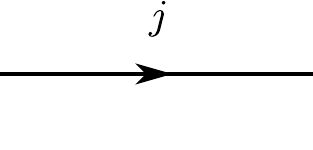}{0.6}
\end{align}
In such drawings each free end of a line carries a magnetic index; the magnetic indices are usually not written explicitly. Contraction of indices is carried out by joining the corresponding lines. The properties of $\epsilon^{(j)}_{mn}$ then imply that the arrow behaves according to the rules
\vspace{-12pt}
\begin{align}
\RealSymb{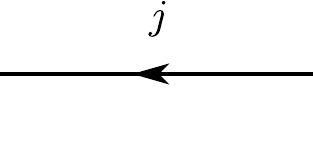}{0.6} &\quad = \quad (-1)^{2j}\RealSymb{fig6-2.pdf}{0.6} \\ 
\RealSymb{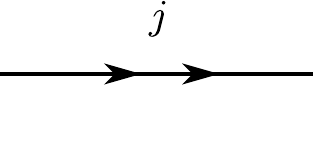}{0.6} &\quad = \quad (-1)^{2j}\RealSymb{fig6-1.pdf}{0.6}
\end{align}
The 3$j$-symbol is represented by three lines joined together at a node (with the free ends of the lines again carrying unwritten magnetic indices):
\be\label{3j-gra}
\threej{j_1}{j_2}{j_3}{m_1}{m_2}{m_3} \quad = \quad \RealSymb{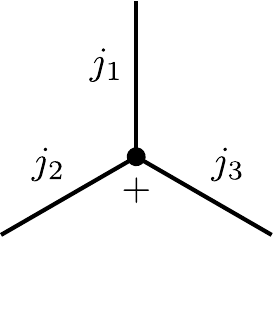}{0.6} 
\ee
Since the 3$j$-symbol is sensitive to non-cyclic permutations of the columns, the ordering of the columns is indicated by a sign at the node; the drawing \eqref{3j-gra} with a minus sign would represent the symbol $\bigl(\begin{smallmatrix} j_1&j_3&j_2 \\ m_1&m_3&m_2 \end{smallmatrix}\bigr)$. From the symmetries of the 3$j$-symbol, we have
\be
\RealSymb{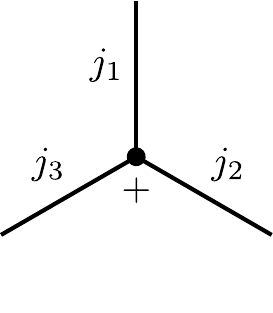}{0.6} \quad = \quad (-1)^{j_1+j_2+j_3}\RealSymb{fig6-5.pdf}{0.6}
\ee
and
\be
\RealSymb{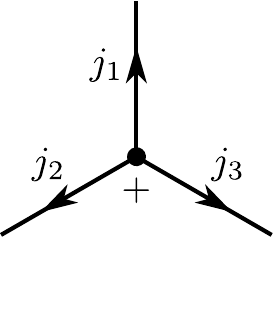}{0.6} \quad = \quad \RealSymb{fig6-5.pdf}{0.6}
\ee
The 3$j$-symbol also satisfies the orthogonality relation
\be
\RealSymb{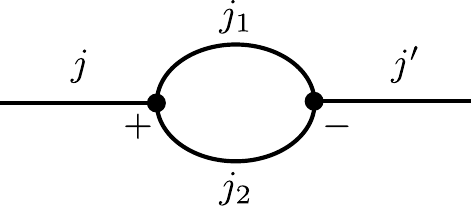}{0.6} \quad = \quad \frac{1}{d_j}\delta_{jj'}\;\RealSymb{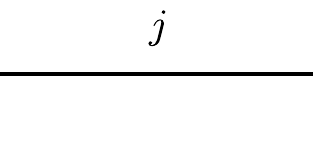}{0.6}
\ee
In graphical form, the four-valent intertwiner \eqref{int4-12} is
\be\label{i12-gra}
\bigl(\iota_{12}^{(k)}\bigr)_{m_1m_2m_3m_4} \quad = \quad \RealSymb{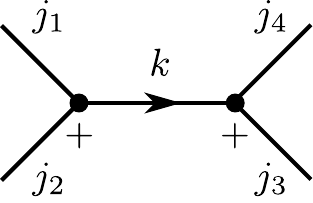}{0.6}
\ee
The graphical representation of the 6$j$-symbol is given by
\be
\sixj{j_1}{j_2}{j_3}{k_1}{k_2}{k_3} \quad = \quad \RealSymb{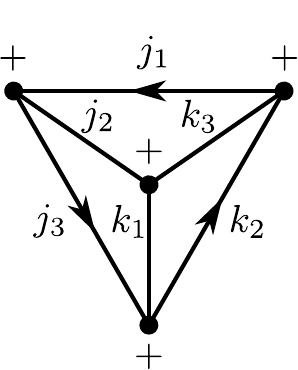}{0.6}
\ee
as can be verified by writing \Eq{chbasis} graphically and contracting both sides with $\iota^{(13)}_{k'}$

The $SU(2)$ generator $\Tau{j}{i}{m}{n}$ is an unnormalized element of the intertwiner space ${\rm Inv}\bigl({\cal H}_1\otimes{\cal H}_j^*\otimes{\cal H}_j\bigr)$. The precise relation is
\vspace{-24pt}
\be\label{tau-gra}
\Tau{j}{i}{m}{n} \quad = \quad iW_j\;\RealSymb{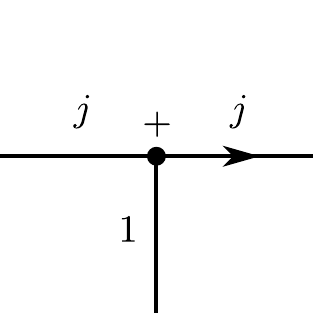}{0.6}
\ee
where $W_j = \sqrt{j(j+1)(2j+1)}$.

We also introduce the graphical notation
\be\label{D-gra}
\D{j}{m}{n}{g} \quad = \quad \RealSymb{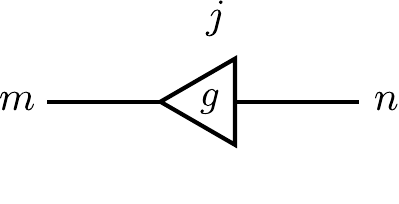}{0.6} 
\ee
for the Wigner matrices. Recalling that $\D{j}{m}{n}{g^{-1}} = \epsilon_{\phantom{m'}}^{(j)mm'}\epsilon^{(j)}_{nn'}\D{j}{n'}{m'}{g}$ we then see that the inverse matrix satisfies the graphical relation  
\be\label{D^-1-gra}
\RealSymb{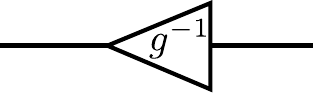}{0.6} \quad = \quad \RealSymb{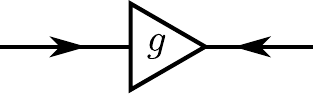}{0.6}
\ee
Relations such as \eqref{h[Ag]} suggest that the index $m$ of $\D{j}{m}{n}{h_e}$ belongs to the endpoint of the edge $e$, while the index $n$ belongs to the beginning point. Therefore the orientation of the triangle in \eqref{D-gra} is consistent with the orientation of the edge in the graphical representation of a spin network state.

\newpage

Using \Eqs{tau-gra} and \eqref{D-gra}, we may cast the action of $J_i^{(x,e)}$ on holonomies into graphical form. From \Eq{J*h_e}, we find
\vspace{-20pt}
\begin{subequations}\label{J*h-gra}
\begin{align}
J_i^{(x,e)}\,\RealSymb{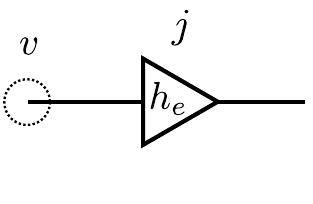}{0.6} &\quad = \quad -W_j\;\RealSymb{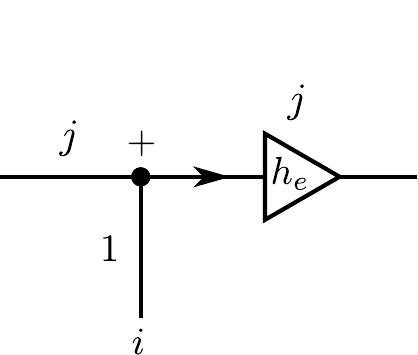}{0.6} \qquad \text{if $e$ begins from $x$} \\
\intertext{and \vspace{-24pt}}
J_i^{(x,e)}\,\RealSymb{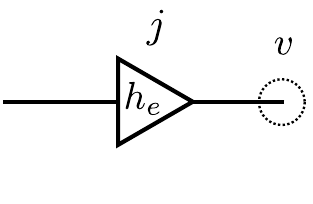}{0.6} &\quad = \quad W_j\;\RealSymb{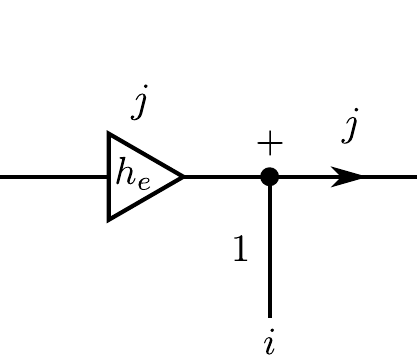}{0.6} \qquad \text{if $e$ ends at $x$.}
\end{align}
\end{subequations}

The seemingly trivial observation that tensors of the space ${\rm Inv}\,\bigl({\cal H}_{j_1}\otimes\cdots\otimes{\cal H}_{j_N}\bigr)$ can be expanded with respect to any basis of the space leads to a set of relations which are so useful in practical calculations that they would deserve to be called the fundamental theorem of graphical calculus. Letting a block with $N$ lines attached denote a general $N$-valent intertwiner, and using the observation for intertwiners of valence two, three and four, we have
\begin{subequations}\label{ftgc}
\begin{align}
\RealSymb{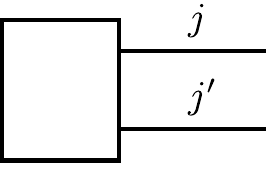}{0.6} &\quad = \quad \delta_{jj'}\frac{1}{d_j}\RealSymb{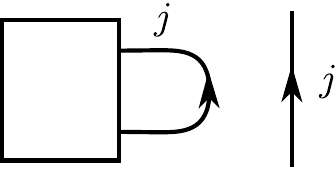}{0.6} \label{ftgc2} \\
\RealSymb{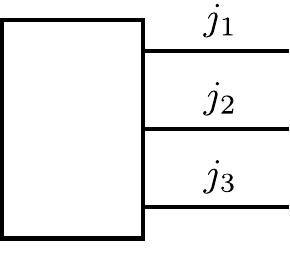}{0.6} &\quad = \quad \RealSymb{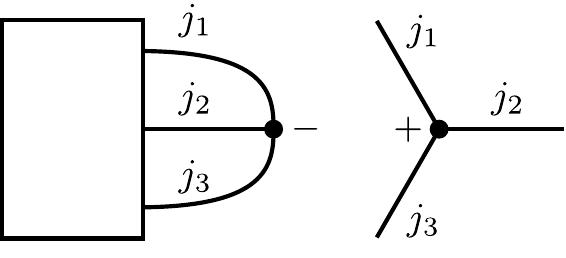}{0.6} \label{ftgc3} \\
\RealSymb{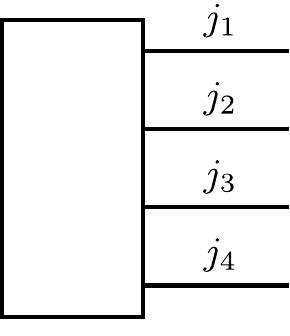}{0.6} &\quad = \quad \sum_x d_x\,\RealSymb{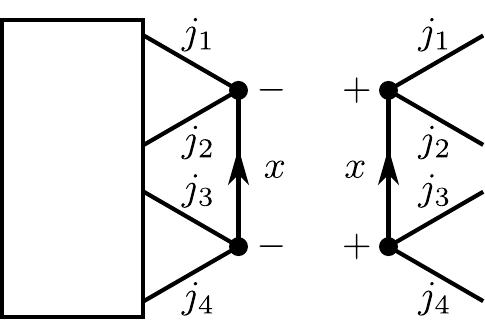}{0.6} \label{ftgc4}
\end{align}
\end{subequations}
and for intertwiners of higher valence an obvious generalization of \Eq{ftgc4} holds. The relations \eqref{ftgc} are normally used to break down complicated graphical expressions into simpler pieces.

\section{Diffeomorphism invariance}\label{ch:diff}

The imposition of the diffeomorphism constraint is a crucial step in the quantization program of loop quantum gravity, since it is at this point that the essential requirement of background independence becomes fully introduced into the formalism. At the technical level, the process of solving the diffeomorphism constraint contains a couple of complications which were absent in the case of the Gauss constraint.

The first complication is that, due to the nature of the scalar product \eqref{cyl-sp} on the kinematical Hilbert space, the infinitesimal generator of diffeomorphisms does not exist as a well-defined operator on ${\cal H}_{\rm kin}$. If the infinitesimal generator did exist, it could be extracted from the operator of finite diffeomorphisms by considering a limit of the kind
\be\label{diff-limit}
\lim_{\epsilon\to 0} \frac{U(\phi_\epsilon)\ket{\Psi_\Gamma} - \ket{\Psi_\Gamma}}{\epsilon}.
\ee
However, this limit does not exist in general, because the states $U(\phi_\epsilon)\ket{\Psi_\Gamma}$ and $\ket{\Psi_\Gamma}$ are orthogonal to each other if the diffeomorphism $\phi_\epsilon$ moves the graph $\Gamma$ even by an infinitesimally small amount. For this reason it is not possible to solve the diffeomorphism constraint by constructing a constraint operator on ${\cal H}_{\rm kin}$ and looking for the states which are annihilated by the operator. Instead, the diffeomorphism invariant states must be sought by using an appropriately constructed group averaging procedure.

The other difficulty is that since diffeomorphisms act on the graph of a cylindrical function, the diffeomorphism invariant states cannot be proper, normalizable elements of the kinematical Hilbert space. In fact, the only diffeomorphism invariant element of ${\cal H}_{\rm kin}$ is the constant function (which can be viewed as a cylindrical function on a graph consisting of no edges and no nodes). Therefore non-trivial diffeomorphism invariant states should be searched for in the dual space $\Cyl^*$ (\ie the space of linear functionals on $\Cyl$). Of course this is not very surprising, as the same situation occurs also in simple examples in quantum mechanics. For example, solutions of the constraint $p_z=0$ are not proper elements of the Hilbert space ${\cal H} = L_2(\R^3)$, but they do belong to the dual space ${\cal H}^*$, since a function of the form $\psi(x,y)$ can be interpreted as a linear functional on ${\cal H}$.

After these preliminary remarks, let us move on to the construction of the group averaging map\footnote{This map was first constructed in \cite{ALMMT}. Our presentation in this chapter follows that of \cite{status} and \cite{Thiemann}.}, which is supposed to map states in ${\cal H}_{\rm kin}$ into diffeomorphism invariant elements of $\Cyl^*$. For a given graph $\Gamma$, a diffeomorphism may either leave the graph completely untouched, map the graph into itself but bring about some non-trivial change such as reversing the orientation of an edge, or move the graph around in $\Sigma$. The first kind of diffeomorphisms should clearly be ignored in the group averaging of a cylindrical function $\Psi_\Gamma$, and it is convenient to carry out the averaging separately with respect to the remaining two kinds of diffeomorphisms.

To be more precise, denote by ${\cal D}_\Gamma$ the group of diffeomorphisms which map $\Gamma$ into itself, and by ${\cal T}_\Gamma$ the subgroup of ${\cal D}_\Gamma$ whose action on $\Gamma$ is completely trivial. Then the (finite) quotient group
\be
{\cal S}_\Gamma = {\cal D}_\Gamma/{\cal T}_\Gamma,
\ee
which we call the group of symmetries of $\Gamma$, comprises the diffeomorphisms of the second kind in the discussion above. Averaging of a cylindrical function $\Psi_\Gamma$ with respect to the symmetries of the graph is then accomplished by using the projection operator $P_\Gamma$, defined as
\be\label{P_Gamma*Psi}
P_\Gamma\ket{\Psi_\Gamma} = \frac{1}{N_{{\cal S}_\Gamma}} \sum_{\phi\in{\cal S}_\Gamma} U(\phi)\ket{\Psi_\Gamma},
\ee
where $N_{S_\Gamma}$ is the number of elements in ${\cal S}_\Gamma$, and the factor of $1/{N_{{\cal S}_\Gamma}}$ is inserted to ensure that the averaging is done with ''total weight'' 1.

It then remains to average the state \eqref{P_Gamma*Psi} with respect to diffeomorphisms which move the graph around in the spatial manifold $\Sigma$. More precisely, the averaging should be done with respect to the different ways that the graph can be transformed by a diffeomorphism in $\Sigma$, since if $\Gamma'$ is a graph related to $\Gamma$ by a diffeomorphism, then there are infinitely many diffeomorphisms for which $\Gamma' = \phi(\Gamma)$ holds. Moreover, since the possible symmetries of the graph are already dealt with in \Eq{P_Gamma*Psi}, the appropriate group for the remaining part of the group averaging is ${\rm Diff}/{\cal D}_\Gamma$, with ${\rm Diff}$ denoting the group of all diffeomorphisms on $\Sigma$. This is an uncountably infinite group, so the result of the averaging will not be a proper cylindrical function, but only an element of $\Cyl^*$. Hence we define
\be\label{P_diff*Psi}
P_{\rm diff}[\Psi_\Gamma] = \sum_{\phi\in{\rm Diff}/{\cal D}_\Gamma} \bra{U(\phi)P_\Gamma\Psi_\Gamma},
\ee
where the uncountable sum on the right-hand side is a well-defined element of $\Cyl^*$, because when acting on a cylindrical function, only a finite number of terms in the sum will be non-vanishing. Note that the operator $P_{\rm diff}$, even though we denote it by the letter $P$, is not a true projection operator; rather, it is a so-called rigging map, which maps elements of $\Cyl$ into (diffeomorphism invariant) elements of $\Cyl^*$. (The action of $P_{\rm diff}$ is extended by linearity to all elements of $\Cyl$.) 

\Eqs{P_Gamma*Psi} and \eqref{P_diff*Psi} define the general solution of the diffeomorphism constraint. The diffeomorphism invariance of the scalar product on ${\cal H}_{\rm kin}$ guarantees that the solution is consistent in the sense that two cylindrical functions which are related to each other by a diffeomorphism give rise to the same diffeomorphism invariant state under the operator $P_{\rm diff}$:
\be
P_{\rm diff}\bigl[U(\phi)\Psi_\Gamma\bigr] = P_{\rm diff}[\Psi_\Gamma].
\ee
Even though the diffeomorphism invariant states are unnormalizable in the scalar product on ${\cal H}_{\rm kin}$, a natural scalar product between them can nevertheless be defined as 
\be\label{diff-sp}
\bbraket{P_{\rm diff}[\Psi]}{P_{\rm diff}[\Psi']} \equiv P_{\rm diff}[\Psi]\bigl(\ket{\Psi'}\bigr),
\ee
where the right-hand side is defined by \Eq{P_diff*Psi} and the scalar product on ${\cal H}_{\rm kin}$.

By applying the operator $P_{\rm diff}$ to all cylindrical functions, one may construct the space of all diffeomorphism invariant ''cylindrical functionals'' (\ie diffeomorphism invariant elements of $\Cyl^*$). In order to obtain simultaneous solutions of the Gauss and diffeomorphism constraints, it suffices to restrict the set of initial cylindrical functions to the gauge invariant subspace of $\Cyl$. Therefore we define the diffeomorphism invariant Hilbert space of loop quantum gravity as
\be
{\cal H}_{\rm diff} = \overline{P_{\rm diff}(\Cyl_G)},
\ee
where $\Cyl_G$ is the space of gauge invariant cylindrical functions, and the completion is with respect to the norm defined by the scalar product \eqref{diff-sp}.

Well-defined operators on the space ${\cal H}_{\rm diff}$ can be obtained from diffeomorphism invariant operators on the kinematical Hilbert space. Any diffeomorphism invariant operator $A$ on ${\cal H}_{\rm kin}$ defines an operator $A^\star$ on ${\cal H}_{\rm diff}$ through the prescription 
\be\label{Astar}
\bigl(A^\star P_{\rm diff}[\Psi]\bigr)\bigl(\ket{\Psi'}\bigr) \equiv P_{\rm diff}[\Psi]\bigl(A\ket{\Psi'}\bigr).
\ee
This definition is consistent with the scalar product on ${\cal H}_{\rm diff}$ in the sense that the operator $A^\star$ is self-adjoint with respect to the scalar product \eqref{diff-sp} precisely when $A$ is a self-adjoint operator on ${\cal H}_{\rm kin}$.

Solving the diffeomorphism constraint is the key step in realizing the crucial requirement of background independence in loop quantum gravity. Loosely speaking, one can think of states in the diffeomorphism invariant Hilbert space as spin networks of the form \eqref{spinnetwork}, but instead of a graph embedded in $\Sigma$, each state is labeled by an equivalence class of graphs under diffeomorphisms -- or a kind of ''semi-abstract'' graph, which is not localized anywhere in $\Sigma$, but which still retains some diffeomorphism invariant information from the original embedded graph (such as the possible knottedness of the graph, and the differential structure of the edges at each node). This kind of diffeomorphism invariant information is the only trace that remains of the spatial manifold $\Sigma$ in the diffeomorphism invariant Hilbert space.

\begin{figure}[t]
	\centering
		\includegraphics[width=0.33\textwidth]{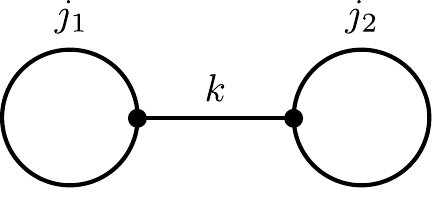}
		\caption{A spin network state with non-obvious transformation properties under diffeomorphisms.}
	\label{fig7-1}
\end{figure}

However, an important caveat that goes with the intuitive picture described above is that not every spin network state has a corresponding state in the diffeomorphism invariant Hilbert space. In particular, some non-trivial spin networks are annihilated by the projection operator $P_\Gamma$. A standard example is the state shown in \Fig{fig7-1}, which consists of two loops connected to each other by a third edge. Using \Eq{D^-1-gra}, one can show that the effect of a diffeomorphism which reverses the orientation of one of the loops while leaving the rest of the graph unchanged is to multiply the state by $(-1)^k$. Therefore this state is mapped to zero by the operator $P_\Gamma$ whenever the spin $k$ is odd; the state survives the averaging with respect to the symmetries of the graph only if $k$ is even. When $k$ is odd, the state of \Fig{fig7-1} does not exist in the diffeomorphism invariant Hilbert space.

Another subtlety related to the structure of the diffeomorphism invariant Hilbert space arises in connection with the question of separability of the space. The fact that diffeomorphism invariant states ''remember'' the differential structure at the nodes of the graph implies that states in ${\cal H}_{\rm diff}$ containing nodes of sufficiently high valence carry continuous, diffeomorphism invariant information, or so-called ''moduli'', at the nodes (see \eg \cite{moduli}). For example, in an $n$-valent node there are $n(n-1)/2$ angles between the tangent vectors of the edges at the node. A diffeomorphism on $\Sigma$ acts on the tangent vectors linearly as a $3\times 3$ -matrix, so there are only 9 free parameters available to change the (continuous) values of these angles. It follows that whenever $n\geq 5$, there exist families of uncountably many graphs which differ from each other only within an infinitesimally small neighborhood of a node, but which nevertheless cannot be transformed into each other by a diffeomorphism. In other words, the space ${\cal H}_{\rm diff}$ contains uncountably many states which are all diffeomorphically inequivalent and therefore orthogonal to each other; hence ${\cal H}_{\rm diff}$ is non-separable.

On the other hand, the continuous ''moduli'' do not seem to have any physical relevance, in that not any known operator in loop quantum gravity is sensitive to their values (as pointed out by Thiemann in \cite{Thiemann-length} and \cite{QSD3}). In practice one may therefore work with a separable subspace of ${\cal H}_{\rm diff}$ obtained by arbitrarily fixing the values of the moduli. Such a subspace would be preserved by all known LQG operators, and the result of any calculation would not depend on the arbitrary choices made in selecting the subspace. In this sense it seems that any set of practically relevant operators will always select a separable subspace (or a ''superselection sector'') of ${\cal H}_{\rm diff}$, even though the entire space ${\cal H}_{\rm diff}$ is non-separable.

\addtocontents{toc}{\protect\newpage}
\section{Geometric operators}\label{ch:geometric}

The basic strategy for constructing operators in loop quantum gravity out of classical functions of the variables $(A_a^i,E^a_i)$ is to re-express the classical function in terms of holonomies and fluxes, which are the elementary variables that can be promoted into well-defined operators in LQG. In particular, operators corresponding to geometric quantities such as areas of surfaces \cite{area, RSvolume} and volumes of regions \cite{ALvolume, RSvolume} in the spatial manifold $\Sigma$ can be constructed in this way. 

\subsection{Area operator}

The area of a two-dimensional surface $S$ on $\Sigma$ is given by $A(S) = \int_S \sqrt h$, with $h$ the determinant of the induced metric on $S$. If the equation of the surface is given in the form $x^a = x^a(\sigma^A)$, where $\sigma^A = (\sigma^1,\sigma^2)$ are coordinates on the surface, then a more explicit expression for the area is
\be\label{A-classical}
A(S) = \int_S d^2\sigma\,\sqrt{\det\biggl(\frac{\partial x(\sigma)}{\partial\sigma^A}\cdot\frac{\partial x(\sigma)}{\partial\sigma^B}\biggr)}.
\ee
To start rewriting this as a function of fluxes, we expand the determinant within the integrand as
\be
\det\biggl(\frac{\partial x(\sigma)}{\partial\sigma^A}\cdot\frac{\partial x(\sigma)}{\partial\sigma^B}\biggr) = (q_{ab}q_{cd} - q_{ad}q_{bc})\frac{\partial x^a}{\partial\sigma^1}\frac{\partial x^b}{\partial\sigma^1}\frac{\partial x^c}{\partial\sigma^2}\frac{\partial x^d}{\partial\sigma^2}.
\ee
Now using the identity $q_{ab}q_{cd} - q_{ad}q_{bc} = \epsilon_{ace}\epsilon_{bdf}(\det q)q^{ef}$ and recalling that the normal of the surface is $n_a = \epsilon_{abc}({\partial x^b}/{\partial\sigma^1})({\partial x^c}/{\partial\sigma^2})$, we find
\be
A(S) = \int_S d^2\sigma\,\sqrt{n_aE^a_in_bE^{bi}}.
\ee
Here the area of $S$ is not yet expressed in terms of fluxes. However, we see that the area of an infinitesimally small surface $S^{(\epsilon)}$ is approximately given by
\be\label{A(Seps)}
A\bigl(S^{(\epsilon)}\bigr) \simeq \sqrt{E_i\bigl(S^{(\epsilon)}\bigr)E^i\bigl(S^{(\epsilon)}\bigr)}.
\ee
In the case of a surface of finite size, this suggests subdividing the surface into a set of infinitesimal surfaces $\{S_\alpha^{(\epsilon)}\}$, whose size is controlled by the parameter $\epsilon$. By doing so, we get to write the area of the surface as
\be\label{A(S)-reg}
A(S) = \lim_{\epsilon\to 0}\, \sum_\alpha \sqrt{E_i\bigl(S_\alpha^{(\epsilon)}\bigr)E^i\bigl(S_\alpha^{(\epsilon)}\bigr)}.
\ee
This concludes the classical preparations for constructing the area operator, since in \Eq{A(S)-reg} the area is expressed as a function of fluxes only, and is therefore in a form which can readily be promoted into an operator in the quantum theory.

In order to derive the action of the area operator on a spin network state, we must study how the operator $E_i\bigl(S_\alpha^{(\epsilon)}\bigr)E^i\bigl(S_\alpha^{(\epsilon)}\bigr)$ acts. As the subdivision of $S$ into the surfaces $\{S_\alpha^{(\epsilon)}\}$ becomes finer and finer, there eventually remain two non-trivial cases to consider: (1) the surface $S_\alpha^{(\epsilon)}$ is intersected transversally by a single edge of the spin network; or (2) a single node of the spin network lies on $S_\alpha^{(\epsilon)}$. In case (1), letting $h_e$ denote the holonomy of the edge which intersects $S_\alpha^{(\epsilon)}$, a simple calculation using \Eqs{E*Psi} and \eqref{J*h_e} shows that
\be
E_i\bigl(S_\alpha^{(\epsilon)}\bigr)E^i\bigl(S_\alpha^{(\epsilon)}\bigr)\D{j}{m}{n}{h_e} = (8\pi\beta G)^2j(j+1)\D{j}{m}{n}{h_e}.
\ee
Therefore in case (1) the spin network is simply an eigenstate of $E_i\bigl(S_a^{(\epsilon)}\bigr)E^i\bigl(S_a^{(\epsilon)}\bigr)$ with eigenvalue $(8\pi\beta G)^2j(j+1)$. Since the eigenvalue is positive, there is no problem with taking the square root in \Eq{A(S)-reg}.

Case (2), where the surface $S_\alpha^{(\epsilon)}$ contains a node $v$ of the spin network, is somewhat more complicated. Using the definition of the flux operator, we find
\be\label{case(2)}
E_i\bigl(S_\alpha^{(\epsilon)}\bigr)E^i\bigl(S_\alpha^{(\epsilon)}\bigr)\bket{\Gamma,j_e,\iota_v}= (4\pi\beta G)^2\sum_{e_I,e_J} \delta(e_I,e_J) J_i^{(v,e_I)}J_{\phantom{i}}^{(v,e_J)i}\bket{\Gamma,j_e,\iota_v}
\ee
where the sum runs over all pairs of edges at the node $v$, and the factor $\delta(e_I,e_J)$ equals $0$ if either $e_I$ or $e_J$ intersects the surface $S_\alpha^{(\epsilon)}$ tangentially at $v$, and otherwise equals $+1$ or $-1$ depending on whether $e_I$ and $e_J$ lie on the same side of the surface or on opposite sides of it. Defining now the operators
\be\label{JuJd}
J_{v,i}^{(u)} \; = \sum_{\text{$e$ above $S$}} J_i^{(v,e)}, \qquad J_{v,i}^{(d)} \; = \sum_{\text{$e$ below $S$}} J_i^{(v,e)},
\ee
we can write the operator on the right-hand side \Eq{case(2)} in the form
\be\label{case(2)operator}
\sum_{e_I,e_J} \delta(e_I,e_J) J_i^{(v,e_I)}J_{\phantom{i}}^{(v,e_J)i} = \bigl(J_v^{(u)} - J_v^{(d)}\bigr)^2 = 2\bigl(J_v^{(u)}\bigr)^2 + 2\bigl(J_v^{(d)}\bigr)^2 - \bigl(J_v^{(u)} + J_v^{(d)}\bigr)^2.
\ee
From this, the eigenvalues of the operator can be read off as
\be
2j_u(j_u+1) + 2j_d(j_d+1) - j_t(j_t+1)
\ee
where the spins $j_u$, $j_d$ and $j_t$ are subject to the Clebsch--Gordan conditions. The eigenvalues are again positive, being the eigenvalues of the manifestly positive operator $\bigl(J_v^{(u)} - J_v^{(d)}\bigr)^2$, so there is no problem in defining the square root in \Eq{A(S)-reg} through the spectral decomposition of the operator. However, in this case a generic spin network state is not necessarily an eigenstate of the area operator (though it is always possible to choose a basis of intertwiners at the node in which the operator \eqref{case(2)operator} is diagonal).

To complete the definition of the area operator, it remains to take the limit $\epsilon\to 0$ in \Eq{A(S)-reg}. This step turns out to be easier than one might have expected, since from the results of the above calculations we can see that the limit is in fact trivial. Once the subdivision $\{S_\alpha^{(\epsilon)}\}$ is fine enough that each of the surfaces $S_\alpha^{(\epsilon)}$ intersects the graph of a spin network at most at a single point, then any further refinement of the subdivision has no effect on the action of the regularized, $\epsilon$-dependent area operator on a spin network state. In other words, the limit $\epsilon\to 0$ of the expression \eqref{A(S)-reg} is reached already at some finite value of $\epsilon$. That a regulator involved in constructing a quantum operator out of a classical function can be removed trivially at the end of the construction is a recurring theme in loop quantum gravity; it is one of the distinctive, powerful features of the background-independent framework of the theory.

In conclusion, we found that the classical area functional \eqref{A-classical} can be promoted into a well-defined operator on the Hilbert space of loop quantum gravity. An explicit expression for the operator can only be given separately within the subspaces associated to every possible graph. We have
\be
A(S)\ket{\Psi_\Gamma} = 4\pi\beta G\sum_{x\in S} \sqrt{\bigl(J_x^{(u)} - J_x^{(d)}\bigr)^2}\,\ket{\Psi_\Gamma}
\ee
with $J_x^{(u)}$ and $J_x^{(d)}$ defined by \Eq{JuJd}. The operator is evidently gauge invariant, though this can also be verified explicitly by checking that $A(S)$ commutes with the Gauss operator $G_v$ of \Eq{G_v}. Since the operator refers to the surface $S$, it is not diffeomorphism invariant, but it does transform covariantly under diffeomorphisms: $U(\phi)A(S) = A\bigl(\phi(S)\bigr)$. The spectrum of the operator can be computed explicitly; the scale of the area eigenvalues is determined by the Planck length $l_P = \sqrt{8\pi G}$ and the Immirzi parameter $\beta$. In particular, the so-called ''main sequence'' of the spectrum, which arises from intersections of the surface $S$ with isolated edges of a spin network, has the form
\be
\lambda(A) = 8\pi\beta G\sum_\alpha \sqrt{j_\alpha(j_\alpha+1)},
\ee
where the sum runs over any arbitrary sequence of spins $\{j_\alpha\}$. Remarkably, the spectrum is discrete (though practically indistinguishable from a continuum at macroscopic values of the area), even though discreteness of geometry is not imposed by hand at any point in the development of the kinematical structure of loop quantum gravity. 

\subsection{Volume operator}

Classically, the volume of a region $R$ in $\Sigma$ is given by
\be\label{V-classical}
V(R) = \int_R d^3x\,\sqrt q = \int_R d^3x\,\sqrt{\biggl|\frac{1}{3!}\epsilon_{abc}\epsilon^{ijk}E^a_iE^b_jE^c_k\biggr|}.
\ee
In broad outline, the steps leading from this expression to the quantum volume operator are identical to those required for the area operator, but there will appear a number of new technical details which were not encountered in the case of the area operator.

As before, the first step is to rewrite the volume as a function of the flux variable. To this end, let us consider an infinitesimal cubical cell $C^{(\epsilon)}$. The volume of the cell can be approximated as
\be\label{V(C)}
V\bigl(C^{(\epsilon)}\bigr) \simeq \sqrt{\biggl|\frac{1}{3!}\epsilon_{abc}\epsilon^{ijk}E_i(S^a)E_j(S^b)E_k(S^c)\biggr|},
\ee
where $(S^1,S^2,S^3)$ is a suitably chosen set of surfaces associated with the cell. We assume that the surfaces are aligned along the coordinate axes, so that on each surface $S^a$ the corresponding coordinate $x^a = {\rm const.}$, and require the areas of the surfaces to satisfy $\sqrt{A(S^1)A(S^2)A(S^3)} = V\bigl(C^{(\epsilon)}\bigr)$, at least to leading order in $\epsilon$. The volume of any region $R$ can then be expressed in terms of fluxes by subdividing $R$ into a set of cells $\{C^{(\epsilon)}_\alpha\}$:
\be\label{V(R)-reg}
V(R) = \lim_{\epsilon\to 0}\, \sum_\alpha V\bigl(C^{(\epsilon)}_\alpha\bigr).
\ee
A vitally important feature of the regularized expression \eqref{V(R)-reg}, just as the expression \eqref{A(S)-reg} for the area, is that it does not depend explicitly on the parameter $\epsilon$, which controls the size of the cells. This is a reflection of the fact that the integrand in \Eq{V-classical} is geometrically a density of weight one, and ultimately guarantees that the regulator can be removed and the expression \eqref{V(R)-reg} promoted into a well-defined operator in the quantum theory.

Compared to the case of the area operator, the first new aspect in the construction of the volume operator is that there is a substantial freedom in choosing the surfaces $\{S^a\}$ associated to the cells $\{C^{(\epsilon)}_\alpha\}$. We will consider in detail only the so-called ''internal regularization'' due to Ashtekar and Lewandowski \cite{ALvolume}, since the resulting operator is the only volume operator that will be used later in this work. In the internal regularization, the surfaces $S^a$ are taken to lie inside their corresponding cell and are bounded by the faces of the cell. Furthermore, the surfaces are adjusted separately to the graph of each spin network state by imposing the following requirements:
\begin{itemize}
\item If a cell $C^{(\epsilon)}_\alpha$ contains a node of the graph, it must lie in the intersection point of the corresponding surfaces $S^a_\alpha$.
\item The surfaces $S^a_\alpha$ of a cell containing a node intersect the graph only at the node.
\item If $C^{(\epsilon)}_\alpha$ does not contain a node, then the graph intersects the surfaces $S^a_\alpha$ at no more than two points.
\end{itemize}
Whenever the first requirement is satisfied, the remaining two can immediately be met by sufficiently refining the subdivision $\{C^{(\epsilon)}_\alpha\}$.

To find the action of the resulting operator on spin networks, we must start by considering (for each cell $C^{(\epsilon)}_\alpha$) the operator
\be
\theta_C = \frac{1}{3!}\epsilon_{abc}\epsilon^{ijk}E_i(S^a)E_j(S^b)E_k(S^c).
\ee
The third requirement and the antisymmetry of $\epsilon_{abc}$ guarantee that the operator $\theta_C$ gives zero when acting on a cell that does not contain a node. If the cell does contain a node, we find (recalling the first two requirements) that the operator acts as
\be\label{theta_C}
\theta_C\bket{\Gamma,j_e,\iota_v} = (8\pi\beta G)^3\frac{1}{48}\sum_{e_I,e_J,e_K} \kappa(e_I,e_J,e_K)\epsilon^{ijk}J_i^{(v,e_I)}J_j^{(v,e_J)}J_k^{(v,e_K)}\bket{\Gamma,j_e,\iota_v},
\ee
where $\kappa(e_I,e_J,e_K) = \epsilon_{abc}\kappa(S_a,e_I)\kappa(S_b,e_J)\kappa(S_c,e_K)$, and each sum runs over all edges at the node, though the antisymmetry of $\kappa(e_I,e_J,e_K)$ implies that non-zero terms arise only when the edges $e_I,e_J,e_K$ are all distinct from each other.

Since the operator $\theta_C$ acts only on the nodes of a spin network as soon as the three requirements above are satisfied, any further refinement of the cells $\{C^{(\epsilon)}_\alpha\}$ has no effect in \Eq{V(R)-reg}, so it is again trivial to take the limit $\epsilon\to 0$. However, the resulting operator is not yet a satisfactory operator, since it retains a dependence on the details of the cells $\{C^{(\epsilon)}_\alpha\}$ and the surfaces $S^a_\alpha$ through the factor $\kappa(e_I,e_J,e_K)$. In order to obtain a background-independent volume operator, one must therefore perform a suitable averaging over the background structures used in the construction, as discussed in detail in \cite{ALvolume}. The conclusion is that such an averaging can be carried out, and the form of the resulting operator is determined uniquely\footnote{Up to an undetermined multiplicative factor, which we will ignore in what follows.} as
\be\label{V(R)}
V(R)\ket{\Psi_\Gamma} = (8\pi\beta G)^{3/2}\sum_{\substack{{v\in\Gamma}\\{\text{within}\,R}}} \sqrt{|q_v|}\,\ket{\Psi_\Gamma},
\ee
where
\be\label{q_v}
q_v = \frac{1}{48}\sum_{e_I,e_J,e_K} \epsilon(e_I,e_J,e_K)\epsilon^{ijk}J_i^{(v,e_I)}J_j^{(v,e_J)}J_k^{(v,e_K)},
\ee
and the ''orientation factor'' $\epsilon(e_I,e_J,e_K)$ equals $+1$ if the triple of tangent vectors $(\dot e_I,\dot e_J,\dot e_K)$ at the node $v$ is positively oriented, $-1$ if it is negatively oriented, and $0$ if the tangent vectors are not linearly independent.

The other well-known construction of the volume operator was introduced by Rovelli and Smolin \cite{RSvolume} and studied further by De Pietri and Rovelli \cite{DPRvolume}. Their construction is based on an ''external regularization'', in which the faces of the cell $C_\alpha^{(\epsilon)}$ are used as the surfaces $S^a_\alpha$. The resulting operator still acts only on the nodes of a spin network, but in order to achieve this, one must start by suitably rewriting the classical expression \eqref{V(C)} in terms of the parallel transported flux variable of \Eq{tr-flux}. (At the classical level this can be done freely when considering an infinitesimal cell, since the holonomies involved in the parallel transported flux reduce to identities in the limit $\epsilon\to 0$.) In the end the externally regularized volume operator takes the form \eqref{V(R)}, but in place of $|q_v|$ one has the operator
\be
q_v^{\rm ext} = \frac{1}{48}\sum_{e_I\neq e_J\neq e_K} \Bigl|\epsilon^{ijk}J_i^{(v,e_I)}J_j^{(v,e_J)}J_k^{(v,e_K)}\Bigr|,
\ee
where the sum runs over all triples of three distinct edges at the node. Therefore the main difference between the internally and externally regularized volume operators (as discussed \eg in \cite{L-volume}) is that the latter is not sensitive to the tangential structure of the edges at the node.

\subsection{Properties of the volume operator}\label{sec:vol-properties}

The volume operator $V(R)$, just as the area operator, is gauge invariant and transforms covariantly under diffeomorphisms. However, the total volume of the spatial manifold $\Sigma$, which we will denote simply by $V$, is diffeomorphism invariant, and is therefore a well-defined operator on the space ${\cal H}_{\rm diff}$.

Due to the factor $\epsilon(e_I,e_J,e_K)$ in \eqref{q_v}, the volume operator gives zero when acting on a bivalent node, or a node of higher valence where the tangent vectors of all the edges are planar. A less obvious property is that every trivalent node of a spin network state is also annihilated by the volume operator. Using the gauge invariance condition $J_i^{(v,e_1)} + J_i^{(v,e_2)} + J_i^{(v,e_3)} = 0$ and the commutation relation \eqref{[J,J]}, we find
\begin{align}
\epsilon^{ijk}J_i^{(v,e_1)}J_j^{(v,e_2)}J_k^{(v,e_3)} &= -\epsilon^{ijk}J_i^{(v,e_1)}J_j^{(v,e_2)}\bigl(J_k^{(v,e_1)} + J_k^{(v,e_2)}\bigr) \notag \\
&= -\frac{1}{2}\epsilon^{ijk}\Bigl(\bigl[J_i^{(v,e_1)},J_k^{(v,e_1)}\bigr]J_j^{(v,e_2)} + J_i^{(v,e_1)}\bigl[J_j^{(v,e_2)},J_k^{(v,e_2)}\bigr]\Bigr) \notag \\
&= i\Bigl(\delta^{jl} J_l^{(v,e_1)}J_j^{(v,e_2)} - \delta^{il}J_i^{(v,e_1)}J_l^{(v,e_2)}\Bigr) = 0.
\end{align}
Therefore a four-valent node is the simplest node on which the volume operator acts in a non-trivial way.

Gauge invariance can also be used to simplify the task of calculating the action of the volume operator even when the result will not be trivial. Let us illustrate this by considering the example of a four-valent node. Introducing the notation
\be
q_v^{(IJK)} = \epsilon^{ijk}J_i^{(v,e_I)}J_j^{(v,e_J)}J_k^{(v,e_K)},
\ee
it would appear that one must compute the action of the operator
\be
\epsilon(e_1,e_2,e_3)q_v^{(123)} + \epsilon(e_1,e_2,e_4)q_v^{(124)} + \epsilon(e_1,e_3,e_4)q_v^{(134)} + \epsilon(e_2,e_3,e_4)q_v^{(234)},
\ee
in which each of the four terms involves a distinct triple of the operators $J_i^{(v,e_I)}$. However, using gauge invariance, we can eliminate one of the angular momentum operators from the calculation by writing $J_i^{(v,e_4)} = -J_i^{(v,e_1)} - J_i^{(v,e_2)} - J_i^{(v,e_3)}$. Hence we find that we can equivalently calculate the action of the simpler operator
\be
\Bigl[\epsilon(e_1,e_2,e_3) - \epsilon(e_1,e_2,e_4) - \epsilon(e_1,e_3,e_4) - \epsilon(e_2,e_3,e_4)\Bigr]q_v^{(123)},
\ee
which depends only on a single triple of the angular momentum operators.

In section \ref{sec:V-el} we will consider in detail the problem of computing the action of the (squared) volume operator in the spin network basis. In the literature, matrix elements of the volume have been calculated \eg in \cite{Thiemann-vol, DPRvolume, BrunnemannThiemann, BrunnemannRideout1, BrunnemannRideout2}. As the simplest non-trivial example, we display here the action of the operator $q_v^{(123)}$ on a four-valent node, in the (unnormalized) basis \eqref{int4-12} of the four-valent intertwiner space:\footnote{\Eq{q-example} can be obtained from the more general result of \Eq{q123-result} by setting the spins $l_1$ and $l_2$ equal to zero and using the equality \cite{Varshalovich}
\[
\sixj{a}{b}{c}{d}{e}{0} = \frac{(-1)^{a+b+c}}{\sqrt{d_ad_b}}\delta_{ae}\delta_{bd}.
\]}
\begin{align}
&q_v^{(123)}\;\RealSymb{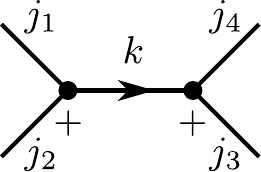}{0.6} \quad = \quad i\sqrt{6}W_{j_1}W_{j_2}W_{j_3}(-1)^{j_3+j_4+k} \notag  \\ 
&\hspace{3cm}\times\sum_x d_x\sixj{j_3}{j_3}{1}{k}{x}{j_4}\ninej{j_1}{j_1}{1}{j_2}{j_2}{1}{k}{x}{1}\;\RealSymb{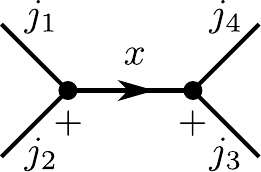}{0.6}\label{q-example}
\end{align}
Here $W_j = \sqrt{j(j+1)(2j+1)}$, and the Wigner 6$j$- and 9$j$-symbols appear in the matrix elements. The properties of the 9$j$-symbol imply that the sum over $x$ runs only over the two values $x=k\pm 1$. Thus, in this case the matrix of the squared volume operator has non-zero entries only in the two diagonals immediately adjacent to the main diagonal.

The spectrum of the volume operator is clearly discrete, but in contrast to the area operator, an analytic expression for the complete set of eigenvalues is not known. The eigenvalue problem of the operator has been solved explicitly only in certain relatively simple special cases (see \eg \cite{BrunnemannRideout1, BrunnemannRideout2}, where the spectra corresponding to nodes of valence up to 7 are studied numerically, and analytical results are also given in the case of a four-valent node).

The basic structure of the area and volume operators is that the quantum area of a surface receives a contribution from each edge of a spin network intersecting the surface, while the quantum volume of a region receives a contribution from each node of a spin network contained inside the region. This naturally suggests a physical interpretation where a spin network is a state of discrete, quantized spatial geometry, consisting of quantized excitations of volume (at the nodes) separated from each other by quantized excitations of area (at the edges). Within this interpretation, the graph of a spin network state is seen as dual to the quantized geometry defined by the state, each node of the graph being dual to an elementary quantum of volume, while each edge is dual to an elementary quantum of area.

After the diffeomorphism constraint is imposed through the operator $P_{\rm diff}$, the excitations of quantum geometry are not localized in any background manifold. Hence the picture is truly background independent, the only physically meaningful information being the relative localization of the quanta of geometry with respect to each other. We emphasize again that this intriguing picture seems to emerge naturally as a result of quantizing general relativity according to the framework of loop quantum gravity. Nowhere in the construction of the theory is discreteness of spatial geometry taken as an assumption or imposed as a postulate.

\subsection{Angle operator}\label{sec:angle}

Since the geometry of the spatial surface $\Sigma$ is completely encoded in the metric $q_{ab}$, any geometric quantity intrinsic to $\Sigma$ can be expressed as a function of the triad $E^a_i$, and could therefore (at least in principle) be quantized following similar steps by which the area and volume operators were constructed. In the remainder of this chapter, we will briefly describe two further geometric operators, which will be used later in this work.

The angle operator $\theta_v^{(e_1,e_2)}$, introduced into loop quantum gravity by Major \cite{Major}, acts on a pair of edges $(e_1,e_2)$ at a node $v$ of a spin network as
\be\label{angle-op}
\theta_v^{(e_1,e_2)} = \cos^{-1}\biggl(\frac{J_i^{(v,e_1)}J_{\phantom{i}}^{(v,e_2)i}}{\bigl|J^{(v,e_1)}\bigr|\bigl|J^{(v,e_2)}\bigr|}\biggr),
\ee
where $\bigl|J^{(v,e)}\bigr| \equiv \sqrt{J_i^{(v,e)}J_{\phantom{i}}^{(v,e)i}}$. As we will see shortly, it is straightforward to determine the eigenvalues and eigenstates of the operator inside $\cos^{-1}$, so the operator \eqref{angle-op} can be defined through its spectral decomposition without any problems.

The motivation for calling \eqref{angle-op} the angle operator is provided by the classical expression
\be\label{angle-class}
C(S_1,S_2) = \frac{E_i(S_1)E^i(S_2)}{\bigl|E(S_1)\bigr|\bigl|E(S_2)\bigr|},
\ee
where again $|E(S)|$ stands for $\sqrt{E_i(S)E^i(S)}$, and whose quantization clearly leads to the operator inside $\cos^{-1}$ in \Eq{angle-op}, when the surfaces are chosen such that each $S_I$ intersects only the edge $e_I$. (Precisely speaking, \eqref{angle-class} must be quantized using the parallel transported flux operator, in order to bring the action of the $J$-operators from the surfaces to the node.) In the limit where the surfaces are infinitesimally small and located infinitesimally close to the point $v$, the fluxes in \eqref{angle-class} can be approximated by $E_i(S) \simeq \epsilon^2E^a_i(v)n_a(v)$, with $\epsilon^2$ the coordinate area of the surface and $n_a$ the normal covector. In this limit, \Eq{angle-class} therefore reduces to
\be
C(S_1,S_2) \simeq \frac{E^a_in_a^{(1)}E^{bi}n_b^{(2)}}{\sqrt{E^a_in_a^{(1)}E^{bi}n_b^{(1)}}\sqrt{E^a_in_a^{(2)}E^{bi}n_b^{(2)}}} = \frac{q^{ab}n_a^{(1)}n_b^{(2)}}{\sqrt{q^{ab}n_a^{(1)}n_b^{(1)}}\sqrt{q^{ab}n_a^{(2)}n_b^{(2)}}},
\ee
which shows that $C(S_1,S_2)$ is a regularized expression for the cosine of the angle between the surfaces $S_1$ and $S_2$.

To determine the eigenvalues and eigenstates of the operator \eqref{angle-op}, it is enough to note that
\be\label{J1.J2}
J_i^{(v,e_1)}J^{(v,e_2)i} = \frac{1}{2}\Bigl(\bigl(J^{(v,e_1)} + J^{(v,e_2)}\bigr)^2 - \bigl(J^{(v,e_1)}\bigr)^2 - \bigl(J^{(v,e_2)}\bigr)^2\Bigr).
\ee
From this it is apparent that $\theta_v^{(e_1,e_2)}$ is diagonal whenever the intertwiner at the node $v$ couples the spins $j_1$ and $j_2$ of the edges $e_1$ and $e_2$ into a definite total spin $k$:
\be\label{angle-eq}
\theta_v^{(e_1,e_2)}\;\RealSymb{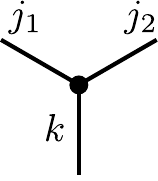}{0.6} \quad = \quad \theta(j_1,j_2,k)\;\RealSymb{fig8-3.pdf}{0.6}
\ee
with the corresponding eigenvalues being
\be\label{angle-eig}
\theta(j_1,j_2,k) = \cos^{-1}\biggl(\frac{k(k+1) - j_1(j_1+1) - j_2(j_2+1)}{2\sqrt{j_1(j_1+1)}\sqrt{j_2(j_2+1)}}\biggr).
\ee
In the ''monochromatic'' case $j_1=j_2\equiv j$, we see that when the total spin is $k=0$ (so that, in a sense, the fluxes of the surfaces dual to the edges point in completely opposite directions), the angle eigenvalue is $\theta(j,j,0)=\pi$, as one would expect. When the total spin has its maximal value $k=2j$ (describing a situation where the fluxes are as parallel as possible), the eigenvalue $\theta(j,j,2j)=\cos^{-1}\bigl(j/(j+1)\bigr)$ is not exactly equal to zero, but does approach zero in the limit of large $j$.

\subsection{Length operator}\label{sec:length}

Two different length operators, due to Thiemann \cite{Thiemann-length} and Bianchi \cite{Bianchi}, have been constructed in loop quantum gravity. The two operators differ from each other not only technically but also conceptually: Thiemann's operator apparently measures the length of the edge of a spin network, while Bianchi's operator measures the length of the curve along which two surfaces dual to two spin network edges intersect each other. We will focus our discussion solely on the operator defined by Bianchi, since it is the only length operator directly relevant to this work.

The starting point for the construction of a length operator is to express the length of a curve $e$ in the spatial manifold $\Sigma$ as
\be\label{length-cl}
L(e) = \int_0^1 ds\,\sqrt{g_{ab}\bigl(e(s)\bigr)\dot e^a(s)\dot e^b(s)} = \int_0^1 ds\,\sqrt{G^i(s)G_i(s)},
\ee
where
\be
G^i(s) = \frac{\epsilon^{ijk}\epsilon_{abc}\dot e^a(s)E^b_jE^c_k}{2\sqrt{\det E}}.
\ee
Through a careful regularization of the expression \eqref{length-cl}, where the curve $e$ is now taken to be the intersection between two surfaces $S_{e_1}$ and $S_{e_2}$ dual to two edges $e_1$ and $e_2$ attached to the same node $v$ of a spin network, Bianchi obtained the operator
\be\label{L_v}
L_v^{(e_1,e_2)} = \frac{1}{2}(8\pi\beta G)^2\sqrt{{\cal V}_v^{-1}Y_i^{(v,e_1,e_2)}Y_{\phantom{i}}^{(v,e_1,e_2)i}{\cal V}_v^{-1}}.
\ee
Here the operator $Y_i^{(v,e_1,e_2)}$ is defined as
\be\label{Y_i}
Y_i^{(v,e_1,e_2)} = \epsilon\downup{i}{jk}J_j^{(v,e_1)}J_k^{(v,e_2)},
\ee
and ${\cal V}_v^{-1}$ is a regularized inverse volume operator of the node $v$. The inverse operator ${\cal V}_v^{-1}$ can be defined by specifying its action on the eigenstates of the standard volume operator\footnote{In this work we depart slightly from Bianchi's construction by using the Ashtekar--Lewandowski volume operator, instead of the Rovelli--Smolin operator used originally by Bianchi, to define the operator ${\cal V}_v^{-1}$.} $V_v$ (restricted to the node $v$) as
\be\label{V^-1}
{\cal V}_v^{-1}\ket v = \begin{cases} v^{-1}\ket v & \text{if $v\neq 0$} \\ 0 & \text{if $v=0$} \end{cases}
\ee
(where $\ket v$ is an eigenstate of $V_v$ with eigenvalue $v$). More generally, one could define ${\cal V}_v^{-1}(\epsilon) = V_v/(V_v^2+\epsilon^2)$, where the parameter $\epsilon$ has a small but finite value. This is known as the so-called Tikhonov regularization of the inverse operator. The operator \eqref{V^-1} results by taking the limit $\epsilon\to 0$ of the operator ${\cal V}_v^{-1}(\epsilon)$.

The two ingredients that go into evaluating the action of the length operator in the spin network basis are the matrix elements of the operators ${\cal V}_v^{-1}$ and $Y_i^{(v,e_1,e_2)}Y_{\phantom i}^{(v,e_1,e_2)i}$. While the matrix elements of ${\cal V}_v^{-1}$ are generally accessible at most through numerical methods, deriving the action of the operator $Y_i^{(v,e_1,e_2)}Y_{\phantom i}^{(v,e_1,e_2)i}$ on a spin network state is a straightforward exercise in angular momentum algebra. (The calculation could alternatively be carried out using the graphical techniques outlined in section \ref{sec:SU2calc}, and described fully in the Appendix.) Using the identity $\epsilon_{ijk}\epsilon^{ilm} = \delta_j^l\delta_k^m - \delta_j^m\delta_k^l$ and the commutation relations of the angular momentum operator, we can express the operator in question in the form
\be\label{Y^2-exp}
Y_i^{(v,e_1,e_2)}Y_{\phantom i}^{(v,e_1,e_2)i} = \bigl(J^{(v,e_1)}\bigr)^2\bigl(J^{(v,e_2)}\bigr)^2 - \bigl(J_i^{(v,e_1)}J_{\phantom{i}}^{(v,e_2)i}\bigr)^2 - J_i^{(v,e_1)}J_{\phantom{i}}^{(v,e_2)i}.
\ee
From this we see that this operator, just as the angle operator, is diagonalized by any intertwiner in which the edges $e_1$ and $e_2$ are coupled to a definite total spin. \mbox{The eigenvalue}
\newpage
\noindent in the diagonal action of the operator,
\be\label{Y^2-eq}
Y_i^{(v,e_1,e_2)}Y_{\phantom i}^{(v,e_1,e_2)i}\;\RealSymb{fig8-3.pdf}{0.6} \quad = \quad \omega(j_1,j_2,k)\;\RealSymb{fig8-3.pdf}{0.6}
\ee
can be read off from \Eq{Y^2-exp}. It is given by
\be\label{Y^2-eig}
\omega(j_1,j_2,k) = j_1(j_1+1)j_2(j_2+1) - C(j_1,j_2,k)^2 - C(j_1,j_2,k),
\ee
where $C(j_1,j_2,k)$ are the eigenvalues of the operator \eqref{J1.J2}, namely
\be
C(j_1,j_2,k) = \frac{k(k+1)-j_1(j_1+1)-j_2(j_2+1)}{2}.
\ee

\section{Coherent states}\label{ch:CS}

An important ingredient in the kinematical structure of any quantum theory is the existence of coherent states, which describe nearly classical configurations of the quantum system. The spin network states of loop quantum gravity are very poor candidates for this task, since some aspects of the quantized geometry described by a spin network state are, in a sense, maximally far from classical. Being an eigenstate of geometric operators such as the area or the volume, a typical spin network state is completely spread with respect to the corresponding conjugate variables. Coherent states for loop quantum gravity at the kinematical level were first introduced by Thiemann \cite{GCS1, ThiemannCS}, building on related work by Hall \cite{Hall}.

\subsection{The complexifier method}

A systematic algorithm for constructing coherent states is provided by the so-called complexifier method \cite{ThiemannCS}. The complexifier ${\cal C}(q,p)$ is a function on the classical phase space, which is assumed to satisfy certain requirements. In particular, the complexifier must be a positive function, and must have a stronger than linear dependence on the momentum variable. By canonically quantizing the classical function ${\cal C}(p,q)$, one obtains the corresponding quantum operator $C$, which we also refer to as the complexifier.

The complexifier is used to construct coherent states in the following way. We start by applying the operator $e^{-C}$ to the delta function of the configuration variable, obtaining the function
\be\label{psi_q0}
\psi_{q_0}(q) = e^{-C}\delta(q,q_0).
\ee
The effect of the operator $e^{-C}$ is to smooth out the delta function, producing a function which has a peak of finite width concentrated around the point $q=q_0$. For example, considering the complexifier ${\cal C} = tp^2/2$ in quantum mechanics, the action of the operator $e^{-tp^2/2}$ on the delta function $\delta(x-x_0) = (2\pi)^{-1}\int dk\,e^{ik(x-x_0)}$ produces the Gaussian function $\psi_{x_0}(x) \sim e^{-(x-x_0)^2/2t}$. The requirements imposed on the complexifier are necessary in order that the function \eqref{psi_q0} can be a proper element of the Hilbert space of the quantum theory.

The functions \eqref{psi_q0} are now made into coherent states by ''complexifying'' the label $q_0$, \ie by analytically extending the functions \eqref{psi_q0} to complex values of $q_0$. Symbolically,
\be\label{psi_z0}
\psi_{z_0}(q) = \Bigl(e^{-C}\delta(q,q_0)\Bigr)\Big|_{q_0\to z_0}.
\ee
The precise way in which the label $q_0$ is ''complexified'' into $z_0$ is specified by the rule
\be\label{z(q,p)}
z(q,p) = \sum_n \frac{i^n}{n!}\bigl\{q,{\cal C}(q,p)\bigr\}^{(n)},
\ee
where $\{q,{\cal C}\}^{(n)}$ is the $n$ times iterated Poisson bracket, \ie $\{q,{\cal C}\}^{(n+1)} = \{\{q,{\cal C}\}^{(n)},{\cal C}\}$ with $\{q,{\cal C}\}^{(0)} = q$. \Eq{z(q,p)} also provides the relation between the label $z$ and the variables of the classical phase space, thereby defining the point in the classical phase space on which the coherent state \eqref{psi_z0} is supposedly peaked. Continuing the example from quantum mechanics, we see that replacing $x_0 \to z_0 = x_0+itp_0$ in the function $e^{-(x-x_0)^2/2t}$ gives the standard Gaussian coherent states $\psi_{(x_0,p_0)}(x) \sim e^{ip_0x}e^{-(x-x_0)^2/2t}$.

The construction of the states \eqref{psi_z0} guarantees that they are eigenstates of an annihilation operator $a$, which is defined through a quantum analog of the classical relation \eqref{z(q,p)} as
\be
a = e^{-C}qe^C = \sum_n \frac{1}{n!}[q,C]^{(n)},
\ee
with $[q,C]^{(n)}$ the $n$ times iterated commutator. Making use of \Eq{psi_z0}, a simple calculation shows that the state $\psi_{z_0}$ is an eigenstate of the operator $a$ with the complex eigenvalue $z_0$:
\be\label{apsi=zpsi}
a\psi_{z_0} = z_0\psi_{z_0}.
\ee
This result provides the main motivation for why the states \eqref{psi_z0} can be expected to be legitimate coherent states. Indeed, the eigenvalue equation \eqref{apsi=zpsi} implies that the states \eqref{psi_z0} are optimally peaked with respect to the Hermitian operators
\be\label{QandP}
Q = \frac{1}{2}(a+a^\dagger), \qquad P = \frac{1}{2i}(a-a^\dagger),
\ee
in the sense that the equality sign in the uncertainty relation
\be
\Delta P \Delta Q \geq \frac{1}{2}\big|\bigl\langle[P,Q]\bigr\rangle\big|
\ee
is reached when the both sides of the inequality are evaluated in the state \eqref{psi_z0}. However, even though it is certainly reasonable to expect the states $\psi_{z_0}$ to also possess other semiclassical properties, such as $\langle {\cal O}(q,p)^2\rangle / \langle {\cal O}(q,p)\rangle^2 - 1 \ll 1$ for a set of ''reasonable'' operators ${\cal O}(q,p)$, such properties are strictly speaking not guaranteed by the construction, and should be checked separately in each case.

\subsection{Heat kernel coherent states}

In order to derive coherent states from the complexifier construction, one must start by considering the form of the delta function on the appropriate Hilbert space. For example, on the gauge-invariant Hilbert space ${\cal H}_G^\Gamma$ associated to a fixed graph $\Gamma$, the corresponding delta function $\delta_\Gamma$ should satisfy
\be
\int d\mu_\Gamma(g_e)\,\delta_\Gamma(g_e,q_e)f(g_e) = f(q_e),
\ee
where $d\mu_\Gamma$ denotes the measure defined by \Eq{cyl-sp} on the space ${\cal H}_G^\Gamma$. It is easy to verify that the expansion of the delta function in the spin network basis of ${\cal H}_G^\Gamma$ is given by
\be
\delta_\Gamma(g_e,q_e) = \sum_{j_e,\iota_v} {\Psi_{\Gamma,\{j_e\},\{\iota_v\}}(q_e)}\overline{\Psi_{\Gamma,\{j_e\},\{\iota_v\}}(g_e)}.
\ee
An explicit expression for coherent states on ${\cal H}_G^\Gamma$ can now be written down, if we assume that the states $\Psi_{\Gamma,\{j_e\},\{\iota_v\}}$ are eigenstates of the complexifier. Denoting the eigenvalues of the complexifier by $\lambda(j_e,\iota_v)$, we have
\be\label{psi_z-HG}
\psi^\Gamma_{\{z_e\}}(g_e) = \sum_{j_e,\iota_v} e^{-\lambda(j_e,\iota_v)}{\Psi_{\Gamma,\{j_e\},\{\iota_v\}}\bigl(z_e(A,E)\bigr)}\overline{\Psi_{\Gamma,\{j_e\},\{\iota_v\}}(g_e)}.
\ee
Clearly the expression \eqref{psi_z-HG} is not of much practical use, unless an explicit expression is available for the eigenvalues of the complexifier. This requirement rules out the volume operator as a possible complexifier, even though the total volume is a diffeomorphism invariant operator, and would therefore appear to be a promising candidate complexifier for constructing coherent states on the diffeomorphism invariant Hilbert space. The spectrum of the volume operator (particularly the frequency of zeros among the eigenvalues) is not even known in sufficient detail to determine whether the expression \eqref{psi_z-HG} converges into a state of finite norm in ${\cal H}_G^\Gamma$.

Among the operators whose spectra are known in closed form, the area operator seems like a natural candidate for a complexifier, due to the simplicity of its action on spin networks -- depending on the surface $S$, the action of the operator $A(S)$ could possibly involve only a single edge of the spin network\footnote{Note, however, that there exist other operators, such as the angle operator or the operator \eqref{Y^2-exp}, whose eigenvalues and eigenstates in the spin network basis are known in explicit form. These operators could therefore be used as prospective complexifiers; however, to the author's best knowledge, coherent states resulting from complexifiers based on such operators have not been studied anywhere.}. Coherent states based on using (a suitable version of) the area operator as the complexifier are commonly known as heat kernel coherent states. They were introduced as coherent states for loop quantum gravity by Thiemann \cite{GCS1}, after having been studied earlier in a different context by Hall \cite{Hall}. Their properties in loop quantum gravity were investigated further in a series of papers by Thiemann and Winkler \cite{GCS2, GCS3, GCS4}.

The specific complexifier chosen by Thiemann in \cite{GCS1} is a variant of the squared area operator, defined in terms of the parallel transported flux operator of \Eq{E_i^p}. To each edge $e$ of a spin network there is associated a corresponding surface $S_e$, which intersects $e$ but does not intersect any other edges of the spin network. Using the surface $S_e$, one then defines a parallel transported flux $E_i^{(x_0)}(S_e)$, in which points on $S_e$ are transported to a fixed point $x_0$ on the edge $e$ through a path which goes within $S_e$ (\eg along a straight line in the coordinates chosen on the surface) until the point where $S_e$ intersects $e$, and from there to $x_0$ along $e$. The complexifier associated to the edge is then defined as
\be
{\cal C}_e = \frac{t}{2}(p_{x_0})^2,
\ee
where $(p_{x_0})_i = (8\pi\beta G)^{-1} E_i^{(x_0)}(S_e)$, and $t$ is a parameter, which turns out to control the peakedness properties of the resulting coherent states. Since the operator corresponding to ${\cal C}_e$ acts only on the edge $e$, it suffices to carry out the construction of coherent states within the Hilbert space ${\cal H}_e = L_2(SU(2))$ of a single edge. After the construction is completed, coherent states on a fixed graph can be obtained at the non-gauge invariant level as tensor products of the single-edge coherent states, and gauge invariant coherent states will be given by projections of such tensor products onto the gauge invariant Hilbert space (see section \ref{sec:GICS}).

The standard expansion of the delta function on ${\cal H}_e$ is given by
\be
\delta(g,g_0) = \sum_j d_j\chi^{(j)}(g_0g^{-1}),
\ee
where $\chi^{(j)}(g) \equiv \Tr D^{(j)}(g)$ is the character of the spin-$j$ representation of $SU(2)$. Since the complexifier operator $C_e$ acts on the spin-$j$ subspace of ${\cal H}$ diagonally, with the eigenvalue $(t/2)j(j+1) \equiv (t/2)\lambda_j$, we immediately find the expression
\be\label{psi_h0}
\psi_{h_0}(g) = \sum_j d_je^{-t\lambda_j/2}\chi^{(j)}(h_0g^{-1})
\ee
for coherent states on ${\cal H}_e$. Here $h_0$, which is the complexification of the $SU(2)$-element $g_0$, is an element of $SL(2,\C)$, as can be verified by computing $h_0$ from \Eq{z(q,p)}. 

In order to evaluate
\be\label{hSL2C}
h = \sum_n \frac{i^n}{n!}\bigl\{h_e,\frac{t}{2}(p_{x_0})^2\bigr\}^{(n)},
\ee
we note that the Poisson bracket of the parallel transported flux with the holonomy of the edge is given by
\be
\{h_e,E_i^{(x_0)}(S_e)\} = 8\pi\beta Gh_{t(e)\leftarrow x_0}\tau_ih_{x_0\leftarrow s(e)},
\ee
where $h_{t(e)\leftarrow x_0}$ is the holonomy along the segment of $e$ from $x_0$ to $t(e)$, and $h_{x_0\leftarrow s(e)}$ is defined similarly. In particular, when $x_0$ is the beginning or ending point of $e$, we have
\be
\{h_e,(p_{s(e)})_i\} = h_e\tau_i, \qquad \bigl\{h_e,(p_{t(e)})_i\bigr\} = \tau_ih_e.
\ee
Relabeling the variables as $h_e\to g_0$, $(p_{s(e)})_i \to (p_0)_i$ and $(p_{t(e)})_i \to (p_0')_i$ in order to switch to the notation used in \cite{paper3}, we find from \Eq{hSL2C} the two equivalent expressions
\vspace{-12pt}
\begin{subequations}\label{h0-decomp}
\begin{align}
h_0 &= g_0e^{t\vec p_0\cdot\vec\sigma/2}, \label{h0-left} \\
h_0 &= e^{t\vec p_0^{\,\prime}\cdot\vec\sigma/2}g_0 \label{h0-right}
\end{align}
\end{subequations}
for the object $h_0$. This confirms that $h_0$ is indeed an element of $SL(2,\C)$, since the expressions \eqref{h0-decomp} are the standard decompositions of an $SL(2,\C)$ element into a $SU(2)$ rotation and a boost. The variables of the two decompositions are related to each other by
\be
\vec p_0^{\,\prime}\cdot\vec\sigma = g_0(\vec p_0\cdot\vec\sigma)g_0^{-1},
\ee
reflecting the relation $E^{t(e)}(S) = h_eE^{s(e)}(S)h_e^{-1}$ between the two flux variables $E^{s(e)}(S)$ and $E^{t(e)}(S)$.

An orthonormal basis in ${\cal H}_e = L_2(SU(2))$ is given by the states $\ket{jmn}$, whose wave functions are defined as
\be\label{H_e-basis}
\braket{g}{jmn} = \sqrt{d_j}\overline{\displaystyle \D{j}{m}{n}{g}}
\ee
(where the complex conjugation on the right-hand side is introduced in order to be consistent with the convention chosen in \cite{paper3}). The basis \eqref{H_e-basis} diagonalizes the left- and right-invariant vector fields \eqref{L_iandR_i}, as indicated by the eigenvalue equations
\begin{subequations}
\begin{align}
L_z\ket{jmn} &= n\ket{jmn} \\
R_z\ket{jmn} &= m\ket{jmn} \\
L^2\ket{jmn} &= j(j+1)\ket{jmn}\\
R^2\ket{jmn} &= j(j+1)\ket{jmn}.
\end{align}
\end{subequations}
From \Eq{psi_h0} we see that the expansion of the coherent state $\psi_{h_0}(g) \equiv \braket{g}{g_0,\vec p_0}$ in this basis reads
\be\label{ket_h0}
\ket{g_0,\vec p_0} = \sum_{jmn} \sqrt{d_j}e^{-t\lambda_j/2}\D{j}{m}{n}{g_0e^{t\vec p_0\cdot\vec\sigma/2}}\ket{jmn}.
\ee

For later use, let us note the behaviour of the states $\ket{g_0,\vec p_0}$ under local $SU(2)$ gauge transformations. We recall that under a transformation described by a gauge function $a(x)\in SU(2)$, the holonomy transforms as
\be
h_e \to a\bigl(t(e)\bigr)h_ea^{-1}\bigl(s(e)\bigr).
\ee
Making the replacement $g\to a_tga_s^{-1}$ in the wave function \eqref{psi_h0}, and choosing either one of the parametrizations \eqref{h0-decomp}, we see that the effect of a gauge transformation on a coherent state consists entirely of a simple change in the labels $g_0$ and $\vec p_0$. For example, in the parametrization \eqref{h0-left}, we have
\be\label{SU2*gp}
\ket{g_0,\vec p_0} \to \bket{a_t^{-1}g_0a_s,R^{-1}(a_s)\vec p_0},
\ee
where the $\R^3$ rotation matrix $R\updown{i}{j}(a)$ is defined by $a(p^i\sigma_i)a^{-1} = \bigl(R\updown{i}{j}(a)p^j\bigr)\sigma_i$. In the other parametrization \eqref{h0-right}, we similarly find
\be\label{SU2*gp'}
\ket{g_0,\vec p_0^{\,\prime}} \to \bket{a_t^{-1}g_0a_s,R^{-1}(a_t)\vec p_0^{\,\prime}}.
\ee
The coherent states $\ket{g_0,\vec p_0}$ therefore remain coherent under gauge transformations, simply becoming peaked on a different point of the classical phase space.

\subsection{Resolution of the identity}

An important property of the coherent states $\ket{g_0,\vec p_0}$ is that they resolve the identity on the space ${\cal H}_e = L_2(SU(2))$,
\be\label{H_e-identity}
\Id = \int d\mu(g,p)\,\ket{g,\vec p}\bra{g,\vec p},
\ee
and therefore provide an overcomplete basis on ${\cal H}_e$. In \Eq{H_e-identity}, the integral is taken over the classical phase space, and integration measure has the factorized form \cite{GCS2}
\be
d\mu(g,p) = dg\,d\nu(p),
\ee
where $dg$ is the Haar measure of $SU(2)$, and the factor involving $\vec p$ is given by
\be
d\nu(p) = d^3p\,e^{-t/4}\biggl(\frac{t}{\pi}\biggr)^{3/2}\frac{\sinh tp}{tp}e^{-tp^2},
\ee
where we denote $p = |\vec p|$.

In order to check that the operator \eqref{H_e-identity} really is the unit operator on ${\cal H}_e$, we compute its matrix elements $\bra{jmn}\Id\ket{j'm'n'}$ in the basis \eqref{H_e-basis}. After inserting the expression \eqref{ket_h0}, the integral over $g$ can be calculated immediately using the orthogonality theorem of the Wigner matrices, and we are left with
\be\label{Id-element}
\bra{jmn}\Id\ket{j'm'n'} = \delta_{jj'}\delta^m_{m'}e^{-t\lambda_j}\int d\nu(p)\,\D{j}{n'}{n}{e^{t\vec p\cdot\vec\sigma}}.
\ee
Let us denote the remaining integral on the right-hand side as $I\updown{n'}{n}$. The key to evaluating the integral is to view it as a tensor in ${\cal H}_j\otimes{\cal H}_{j}^*$, and observe that the rotational invariance of the measure $d\nu(p)$ implies that $I\updown{n'}{n}$ is invariant under the action of $SU(2)$:
\be
\D{j}{m'}{n'}{g^{-1}}I\updown{n'}{n}\D{j}{n}{m}{g} = I\updown{m'}{m}.
\ee
Hence $I\updown{n'}{n}$ is an element of the space ${\rm Inv}\,({\cal H}_j\otimes{\cal H}_{j}^*)$, and so must be proportional to $\delta_n^{n'}$, which is the only invariant tensor carrying one upper and one lower index. The coefficient of proportionality in $I\updown{n'}{n} = c(j)\delta_n^{n'}$ can be determined by contracting both sides with $\delta_{n'}^n$. This leads to
\be
c(j) = \frac{1}{d_j}\int d\nu(p)\,\chi^{(j)}(e^{t\vec p\cdot\vec\sigma}) = \frac{1}{d_j}\int d\nu(p)\,\frac{\sinh d_jtp}{\sinh tp} = e^{t\lambda_j},
\ee
where the trace $\chi^{(j)}(e^{t\vec p\cdot\vec\sigma})$ was evaluated in the basis where $D^{(j)}(e^{t\vec p\cdot\vec\sigma})$ is diagonal, with eigenvalues $e^{tjp}$, $e^{t(j-1)p}$, $\dots$, $e^{-tjp}$. Putting everything together, we have shown that
\be
\bra{jmn}\Id\ket{j'm'n'} = \delta_{jj'}\delta^m_{m'}\delta_n^{n'},
\ee
from which we can conclude that \eqref{H_e-identity} is indeed a correct expression for the identity operator on ${\cal H}_e$.

\subsection{Peakedness properties}\label{peakedness}

As we have seen, the construction of the heat kernel coherent states \eqref{psi_h0} guarantees that they are sharply peaked with respect to the operators \eqref{QandP}, where $a$ is now given by a quantization of the classical variable $h = ge^{t\vec p\cdot\vec\sigma/2}$. Therefore it is not immediately obvious (though it is certainly very plausible) that the states are also properly peaked on the holonomy and the flux, which are the operators directly relevant to loop quantum gravity. The peakedness properties of the states \eqref{psi_h0} with respect to the holonomy and the flux were established by means of direct calculations by Thiemann and Winkler \cite{GCS2}. In this section we will briefly summarize their results.

The possible peakedness of the states \eqref{psi_h0} with respect to the holonomy is described by the probability distribution in ''holonomy space'',
\be\label{rho(g)}
\rho_{h_0}(g) = \frac{\bigl|\psi_{h_0}(g)\bigr|^2}{\braket{\psi_{h_0}}{\psi_{h_0}}}
\ee
(where it is necessary to divide by the norm $\braket{\psi_{h_0}}{\psi_{h_0}}$, since the state $\ket{\psi_{h_0}}$ is in general not normalized). If the state $\ket{\psi_{h_0}}$ is properly peaked on the holonomy, the function \eqref{rho(g)} should have a sharp peak concentrated around the point $g=g_0$ (we recall that $h_0 = g_0e^{t\vec p_0\cdot\vec\sigma/2}$). The observation
\be
\psi_{h_0}(g) = \psi_{e^{\vec p_0\cdot\vec\sigma/2}}(g_0^{-1}g)
\ee
allows one to replace the problem of showing that \eqref{rho(g)} is peaked on $g=g_0$ with the equivalent problem of showing that
\be
\rho_{e^{\vec p_0\cdot\vec\sigma/2}}(g)
\ee
is peaked on $g=\Id$ independently of the value of $\vec p_0$. By going through a long and tedious calculation, which we have little hope of even outlining in the space available here\footnote{We merely note that the essential tool in Thiemann and Winkler's calculation is the Poisson summation formula
\[ \sum_{n=-\infty}^\infty f(tn) = \frac{2\pi}{t}\sum_{n=-\infty}^\infty \tilde f\biggl(\frac{2\pi n}{t}\biggr),\]
where $\tilde f$ denotes the Fourier transform of $f$. The importance of the Poisson summation formula to the present problem is that it allows one to convert a sum like \eqref{psi_h0}, which converges extremely slowly in the limit $t\to 0$, into a sum which typically is converging very rapidly when $t\to 0$, and which can be approximated very well by keeping only the leading term.}, Thiemann and Winkler managed to show that this is indeed the case. The width of the peak is characterized by the parameter $t$, with the peak becoming sharp in the limit $t\to 0$.

In order to study the peakedness of the state $\ket{\psi_{h_0}}$ with respect to flux operators, the state should be expanded in the eigenbasis \eqref{H_e-basis} of the left- and right-invariant vector fields as
\be
\ket{\psi_{h_0}} = \sum_{jmn} \psi_{h_0}(jmn)\ket{jmn},
\ee
where the coefficients of the expansion are given by
\be
\psi_{h_0}(jmn) = \sqrt{d_j}e^{-t\lambda_j/2}\D{j}{m}{n}{h_0}.
\ee
The probability of the state $\ket{\psi_{h_0}}$ to have a specific momentum configuration is then described by the (discrete) probability distribution
\be\label{rho(jmn)}
\rho_{h_0}(jmn) = \frac{\bigl|\psi_{h_0}(jmn)|^2}{\braket{\psi_{h_0}}{\psi_{h_0}}}.
\ee
By making use of estimates based on the explicit expression \eqref{WignerD-expl} for the matrix elements $\D{j}{m}{n}{h_0}$, Thiemann and Winkler were able to obtain the result\footnote{Precisely speaking, \Eq{rho(jmn)-est} is valid only when neither $p_{0z}/p_0$ nor $p_{0z}'/p_0$ is too close to 1. However, a very similar estimation holds in the case $p_{0z}/p_0\simeq 1$ or $p_{0z}'/p_0\simeq 1$.}
\be\label{rho(jmn)-est}
\hspace{0.5cm}\rho_{h_0}(jmn) \lesssim \frac{\sqrt t}{4\sqrt\pi p_0}\exp\biggl[-t(j+\half-p_0)^2  - \frac{j}{2}\frac{(m/j - p_{0z}'/p_0)^2}{1-(p_{0z}'/p_0)^2} - \frac{j}{2}\frac{(n/j - p_{0z}/p_0)^2}{1 - (p_{0z}/p_0)^2}\biggr],\hspace{0.5cm}
\ee
where $p_{0z}$ and $p_{0z}'$ refer to the two decompositions \eqref{h0-decomp} of the $SL(2,\C)$ element $h_0$. The peakedness properties of the state $\ket{\psi_{h_0}}$ with respect to flux operators can be read off from \Eq{rho(jmn)-est}. We see that the probability distribution \eqref{rho(jmn)} is peaked on the values
\be
j = p_0-\frac{1}{2}, \qquad m \simeq p_{0z}', \qquad n \simeq p_{0z},
\ee
and the peak becomes sharp in the limit of large $p_0$ -- specifically, when $p_0\gg 1/\sqrt t$ (the value of $t$ being fixed, for example, by the requirement that the coherent state is sufficiently well peaked on the holonomy).

Thiemann and Winkler also showed that the overlap function
\be
\Delta(h_1,h_2) = \frac{\bigl|\braket{\psi_{h_1}}{\psi_{h_2}}\bigr|^2}{\braket{\psi_{h_1}}{\psi_{h_1}}\braket{\psi_{h_2}}{\psi_{h_2}}}
\ee
is peaked on $h_1=h_2$, with the peak falling off exponentially fast around the maximum, at least in the limit of small $t$. Figuratively speaking, if we view the coherent state $\ket{\psi_{h_0}} \equiv \ket{g_0,\vec p_0}$ as a state-vector valued function on the classical phase space, then the peakedness of the overlap function means that $\ket{g_0,\vec p_0}$ differs significantly from zero only within a small neighborhood around the point $(g_0,\vec p_0)$.

\subsection{Gauge invariant coherent states}\label{sec:GICS}

So far our discussion of coherent states has been restricted to a single spin network edge. However, the generalization of the construction to coherent states on a fixed graph is rather trivial, at least at the level of non-gauge invariant states. Coherent states based on a fixed graph $\Gamma$ are given simply by tensor products of the single-edge coherent states \eqref{psi_h0} over the edges of the graph:
\be\label{phi_h}
\phi^\Gamma_{\{h_e\}}(g_{e_1},\dots,g_{e_N}) = \prod_{e\in\Gamma} \sum_{j_e} d_{j_e} e^{-t_e\lambda_{j_e}/2}\chi^{(j_e)}(h_eg_e^{-1}).
\ee
Gauge invariant coherent states can be constructed by group averaging the tensor product states \eqref{phi_h} with respect to gauge transformations at each node of the graph:
\be\label{Phi_h-int-first}
\Phi^\Gamma_{\{h_e\}}(g_{e_1},\dots,g_{e_N}) = \int da_1\cdots da_M\,\phi^\Gamma_{\{h_e\}}\bigl(a_{t(e_1)}g_{e_1}a_{s(e_1)}^{-1},\dots,a_{t(e_N)}g_{e_N}a_{s(e_N)}^{-1}\bigr).
\ee
Inserting the expression \eqref{phi_h}, we obtain
\begin{align}
\Phi^\Gamma_{\{h_e\}}(g_{e_1},\dots,g_{e_N})&= \prod_{e\in\Gamma}\biggl(\sum_{j_e} d_{j_e} e^{-t_e\lambda_{j_e}/2}\D{j_e}{m_e}{n_e}{h_e}\D{j_e}{m_e'}{n_e'}{g_e^{-1}}\biggr) \notag \\
&\quad\times \int da_1\cdots da_M\,\prod_{e\in\Gamma} \biggl(\Db{j_e}{n_e}{m_e'}{a_{s(e)}}\Db{j_e}{n_e'}{m_e}{a_{t(e)}^{-1}}\biggr).\label{Phi_h-int}
\end{align}
At each node of the graph we now have an integral of the form
\be
\int da_v\,\biggl(\prod_{\text{$e$ out}}\D{j_e}{n_e}{m_e'}{a_v}\biggr)\biggl(\prod_{\text{$e$ in}}\D{j_e}{n_e'}{m_e}{a_v^{-1}}\biggr),
\ee
which, viewed as an $SU(2)$ tensor, is essentially a normalized projection operator onto the intertwiner space of the node. Therefore the integral can be expressed as
\be
\sum_\iota \overline{\displaystyle \iota\updown{n_1\cdots n_O}{m_1\cdots m_I}}\iota\updown{n_1'\cdots n_I'}{m_1'\cdots m_O'},
\ee
where the sum runs over any orthonormal basis of the intertwiner space, and the indices $m_1,\dots,m_I$ refer to edges coming in to the node, while $n_1,\dots,n_O$ refer to edges going out of the node. Going back to \Eq{Phi_h-int}, we see that the gauge invariant coherent states can be written in the form
\be\label{Phi_h}
\Phi^\Gamma_{\{h_e\}}(g_e) = \sum_{\{j_e\},\{\iota_v\}} \biggl(\prod_e e^{-t_e\lambda_{j_e/2}} \biggr){\Psi_{\Gamma,\{j_e\},\{\iota_v\}}(h_e)}\overline{\Psi_{\Gamma,\{j_e\},\{\iota_v\}}(g_e)},
\ee
where $\Psi_{\Gamma,\{j_e\},\{\iota_v\}}(g_e)$ are the standard spin network states on the graph $\Gamma$.

It is not immediately obvious how the group averaging operation affects the peakedness properties of the coherent states. In their paper on peakedness proofs \cite{GCS2}, Thiemann and Winkler argue that peakedness of the gauge invariant coherent states with respect to holonomies, and of the overlap function of gauge invariant coherent states, follow from the corresponding properties of the non-gauge invariant coherent states. The overlap between gauge invariant coherent states has also been studied by Thiemann and Bahr \cite{BahrThiemann}.

The situation with respect to flux operators becomes the most transparent through a result obtained by Bianchi, Magliaro and Perini \cite{BMP}, which establishes a relation between the gauge-invariant coherent states \eqref{Phi_h} and the so-called Livine--Speziale coherent intertwiners \cite{LivineSpeziale}. The derivation of this result is based on the alternative decomposition
\be\label{h=nzn}
h = g(\vec n)e^{-iz\sigma_z/2}g^{-1}(\vec n')
\ee
of the $SL(2,\C)$ element $h$; here $z = \xi + it\eta$ is an arbitrary complex number, and $g(\vec n)$ is an $SU(2)$ element of the specific form $e^{-i\vec m(\vec n)\cdot\vec\sigma/2}$, where the rotation vector $\vec m(\vec n)$ lies in the $xy$-plane. This decomposition is particularly suited for discussing the limit of large $p_0$, which turns out to correspond to ${\rm Im}\, z\gg 1$ in the parametrization of \Eq{h=nzn}. When ${\rm Im}\, z$ is large, the matrix elements of $D^{(j)}(e^{-iz\sigma_z/2})$ are dominated by the largest, $jj$-matrix element:
\be\label{BMP-appr}
\D{j}{m}{n}{e^{-iz\sigma_z/2}} = \delta^m_n e^{-imz} \simeq \delta_{mj}\delta_{nj}e^{(t\eta - i\xi)j}.
\ee
Equivalently, the matrix $D^{(j)}(e^{-iz\sigma_z/2})$ is approximated by
\be
D^{(j)}(e^{-iz\sigma_z/2}) \simeq e^{(t\eta-i\xi)j}\ket{jj}\bra{jj},
\ee
with the correction being exponentially small relative to the leading term.

Under the approximation \eqref{BMP-appr}, the trace in \Eq{phi_h} becomes
\be\label{chi-approx}
\chi^{(j)}(hg^{-1}) \simeq e^{(t\eta-i\xi)j}\chi^{(j)}\bigl(\ket{j\vec n}\bra{j\vec n'}g^{-1}\bigr),
\ee
with $\ket{j\vec n} = g(\vec n)\ket{jj}$ the standard coherent states of angular momentum \cite{Radcliffe, Perelomov1972}. When \Eq{chi-approx} is used in \Eqs{phi_h} and \eqref{Phi_h-int-first}, the group integration at each node produces the coherent intertwiner
\be
\bket{\iota_v(\{\vec n_{e_{\rm in}}\},\{\vec n_{e_{\rm out}}'\})} = \int da_v\,\biggl(\prod_{\text{$e$ in}}a_v^{-1}\ket{j_e\vec n_e}\biggr)\biggl(\prod_{\text{$e$ out}} \bra{j_e\vec n_e'}a_v\biggr),
\ee
where the ket on the left-hand side should be understood as an element of the intertwiner space ${\rm Inv}\,\bigl[\bigl(\otimes_{\text{$e$ in}} {\cal H}_{j_e}\bigr)\bigl(\otimes_{\text{$e$ out}} {\cal H}_{j_e}^*\bigr)\bigr]$. In this way one finds that the gauge invariant coherent states can be expressed as superpositions of spin network states with coherent intertwiners at the nodes:
\be\label{Phi_nnz}
\Phi^\Gamma_{\{\vec n_e,\vec n_e',z_e\}}(g_e) \simeq \sum_{j_e}\biggl(\prod_{e\in\Gamma} \sqrt{d_{j_e}}e^{-t_e(j_e-j_e^{(0)})^2/2}e^{-i\xi_ej_e}\biggr)\overline{\Psi_{\Gamma,\{j_e\},\{\iota_v(\vec n_v,\vec n_v')\}}(g_e)},
\ee
where $j_e^{(0)} = \eta_e - \half$. In particular, this result seems to establish peakedness of the gauge invariant coherent states with respect to flux operators, since the Livine--Speziale intertwiners are well known to have good semiclassical properties with respect to the fluxes.

\newpage

\section{Coherent state operators}\label{ch:CSoperators}
\newcommand{\pp}{\vec p^{\,\prime}}

Our overview of the kinematical framework of loop quantum gravity will be concluded in this chapter, in which we show how the complexifier coherent states introduced in the previous chapter can be used to set up an alternative, non-canonical prescription for systematically defining kinematical operators (such as holonomy, flux, area and volume) in loop quantum gravity. The results that are reviewed in this chapter were obtained by the author in collaboration with Emanuele Alesci, Andrea Dapor, Jerzy Lewandowski and Jan Sikorski, and have been published in the article \cite{paper3}.

The work presented in this chapter is based on the ideas of so-called ''coherent state quantization'' (see \eg \cite{Gazeau-book} for an introduction), and was inspired by the earlier work of Gazeau and collaborators, who studied the construction of operators from coherent states in the context of quantum mechanics \cite{Gazeau2} and quantum cosmology \cite{Gazeau3, Gazeau4}. However, similar ideas have been discussed in the literature already much earlier, particularly by Berezin \cite{Berezin} and Klauder \cite{Klauder, KlauderSkagerstam}.

It should be pointed out that the usual terminology of ''coherent state quantization'' is somewhat of a misnomer, since it might give the impression that one is performing an entirely different quantization of the classical theory. In reality, coherent state quantization simply provides a mechanism by which coherent states labeled by points on a classical phase space can be used to promote functions on the classical phase space into operators on the Hilbert space spanned by the coherent states. In other words, the Hilbert space of the quantum theory is already given, and is the same space on which one would perform a canonical quantization of the classical theory; one merely has a different, non-canonical way of associating quantum operators to classical phase space functions. For this reason, we prefer the expression ''coherent state operators'' over the potentially misleading ''coherent state quantization''.

Coherent state operators have several definite advantages in comparison to their canonical counterparts. Most notably, once a family of coherent states is chosen, the procedure associates a unique and unambiguous operator to every classical function; there are no ordering ambiguities which are often encountered in canonical quantization. Moreover, certain properties of the classical function are automatically passed to the corresponding coherent state operator, in contrast to the situation with canonical quantization. For example, the operator corresponding to a classical function which is real-valued, or positive, is guaranteed to be symmetric, or positive-definite. A feature of the construction which one might consider a disadvantage is that the commutator of coherent state operators generally does not exactly reproduce the Poisson bracket of the corresponding classical functions, even if one considers just the elementary ''position'' and ''momentum'' operators.

In what follows, we will start by presenting the general recipe for constructing coherent state operators in loop quantum gravity, using the complexifier coherent states of the previous chapter. The nature of the coherent states implies that the operators obtained from them will be restricted to the Hilbert space of a single, fixed spin network graph. Afterwards we will establish some general properties of the resulting operators, and present a number of concrete examples of the construction. We will consider in detail both the elementary holonomy and flux operators, and the geometric operators of area, angle and volume. The discussion will then be concluded with some closing remarks.

\newpage

\subsection{Construction of coherent state operators}\label{CS-construction}

The construction by which coherent states can be used to associate quantum operators to classical phase space functions is based on the resolution of identity in terms of the coherent states, which involves an integration over the appropriate classical phase space. Restricting our attention first to a single spin network edge, we have the resolution of identity
\be\label{CS-Id}
\Id = \int d\mu(g,p)\,\ket{g,\vec p}\bra{g,\vec p}
\ee
on the Hilbert space of the edge. The object on the right-hand side can be viewed as an integral over the classical phase space associated to the edge\footnote{A detailed discussion of the classical phase space of loop quantum gravity can be found in \cite{Geiller_et_al}.}, the variables $g$ and $\vec p$ being interpreted respectively as the holonomy of $A$ along the edge, and the flux of $E$ through a surface dual to the edge. In fact, the behaviour of the variable $\vec p$ under $SU(2)$ gauge transformations, given by \Eq{SU2*gp}, suggests that $\vec p$ should precisely speaking be interpreted as the parallel transported flux variable, where the parallel transport is taken to the beginning point of the edge.

Given now any function $f(g,\vec p)$ on the corresponding classical phase space, we can define a coherent state operator representing the function by inserting $f(g,\vec p)$ into the integral in \Eq{CS-Id}. In this way we obtain the operator
\be\label{O_f}
{\cal O}_f = \int d\mu(g,p)\,f(g,\vec p)\ket{g,\vec p}\bra{g,\vec p}.
\ee
This construction associates a unique operator on the single-edge Hilbert space to every function on the classical phase space of the edge.

It is straightforward to generalize the prescription \eqref{O_f} to obtain operators on the Hilbert space of a fixed graph, which has the structure of a tensor product of Hilbert spaces associated to each edge of the graph. In particular, the identity operator on the Hilbert space of a graph is resolved by the tensor product states $\ket{\{g_e\},\{\vec p_e\}} \equiv \otimes_e \ket{g_e,\vec p_e}$ (with respect to the measure $\prod_e d\mu(g_e,\vec p_e)$). To any function $f(\{g_e\},\{\vec p_e\})$ on the classical phase space of the graph, we therefore associate the operator
\be\label{O_f N}
{\cal O}_f = \int d\mu(g_1,p_1)\cdots d\mu(g_N,p_N)\,f\bigl(\{g_e\},\{\vec p_e\}\bigr)\,\bket{\{g_e\},\{\vec p_e\}}\bbra{\{g_e\},\{\vec p_e\}}.
\ee
In practice a slightly modified version of the definition \eqref{O_f N} must be used in order to obtain an operator which accounts correctly for the orientation of the graph, as we will now explain.

The classical interpretation of the variable $\vec p$ as a parallel transported flux indicates that $\vec p$ belongs to the beginning point of the corresponding edge. Similarly, the variable $\pp$ corresponding to the decomposition $h = e^{t\pp\cdot\vec\sigma/2}g$ is interpreted as the flux parallel transported to the endpoint of the edge, and therefore belongs to the endpoint. This interpretation is respected by the operator \eqref{O_f N}, in the sense that whenever one takes a $\vec p_e$ in the classical function $f(\{g_e\},\{\vec p_e\})$ and the corresponding coherent state $\ket{g_e,\vec p_e}$, the resulting operator will act on the beginning point of the edge $e$, whereas if one takes a $\pp_e$ for the same edge, one gets an operator which acts on the endpoint of the edge. (These claims are confirmed by the calculations made later in this chapter; see in particular \Eqs{CS-p_i} and \eqref{CS-p'_i}.) In particular this means that if one wants to construct an operator (such as the volume operator) which acts on a single node of a spin network state, one should put in \Eq{O_f N} a $\vec p_e$ for every edge oriented away from the node, and a $\pp_e$ for every edge oriented into the node.

At this point one might be inclined to ask whether it would not be better to take the gauge-invariant coherent states of \Eq{Phi_h}, rather than the simple tensor product states used in \Eq{O_f N}, as the starting point for constructing coherent state operators, considering that practical calculations in loop quantum gravity are usually made at the gauge-invariant level, and not in the kinematical Hilbert space. However, it is straightforward to see that if one uses the operator to act exclusively on gauge-invariant states, then there will be no difference between the operator \eqref{O_f N} and the operator  
\be\label{O_f g}
{\cal O}_f^{[g]} = \int d\mu(g_1,p_1)\cdots d\mu(g_N,p_N)\,f\bigl(\{g_e\},\{\vec p_e\}\bigr)\,\bket{\Phi^\Gamma_{\{g_e,\vec p_e\}}}\bbra{\Phi^\Gamma_{\{g_e,\vec p_e\}}}
\ee
constructed from the gauge invariant coherent states \eqref{Phi_h}. Since the states $\ket{\Phi^\Gamma_{\{g_e,\vec p_e\}}}$ are obtained simply by applying the gauge invariant projector $P_G$ to the tensor product states $\ket{\{g_e\},\{\vec p_e\}}$, it follows immediately that if $\ket\Phi$ and $\ket\Psi$ are any two gauge invariant states, the operators \eqref{O_f N} and \eqref{O_f g} have the same matrix elements between these states:
\be
\bra\Phi{\cal O}_f^{[g]}\ket\Psi =  \bra\Phi{\cal O}_f\ket\Psi.
\ee
This being the case, it is actually preferable to use the non-gauge invariant tensor product states chosen in \Eq{O_f N}, since they will be technically much easier to handle.

\subsection{General properties}\label{CS-properties}

Before moving on to consider concrete examples of coherent state operators in loop quantum gravity, we will establish some general properties of operators constructed in this way. Let us first consider the behaviour of the operator \eqref{O_f N} under gauge transformations. In particular, the result of this calculation will show that it is possible to obtain gauge invariant operators from \Eq{O_f N}, even though the coherent states used to construct the operator are not gauge invariant.

In the general case, where we may have associated either a $\vec p$ or $\pp$ to each edge of the graph, it is convenient to separate the two types of momentum variables, writing the operator \eqref{O_f N} as
\be
{\cal O}_f = \int d\mu(g_1,p_1)\cdots d\mu(g_N,p_N)\,f\bigl(\{g_e\},\{\vec p_e\},\{\pp_e\}\bigr)\,\bket{\{g_e\},\{\vec p_e\},\{\pp_e\}}\bbra{\{g_e\},\{\vec p_e\},\{\pp_e\}}.
\ee
Letting $U(a)$ denote the action of a local $SU(2)$ gauge transformation defined by the function $a(x)\in SU(2)$, and recalling the action of a gauge transformation on the coherent states from \Eqs{SU2*gp} and \eqref{SU2*gp'}, we find
\begin{align}
&U^{-1}(a){\cal O}_fU(a) = \int d\mu(g_1,p_1)\cdots d\mu(g_N,p_N) \notag \\
&\times f\bigl(\{a_{t(e)}^{-1}g_ea_{s(e)}\},\{R^{-1}(a_{s(e)})\vec p_e\},\{R^{-1}(a_{t(e)})\pp_e\}\bigr)\,\bket{\{g_e\},\{\vec p_e\},\{\pp_e\}}\bbra{\{g_e\},\{\vec p_e\},\{\pp_e\}},
\end{align}
where the $SU(2)$ invariance of the Haar measure and the rotational invariance of the measure $d\nu(p)$ was used to move the action of the gauge transformation from the coherent states to the function $f$. Therefore we see that it is indeed possible for the operator ${\cal O}_f$ to be gauge invariant, the condition for its gauge invariance being simply that the function $f(\{g_e\},\{\vec p_e\},\{\pp_e\})$ be invariant under a gauge transformation of its arguments.

Another general property of the operator \eqref{O_f N} concerns the case where the function $f(\{g_e\},\{\vec p_e\})$ is strictly positive everywhere, except possibly in a set of measure zero with respect to the measure $d\mu(g_1,p_1)\cdots d\mu(g_N,p_N)$. Then all the eigenvalues of ${\cal O}_f$ will be strictly positive, at least if the corresponding eigenstates are proper, non-distributional states.

To prove this statement, let $\ket\lambda$ be a proper, normalizable eigenstate of ${\cal O}_f$. Then the eigenvalue can be written as
\be\label{O_f lambda}
\lambda = \bra\lambda{\cal O}_f\ket\lambda = \int d\mu(g_1,p_1)\cdots d\mu(g_N,p_N)\,f\bigl(\{g_e\},\{\vec p_e\}\bigr)\,\bigl|\bigl\langle \{g_e\},\{\vec p_e\}\big|\lambda\bigr\rangle\bigr|^2.
\ee
We see immediately that the eigenvalue is non-negative. The only way in which $\lambda$ can be equal to zero is that $\braket{\{g_e\},\{\vec p_e\}}{\lambda}=0$ everywhere except in a set of measure zero, since we have assumed that the set where $f(\{g_e\},\{\vec p_e\})\leq 0$ is of measure zero. However, the condition for the eigenstate to be normalized,
\be
\braket\lambda\lambda = \int d\mu(g_1,p_1)\cdots d\mu(g_N,p_N)\,\bigl|\bigl\langle \{g_e\},\{\vec p_e\}\big|\lambda\bigr\rangle\bigr|^2 = 1
\ee
together with the assumption that $\ket\lambda$ is a non-distributional state, so that the projections $\braket{\{g_e\},\{\vec p_e\}}{\lambda}$ have finite values, is sufficient to ensure that $\braket{\{g_e\},\{\vec p_e\}}{\lambda}\neq 0$ in a set of positive measure. It follows that the integral on the right-hand side of \Eq{O_f lambda} is strictly positive, and therefore $\lambda>0$.

\subsection{Holonomy operator}

We will now study in detail the coherent state operators corresponding to the elementary holonomy and flux operators of loop quantum gravity. Starting with the holonomy, we take the function $f(g,\vec p)$ in \Eq{O_f} to be the Wigner matrix $\D{j}{m}{n}{g}$, thereby obtaining the operator
\be
{\cal O}_{\D{j}{m}{n}{g}} = \int d\mu(g,p)\,\D{j}{m}{n}{g}\ket{g,\vec p}\bra{g,\vec p}.
\ee
In order to understand the action of this operator, we compute its matrix elements between two states $\ket{jmn}$ of the basis \eqref{H_e-basis}. Using \Eq{ket_h0} for the components of the coherent states with respect to the basis \eqref{H_e-basis}, we get
\begin{align}
\bra{j_1m_1n_1}{\cal O}_{\D{j}{m}{n}{g}}\ket{j_2m_2n_2} &= \sqrt{d_{j_1}d_{j_2}}e^{-t(\lambda_{j_1}+\lambda_{j_2})/2} \notag \\
&\times\displaystyle{\int dg\,\D{j_1}{m_1}{\mu_1}{g}\D{j}{m}{n}{g}\overline{\displaystyle \D{j_2}{m_2}{\mu_2}{g}}} \notag \\
&\times\displaystyle{\int d\nu(p)\,\D{j_1}{\mu_1}{n_1}{e^{t\vec p\cdot\vec\sigma/2}}\overline{\displaystyle \D{j_2}{\mu_2}{n_2}{e^{t\vec p\cdot\vec\sigma/2}}}}.\label{CSholonomy-1}
\end{align}
Here the integral over $g$ gives
\be\label{CSholonomy-g-int}
\int dg\,\D{j_1}{m_1}{\mu_1}{g}\D{j}{m}{n}{g}\overline{\displaystyle \D{j_2}{m_2}{\mu_2}{g}} = \frac{1}{d_{j_2}}\CGi{j_1j}{j_2}{m_1m}{m_2}\CG{j_1j}{j_2}{\mu_1n}{\mu_2}.
\ee
To evaluate the integral over $\vec p$, we start by applying the Clebsch--Gordan series
\be\label{SL2C-series}
\D{j_1}{m_1}{n_1}{h}\D{j_2}{m_2}{n_2}{h} = \sum_{jmn} \CGi{j_1j_2}{j}{m_1m_2}{m}\CG{j_1j_2}{j}{n_1n_2}{n}\D{j}{m}{n}{h},
\ee
which is valid also in the case that $h$ is an element of $SL(2,\C)$, even though it is better known as a relation between $SU(2)$ representation matrices.\footnote{To prove this statement, we recall that a general element of $SL(2,\C)$ can be written in the form $h = ge^{-i\xi\sigma_z}g'$, where $g$ and $g'\in SU(2)$, and $e^{-i\xi\sigma_z}$ is a diagonal matrix. Using this decomposition, we have
\vspace{-6pt}
\begin{align*}
\D{j_1}{m_1}{n_1}{h}\D{j_2}{m_2}{n_2}{h} &= \D{j_1}{m_1}{\mu_1}{g}\D{j_1}{\mu_1}{\nu_1}{e^{-i\xi\sigma_z}}\D{j_1}{\nu_1}{n_1}{g'} \\
&\times \D{j_2}{m_2}{\mu_2}{g}\D{j_2}{\mu_2}{\nu_2}{e^{-i\xi\sigma_z}}\D{j_2}{\nu_2}{n_2}{g'}.
\end{align*}
Here we can use the Clebsch--Gordan series of $SU(2)$ to couple the representation matrices of $g$ and those of $g'$. Inserting also $\D{j}{\mu}{\nu}{e^{-i\xi\sigma_z}} = \delta^\mu_\nu e^{-i\mu\xi}$, we obtain
\begin{align*}
\sum_{\mu_1\mu_2}e^{-i(\mu_1+\mu_2)\xi}\sum_{jmn}\CGi{j_1j_2}{j}{m_1m_2}{m}&\CG{j_1j_2}{j}{\mu_1\mu_2}{n}\D{j}{m}{n}{g} \\ 
&\times\sum_{j'm'n'}\CGi{j_1j_2}{j'}{\mu_1\mu_2}{m'}\CG{j_1j_2}{j'}{n_1n_2}{n'}\D{j'}{m'}{n'}{g'}.
\end{align*}
Now the condition $\mu_1+\mu_2=m'$ on one of the Clebsch--Gordan coefficients allows us to replace $e^{-i(\mu_1+\mu_2)\xi}$ with $e^{-im'\xi}$, after which the sum over $\mu_1$ and $\mu_2$ can be performed using the orthogonality properties of Clebsch--Gordan coefficients. This leads to
\[
\D{j_1}{m_1}{n_1}{h}\D{j_2}{m_2}{n_2}{h} = \sum_{jmnn'} \CGi{j_1j_2}{j}{m_1m_2}{m}\CG{j_1j_2}{j}{n_1n_2}{n'}\D{j}{m}{n}{g}e^{-in\xi}\D{j}{n}{n'}{g'},
\]
which completes the proof of \Eq{SL2C-series}, since $\sum_n \D{j}{m}{n}{g}e^{-in\xi}\D{j}{n}{n'}{g'} = \D{j}{m}{n'}{h}$.} Noting that the matrix $e^{t\vec p\cdot\vec\sigma/2}$ is Hermitian, and using \Eq{SL2C-series}, we can write the integrand in \Eq{CSholonomy-1} as
\be\label{CSholonomy-p-integrand}
\D{j_1}{\mu_1}{n_1}{e^{t\vec p\cdot\vec\sigma/2}}\overline{\displaystyle \D{j_2}{\mu_2}{n_2}{e^{t\vec p\cdot\vec\sigma/2}}} = \sum_{k\mu\nu} \CGi{j_1j_2}{k}{\mu_1n_2}{\mu}\CG{j_1j_2}{k}{n_1\mu_2}{\nu}\D{k}{\mu}{\nu}{e^{t\vec p\cdot\vec\sigma/2}}.
\ee
It then remains to consider the integral
\be\label{CSholonomy-p-int}
\int d\nu(p)\,\D{k}{\mu}{\nu}{e^{t\vec p\cdot\vec\sigma/2}}.
\ee
This integral is of the same type as the integral in \Eq{Id-element}, and so we can repeat the arguments given below \Eq{Id-element} to show that it is proportional to $\delta^\mu_\nu$, with the coefficient of proportionality given by $1/d_k$ times the trace of the integral. The integral \eqref{CSholonomy-p-int} is therefore equal to
\be
\delta^\mu_\nu\frac{1}{d_k}\int d\nu(p)\,\chi^{(k)}(e^{t\vec p\cdot\vec\sigma/2}) = \delta^\mu_\nu\frac{1}{d_k}e^{t(\lambda_k/4-1/8)}\biggl(d_k\cosh\frac{d_kt}{8} + \sinh\frac{d_kt}{8}\biggr).
\ee
When \Eqs{CSholonomy-g-int} and \eqref{CSholonomy-p-integrand} are now inserted back into \Eq{CSholonomy-1}, there appears a partial contraction of three Clebsch--Gordan coefficients. Using the graphical representation of the Clebsch--Gordan coefficient from \Eq{clebsch g}, a simple graphical calculation shows that this contraction is equal to
\be
\CG{j_1j}{j_2}{\mu_1n}{\mu_2}\CGi{j_1j_2}{k}{\mu_1n_2}{\mu}\CG{j_1j_2}{k}{n_1\mu_2}{\mu} = d_k(-1)^{2k}\sixj{j_1}{j_2}{j}{j_1}{j_2}{k}\CG{j_1j}{j_2}{n_1n}{n_2}.
\ee
Hence we have shown that the matrix elements of the coherent state holonomy operator are given by
\be\label{CS-Dmn}
\bra{j_1m_1n_1}{\cal O}_{\D{j}{m}{n}{g}}\ket{j_2m_2n_2} = H_t(j_1,j_2,j)\sqrt{\frac{d_{j_1}}{d_{j_2}}}\CGi{j_1j}{j_2}{m_1m}{m_2}\CG{j_1j}{j_2}{n_1n}{n_2},
\ee
where
\be\label{H_t}
H_t(j_1,j_2,j) = \sum_k (-1)^{2k}e^{t(\lambda_k/4-\lambda_{j_1}/2-\lambda_{j_2}/2-1/8)}\biggl(d_k\cosh\frac{d_kt}{8} + \sinh\frac{d_kt}{8}\biggr)\sixj{j_1}{j_2}{j}{j_1}{j_2}{k}.
\ee
The result of \Eq{CS-Dmn} may be compared against the matrix elements of the canonical holonomy operator, which read
\begin{align}
\bra{j_1m_1n_1}{D^{(j)m}}_{n}\ket{j_2m_2n_2} &= \sqrt{d_{j_1}d_{j_2}}\int dg\,\D{j_1}{m_1}{n_1}{g}\D{j}{m}{n}{g}\overline{\displaystyle \D{j_2}{m_2}{n_2}{g}} \notag \\
&=\sqrt{\frac{d_{j_1}}{d_{j_2}}}\CGi{j_1j}{j_2}{m_1m}{m_2}\CG{j_1j}{j_2}{n_1n}{n_2}.
\end{align}
We see that the matrix elements of the coherent state operator are related to those of the canonical operator by the factor \eqref{H_t}:
\be
\bra{j_1m_1n_1}{\cal O}_{\D{j}{m}{n}{g}}\ket{j_2m_2n_2} = H_t(j_1,j_2,j)\bra{j_1m_1n_1}{D^{(j)m}}_{n}\ket{j_2m_2n_2}.
\ee
In the limit $t\to 0$ -- where, as we recall from section \ref{peakedness}, the coherent state $\psi_{(g_0,\vec p_0)}(g)$ becomes sharply peaked on $g=g_0$ -- the function $H_t(j_1,j_2,j)$ becomes
\be
H_{t\to 0}(j_1,j_2,j) = \sum_k d_k(-1)^{2k}\sixj{j_1}{j_2}{j}{j_1}{j_2}{k} = 1.
\ee
This shows that the coherent state holonomy operator reduces to the canonical holonomy operator in the limit in which the states used to construct the coherent state operator are sharply peaked with respect to the holonomy.

\subsection{Left- and right-invariant vector fields}

The variable $\vec p$, corresponding to the decomposition $h = ge^{t\vec p\cdot\vec\sigma/2}$, is invariant under left multiplication by $SU(2)$, and is therefore a natural candidate for the classical function from which the coherent state operator corresponding to the left-invariant vector field could be obtained. Let us therefore consider the operator
\be\label{O_pi}
{\cal O}_{p_i} = \frac{1}{2}\int d\mu(g,p)\,p_i\ket{g,\vec p}\bra{g,\vec p}.
\ee
As with the holonomy operator, we proceed to compute the matrix elements of this operator in the basis $\ket{jmn}$. We have
\begin{align}
\bra{j_1m_1n_1}{\cal O}_{p_i}\ket{j_2m_2n_2} &= \frac{1}{2}\sqrt{d_{j_1}d_{j_2}}e^{-t(\lambda_{j_1}+\lambda_{j_2})/2}\displaystyle{\int dg\,\D{j_1}{m_1}{\mu_1}{g}\overline{\displaystyle \D{j_2}{m_2}{\mu_2}{g}}} \notag \\
&\times\displaystyle{\int d\nu(p)\,p_i\D{j_1}{\mu_1}{n_1}{e^{t\vec p\cdot\vec\sigma/2}}\overline{\displaystyle \D{j_2}{\mu_2}{n_2}{e^{t\vec p\cdot\vec\sigma/2}}}},
\end{align}
where the integral over the group immediately gives $(1/d_{j_1})\delta_{j_1j_2}\delta^{m_1}_{m_2}\delta^{\mu_2}_{\mu_1}$. We are then left with
\be\label{O_pi-el}
\bra{j_1m_1n_1}{\cal O}_{p_i}\ket{j_2m_2n_2} = \frac{1}{2}e^{-t\lambda_{j_1}}\delta_{j_1j_2}\delta^{m_1}_{m_2}\int d\nu(p)\,p_i\D{j_1}{n_2}{n_1}{e^{t\vec p\cdot\vec\sigma}}.
\ee
In order to compute this integral, it is convenient to express the variable $p_i$ in terms of objects compatible with $SU(2)$ recoupling theory by writing
\be
p_i = \frac{p}{2\sinh tp}\Tr\bigl(\sigma_ie^{t\vec p\cdot\vec\sigma}\bigr).
\ee
In this way we obtain
\begin{align}
&\int d\nu(p)\,p_i\D{j_1}{n_2}{n_1}{e^{t\vec p\cdot\vec\sigma}} = (\sigma_i){}\updown{A}{B}\int d\nu(p)\,\frac{p}{2\sinh tp}\D{1/2}{B}{A}{e^{t\vec p\cdot\vec\sigma}}\D{j_1}{n_2}{n_1}{e^{t\vec p\cdot\vec\sigma}} \notag \\
&= \frac{1}{2}(\sigma_i){}\updown{A}{B}\sum_{jmn}\CGi{j_1\half}{j}{n_2B}{m}\CG{j_1\half}{j}{n_1A}{n}\int d\nu(p)\,\frac{p}{\sinh tp}\D{j}{m}{n}{e^{t\vec p\cdot\vec\sigma}},\label{CSflux-1}
\end{align}
where \Eq{SL2C-series} was used to couple the $SL(2,\C)$ representation matrices. By the same reasoning as before, the remaining integral is equal to
\begin{align}
\int d\nu(p)\,\frac{p}{\sinh tp}\D{j}{m}{n}{e^{t\vec p\cdot\vec\sigma}} &= \delta^m_n\frac{1}{d_j}\int d\nu(p)\,\frac{p}{\sinh tp}\chi^{(j)}(e^{t\vec p\cdot\vec\sigma}) \notag \\
&= \delta^m_n\frac{1}{d_j}\frac{2}{t}e^{-t/4}\sum_{s=-j}^j\bigl(1+2s^2t\bigr)e^{s^2t}.
\end{align}
In \Eq{CSflux-1} we then have the contraction
\be
(\sigma_i){}\updown{A}{B}\CGi{j_1\half}{j}{n_2B}{m}\CG{j_1\half}{j}{n_1A}{m} = \sqrt{\frac{3}{2}}\frac{d_j}{\sqrt{d_{j_1}\lambda_{j_1}}}(-1)^{j+j_1+1/2}\sixj{\half}{\half}{1}{j_1}{j_1}{j}\Sigm{j_1}{i}{n_2}{n_1}
\ee
which is again convenient to evaluate graphically, recalling the graphical representation \eqref{tau=C g} for the $SU(2)$ generators. At this point we have found
\begin{align}
\bra{j_1m_1n_1}{\cal O}_{p_i}\ket{j_2m_2n_2} &= \delta_{j_1j_2}\delta^{m_1}_{m_2}\Sigm{j_1}{i}{n_2}{n_1} \notag \\
&\times\sqrt{\frac{3}{2}}\frac{e^{-t(\lambda_{j_1}+1/4)}}{2t\sqrt{d_{j_1}\lambda_{j_1}}}\sum_j (-1)^{j+j_1+1/2}\sixj{\half}{\half}{1}{j_1}{j_1}{j}S_t(j),\label{CSflux-2}
\end{align}
where we introduced the abbreviation
\be
S_t(j) = \sum_{s=-j}^j\bigl(1+2s^2t\bigr)e^{s^2t}.
\ee
This result can be simplified further by performing the sum over $j$, which runs over the two values $j=j_1\pm\half$. Inserting the explicit expressions for the 6$j$-symbols involved in the sum,
\be
\sixj{\half}{\half}{1}{j}{j}{j-\half} = \frac{(-1)^{2j+1}}{\sqrt{6d_j}}\sqrt{\frac{j+1}{j}}, \qquad \sixj{\half}{\half}{1}{j}{j}{j+\half} = \frac{(-1)^{2j+1}}{\sqrt{6d_j}}\sqrt{\frac{j}{j+1}},
\ee
we conclude that the action of the coherent state left-invariant vector field can be expressed in the form
\be\label{CS-p_i}
{\cal O}_{p_i}\ket{jmn} = \frac{1}{2}F_t(j)\Sigm{j}{i}{\nu}{n}\ket{jm\nu},
\ee
where
\begin{align}
F_t(j) &= \frac{1}{2td_j\lambda_j}e^{-t(\lambda_{j_1}+1/4)}\biggl(j\Bigl[S_t(j+\half)-S_t(j-\half)\Bigr] - S_t(j-\half)\biggr)\notag \\
&= \frac{1}{2td_j\lambda_j}\biggl(j\bigl(2+td_j^2\bigr) - e^{-t(j+1/2)^2}\sum_{s=-j+1/2}^{j-1/2}\bigl(1+2s^2t\bigr)e^{s^2t}\biggr).
\end{align}
For comparison, the action of the canonical left-invariant vector field of \Eq{L_iandR_i} on the states $\ket{jmn}$ is given by
\be
L_i\ket{jmn} = \frac{1}{2}\Sigm{j}{i}{\nu}{n}\ket{jm\nu}.
\ee

Hence we see that the coherent state operator is again related to the corresponding canonical operator simply by a multiplicative factor:
\be
{\cal O}_{p_i}\ket{jmn} = F_t(j)L_i\ket{jmn}.
\ee
However, in this case the canonical operator is not recovered from the coherent state operator by taking a limit with $t$, but rather by letting the coherent state operator act on a state $\ket{jmn}$ having a sufficiently large value of $j$. Since the large-$j$ behaviour of the multiplicative factor is $F_t(j) = 1 + {\cal O}(j^{-1})$ (independently of the value of $t$), the \mbox{operator ${\cal O}_{p_i}$} behaves approximately like the canonical operator $L_i$ when it is restricted to the large-$j$ sector of the space spanned by the states $\ket{jmn}$. This seems to be consistent with the discussion of section \ref{peakedness}, according to which the coherent states $\ket{g_0,\vec p_0}$ become sharply peaked on the momentum variable in the limit of large $p_0$.

The coherent state operator corresponding to the right-invariant vector field can be derived by taking the variable $\pp$ of the decomposition $h = e^{t\pp\cdot\vec\sigma/2}g$ as the classical function from which the operator is constructed. A calculation entirely similar to the one made above for the left-invariant vector field shows that the operator
\be\label{CS-p'_i}
{\cal O}_{p'_i} = -\frac{1}{2}\int d\mu(g,p')\,p'_i\ket{g,\pp}\bra{g,\pp}
\ee
acts on the states $\ket{jmn}$ as
\be
{\cal O}_{p'_i}\ket{jmn} = -\frac{1}{2}\Sigm{j}{i}{\mu}{m}\ket{j\mu n}.
\ee
Equivalently,
\be
{\cal O}_{p'_i}\ket{jmn} = F_t(j)R_i\ket{jmn},
\ee
where $R_i$ is the canonical right-invariant vector field.

Now that the action of the coherent state operators corresponding to holonomies and fluxes has been computed in explicit form, we may address the question of whether the coherent state prescription has produced a genuinely non-canonical quantization of holonomies and fluxes, or whether we have merely obtained a non-standard representation of the canonical holonomy-flux algebra. To this end, let us look at the commutator between the operators ${\cal O}_{p_i}$ and ${\cal O}_{\D{j}{m}{n}{g}}$. Inserting a resolution of identity in the states $\ket{jmn}$, and using the matrix elements given by \Eqs{CS-Dmn} and \eqref{CS-p_i}, one can show that
\begin{align}
\bra{j_1m_1n_1}\bigl[{\cal O}_{p_i},{\cal O}&_{\D{j}{m}{n}{g}}\bigr]\ket{j_2m_2n_2} = \frac{1}{2}\Sigm{j}{i}{\mu}{n}\bra{j_1m_1n_1}{\cal O}_{\D{j}{m}{\mu}{g}}\ket{j_2m_2n_2} \notag \\
&+\frac{1}{2}\bigl(F_t(j_1)-1\bigr)\Sigm{j_1}{i}{\mu}{n_1}\bra{j_1m_1\mu}{\cal O}_{\D{j}{m}{n}{g}}\ket{j_2m_2n_2} \notag \\
&-\frac{1}{2}\bigl(F_t(j_2)-1\bigr)\Sigm{j_2}{i}{n_2}{\mu}\bra{j_1m_1n_1}{\cal O}_{\D{j}{m}{n}{g}}\ket{j_2m_2\mu}.
\end{align}
The first term on the right-hand side matches with the commutator of the canonical operators, which reads
\be
\bigl[L_i,{D^{(j)m}}_{n}\bigr] = \frac{1}{2}\Sigm{j}{i}{\mu}{n}{D^{(j)m}}_{\mu}.
\ee
The presence of additional terms means that the algebra of the coherent state operators is indeed inequivalent to the algebra of canonical holonomies and fluxes -- however, the canonical algebra is approximately reproduced if the coherent state operators are restricted to act on the large-$j$ sector of the space spanned by the states $\ket{jmn}$. 

\subsection{Geometric operators}

In this section we continue our series of examples by discussing the coherent state counterparts of some of the geometric operators of loop quantum gravity that were introduced in Chapter \ref{ch:geometric}.

\subsubsection*{Area operator}

Classically, the area of a surface $S$ can be expressed as the ''length'' of the associated flux variable, \ie $A(S) = \sqrt{E^i(S)E_i(S)}$. Consequently, we take the operator 
\be\label{O_p}
{\cal O}_{|\vec p|} = \int d\mu(g,p)\,p\ket{g,\vec p}\bra{g,\vec p}
\ee
as the coherent state operator describing the area dual to a spin network edge. Following the calculation that lead from \Eq{O_pi} to \Eq{O_pi-el}, we see that the operator \eqref{O_p} has the matrix elements
\be
\bra{j_1m_1n_1}{\cal O}_{|\vec p|}\ket{j_2m_2n_2} = e^{-t\lambda_{j_1}}\delta_{j_1j_2}\delta^{m_1}_{m_2}\int d\nu(p)\,p\D{j_1}{n_2}{n_1}{e^{t\vec p\cdot\vec\sigma}}.
\ee
Here the integral is of the by now familiar form, and is proportional to $\delta^{n_2}_{n_1}$. Thus we conclude that the states $\ket{jmn}$ are eigenstates of the coherent state area operator
\be
{\cal O}_{|\vec p|}\ket{jmn} = \alpha(j)\ket{jmn},
\ee
with the eigenvalues given by
\begin{align}
\alpha(j) &= e^{-t\lambda_j}\frac{1}{d_j}\int d\nu(p)\,p\chi^{(j)}(e^{t\vec p\cdot\vec\sigma}) \notag \\
&= \biggl(\frac{d_j}{2} + \frac{1}{td_j}\biggr)\,{\rm erf}\,\biggl(\frac{\sqrt t}{2}d_j\biggr) + \frac{1}{\sqrt{\pi t}}e^{-t(\lambda_j+1/4)} \label{alpha(j)},
\end{align}
where the error function is defined as ${\rm erf}\,x = \frac{2}{\sqrt\pi}\int_0^x dt\,e^{-t^2}$.

\begin{figure}[p]
	\centering
		\includegraphics[width=0.7\textwidth]{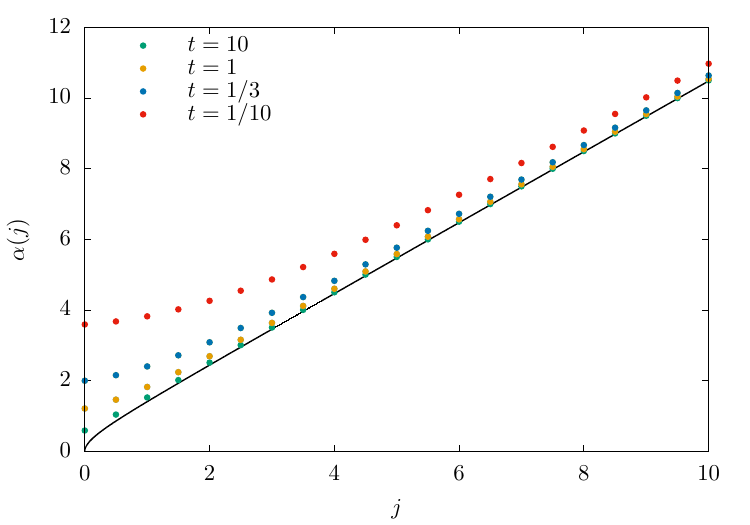}
		\caption{Eigenvalues of the coherent state area operator as a function of $j$, for various values of the parameter $t$. For comparison, the eigenvalues of the canonical area operator are shown by the solid line. As $j$ grows, the eigenvalues of the coherent state operator converge to the canonical eigenvalues.}
	\label{areaplot}
\end{figure}

The eigenvalues \eqref{alpha(j)} are plotted in \Fig{areaplot} as a function of $j$ for various values of the parameter $t$. For comparison, the eigenvalues $\sqrt{j(j+1)}$ of the canonical area operator are also shown. We see that as $j$ increases, the eigenvalues of the coherent state operator approach those of the canonical operator, the convergence being faster for larger values of $t$. Indeed, for large values of $j$ the eigenvalue \eqref{alpha(j)} has the asymptotic behaviour
\be
\alpha(j) = j + \frac{1}{2} + {\cal O}(j^{-1}),
\ee
showing that the eigenvalues of the coherent state operator agree with the canonical eigenvalues in the limit of large $j$. For smaller values of $j$, the eigenvalues $\alpha(j)$ deviate noticeably from the canonical eigenvalues, especially if the value of $t$ is not very large, so that the coherent states used to construct the operator \eqref{O_p} are not very sharply peaked on the flux variable. The discrepancy is the most drastic for $j=0$, since $\alpha(0)>\half$ for all values of $t$, and so even an edge of spin zero carries a non-zero area according to the coherent state area operator\footnote{As could have been expected, keeping in mind the theorem proven in section \ref{CS-properties}.}.

\subsubsection*{Angle operator}

Taking the function $f(\{g_e\},\{\vec p_e\})$ in \Eq{O_f N} to be the angle between two flux vectors, we obtain the coherent state angle operator
\be\label{O_theta}
{\cal O}_{\theta(\vec p_1,\vec p_2)} = \int d\mu(g_1,p_1)\,d\mu(g_2,p_2)\,\cos^{-1}\biggl(\frac{\vec p_1\cdot\vec p_2}{|\vec p_1||\vec p_2|}\biggr)\,\ket{g_1,\vec p_1;g_2,\vec p_2}\bra{g_1,\vec p_1;g_2,\vec p_2},
\ee
which acts on a pair of edges sharing a node in a spin network state. Recalling the discussion after \Eq{O_f N}, we have assumed that both edges are oriented out of the node. If an edge is oriented into the node, the corresponding variable $\vec p_e$ should be replaced with $\pp_e$ in \Eq{O_theta}.

Let us study the action of the operator \eqref{O_theta} on a state of the form
\be\label{CS-state12}
\Biggl| \RealSymb{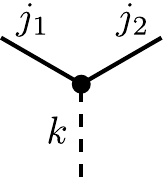}{0.5} \Biggr\rangle \; \equiv \; \sqrt{d_k}\threej{j_1}{j_2}{k}{n_1}{n_2}{\mu}\ket{j_1m_1n_1}\ket{j_2m_2n_2},
\ee
where the two edges are coupled into an internal spin $k$ by the intertwiner at the node. Taking the matrix element of the operator between two such states, and performing the immediate integrations over $SU(2)$, we get
\begin{align}
\Biggl\langle \RealSymb{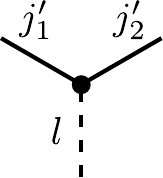}{0.5} \Biggr|&{\cal O}_{\theta(\vec p_1,\vec p_2)} \Biggl| \RealSymb{fig9-state12.pdf}{0.5} \Biggr\rangle \notag \\
&=\delta_{j_1j_1'}\delta_{j_2j_2'}\delta^{m_1}_{m'_1}\delta^{m_2}_{m'_2}\sqrt{d_kd_l}e^{-t(\lambda_{j_1}+\lambda_{j_2})}\threej{j_1}{j_2}{k}{\mu_1}{\mu_2}{\mu}\threej{j_1}{j_2}{l}{\nu_1}{\nu_2}{\nu} \notag \\
&\times\int d\nu(p_1)\,d\nu(p_2)\,\cos^{-1}\biggl(\frac{\vec p_1\cdot\vec p_2}{|\vec p_1||\vec p_2|}\biggr)\D{j_1}{\mu_1}{\nu_1}{e^{t\vec p_1\cdot\vec\sigma}}\D{j_2}{\mu_2}{\nu_2}{e^{t\vec p_2\cdot\vec\sigma}}.\label{CS-angle-1}
\end{align}
The integral appearing here -- let us denote it by $I(j_1,j_2)^{\mu_1\mu_2}_{\nu_1\nu_2}$ -- is of a form not encountered so far, but it can be dealt with by a generalization of the reasoning used to evaluate integrals earlier in this chapter. The rotational invariance of the angle function implies that the integral is invariant under the action of $SU(2)$ on its indices, in the sense that
\be
I(j_1,j_2)^{\mu_1\mu_2}_{\nu_1\nu_2} = \D{j_1}{n_1}{\nu_1}{g}\D{j_2}{n_2}{\nu_2}{g}\D{j_1}{\mu_1}{m_1}{g^{-1}}\D{j_2}{\mu_2}{m_2}{g^{-1}}I(j_1,j_2)^{m_1m_2}_{n_1n_2}.
\ee
This means that the integral, viewed as an $SU(2)$ tensor, is an element of the intertwiner space ${\rm Inv}\,\bigl({\cal H}_{j_1}\otimes{\cal H}_{j_2}\otimes{\cal H}_{j_1}^*\otimes{\cal H}_{j_2}^*\bigr)$. Expanding $I(j_1,j_2)^{\mu_1\mu_2}_{\nu_1\nu_2}$ in a suitable basis of this space, we can write
\be\label{I12-expanded}
I(j_1,j_2)^{\mu_1\mu_2}_{\nu_1\nu_2} = \sum_x d_x\,I(j_1,j_2,x)(\iota_x)^{\mu_1\mu_2}_{\nu_1\nu_2},
\ee
where the basis intertwiners are
\be
(\iota_x)^{\mu_1\mu_2}_{\nu_1\nu_2} = (\iota^{(j_1j_2x)})^{\mu_1\mu_2\alpha}(\iota^{(j_1j_2x)})_{\nu_1\nu_2\alpha} = \threej{j_1}{j_2}{x}{\mu_1}{\mu_2}{\alpha}\threej{j_1}{j_2}{x}{\nu_1}{\nu_2}{\alpha},
\ee
and therefore
\be
I(j_1,j_2,x) = I(j_1,j_2)^{\mu_1\mu_2}_{\nu_1\nu_2}\threej{j_1}{j_2}{x}{\mu_1}{\mu_2}{\alpha}\threej{j_1}{j_2}{x}{\nu_1}{\nu_2}{\alpha}.
\ee
Inserting now \Eq{I12-expanded} back into \Eq{CS-angle-1}, and using the orthogonality relation \eqref{3j-orth-mm} of the 3$j$-symbols, we find
\be
\Biggl\langle \RealSymb{fig9-state12p.pdf}{0.5} \Biggr|{\cal O}_{\theta(\vec p_1,\vec p_2)} \Biggl| \RealSymb{fig9-state12.pdf}{0.5} \Biggr\rangle  \; = \; \delta_{j_1j_1'}\delta_{j_2j_2'}\delta_{kl}\delta^{m_1}_{m_2}\delta^{m_2}_{n_2}\delta^\mu_\nu e^{-t(\lambda_{j_1}+\lambda_{j_2})}I(j_1,j_2,k).
\ee
Hence we see that the coherent state angle operator (just as its canonical counterpart) acts diagonally on the state \eqref{CS-state12}:
\be
{\cal O}_{\theta(\vec p_1,\vec p_2)} \Biggl| \RealSymb{fig9-state12.pdf}{0.5} \Biggr\rangle \; = \; \theta(j_1,j_2,k)\Biggl| \RealSymb{fig9-state12.pdf}{0.5} \Biggr\rangle 
\ee
where the eigenvalue is
\begin{align}
\theta(j_1,j_2,k) &= e^{-t(\lambda_{j_1}+\lambda_{j_2})}\int d\nu(p_1)\,d\nu(p_2)\,\cos^{-1}\biggl(\frac{\vec p_1\cdot\vec p_2}{|\vec p_1||\vec p_2|}\biggr) \notag \\
&\times\threej{j_1}{j_2}{k}{m_1}{m_2}{\mu}\D{j_1}{m_1}{n_1}{e^{t\vec p_1\cdot\vec\sigma}}\D{j_2}{m_2}{n_2}{e^{t\vec p_2\cdot\vec\sigma}}\threej{j_1}{j_2}{k}{n_1}{n_2}{\mu}.\label{CSangle-eig}
\end{align}
It is not clear how one could try to derive an explicit, analytical expression for the eigenvalues, so the integral \eqref{CSangle-eig} was evaluated numerically in order to compare the coherent state angle operator with the canonical operator, whose eigenvalue on the state \eqref{CS-state12} is
\be
\theta_{\rm can}(j_1,j_2,k) = \cos^{-1}\biggl(\frac{k(k+1) - j_1(j_1+1) - j_2(j_2+1)}{2\sqrt{j_1(j_1+1)}\sqrt{j_2(j_2+1)}}\biggr).
\ee

\begin{figure}[p]
	\centering
		\includegraphics[width=0.7\textwidth]{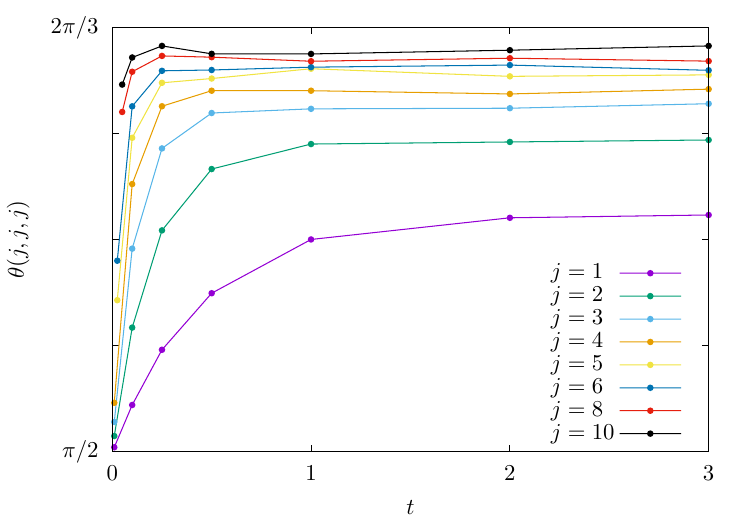}
		\caption{Numerical evaluation of the ''equilateral'' eigenvalues of the coherent state angle operator. The corresponding eigenvalue of the canonical angle operator is equal to $2\pi/3$ independently of the value of $j$. The eigenvalues of the coherent state operator again approach the canonical eigenvalue with increasing $j$.}
	\label{angleplot}
\end{figure}

In \Fig{angleplot} we show the results of a numerical calculation in the ''equilateral'' case, in which $j_1=j_2=k \equiv j$, and the canonical eigenvalue is equal to $\theta_{\rm can}(j,j,j) = 2\pi/3$ independently of the value of $j$. The eigenvalues for various values of $j$ are plotted as a function of the parameter $t$. We see that the eigenvalues converge to certain limiting values with increasing $t$, the limiting value being reached faster for larger values of $j$. As a function of $j$, the eigenvalues of the coherent state operator seem to approach the canonical eigenvalue as the value of $j$ grows. The small fluctuations in the eigenvalues, which are seen in the plot for larger values of $j$, can be attributed to numerical error and are not a genuine feature of the data.

\begin{figure}[p]
	\centering
		\includegraphics[width=0.75\textwidth]{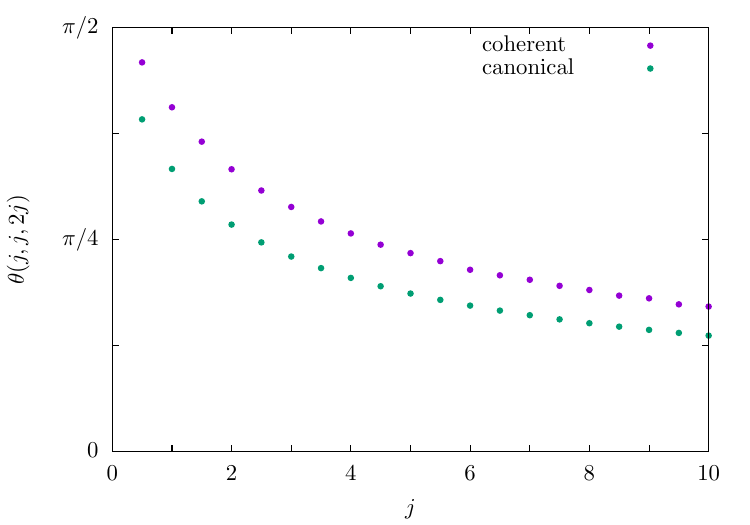}
		\caption{Numerical evaluation of the ''degenerate'' eigenvalues of the coherent state angle operator as a function of $j$ (for $t=3$). The corresponding canonical eigenvalues are also shown for comparison. The two sets of eigenvalues behave in a similar way as a function of $j$, but the eigenvalues of the coherent state operator approach the canonical eigenvalues much more slowly than in the equilateral case.}
	\label{angleplot-2}
\end{figure}

\begin{figure}[p]
	\centering
		\includegraphics[width=0.75\textwidth]{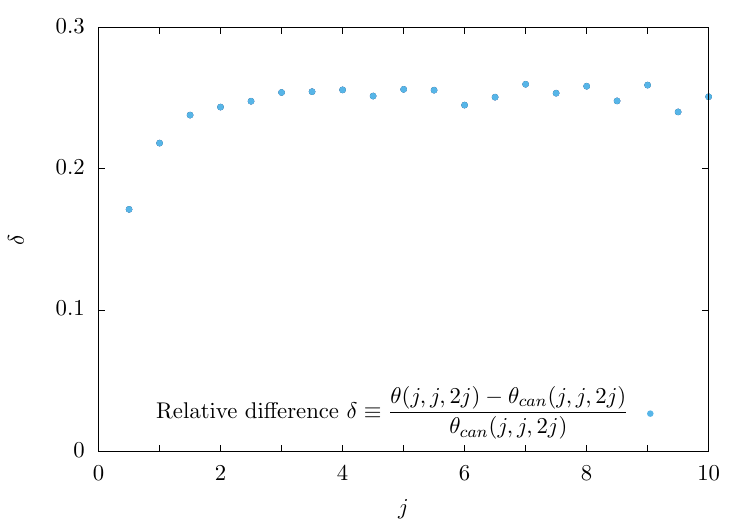}
		\caption{The relative difference between the eigenvalues of the canonical and coherent state angle operators in the degenerate case. As $j$ increases, $\delta$ seems to remain roughly constant, being approximately equal to $25\,\%$. This suggests that the eigenvalues of the coherent state operator (like those of the canonical operator) could approach zero in the limit of large $j$.}
	\label{angle-diff}
\end{figure}

In \Fig{angleplot-2} we have the ''degenerate'' case, where $j_1=j_2\equiv j$, and $k=2j$, with the value of $t$ being fixed to $t=3$. The canonical eigenvalue is now given by $\theta_{\rm can}(j,j,2j) = \cos^{-1}[j/(j+1)]$. We see that the eigenvalues of the coherent state operator behave as a function of $j$ in a similar way as the eigenvalues of the canonical operator, but with increasing $j$, the coherent state eigenvalues approach the canonical eigenvalues much more slowly than in the equilateral case. Nevertheless, the relative difference between the two sets of eigenvalues, shown in \Fig{angle-diff}, seems to remaing roughly constant as $j$ increases, so it seems plausible that the eigenvalues of the coherent state operator would approach zero, and hence converge to the canonical eigenvalues, in the limit of large $j$. 

\subsubsection*{Volume operator}

A coherent state volume operator associated to an $N$-valent spin network node can be constructed as
\begin{align}
{\cal O}_{V_N(\vec p_1,\cdots,\vec p_N)} = \int &d\mu(g_1,p_1)\cdots d\mu(g_N,p_N) \notag \\
&\times V_N(\vec p_1,\cdots,\vec p_N)\,\ket{g_1,\vec p_1;\dots;g_N,\vec p_N}\bra{g_1,\vec p_1;\dots;g_N,\vec p_N},\label{O_V}
\end{align}
where $V_N(\vec p_1,\cdots,\vec p_N)$ is some volume function defined by the vectors $\vec p_1,\dots,\vec p_N$. Naturally the remarks made in section \ref{CS-construction} on the relation between the variables $\vec p$ and the orientation of the graph should be kept in mind also here, and for each edge with an incoming orientation, the corresponding $\vec p$ in \Eq{O_V} should be understood as a $\pp$. It should also be noted that, pending the choice of the function $V_N(\vec p_1,\cdots,\vec p_N)$, \Eq{O_V} defines the coherent state volume operator in an explicit form, in contrast to the situation with the canonical volume operator, which can be defined only implicitly, as the square root of another operator for which an explicit expression can be written down.

Computing the matrix elements of the operator \eqref{O_V} in the intertwiner space of the node proceeds along the same lines as the calculation made above for the angle operator, and leads to the result
\begin{align}
&\bbra{\iota_1^{(j_1\cdots j_N)}}{\cal O}_{V_N(\vec p_1,\dots,\vec p_N)}\bket{\iota_2^{(j_1\cdots j_N)}} \notag \\
&= e^{-t(\lambda_{j_1}+\dots+\lambda_{j_N})}\int d\mu(g_1,p_1)\cdots d\mu(g_N,p_N)\,V_N(\vec p_1,\cdots,\vec p_N) \notag \\
&\times \bigl(\iota_1^{(j_1\cdots j_N)}\bigr)_{m_1\cdots m_N}\D{j_1}{m_1}{n_1}{e^{\vec p_1\cdot\vec\sigma}}\cdots\D{j_N}{m_N}{n_N}{e^{\vec p_N\cdot\vec\sigma}}\bigl(\iota_2^{(j_1\cdots j_N)}\bigr)^{n_1\cdots n_N}.\label{O_V-el}
\end{align}
In particular, a three-valent spin network node, whose intertwiner space is one-dimen-sional, is an eigenstate of the operator ${\cal O}_{V_3(\vec p_1,\vec p_2,\vec p_3)}$. The eigenvalue is given by \Eq{O_V-el} with $N=3$, and $\iota_1$ and $\iota_2$ equal to the 3$j$-symbol with spins $j_1$, $j_2$ and $j_3$. Recalling the theorem of section \ref{CS-properties}, one should not expect the eigenvalue to be zero, unless the volume function $V_3(\vec p_1,\vec p_2,\vec p_3)$ vanishes identically.

The main obstacle against obtaining a satisfactory volume operator from the definition \eqref{O_V} is due to the fact that the integrations in \Eq{O_V} place no restriction on the vectors $\vec p_1,\dots,\vec p_N$, and consequently it seems difficult to choose the volume function $V_N(\vec p_1,\cdots,\vec p_N)$ in a way that would be compatible with the usual geometrical picture associated with a spin network state, in which an $N$-valent spin network node is interpreted as an $N$-faced ''quantum polyhedron''. If one were willing to give up this interpretation, one could simply choose $V_N(\vec p_1,\cdots,\vec p_N)$ to be the volume of the $(N+1)$-faced polyhedron spanned by the vectors $\vec p_1,\cdots,\vec p_N$ (in the sense that the area and unit normal of the $i$-th face are given respectively by $|\vec p_i|$ and $\vec p_i/|\vec p_i|$). Otherwise one might have to look for a modified resolution of identity, which would express the unit operator on the Hilbert space of a graph in such a way that the condition $\vec p_1+\dots+\vec p_N = 0$ would be enforced at every node of the graph, since the polyhedron spanned by the vectors $\vec p_1,\cdots,\vec p_N$ would then have only $N$ faces.

A very similar situation arises with the Livine--Speziale coherent intertwiners \cite{LivineSpeziale}
\be\label{LS-intertwiner}
\bket{\iota({\vec n_1,\dots,\vec n_N})} \equiv \int dg\,D^{(j_1)}(g)\ket{j_1\vec n_1}\otimes\cdots\otimes D^{(j_N)}(g)\ket{j_N\vec n_N}.
\ee
Starting with the resolution of identity on the space ${\cal H}_j$ in terms of the states $\ket{j\vec n}$, it is immediate to see that the identity operator on the $N$-valent intertwiner space ${\rm Inv}\,\bigl({\cal H}_{j_1}\otimes\cdots\otimes{\cal H}_{j_N}\bigr)$ can be resolved in terms of the intertwiners \eqref{LS-intertwiner} as
\be\label{LS-Id}
\Id_{{\rm Inv}({\cal H}_{j_1}\otimes\cdots\otimes{\cal H}_{j_N})} = \int \biggl(\prod_a d_{j_i}\frac{d^2n_i}{4\pi}\biggr)\,\bket{{\iota({\vec n_1,\dots,\vec n_N})}}\bbra{{\iota({\vec n_1,\dots,\vec n_N})}}.
\ee
In principle, \Eq{LS-Id} could be taken as a starting point for constructing coherent state operators representing geometrical quantities described by functions of the vectors $\vec n_1,\dots,\vec n_N$. However, it would again be difficult to define a reasonable volume operator in this way, since the integral in \Eq{LS-Id} is taken unrestricted over all configurations of the vectors $\vec n_1,\dots,\vec n_N$.

On the other hand, in the article \cite{ConradyFreidel} Conrady and Freidel obtained the remarkable and certainly not obvious result that the unit operator on the four-valent\footnote{Even though Conrady and Freidel proved \Eq{CF-Id} only for four-valent intertwiners, it seems reasonable to expect that a similar result would hold in the $N$-valent case as well.} intertwiner space ${\rm Inv}\,\bigl({\cal H}_{j_1}\otimes\cdots\otimes{\cal H}_{j_4}\bigr)$ can be expressed in the alternate form
\be\label{CF-Id}
\Id_{{\rm Inv}({\cal H}_{j_1}\otimes\cdots\otimes{\cal H}_{j_4})} = \int d\rho(\vec n_1,\dots,\vec n_4)\,\delta\biggl(\sum_i j_i\vec n_i\biggr)\,\bket{{\iota({\vec n_1,\dots,\vec n_4})}}\bbra{{\iota({\vec n_1,\dots,\vec n_N})}},
\ee
where the integration measure $d\rho(\vec n_1,\dots,\vec n_4)$ is a non-trivial function of the $j_i$'s and $\vec n_i$'s. The key feature of the result \eqref{CF-Id} is the delta function, which ensures that the identity operator is resolved entirely in terms of coherent intertwiners satisfying the so-called closure condition $\sum_i j_i\vec n_i = 0$. When the closure condition is satisfied, the vectors $\vec n_1,\dots,\vec n_N$ can be interpreted as the unit normals of an $N$-faced polyhedron, while the spins $j_i$ are interpreted as the areas of the faces. It is then clear how a coherent state volume operator consistent with the geometrical interpretation of a spin network node could be constructed from \Eq{CF-Id}, and in fact such an operator has been considered in \cite{polyhedra}.

In the same way, it seems that in order to obtain a satisfactory volume operator from Thiemann's complexifier coherent states, one would likely have to start by deriving a resolution of identity analogous to \Eq{CF-Id}, where a closure condition for the $\vec p$-vectors would be implemented at every node of the graph, rather than trying to construct a volume operator from the simple resolution of identity which was used to write down \Eq{O_f N}. 

\subsection{Discussion}

In all the examples considered in this chapter, the analysis of a given coherent state operator has been significantly more complicated than that of the corresponding canonical operator. Therefore one might be inclined to write off the idea of considering coherent state operators altogether as unpromising, and unlikely to lead to anything practically useful. However, in the author's opinion, the apparent complexity of coherent state operators is most of all a reflection of the fact that the technology associated with coherent states in loop quantum gravity is still relatively poorly developed. In particular, the problem of devising effective tools for computations involving Thiemann's coherent states has not received very much attention in the literature, whereas the proper techniques for performing calculations in the usual spin network basis are already well known.

Some of the limitations of the construction of coherent state operators given in this chapter can likewise be attributed to the analogous limitations of the coherent states taken as the starting point for the construction. For instance, the operators constructed in this chapter are restricted to the Hilbert space of a fixed graph. In general, they cannot even be consistently extended to operators defined on the entire (kinematical) Hilbert space of loop quantum gravity, as shown very clearly by the example of the area operator, which associates a non-zero area even to a spin network edge carrying spin zero. However, it is not very surprising that this is the best one can do, if the coherent states from which the operators are constructed are themselves defined on a fixed graph. Once coherent states defined over a superposition of graphs are developed, we expect that they could be used to construct coherent state operators which are not restricted to a fixed graph, and which could act on spin network states in a graph-changing manner. For now the utility of coherent state operators in loop quantum gravity is certainly limited by the mediocre selection of coherent states available to be used in their construction.

So far we have studied coherent state operators by deriving exact expressions for their action on the spin network basis, essentially considering the coherent state operators as fundamental operators alternative to the ones obtained through canonical quantization. A potentially more fruitful point of view, suggested by the fact that coherent state operators seem to converge to the corresponding canonical operators in the limit of large spins, could be to consider coherent state operators merely as a technical tool providing some sort of a semiclassical approximation to the fundamental canonical operators. On a technical level, the action of the (for now hypothetical) coherent state volume operator on a suitable semiclassical state might well be easier to handle than that of the canonical volume operator, since the integrations involved in the matrix elements of the coherent state operator could possibly be dealt with by saddle point techniques based on the peakedness properties of the semiclassical state, whereas computing the matrix elements of the canonical operator would {\em a priori} involve extracting the square root of a large matrix.

\begin{nopage}

\sis{Part 2. Dynamics of LQG with scalar field as a physical time variable}

$\phantom{x}$

\vspace{96pt}

\noindent\textbf{\LARGE Part 2 \\}

\vspace{24pt}

\noindent\textbf{\huge Dynamics of loop quantum gravity \vspace{10pt} \\
\huge with scalar field as a physical time variable} 

\vspace{72pt}

\begin{quote}

Some people are so much impressed by the difficulties of passing over from Hamiltonian classical mechanics to quantum mechanics that they think that maybe the whole method of working from Hamiltonian classical theory is a bad method. Particularly in the last few years people have been setting up alternative methods for getting quantum field theories. They have made quite considerable progress on these lines. (\ldots) Still I feel that these alternative methods (\ldots) will not lead to a final solution to the problem. I feel that there will always be something missing from them which we can only get by working from a Hamiltonian, or maybe from some generalization of the concept of a Hamiltonian. So I take the point of view that the Hamiltonian is really very important for quantum theory.

\begin{flushright}
-- P.\@\xspace A.\@\xspace M.\@\xspace Dirac
\end{flushright}
\end{quote}

\end{nopage}

\section{Dynamics in loop quantum gravity}\label{ch:dynamics}

\subsection{The problem of time}

The spin network states of loop quantum gravity provide a mathematically consistent and physically appealing kinematical description of the quantized gravitational field. The physical picture associated to spin network states describes a quantum state of the gravitational field as a state of discrete, quantized spatial geometry, formed by elementary quanta of volume which are separated from each other by elementary quanta of area. This is a compelling realization of the idea, fundamental to general relativity, that the gravitational field and the geometry of spacetime are essentially the same physical object.

Before the structure of the quantum theory can be considered complete, the crucial element which must still be introduced is a prescription which specifies the dynamics of the quantum states of the theory. Aside from any possible technical difficulties which might be involved in the definition of dynamics, the issue is made conceptually complicated by the so-called ''problem of time'' in general relativity. In general relativity, there is no time variable possessing any absolute, physically distinguished meaning. Consequently, in the canonical formulation of the theory, there is no Hamiltonian which would govern the dynamics of the theory by generating time evolution with respect to such a time variable. Instead, the canonical Hamiltonian is merely a linear combination of constraints, so all physical (gauge-invariant) observables simply commute with the Hamiltonian. At a first sight, it would therefore seem that the observables are ''frozen'' and there is no dynamics, nor any notion of ''time'', in the formalism.

The resolution of this apparent paradox is that in general relativity, as in any generally covariant theory, the physically significant information which can be extracted from the theory is not contained in the dependence of a physically measurable quantity on the spacetime coordinates, which are in general essentially arbitrary and physically meaningless. Rather, it is contained in the relations between a measurable quantity and other measurable quantities. In particular, the dynamics of the theory is not revealed by the way in which a single measurable quantity changes with respect to an arbitrary time coordinate, but in the way in which the different measurable quantities of the theory change in relation to one another.

As an example (adapted from the examples discussed in \cite{partial}), consider some measurable physical quantity in general relativity, such as the angular position of the Sun in the sky (say, relative to the telescope of a certain observatory). Once a specific choice of coordinates is made, the angles $\bigl(\theta_{\rm Sun}(t),\phi_{\rm Sun}(t)\bigr)$ can (at least in principle) be computed from the Einstein equations as a function of the chosen time coordinate. However, such a result cannot really be regarded as a physical prediction, considering that the choice of $t$ is practically arbitrary, and in general a procedure by which the value of the time coordinate could be measured might not be available.

In order to obtain a genuine prediction about measurable physics from the formalism of general relativity, we must extend our considerations to include another physical quantity, such as the angular position of the Moon in the sky. After the angles $\bigl(\theta_{\rm Moon}(t),\phi_{\rm Moon}(t)\bigr)$ as a function of $t$ have been computed, the time coordinate can be eliminated (at least locally, within a certain range of values of $t$) to express the position of one of the celestial bodies as a (possibly implicit) function of the position of the other. 

The relation between the position of the Sun and that of the Moon is therefore a quantity which can be determined from the equations of general relativity. In other words, a valid physical question, whose answer can be predicted by the theory, is: What will be the position of the Sun, given that the position of the Moon is observed to have a certain value?\footnote{In Rovelli's terminology \cite{Rovelli, partial}, quantities which can be measured but whose value cannot be predicted (such as the position of the Sun or the Moon) are called partial observables. Measurable quantities which can be predicted from the theory (such as the position of the Sun relative to the position of the Moon) are called complete observables.} This prediction can naturally be formulated in a way which makes no reference to any arbitrarily chosen system of spacetime coordinates. The only role of the coordinates is to serve as an auxiliary bookkeeping device, facilitating the derivation of the eventual coordinate-free prediction.

\subsection{Dynamics in the fully constrained theory}

Coming back to the issue of dynamics in the quantum theory, the approach most in line with Dirac's ideas on quantizing a generally covariant theory would begin with trying to promote the Hamiltonian constraint into an operator on a suitable Hilbert space of loop quantum gravity. The constraint operator serves a dual purpose: On one hand, the states which are annihilated by the constraint form the Hilbert space of physical states; on the other hand, the constraint operator also selects the physical (\ie gauge invariant) observables of the theory. If the physical Hilbert space can be constructed, information about dynamics will be encoded in the scalar product on this space, which defines transition amplitudes between physical states.

As an example illustrating how dynamical information can be contained in transition amplitudes, let us consider the setting of a calculation which has been performed in the spin foam formulation of loop quantum gravity \cite{SFC1, SFC2}. Suppose we have a family of semiclassical states $\ket{\Psi_{(a,\dot a)}}$, each of which is peaked on a homogeneous and isotropic spatial geometry characterized by certain values of the scale factor $a$ and the ''velocity'' $\dot a$. Denoting by $W$ the operator defining transition amplitudes, the amplitude
\be
W(a,\dot a;a_0,\dot a_0) \equiv \bra{\Psi_{(a,\dot a)}}W\ket{\Psi_{(a_0,\dot a_0)}}
\ee
describes the probability of obtaining a state peaked on the values $(a,\dot a)$ from an initial state which is peaked on the values $(a_0,\dot a_0)$. Considering $W(a,\dot a;a_0,\dot a_0)$ as a function of $a$ and $\dot a$ for fixed values of $a_0$ and $\dot a_0$, one could argue that the classical dynamics of a homogeneous and isotropic spatial geometry is correctly recovered from the quantum theory if the magnitude of $W(a,\dot a;a_0,\dot a_0)$ is suppressed unless the point $(a,\dot a)$ is on (or at least very close to) the classical trajectory resulting from the evolution of the initial values $(a_0,\dot a_0)$. (This is the conclusion which was found in \cite{SFC2}.)

In the above example, information about dynamics is again given by the relation of the variables $a$ and $\dot a$ to one another, as opposed to the evolution of these variables with respect to a physically meaningless time parameter; in fact, such a time parameter is entirely absent from the calculation.

In loop quantum gravity, the construction of the Hamiltonian constraint as a concrete and well-defined operator has been accomplished originally by Thiemann \cite{QSD}, and later by other authors \cite{paper2, AlesciC, GaulRovelli, YangMa} following the example set by Thiemann in varying degrees of closeness (see Chapters \ref{ch:Thiemann} and \ref{ch:constraint}). This is already a considerable achievement, in light of the complicated form of the classical Hamiltonian constraint as a function of the Ashtekar variables. However, the complexity of the classical constraint functional is invariably reflected in the complexity of the corresponding quantum operator, and consequently the task of deriving solutions of the constraint operator in explicit form, not to mention uncovering the physical interpretation of the resulting states, has proven to be extremely challenging. In fully concrete terms, very little is known about the physical Hilbert space (or the physical scalar product thereon) defined by the kernel of any of the constraint operators available in the literature, even though the structure of the kernel of a given constraint operator can often be described to some extent on a qualitative level.

\subsection{Deparametrization}

The practical difficulties encountered with the Hamiltonian constraint operator have motivated the search for alternative ways to deal with the problem of dynamics in loop quantum gravity. Within the canonical formalism, one of the main alternatives to working with the Hamiltonian constraint is the so-called method of deparametrization, in which one of the degrees of freedom contained in the theory is assigned the role of a physical, relational time variable, with respect to which the dynamics of the remaining degrees of freedom is described.

It is possible to deparametrize general relativity ''intrinsically'', in terms of geometric quantities such as proper distances and angles (see \eg \cite{Jedrek.et.al, Jedrek.et.al.add, Norbert.Jedrek}), but the resulting formalism is rather cumbersome and difficult to handle in practice. A more tractable approach is to consider gravity coupled with a matter field -- in practice, a suitable scalar field -- which is used as a physical time variable for the dynamics of the gravitational field.

In a deparametrized model of gravity, the Hamiltonian constraint is essentially traded for a physical Hamiltonian, which generates time evolution with respect to the time variable provided by the scalar field. After the scalar field has been singled out as the time variable, reparametrization of the time coordinate is no longer a gauge invariance of the theory. In the quantum theory, the physical Hilbert space is therefore the space of diffeomorphism invariant states, and the dynamics of states in ${\cal H}_{\rm diff}$ is governed by the Schr\"odinger equation
\be
i\frac{d}{dt}\ket\Psi = H_{\rm phys}\ket\Psi,
\ee
in which $t$ is the time defined by the scalar field, and a physical Hamiltonian operator acts as the generator of time evolution. The issues of extracting solutions of the Hamiltonian constraint and interpreting them physically, and determining the scalar product on the physical Hilbert space, are entirely bypassed in models of loop quantum gravity where the dynamics is defined through deparametrization.

The use of scalar fields as reference fields for general relativity was first considered in a series of articles by Kucha\v{r} and collaborators \cite{BrownKuchar, KucharBicak, KucharTorre91, KucharTorre93}. In the context of loop quantum gravity, a model in which a massless Klein--Gordon field is used as a physical time variable was proposed by Rovelli and Smolin \cite{RovelliSmolin93}. The quantum theory of the model has been developed in more detail in \cite{quantized} and \cite{paper1}, while the classical structure of the model was studied by Kucha\v{r} and Romano in \cite{KucharRomano}. In addition, a non-rotational dust field as a time variable for loop quantum gravity has been considered by Husain and Paw\l owski \cite{HusainPawlowski, HusainPawlowski2}. A comprehensive catalog of deparametrized models, including models where three additional scalar fields are introduced to serve as a spatial reference system, is given by Giesel and Thiemann in \cite{GieselThiemann}.

\newpage

In this work, deparametrization by means of a scalar field will be considered as the main approach for tackling the problem of dynamics in loop quantum gravity. Since this point of view is motivated entirely by practical considerations, we will focus our entire attention on specific models containing a single scalar field, whose purpose is to act a physical time variable. These are the free scalar field model proposed by Rovelli and Smolin, and the non-rotational dust model due to Brown and Kucha\v{r}, and Husain and Paw\l owski. Models in which four scalar fields are used to deparametrize the entire spacetime will not be considered in this work.

\subsection{Other approaches to dynamics}

To conclude this chapter, we will briefly mention some of the other approaches that have been developed in an attempt to give an adequate definition of the dynamics for loop quantum gravity. Staying within the context of the canonical theory, Thiemann has proposed the so-called master constraint programme \cite{master}, in which the infinite number of Hamiltonian constraints $C(x)=0$ are replaced with the single ''master constraint''
\be
M = \int d^3x\,\frac{C^2}{\sqrt q},
\ee
the vanishing of which is equivalent to all the constraints $C(x)$ vanishing one-by-one. The master constraint carries no dependence on the lapse function, and its constraint algebra is significantly simpler than the algebra of the standard Hamiltonian constraint. Due to these features, the introduction of the master constraint can potentially clear up some issues of a formal, mathematical nature. However, it seems unlikely that the practical problems one faces when trying to extract solutions of the Hamiltonian constraint would be any less severe in the case of the master constraint.

A notable and highly prominent framework for the dynamics of loop quantum gravity is the spin foam formalism, also often referred to as covariant loop quantum gravity (see \cite{Perez, Zakopane, RovelliVidotto} and references therein). The dynamics of the covariant theory is defined, broadly speaking, by performing a concrete implementation of a particular version of the path integral for general relativity, thereby providing a prescription for associating transition amplitudes to spin network states.

Technically, the spin foam formalism assigns an amplitude to every colored two-complex (the faces and edges of the two-complex being ''colored'' respectively with spins and intertwiners). The two-complex arises as the dual of a simplicial decomposition of the spacetime manifold, which is introduced as a tool for regularizing the path integral. If the boundary of the two-complex consists of two disjoint graphs, the corresponding amplitude can be interpreted as a transition amplitude between the spin network states defined by the two boundary graphs together with the associated spins and intertwiners. The two-complex itself can then be interpreted as a sort of Feynman diagram describing a particular ''history'' of a spin network state (in the same sense in which a Feynman diagram in conventional quantum field theory describes a history of particles).

The group field theory approach \cite{Oriti} pioneered by Oriti represents a more significant departure from the conventional framework of loop quantum gravity. On the kinematical level, the state space considered in this approach is similar to the kinematical Hilbert space of loop quantum gravity, but there are a number of significant differences. The dynamics is specified by using a Feynman expansion of the group field theory to assign transition amplitudes to the kinematical states.

\section{Classical theory of deparametrized models}\label{ch:deparametrized}

In this chapter we venture briefly into the classical theory of the deparametrized models of general relativity which will be used later in this work to provide a definition of the dynamics for loop quantum gravity. The two models which will be considered are the free scalar field model proposed by Rovelli and Smolin \cite{RovelliSmolin93} (and developed further in \cite{quantized} and \cite{paper1}), and the non-rotational dust model, which was introduced classically by Brown and Kucha\v{r} \cite{BrownKuchar}, and first studied in the context of loop quantum gravity by Husain and Paw\l owski \cite{HusainPawlowski}. Here we do not attempt to give a comprehensive discussion of the classical theory of these models; the only goal of this chapter is to establish the form of the physical Hamiltonian which generates the dynamics of the gravitational field with respect to the relational time provided by the reference scalar field, and whose quantization will lead to the operator governing the dynamics in the quantum theory.

\subsection{Free scalar field model}

\newcommand{\gr}{{\rm gr}}

The system formed by the gravitational field and a minimally coupled massless Klein--Gordon field is described by the action
\be
S = \frac{1}{16\pi G}\int d^4x\,\sqrt{-g}R - \frac{1}{2}\int d^4x\,\sqrt{-g}g^{\mu\nu}\partial_\mu\phi\partial_\nu\phi.
\ee
After a standard 3+1 decomposition in the ADM variables is performed, the action takes the form
\be
S = \int dt\int d^3x\,\Bigl(p^{ab}\dot q_{ab} + \pi\dot\phi - NC - N^aC_a\Bigr),
\ee
where $\pi = \sqrt q\dot\phi$ is the canonical momentum of the scalar field. Here the diffeomorphism and Hamiltonian constraints are
\be\label{C_a-KG}
C_a = C_a^\gr + \pi\partial_a\phi
\ee
and
\be\label{C-KG}
C = C_\gr + \frac{\pi^2}{2\sqrt q} + \frac{1}{2}\sqrt q q^{ab}\partial_a\phi\partial_b\phi
\ee
with $C_a^\gr$ and $C_\gr$ denoting the gravitational contributions to the constraints, as given by \Eqs{C_a-ADM} and \eqref{C-ADM}.

In practical terms, the physical Hamiltonian, which generates the dynamics of the gravitational field with respect to the internal time provided by the scalar field, is now determined by solving the system of constraints \eqref{C_a-KG}--\eqref{C-KG} for the momentum of the scalar field, and taking the resulting expression as the physical Hamiltonian density. That this procedure indeed correctly gives the physical Hamiltonian can be justified in various ways. In \cite{quantized}, the authors considered the quantization of the classical theory defined by the diffeomorphism constraint \eqref{C_a-KG} and the modified Hamiltonian constraint\footnote{If the Ashtekar variables are used to describe the gravitational field, the Gauss constraint must naturally be included in the system of constraints.}
\be\label{C'-KG}
C' = \pi - h,
\ee
where
\be\label{h-KG}
h = \sqrt{-qC_\gr + \sqrt q\sqrt{C_\gr^2 - q^{ab}C^\gr_a C^\gr_b}}.
\ee
When taking the square roots in \Eq{h-KG}, the signs have been chosen according to what seems to be the unique choice leading to a positive-definite and non-trivial Hamiltonian operator in the quantum theory. In the region of phase space corresponding to the sign choices made in \Eq{h-KG}, the constraint \eqref{C'-KG} is equivalent to the original Hamiltonian constraint \eqref{C-KG}.

It was then shown in \cite{quantized} that the quantum observables defined by the constraint \eqref{C'-KG} (together with the Gauss and diffeomorphism constraints) are essentially certain operators on ${\cal H}_{\rm diff}$ which are labeled by values of the scalar field $\phi$. If the scalar field is interpreted as a physical time variable, the dynamics of an observable ${\cal O}_\phi$ is generated by the Heisenberg-like equation
\be
i\frac{d}{d\phi}{\cal O}_\phi = [{\cal O}_\phi,H_{\rm phys}],
\ee
where the physical Hamiltonian $H_{\rm phys}$ is an operator constructed by quantizing the classical functional \eqref{h-KG}. In practice, on grounds of the reasoning that the physical Hamiltonian will act in the diffeomorphism invariant Hilbert space, the factor involving the diffeomorphism constraint in the classical expression \eqref{h-KG} is discarded, and
\be\label{hphys-KG}
h_{\rm phys} = \int d^3x\,\sqrt{-2\sqrt q C_\gr}
\ee
is taken as the classical functional to be quantized in order to define the physical Hamiltonian for the free scalar field model.

An alternative way of seeing that \eqref{hphys-KG} is the generator of dynamics with respect to the scalar field time can be derived by considering the time evolution of the gravitational variables $(q_{ab},p^{ab})$ under the canonical Hamiltonian
\be
H = \int d^3x\,\bigl(NC + N^aC_a\bigr),
\ee
which generates evolution with respect to the time coordinate $t$ corresponding to the foliation defined by the lapse $N$ and the shift $N^a$. Considering first an arbitrary, unspecified foliation, the evolution of a functional $F(q_{ab},p^{ab})$ is given by the equation
\be\label{dF/dt KG}
\frac{d}{dt}F(q_{ab},p^{ab}) = \{F,H\} = \int d^3x\,\biggl[N\Bigl(\{F,C_\gr\} + \frac{\pi^2}{2}\{F,\frac{1}{\sqrt q}\}\Bigr) + N^a\{F,C^\gr_a\}\biggr].
\ee
Let us now specialize this general result to the case in which the scalar field is taken as the time variable, by introducing the following assumptions:
\begin{itemize}
\item The spatial surfaces $\Sigma$ are chosen to be the surfaces on which the scalar field $\phi$ takes constant values. (In particular, this choice implies that $\partial_a\phi = 0$.)
\item The value of the lapse function is set to $N=\sqrt q/\pi$, this value being derived from the requirement that the coordinate choice $t=\phi$ is preserved under time evolution.
\item We also assume the constraints \eqref{C_a-KG} and \eqref{C-KG} to be satisfied, so that the expression \eqref{h-KG} (where now $C_a^\gr=0$) can be substituted for the value of the scalar field momentum $\pi$.
\end{itemize}
If we additionally restrict ourselves to a diffeomorphism invariant observable $F$, so that $F$ commutes with the diffeomorphism constraint, then a short calculation shows that \Eq{dF/dt KG} is equivalent to the equation 
\be\label{dF/dt'}
\frac{d}{dt}F(q_{ab},p^{ab}) = \{F,h_{\rm phys}\},
\ee
where $t\equiv\phi$, and $h_{\rm phys}$ is the functional given by \Eq{hphys-KG}. 

Under the point of view summarized by \Eq{dF/dt'}, the scalar field is chosen as a time variable already at the classical level, and the physical Hamiltonian \eqref{hphys-KG} is considered as the generator of time evolution (for diffeomorphism invariant observables) in the classical theory. This is to be contrasted with the point of view in which the classical theory is regarded as a fully constrained theory, with the Hamiltonian constraint given by \Eq{C'-KG}. The choice of the scalar field as a time variable is then introduced at the level of the quantum theory, and the physical Hamiltonian emerges as the operator governing the dynamics of the observables selected by the constraint \eqref{C'-KG}. As the structure of the resulting quantum theory will be exactly the same in either case, the question of which point of view is preferable is ultimately a matter of taste, and seems to have little practical importance.

\subsection{Non-rotational dust model}

The model of gravity coupled to non-rotational dust \cite{BrownKuchar, HusainPawlowski} is described by the action
\be\label{S_g+S_dust}
S = \frac{1}{16\pi G}\int d^4x\,\sqrt{-g}R - \frac{1}{2}\int d^4x\,\sqrt{-g}M\bigl(g^{\mu\nu}\partial_\mu T\partial_\nu T + 1\bigr).
\ee
The variable $M$ acts essentially as a Lagrange multiplier; it enforces the gradient $\partial_\mu T$ to be timelike, \ie the surfaces of constant $T$ to be spacelike\footnote{In the dust model, the constant surfaces of the dust field are therefore guaranteed to be suitable for use as the spatial surfaces of a 3+1 decomposition of spacetime. Strictly speaking, no similar guarantee holds for the constant surfaces of the Klein--Gordon field in the free scalar field model.}. The physical interpretation of the field $T$ as a dust field is revealed by the energy-momentum tensor of the field,
\be
T_{\mu\nu} = Mu_\mu u_\nu + \frac{M}{2}g_{\mu\nu}\bigl(1 + g_{\alpha\beta}u^\alpha u^\beta\bigr)
\ee
(with $u_\mu = \partial_\mu T$), which has the form corresponding to pressureless dust.

Upon a 3+1 decomposition, using again the ADM variables to describe the gravitational field, the action of the dust field is transformed into
\be\label{S_dust}
S_{\rm dust} = \int dt\int d^3x\,\frac{M}{2N}\sqrt q\Bigl[\bigl(\dot T + N^a\partial_a T\bigr)^2 - N^2\bigl(q^{ab}\partial_a T\partial_b T + 1\bigr)\Bigr].
\ee
From this we can read off the canonical momentum of the dust field as
\be
p_T = \frac{\delta L_{\rm dust}}{\delta\dot T} = \frac{M}{N}\sqrt q\bigl(\dot T + N^a\partial_a T\bigr).
\ee
The action \eqref{S_dust} can then be cast in the form
\be
S_{\rm dust} = \int dt\int d^3x\,\Bigl(\dot T p_T - NC_{\rm dust} - N^aC_a^{\rm dust}\Bigr),
\ee
with the contributions of the dust field to the constraints given by
\be\label{C_a-dust}
C_a^{\rm dust} = p_T\partial_a T
\ee
and
\be\label{C-dust}
C_{\rm dust} = \frac{p_T^2}{2M\sqrt q} + \frac{M}{2}\sqrt q\bigl(q^{ab}\partial_a T\partial_b T + 1\bigr).
\ee
Before proceeding further with the canonical analysis of the model, we choose at this point to impose the choice of the dust field as the time coordinate, following the approach proposed in \cite{Jedrek} as an improvement over the treatment of \cite{HusainPawlowski}. The coordinate choice $t=T$ immediately implies that $\dot T=1$ and $\partial_a T = 0$. Moreover, by requiring that the condition $t=T$ is preserved under time evolution, one finds that the value of the lapse function is fixed to $N=1$ (while no restriction is placed on the shift vector). Under these assumptions, the momentum of the dust field becomes
\be
p_T = M\sqrt q.
\ee
When this relation is used to eliminate the non-dynamical field $M$ in favour of $p_T$ and $\sqrt q$, the constraint \eqref{C-dust} reduces to $C_{\rm dust} = p_T$. (The diffeomorphism constraint \eqref{C_a-dust} vanishes identically when the value of $T$ is constant on each spatial surface.) Going then back to the action \eqref{S_g+S_dust} for the entire model, we obtain
\be\label{S'}
S = \int dt \int d^3x\,\Bigl(p^{ab}\dot q_{ab} - C_\gr - N^aC_a^\gr\Bigr),
\ee
which can be seen as a kind of gauge-fixed action for the gravitational field, when the role of a physical time variable has been assigned to the dust field. Since the time variable has been fixed, reparametrization of the time coordinate is not a gauge symmetry of the model; rather, $C_\gr$ in \Eq{S'} is interpreted as a generator of time evolution for the gravitational variables with respect to the time defined by the dust field. (Again, one can alternatively come to this conclusion by implementing the choice $t=T$ in the analog of \Eq{dF/dt KG} in the dust model.)

We have thus found that the physical Hamiltonian for the non-rotational dust model is given by the functional
\be
h_{\rm phys} = \int d^3x\,C_\gr.
\ee
From this we see the central advantage of the dust model over the free scalar field model considered in the previous section: No square root is involved in the classical expression for the physical Hamiltonian in the dust model. Accordingly, when it comes to the construction of the quantum theory, the physical Hamiltonian for the dust model can be defined as an operator whose action on spin network states can be computed in explicit form even without knowing anything about the spectrum of the operator. In contrast, in the case of the scalar field model the square root in the classical expression \eqref{hphys-KG} carries over to the Hamiltonian operator in the quantum theory, and so a knowledge of the spectral decomposition of the Hamiltonian is required in order to resolve the square root and make the action of the Hamiltonian explicitly computable.

\newpage

\section{Thiemann's Hamiltonian}\label{ch:Thiemann}

This chapter is devoted to a presentation of Thiemann's construction of the Hamiltonian constraint operator. The construction of a mathematically well-defined and explicitly formulated constraint operator is already a remarkable achievement, considering the complicated and highly non-polynomial dependence of the classical constraint functional on the elementary Ashtekar variables. Thiemann's work therefore represents an important milestone in the history of loop quantum gravity.

Our reasons for giving a detailed review of Thiemann's Hamiltonian are twofold. Firstly, Thiemann's work, with its subtle rewritings of classical expressions in terms of objects which can be consistently quantized, and its careful construction of a regularization which leads in the end to a well-defined quantum operator, serves as a prototype and a model example on how operators can be constructed in loop quantum gravity. Secondly, we will establish that in spite of its great importance in the fully constrained theory, Thiemann's constraint operator is not a suitable candidate for the physical Hamiltonian in deparametrized models of loop quantum gravity, and we will therefore have to search for a more appropriate Hamiltonian in the deparametrized context. The construction of an adequate physical Hamiltonian is one of the central results of this work, and will be taken up in the next chapter.

\subsection{Classical preparations}

Thiemann's construction \cite{QSD} (see also \cite{Thiemann}) is based on the classical expression \eqref{C-AE-1} for the constraint. The smeared constraint $C(N) = \int d^3x\,NC$ has the form
\be\label{C(N)}
C(N) = C_E(N) - (1+\beta^2)C_L(N),
\ee
where
\be\label{C_E(N)}
C_E(N) = \frac{1}{16\pi G}\int d^3x\,N\frac{\epsilon\updown{ij}{k}E^a_iE^b_jF_{ab}^k}{\sqrt{\det E}}
\ee
is called the Euclidean part of the constraint (since \eqref{C_E(N)} could be taken as the entire Hamiltonian constraint in the case of general relativity with Euclidean signature), and the remainder
\be\label{C_L(N)}
C_L(N) = \frac{1}{16\pi G}\int d^3x\,N\frac{E^a_iE^b_j}{\sqrt{\det E}}(K_a^iK_b^j - K_a^jK_b^i)
\ee
is called the Lorentzian part.

At a first sight \Eqs{C(N)}--\eqref{C_L(N)} may seem to suffer from certain concrete difficulties, making them a questionable starting point for the quantization of the Hamiltonian constraint. The inverse volume element appearing in \Eqs{C_E(N)} and \eqref{C_L(N)} cannot be quantized in a straightforward way, unless one is willing to employ a quantization of the type \eqref{V^-1} to circumvent the fact that the volume operator has a large kernel and is therefore not invertible. Furthermore, even though the curvature $F_{ab}^k$ in the Euclidean part can be quantized by relating it to a holonomy around a small loop, as indicated by \Eq{h-loop}, it is not immediately clear how to quantize the extrinsic curvature terms in the Lorentzian part, since no obvious operator corresponding to $K_a^i$ is available in the quantum theory.

An ingenious way to overcome these difficulties was discovered by Thiemann, who used a series of classical identities -- now commonly referred to as Thiemann's tricks -- to manipulate the expressions \eqref{C_E(N)} and \eqref{C_L(N)} into a form suitable for quantization. First of all, Thiemann observed that the factors of $1/\sqrt{\det E}$ can be eliminated by means of the Poisson bracket relation
\be\label{A,V}
\{A_a^i,V\} = 8\pi\beta G\frac{\delta V}{\delta E^a_i} = (8\pi\beta G)\frac{1}{4}\epsilon_{abc}\epsilon^{ijk}\frac{E^b_jE^c_k}{\sqrt{\det E}}.
\ee
Using \Eq{A,V}, the Euclidean part of the constraint can be rewritten as
\be\label{C_E-brac}
C_E(N) = \frac{1}{(8\pi G)^2\beta}\int d^3x\,N\epsilon^{abc}F_{ab}^i\{A_{ci},V\},
\ee
which provides an appropriate starting point for quantization, since all the elements on the right-hand side now correspond to some definite operators in loop quantum gravity. In analogy with the quantization of the curvature of the Ashtekar connection, the connection itself can be quantized by relating it to the holonomy over a short line segment; then the Poisson bracket can simply be replaced with the commutator of the corresponding operators in the quantum theory.

While the inverse volume element can be eliminated also in the Lorentzian part with the help of \Eq{A,V}, further manipulations are needed in order to deal with the factor involving the extrinsic curvature. Thiemann's treatment of this factor consists of two steps, the first of which is to observe that the integrated trace of the extrinsic curvature,
\be
K = \int d^3x\,K_a^iE^a_i,
\ee
satisfies the identity
\be\label{A,K}
\{A_a^i,K\} = 8\pi\beta GK_a^i.
\ee
In verifying this, one should recall from Chapter 1 that the variation $E^a_i\delta\Gamma_a^i$ is a total derivative, from which it follows that $\int d^3x\,\{A_a^i,\Gamma_b^j\}E^b_j=0$. Inserting \Eqs{A,V} and \eqref{A,K} into \Eq{C_L(N)}, the Lorentzian part takes the form
\be
C_L(N) = \frac{1}{(8\pi G)^4\beta^3}\int d^3x\,N\epsilon^{abc}\epsilon_{ijk}\{A_a^i,K\}\{A_b^j,K\}\{A_c^k,V\},
\ee
and it remains to explain how to promote $K$ into a loop quantum gravity operator. The last one in Thiemann's sequence of tricks accomplishes this by relating $K$ to the Poisson bracket $\{C_E(1),V\}$, both of whose arguments are available as operators in loop quantum gravity (assuming, of course, that the operator corresponding to the Euclidean constraint has been successfully constructed). We shall take a moment to derive the precise relation, since several different versions of it, with varying numerical prefactors, can be found in the literature.

The Poisson bracket in question is
\be\label{C,V}
\{C_E(1),V\} = 8\pi\beta G\int d^3x\,\frac{\delta C_E(1)}{\delta A_a^i(x)}\frac{\delta V}{\delta E^a_i(x)}.
\ee
From \Eq{C_E(N)} we compute the first factor of the integrand as
\be\label{dC/dA}
\frac{\delta C_E(1)}{\delta A_a^i(x)} = \frac{1}{8\pi G}\biggl[\epsilon\downup{i}{jk}\partial_b\biggl(\frac{E^a_jE^b_k}{\sqrt{\det E}}\biggr) + \frac{E^a_iE^b_j - E^a_jE^b_i}{\sqrt{\det E}}A_b^j\biggr].
\ee
For the present calculation, the functional derivative of the volume, given by \Eq{A,V}, is better expressed as
\be\label{dV/dE}
\frac{\delta V}{\delta E^a_i(x)} = \frac{1}{2}\sqrt{\det E}E_a^i,
\ee
where $E_a^i$ is the inverse of $E^a_i$ (\ie $E_a^i = e_a^i/\sqrt q$). Then the contraction of \eqref{dV/dE} with the second term of \Eq{dC/dA} immediately gives $(8\pi G)^{-1}E^a_iA_a^i$. Inserting \eqref{dV/dE} and the first term of \eqref{dC/dA} into \Eq{C,V}, and performing an integration by parts, we obtain
\be\label{int GE}
\frac{\beta}{2}\int d^3x\,\epsilon\downup{i}{jk}\partial_b\biggl(\frac{E^a_jE^b_k}{\sqrt{\det E}}\biggr)\sqrt{\det E}E_a^i = -\frac{\beta}{2}\int d^3x\,\epsilon\downup{i}{jk}E^a_jE^b_k\partial_bE_a^i.
\ee
On the other hand, contracting \Eq{Gamma_a^i} with $E^a_i$, we see that $E^a_i\Gamma_a^i = \half\epsilon\downup{i}{jk}E^a_jE^b_k\partial_bE_a^i$, and so \eqref{int GE} is equal to $-\beta\int d^3x\,\Gamma_a^iE^a_i$. All in all, we therefore have
\be\label{C_E,V}
\{C_E(1),V\} = \beta\int d^3x\,(A_a^i-\Gamma_a^i)E^a_i = \beta^2K,
\ee
which is the relation we were looking to prove. Using it, we finally get to express the Lorentzian part of the constraint in the form
\be\label{C_L-brac}
C_L(N) = \frac{1}{(8\pi G)^4\beta^7}\int d^3x\,N\epsilon^{abc}\epsilon_{ijk}\bigl\{A_a^i,\{C_E(1),V\}\bigr\}\bigl\{A_b^j,\{C_E(1),V\}\bigr\}\{A_c^k,V\},
\ee
which carries no essential obstruction against promoting it into a well-defined operator (though from a practical point of view one might complain that the form of \Eq{C_L-brac} is perhaps more complicated than one would have hoped.)

\subsection{Regularization and quantization}\label{sec:Thiemann-reg}

In order to express the Ashtekar connection and its curvature in \Eqs{C_E-brac} and \eqref{C_L-brac} in terms of holonomies, we (following Thiemann) consider a tetrahedral triangulation of the spatial manifold $\Sigma$, though more general partitions of $\Sigma$ could certainly be taken as an alternative starting point for regularization. For each tetrahedron $\Delta$ of the triangulation, we pick one of its vertices $v_\Delta$, denote by $s_I(\Delta)$ $(I=1,2,3)$ the three edges emerging from $v_\Delta$, and let $\alpha_{IJ}(\Delta)$ denote the triangular loop spanned by the two edges $s_I(\Delta)$ and $s_J^{-1}(\Delta)$. Assuming for simplicity that each $s_I$ has coordinate length $\epsilon$, the holonomies associated to the edge $s_I$ and the loop $\alpha_{IJ}$ can be expanded as
\begin{align}
h_{s_I} &= 1 - \epsilon\dot s_I^aA_a^i\tau_i + {\cal O}(\epsilon^2), \\
h_{\alpha_{IJ}} &= 1 - \frac{\epsilon^2}{2}\dot s_I^a\dot s_J^bF_{ab}^i\tau_i + {\cal O}(\epsilon^3).
\end{align}
Using these, it is straightforward to check that the expression
\be\label{C_E-reg}
C_E(N) \simeq \frac{1}{(8\pi G)^2\beta} \frac{-2}{3}\sum_\Delta N(v_\Delta)\epsilon^{IJK} \Tr\Bigl(h_{\alpha_{IJ}(\Delta)}h_{s_K(\Delta)}^{-1}\bigl\{h_{s_K(\Delta)},V(\Delta)\bigr\}\Bigr),
\ee
approximates the Euclidean part \eqref{C_E-brac}, in the sense that the sum on the right-hand side converges to \eqref{C_E-brac} in the limit of an arbitrarily fine triangulation, independently of the details of the triangulation. The factor of $-2/3$ arises from the normalization of the $SU(2)$ generators, $\Tr(\tau_i\tau_j) = -\half\delta_{ij}$, and from the fact that $\epsilon^3\epsilon^{IJK}\dot s_I^a\dot s_J^b\dot s_K^c$ is 6 times the coordinate volume of the tetrahedron $\Delta$. At the classical level, the factor $h_{s_K(\Delta)}^{-1}$ plays no role in the convergence of \eqref{C_E-reg} to the continuum expression \eqref{C_E-brac}; it is inserted to ensure that the operator arising from \Eq{C_E-reg} will be gauge invariant. 

Similarly, the Lorentzian part \eqref{C_L-brac} is approximated by
\begin{align}
C_L(N) \simeq \frac{1}{(8\pi G)^4\beta^7} \frac{2}{3}\sum_\Delta &N(v_\Delta)\epsilon^{IJK}\Tr\Bigl(h_{s_I(\Delta)}^{-1}\bigl\{h_{s_I(\Delta)},\{C_E(1),V(\Delta)\}\bigr\} \notag \\
&\times h_{s_J(\Delta)}^{-1}\bigl\{h_{s_J(\Delta)},\{C_E(1),V(\Delta)\}\bigr\} h_{s_K(\Delta)}^{-1}\bigl\{h_{s_K(\Delta)},V(\Delta)\bigr\}\Bigr), \label{C_L-reg}
\end{align} 
where the numerical coefficient is now adjusted to cancel the factor of $-\tfrac{1}{4}$ in the relation $\Tr(\tau_i\tau_j\tau_k) = -\tfrac{1}{4}\epsilon_{ijk}$. We point out that the regularized expressions \eqref{C_E-reg} and \eqref{C_L-reg} again have no explicit dependence on the regularization parameter $\epsilon$.

Before \Eqs{C_E-reg} and \eqref{C_L-reg} can be promoted into operators, the triangulation must be adjusted to the graph of the spin network state on which the operators are to act. By ordering the volume operator (or at least one factor of the volume operator, in the case of the Lorentzian part) to the right, the resulting operators can be made to act only on the nodes of the spin network. It is therefore sufficient to specify the details of the triangulation only in a neighborhood of each node. 

Fixing the node as the point $v_\Delta$, for each triple of edges $(e_I,e_J,e_K)$ at the node one considers a tetrahedron spanned by segments $(s_I,s_J,s_K)$ of the edges, and constructs seven additional ''virtual'' tetrahedra (as described in great detail in \cite{QSD}) so that the eight tetrahedra together triangulate a neighborhood of the node. With this choice of triangulation, the action of the regularized Euclidean constraint on a state based on a graph $\Gamma$ is given by
\be\label{C_E^Delta}
C_E^{(\Delta)}(N)\ket{\Psi_\Gamma} = \frac{2i}{3(8\pi G)^2\beta}\sum_{v\in\Gamma} N(v)\frac{8}{E(v)} \sum_{\Delta(v)} \epsilon^{IJK}\Tr\Bigl(h_{\alpha_{IJ}(\Delta)}h_{s_K(\Delta)}^{-1}\bigl[h_{s_K(\Delta)},V_v\bigr]\Bigr)\ket{\Psi_\Gamma},
\ee
where the factor of 8 appears because one must sum over the 8 tetrahedra to cover an entire neighborhood of the node. Furthermore, $E(v) = \bigl(\begin{smallmatrix} n(v)\\3 \end{smallmatrix}\bigr)$, with $n(v)$ the valence of the node $v$, is the number of different triples of edges at $v$, and the division by $E(v)$ is to compensate for the fact that an independent triangulation is constructed for each triple of edges. In the same way, the regularized Lorentzian part acts as
\begin{align}
C_L^{(\Delta)}(N)\ket{\Psi_\Gamma} = \frac{2i}{3(8\pi G)^4\beta^7}&\sum_{v\in\Gamma} N(v)\frac{8}{E(v)} \sum_{\Delta(v)}\epsilon^{IJK}\Tr\Bigl(h_{s_I(\Delta)}^{-1}\bigl[h_{s_I(\Delta)},[C_E^{(\Delta')}(1),V_v]\bigr] \notag \\
&\times h_{s_J(\Delta)}^{-1}\bigl[h_{s_J(\Delta)},[C_E^{(\Delta)}(1),V_v]\bigr] h_{s_K(\Delta)}^{-1}\bigl[h_{s_K(\Delta)},V_v\bigr]\Bigr)\ket{\Psi_\Gamma}, \label{C_L^Delta}
\end{align}
where the notation $\Delta'$ indicates that the triangulation used to define the second action of the Euclidean operator should be adapted to the graph created by the first action of $C_E^{(\Delta)}(1)$ on the state $\ket{\Psi_\Gamma}$. 

\newpage

In \Eqs{C_E^Delta} and \eqref{C_L^Delta}, we have regularized the connection and the curvature using holonomies in the fundamental representation, but there is of course room to generalize the construction in various ways. For example, one could use holonomies in a higher irreducible representation (first considered in \cite{GaulRovelli}) or linear combinations of holonomies in different representations \cite{LinComb}, or even allow the representation of the holonomy $h_{\alpha_{IJ}}$ to depend on the spins carried by the edges $e_I$ and $e_J$ on which the loop $\alpha_{IJ}$ is attached \cite{YangMa, Cong}.

As they stand, the operators defined by \Eqs{C_E^Delta} and \eqref{C_L^Delta} still depend on the triangulation $\Delta$. To complete the construction of a uniquely defined constraint operator, one would therefore like to remove the regulator by taking the limit $\epsilon\to 0$. However, if one attempts to take this limit in some way in the kinematical Hilbert space, one immediately encounters the problem already familiar from \Eqs{A-op-limit} and \eqref{diff-limit}. For example, no meaningful limit of $C_E^{(\Delta)}(N)$ (or $C_L^{(\Delta)}(N)$) can be defined by considering limits of matrix elements of the operator in the spin network basis, since a matrix element of the form $\bra{\Phi_{\Gamma'}}C^{(\Delta)}(N)\ket{\Psi_\Gamma}$ will be different from zero for at most one value of $\epsilon$.

At a first sight the situation seems to be more promising in the diffeomorphism invariant Hilbert space, as the triangulations used to obtain \Eqs{C_E^Delta} and \eqref{C_L^Delta} can be constructed in such a way that the loops $\alpha_{IJ}(\Delta)$ and $\alpha_{IJ}(\Delta')$, associated to triangulations characterized by two different values $\epsilon$ and $\epsilon'$ of the regularization parameter, are diffeomorphically equivalent with each other. In other words, there always exists a diffeomorphism $\phi$ such that $C^{(\Delta')}(N)\ket{\Psi_\Gamma} = U(\phi)C^{(\Delta)}(N)\ket{\Psi_\Gamma}$. (See again \cite{QSD} for the details on how this can be achieved.) Then, if $\xi\in{\cal H}_{\rm diff}$ is any diffeomorphism invariant state, the number $\xi\bigl(C^{(\Delta)}(N)\ket{\Psi_\Gamma}\bigr)$ is in fact independent of $\epsilon$, so if understood in this sense, the limit $\epsilon\to 0$ becomes completely trivial\footnote{To be more precise, the limit $\epsilon\to 0$ is defined according to the following criterion \cite{Thiemann}: For a given $\delta>0$, there must exist an $\epsilon(\delta)$ such that
\[
\bigl|\xi\bigl(C^{(\Delta_2)}(N)\ket{\Psi_\Gamma}\bigr) - \xi\bigl(C^{(\Delta_1)}(N)\ket{\Psi_\Gamma}\bigr)\bigr| < \delta
\] 
whenever $|\epsilon_2-\epsilon_1|<\epsilon(\delta)$. If this requirement is satisfied, the limit operator can be defined simply as the operator $C^{(\Delta)}(N)$ for an arbitrary but fixed value of $\epsilon$. This way of defining the limit $\epsilon\to 0$ is often referred to as the uniform Rovelli--Smolin topology. \label{URST}}. However, due to the presence of the lapse function, $C(N)$ itself cannot be consistently defined as an operator on ${\cal H}_{\rm diff}$ -- at most one can define a non-diffeomorphism invariant operator $C(N): {\cal H}_{\rm diff} \to {\rm Cyl}^*$.

Finally, a point of view advocated by Thiemann in \cite{Thiemann}, which completely bypasses the question of taking the limit $\epsilon\to 0$, is to consider $\epsilon$ as a fixed parameter (by choosing once and for all a triangulation for each possible graph), and view the operator defined by \Eqs{C_E^Delta} and \eqref{C_L^Delta} as a kind of an effective operator. When evaluated on semiclassical states, the effective operator is supposed to reproduce the regularized classical expressions \eqref{C_E-reg} and \eqref{C_L-reg}, rather than the continuum expressions \eqref{C_E-brac} and \eqref{C_L-brac}. In this approach the continuum limit would correspond to evaluating the operator on states based on finer and finer graphs, as opposed to taking the regularization parameter to zero on a single, fixed graph.

\subsection{Action of the operator} 

The operator defined by \Eqs{C_E^Delta} and \eqref{C_L^Delta} acts by creating new nodes and edges around the previously existing nodes of a spin network state. As an illustration, the action of the Euclidean part \eqref{C_E^Delta} on a three-valent node has the schematic form
\be\label{C_E-drawing}
C_E\RealSymb{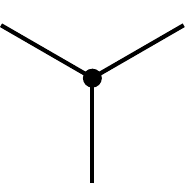}{0.6} \quad = \quad c_1\RealSymb{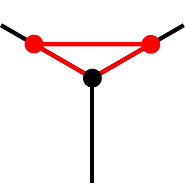}{0.6} \quad + \quad c_2\RealSymb{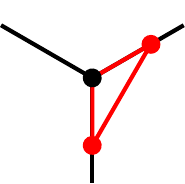}{0.6} \quad + \quad c_3\RealSymb{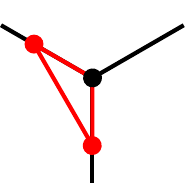}{0.6}
\ee
On the segments where the loop inserted by the operator overlaps an edge of the spin network, the holonomies should be coupled using the Clebsch--Gordan series \eqref{CGser}. For example, if the loop carries spin $\half$, then the action of the loop on an edge of spin $j$ produces two terms, in which the edge has spin $j-\half$ and $j+\half$. Therefore each drawing on the right-hand side of \Eq{C_E-drawing} actually represents multiple terms (four, in the case where the loop has spin $\half$), corresponding to the different combinations of spins that result from coupling the overlapping holonomies.

The Lorentzian part \eqref{C_L^Delta} contains two factors of the Euclidean part, and hence acts by creating two loops, producing terms of the form
\be\label{C_L-drawing}
C_L\RealSymb{fig13-1.pdf}{0.6} \quad = \quad c_{11}\RealSymb{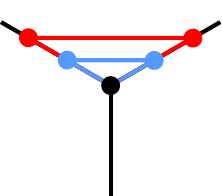}{0.6} \quad + \quad c_{12}\RealSymb{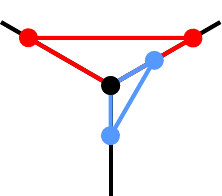}{0.6} \quad + \; \dots
\ee
Since the tetrahedra used to regularize the second factor of $C_E$ in \Eq{C_L^Delta} are contained inside the tetrahedra of the first factor of $C_E$, the second action of the Euclidean operator is responsible for creating the ''smaller'' loops in \Eq{C_L-drawing}. Both the Euclidean and Lorentzian operators naturally also act on the intertwiner at the node. In the case of a three-valent node, the change in intertwiner is of course rather trivial, but if the valence of the node is higher, the operators generally act in the intertwiner space in a non-trivial way.

The nodes created by the action of the Hamiltonian constraint are three-valent and planar. If the constraint operator is acting on a state which contains such nodes, the operator will not act on them, provided that the internally regularized Ashtekar--Lewandowski volume operator is used in \Eqs{C_E^Delta} and \eqref{C_L^Delta}. This is because both the Euclidean and the Lorentzian part have a volume operator ordered to the right, and any planar node is annihilated by the Ashtekar--Lewandowski volume operator\footnote{The action of the volume operator does not give zero just because the node is three-valent, since there are terms in \Eqs{C_E^Delta} and \eqref{C_L^Delta} in which the volume acts on a non-gauge invariant node, and the action of the volume on a non-gauge invariant three-valent node is equivalent to that on a four-valent gauge invariant node. Using the externally regularized volume operator (which is insensitive to the differential structure of the node) in \Eqs{C_E^Delta} and \eqref{C_L^Delta} would therefore give a constraint operator which does act on the nodes that it creates.}. It follows that if one repeatedly acts with the Hamiltonian constraint on a given state, each action of the operator will only act on the nodes present in the original state; the nodes created by the successive actions of the constraint will remain untouched. This property turns out to play a crucial role in ensuring that the commutator of the Hamiltonian constraint with itself is free of anomalies.

The matrix elements of the Hamiltonian constraint in the spin network basis, represented by the numbers $c_I$ and $c_{IJ}$ in \Eqs{C_E-drawing} and \eqref{C_L-drawing}, can be computed explicitly for nodes of sufficiently low valence, up to the fact that the matrix elements of the volume operator are not available in explicit form. The first such computations were done in \cite{Borissov_et_al} and \cite{GaulRovelli}, where the action of the Euclidean part on a three-valent node was derived using a somewhat complicated diagrammatic formalism. The more powerful graphical methods presented in the Appendix were utilized to calculate the matrix elements of the Euclidean part for three- and four-valent nodes in \cite{AlesciThiemannZipfel}, and those of the entire constraint, including the Lorentzian part, for three-valent nodes in \cite{AlesciLiegenerZipfel}.

\subsection{The constraint algebra}

An essential consistency requirement in the quantization of constrained theories concerns the commutators between the constraint operators in the quantum theory. Ideally one would like the commutators among the constraint operators to reproduce the algebra of the corresponding Poisson brackets between the classical constraints, but at the least the commutator between any two constraints should again be proportional to one of the constraints (or possibly a linear combination of them). If this were not the case, one should still require for consistency that physical states be annihilated by the commutator. In effect, the commutator would generate a new, independent constraint, and the space of solutions could not be consistently determined merely by solving the original set of constraints. Then one would naturally suspect that the set of physical states obtained in this way might be too small.

The commutator algebra of Thiemann's Hamiltonian constraint was investigated in \cite{QSD} and \cite{QSD3}; here we will summarize the results following the discussion in \cite{Thiemann}. Since the infinitesimal generator of diffeomorphisms is not available, the commutator between a diffeomorphism constraint and a Hamiltonian constraint should be tested in its finite, exponentiated form as
\be\label{phiCphi}
U^{-1}(\phi)C(N)U(\phi) = C(\phi^* N).
\ee
However, a straightforward calculation (which can be done on general grounds, without having to descend down to the detailed action of the Hamiltonian constraint in the spin network basis) shows that instead of the expected relation \eqref{phiCphi}, one has
\be
U^{-1}(\phi)C(N)U(\phi)\ket{\Psi_\Gamma} = U^{-1}(\varphi_{\phi,\Gamma})C(\phi^* N)U(\varphi_{\phi,\Gamma})\ket{\Psi_\Gamma}.
\ee
The diffeomorphism operators on the right-hand side compensate for the fact that action of the operator $C(N)$ on the left-hand side is defined through a triangulation adapted to the graph $\phi(\Gamma)$, and in general this triangulation is not necessarily mapped under $\phi^{-1}$ into the triangulation by which $C(N)$ is supposed to act on a state based on $\Gamma$ (though it certainly can be mapped into the correct triangulation by applying a further diffeomorphism).

Thus the conclusion is that the relation \eqref{phiCphi} is not precisely satisfied -- however, the discrepancy is harmless since it is given solely in terms of diffeomorphism operators, and hence is not visible at the diffeomorphism invariant level. In particular, \Eq{phiCphi} is a valid equality between operators in the sense of the uniform Rovelli--Smolin topology of footnote \ref{URST}. (Figuratively speaking, we have here an example of the case where the commutator between two constraints is again a constraint, but not the exact constraint one would expect based on the classical algebra of the constraints.) 

As to the commutator between two Hamiltonian constraints, a direct calculation -- a central ingredient of which is the property that the Hamiltonian constraint does not act on the nodes created by a previous action of the constraint -- produces a result of the form 
\be\label{[C,C]Th}
[C(M),C(N)]\ket{\Psi_\Gamma} = \frac{1}{2}\sum_{v\neq v'\in\Gamma} \bigl(M(v)N(v') - M(v')N(v)\bigr)\bigl(U(\phi_{v,v'}) - U(\phi_{v',v})\bigr)C_vC_{v'}\ket{\Psi_\Gamma},
\ee
where $C_v$ denotes the part of the constraint that acts on the node $v$, \ie the coefficient of $N(v)$ in \Eqs{C_E^Delta} and \eqref{C_L^Delta}. Since the right-hand side involves the difference between two diffeomorphism operators, the commutator is identically zero at the diffeomorphism invariant level (according to the criterion of footnote \ref{URST}). Therefore the algebra of commutators is consistent, provided that the Ashtekar--Lewandowski volume operator, rather than the Rovelli--Smolin operator, is used to construct the Hamiltonian constraint. 

The appearance of a difference between diffeomorphisms in \Eq{[C,C]Th} further suggests that the right-hand side of \eqref{[C,C]Th} might have something to do with the right-hand side of the classical Poisson bracket \eqref{C,C ADM}. In \cite{QSD3} it was indeed shown how the right-hand side of \eqref{[C,C]Th} can be interpreted as a quantization of the function $C_a\bigl(q^{ab}(M\partial_b N - N\partial_b M)\bigr)$.

\subsection{Symmetrizing the constraint operator}

The operator considered in this chapter is evidently not symmetric, since it acts by creating loops, while a symmetric operator should be a sum (or some other symmetric combination) of an operator that creates loops and one that removes them. However, as far as the Hamiltonian constraint is concerned, it is strictly speaking not necessary for the operator to be symmetric. Even though real-valued classical observables should normally be promoted to self-adjoint operators upon quantization, the Hamiltonian constraint is not really an ''observable'', at least in the literal sense of being a quantity whose value can be observed by making measurements. In the quantum theory, the Hamiltonian constraint only plays a role in determining the physical states and physical observables, and therefore the mathematical consistency of the theory is not necessarily endangered even if the constraint operator is not symmetric. From this perspective, a sufficient requirement for a consistent constraint operator is that the eigenvalue zero be contained in its spectrum.

It has even been argued in the literature that in the case of self-adjoint constraint operators, there is a risk of the commutator algebra of the constraints becoming anomalous, and therefore it is better for constraint operators to not be symmetric \cite{ref18ofpaper2, Thiemann}. However, in the author's opinion it is not clear how seriously such arguments should be taken in the context of loop quantum gravity. The usual argument calling for the constraint operators to be non-symmetric (see \eg Chapter 30 of \cite{Thiemann}) assumes that the classical Poisson brackets $\{C_\alpha,C_\beta\} = f\downup{\alpha\beta}{\gamma}C_\gamma$ (where the Greek indices enumerate the constraints, and $f\downup{\alpha\beta}{\gamma}$ are the structure functions of the constraint algebra) are promoted in the quantum theory into the formally identical commutation relations $[C_\alpha,C_\beta] = if\downup{\alpha\beta}{\gamma}C_\gamma$, or if the operators involved in the equation are self-adjoint,
\be
[C_\alpha,C_\beta] = \frac{i}{2}\bigl(f\downup{\alpha\beta}{\gamma}C_\gamma + C_\gamma f\downup{\alpha\beta}{\gamma}\bigr).
\ee
Here a possible anomaly may clearly arise from the second term on the right. However, as we have seen, in loop quantum gravity the commutation relations between the constraints are not exactly identical to the classical Poisson brackets, especially for relations involving the diffeomorphism constraint. Therefore it would seem better to try to resolve the question of anomalies separately in each case, regardless of whether the constraint operators are symmetric or not, rather than categorically refuse to consider symmetric constraint operators merely on grounds of the above formal argument.

On the other hand, the Hamiltonian constraint is relevant not only as a constraint operator in the fully constrained theory, but also as a building block for constructing the physical Hamiltonian operator in deparametrized models. In the deparametrized context it is of course vitally important for the physical Hamiltonian to be symmetric (and eventually self-adjoint) in order to ensure that the time evolution generated by it is unitary. Hence the viability of using Thiemann's Hamiltonian as the physical Hamiltonian in a deparametrized model depends on whether it can be symmetrized. This, in turn, boils down to the question of whether the adjoint operator of Thiemann's Hamiltonian exists as a densely defined operator. If the adjoint operator $C^\dagger(N)$ were available, it would act by removing loops, and a symmetric Hamiltonian could be constructed as $\half\bigl(C(N)+C^\dagger(N)\bigr)$, or $\sqrt{C^\dagger(N)C(N)}$, or any other symmetric combination of $C(N)$ and its adjoint.

It turns out that any operator which acts by adding loops of fixed spin in the way indicated by \Eqs{C_E-drawing} and \eqref{C_L-drawing} cannot possess a densely defined adjoint operator, neither on the kinematical Hilbert space nor at the diffeomorphism invariant level. The problematic situation is encountered whenever the operator acts on an edge whose spin matches the spin of the loops created by the operator. When the holonomies on the segment where the loop overlaps the edge are coupled using the Clebsch--Gordan series, the total spin zero will appear among the spins resulting from the coupling. In the corresponding spin network state, a segment of the edge will effectively have been erased. For example, when the loop and the edge both carry spin $1/2$, the action of the Euclidean operator has the structure
\be
C_E \RealSymb{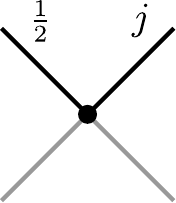}{0.6} \quad = \quad \RealSymb{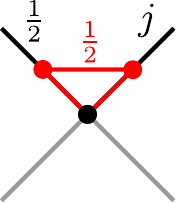}{0.6} \quad = \quad \RealSymb{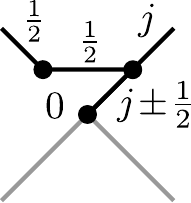}{0.6} \quad + \quad \RealSymb{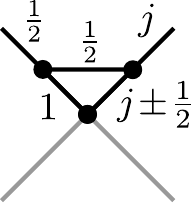}{0.6}   
\ee
In the first term on the right-hand side, all information about the differential relations between the erased edge and the remaining edges at the node has been lost. The same term also arises from the action of the Euclidean operator on different states, which differ from each other only in the way in which the edge whose segment is erased meets the other edges at the node:
\begin{align}
C_E \RealSymb{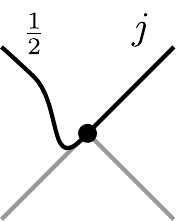}{0.6} \quad &= \quad  \RealSymb{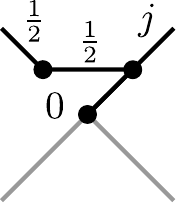}{0.6} \quad + \quad \RealSymb{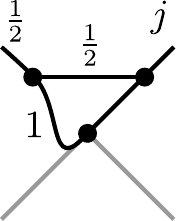}{0.6} \\
C_E \RealSymb{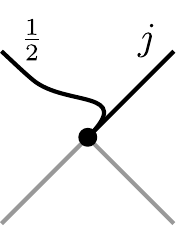}{0.6} \quad &= \quad  \RealSymb{fig13-8b.pdf}{0.6} \quad + \quad \RealSymb{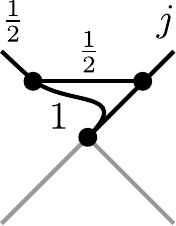}{0.6} \\
&\text{etc.} \notag
\end{align}
It follows that the action of the adjoint operator $C_E^\dagger$ on the state in which the edge has been erased is ill-defined, because -- so to say -- the adjoint operator ''does not know how it should put back the erased edge''. If one tries to act with the adjoint operator on this state, one would obtain an infinite, unnormalizable linear combination of states, in which the missing edge has been re-inserted in all possible ways:
\be\label{adjoint}
C_E^\dagger \RealSymb{fig13-8b.pdf}{0.6} \quad = \quad \RealSymb{fig13-7a.pdf}{0.6} \quad + \quad \RealSymb{fig13-10.pdf}{0.6} \quad + \quad \RealSymb{fig13-12.pdf}{0.6} \quad + \; \dots
\ee
In the kinematical Hilbert space, there are certainly uncountably infinite different ways to put back the erased edge. Even in the diffeomorphism invariant Hilbert space, the number of inequivalent ways in which the missing edge can be reintroduced is at least countably infinite, corresponding to the different orders of tangentiality (see section \ref{sec:special}) between the erased edge and the other edges at the node. If the valence of the node is sufficiently high, the number of terms on the right-hand side of \Eq{adjoint} could be uncountably infinite even at the diffeomorphism invariant level.

These considerations show that the adjoint of Thiemann's Hamiltonian is not available as a densely defined operator, since its action is ill-defined on a large class of spin network states. For this reason, it is not possible to obtain a symmetric operator from Thiemann's construction, and so the operator is not suitable to be used in constructing physical Hamiltonians for deparametrized models of loop quantum gravity. For deparametrized models, a loop assignment different from the one illustrated by \Eqs{C_E-drawing} and \eqref{C_L-drawing} is needed in order to ensure that a satisfactory adjoint operator exists and a symmetric physical Hamiltonian can be constructed. (On the other hand, we emphasize that Thiemann's operator is of course fully adequate as the Hamiltonian constraint operator, which is the purpose for which it was originally designed.)

\section[Physical Hamiltonian for LQG coupled to a free scalar field]{Physical Hamiltonian for loop quantum gravity \\ coupled to a free scalar field}\label{ch:Hphys}

In this chapter we will present an explicit construction of a physical Hamiltonian operator for loop quantum gravity deparametrized with respect to the free Klein--Gordon field. In comparison to Hamiltonians proposed in the literature so far, our construction is characterized by two important, essentially new features:
\begin{itemize}
\item A new regularization is used to quantize the Euclidean part of the Hamiltonian. This results in an operator which can be symmetrized, and which can therefore be a mathematically consistent candidate for the generator of physical time evolution.
\item The usual Lorentzian part of the Hamiltonian is traded for the scalar curvature of the spatial surface by using the expression \eqref{C-AE-2} for the Hamiltonian constraint. This leads to a considerable practical simplification in the properties of the resulting quantum operator.
\end{itemize}
The work on which this chapter is based was carried out by the author in collaboration with Emanuele Alesci, Mehdi Assanioussi and Jerzy Lewandowski, and has been published in the article \cite{paper1}. It can be regarded as a precise and concrete realization of the model first proposed by Rovelli and Smolin in \cite{RovelliSmolin93} and completed on a partially formal level in \cite{quantized}. 

The material in this chapter is organized as follows. The general strategy which we will adopt to define the physical Hamiltonian operator is outlined in section \ref{sec:strategy}. Before descending down to the technicalities involved in constructing the operator, we will explain the key ideas essential to our construction in sections \ref{sec:special} and \ref{sec:curvature}. The precise details of regularizing the classical expression for the physical Hamiltonian and constructing the corresponding quantum operator are then given in sections \ref{sec:reg-E} and \ref{sec:reg-R}. The outcome of the construction is summarized in the concluding section \ref{sec:Hphys-summ}.

\subsection{The general strategy}\label{sec:strategy}

In the classical theory of gravity coupled to a free scalar field, evolution of diffeomorphism invariant observables with respect to the relational time provided by the scalar field is generated by the physical Hamiltonian
\be\label{hphys-1}
h_{\rm phys} = \int d^3x\,\sqrt{-2\sqrt q C_{\rm gr}}.
\ee
Accordingly, the dynamics of the quantum theory will be governed by an operator obtained by quantizing the function \eqref{hphys-1}. In principle, the Hamiltonian \eqref{hphys-1} could be quantized (roughly speaking) by promoting the factor $\sqrt q$ into the volume operator, and using an appropriate modification of Thiemann's construction to quantize the factor $C_{\rm gr}$. However, an essential ingredient of our construction is to manipulate the classical expression \eqref{hphys-1} in a beneficial way before proceeding to quantize it. When \Eq{C-AE-2}, instead of the more commonly used expression \eqref{C-AE-1} for the gravitational Hamiltonian constraint, is inserted into \Eq{hphys-1}, the factor inside the square root takes the form
\be\label{hphys-2}
-2\sqrt q C_{\rm gr} = \frac{1}{8\pi G\beta^2}\Bigl(\epsilon\updown{ij}{k}E^a_iE^b_jF_{ab}^k + (1+\beta^2)(\det E){}^{(3)}\!R\Bigr).
\ee
We see that the problematic factor of $1/\sqrt{\det E}$ has been cancelled in the first term, and in fact a similar cancellation occurs in the second term after it has been regularized by expressing the Ricci scalar in terms of the elementary variables of loop quantum gravity.

A few words about terminology: We will keep referring to the first term on the right-hand side of \Eq{hphys-2} as the Euclidean term, or Euclidean part, since up to a simple multiplicative factor it is identical to the Euclidean part of the Hamiltonian constraint. The second term in \Eq{hphys-2} will be referred to as the curvature term, or curvature part, in order to emphasize that it is genuinely different from the Lorentzian part of the usual expression \eqref{C-AE-1}. This difference is reflected accordingly in the properties of the corresponding quantum operators, as the curvature term in \Eq{hphys-2} can be quantized in such a way that the structure of the resulting operator is extraordinarily simple in comparison to the Lorentzian part of Thiemann's Hamiltonian. Replacing the Lorentzian term with the curvature term is therefore far from being a mere academic exercise -- rather, it offers a very concrete technical advantage, as measured by the degree of complexity of the resulting quantum operator.

The function defined by \Eqs{hphys-1} and \eqref{hphys-2} has the form
\be\label{int a+b}
\int d^3x\,\sqrt{a(x) + b(x)},
\ee
where $a(x)$ and $b(x)$ are functions of the elementary fields $A_a^i$ and $E^a_i$. In order to construct the physical Hamiltonian operator, we must therefore consider how to regularize and quantize an expression of this type. By making use of a cellular decomposition of the spatial manifold $\Sigma$, whose cells are denoted by $\Delta$, the integral \eqref{int a+b} can be approximated as
\be\label{int a+b reg}
\int d^3x\,\sqrt{a(x) + b(x)} \simeq \sum_\Delta \sqrt{\biggl(\int_\Delta d^3x\,\sqrt{a(x)}\biggr)^2 + \biggl(\int_\Delta d^3x\,\sqrt{b(x)}\biggr)^2}.
\ee
This expression suggests how to define an operator corresponding to \eqref{int a+b}, provided that the integrals $\int_\Delta d^3x\,\sqrt{a(x)}$ and $\int_\Delta d^3x\,\sqrt{b(x)}$ can be given a meaning as well-defined operators.

Note, however, that for the square root in \Eq{int a+b} to be well-defined, it is sufficient that the sum $a(x)+b(x)$ is positive definite. The functions $a(x)$ and $b(x)$ do not necessarily have to be separately positive-definite, even though this might seem to be suggested by the right-hand side of \Eq{int a+b reg}. Therefore \Eq{int a+b reg} should be considered as a partially formal expression, its only purpose being to schematically indicate the way in which we derive a quantum operator starting from the classical physical Hamiltonian given by \Eqs{hphys-1} and \eqref{hphys-2}.

If the operators $\sqrt{a(x)}$ and $\sqrt{b(x)}$ are themselves available in the form\footnote{Even though we have not used such notation so far in this work, most of the operators one commonly encounters in loop quantum gravity can nevertheless be expressed in this form. To give a concrete example, the volume operator associated to a region $R$ can be written as
\[
V(R) = \int_R d^3x\,\sqrt{q(x)},
\]
which is formally identical with the classical volume of the region, and where the ''volume element operator'' is
\[
	\sqrt{q(x)} = \sum_{v\in\Sigma} \delta^{(3)}(x,v)\sqrt{|q_v|},
\]
with $q_v$ given by \Eq{q_v}. It is straightforward to check that when the operator defined above acts on a state based on a given graph, the more conventional expression
\[
V(R)\ket{\Psi_\Gamma} = \sum_{v\in\Gamma\cap R} \sqrt{|q_v|}\,\ket{\Psi_\Gamma} 
\]
for the volume operator is recovered.}
\begin{align}
\sqrt{a(x)} &= \sum_{v\in\Sigma} \delta(x,v) \sqrt{a_v}, \label{a(x)op} \\
\sqrt{b(x)} &= \sum_{v\in\Sigma} \delta(x,v) \sqrt{b_v}, \label{b(x)op}
\end{align}
where $a_v$ and $b_v$ act only on the nodes of a spin network state, then the function \eqref{int a+b} can be quantized immediately by inserting \Eqs{a(x)op} and \eqref{b(x)op} into the right-hand side of \Eq{int a+b reg}. When deriving the action of the operator on a spin network state, the cellular decomposition used in \Eq{int a+b reg} has to be refined only so far that each cell $\Delta$ contains at most a single node of the spin network. Any further refinement will not make a difference to the action of the operator, which is given by
\be
\int d^3x\,\widehat{\sqrt{a(x) + b(x)}}\,\ket{\Psi_\Gamma} = \sum_{v\in\Gamma} \sqrt{a_v + b_v}\,\ket{\Psi_\Gamma}.
\ee

\subsection{Special loops}\label{sec:special}

The reason why Thiemann's Hamiltonian is not a suitable candidate for the physical Hamiltonian in the deparametrized context is that the adjoint of the Hamiltonian is not a densely defined operator, making it impossible to symmetrize Thiemann's operator and to eventually obtain a self-adjoint physical Hamiltonian. This is the main obstacle which must be overcome in our construction of the physical Hamiltonian.

The problems with the adjoint of Thiemann's Hamiltonian are ultimately caused by the fact that the loops created by the operator overlap partially with the edges to which they are attached. In order to avoid such problems in our construction, we must therefore adopt a loop assignment in which the loops do not overlap with any edges. However, it then becomes a non-trivial question whether one can tell apart the loop from other edges or loops at the node  -- in particular, whether loops attached to different pairs of edges at the same node, or successive loops attached to the same pair of edges by repeated actions of the Hamiltonian, can be distinguished from each other.

It turns out that all these questions can be answered in the affirmative by making use of a diffeomorphism invariant notion of order of tangentiality between two edges sharing a node. This is defined as follows. Suppose $e_1$ and $e_2$ are two (analytic) edges incident at a node $v$. Then one can choose coordinates in a neighborhood of $v$ such that the edges are parametrized as
\begin{align}
e_1(t) &= (t,0,0), \\
e_2(t) &= \bigl(t,0,f(t)\bigr),
\end{align}
and the node is at $t=0$. If the function $f(t)$ and its first $n$ derivatives vanish at $t=0$, while the $(n+1)$-th derivative does not, we say that $e_1$ and $e_2$ are tangent to each other at order $n$ at the node $v$.

When defining the Euclidean part of the Hamiltonian, the order of tangentiality between the loop created by the operator and the edges to which it is attached is essentially a free parameter in the construction. By exploiting this freedom, we can define the operator in such a way that the loop attached to a given pair of edges is perfectly distinguishable from any other loop which was present at the node before the new loop was attached. In short, this is done by specifying a sufficiently high order of tangentiality between a loop $\alpha_{IJ}$ and the two edges $e_I$ and $e_J$ to which it is attached. For example, denote by $T(e)$ the highest order of tangentiality between an edge $e$ and the other edges at the node before the new loop was introduced. If the loop $\alpha_{IJ}$ is tangent to the edges $e_I$ and $e_J$ (at least) at orders $T(e_I)+1$ and $T(e_J)+1$ respectively, then the loop is clearly distinguished by the fact that all other edges at the node are tangent to $e_I$ and $e_J$ at a lower order than the loop.

Following the terminology introduced in \cite{paper1}, we refer to loops of the type described above as special loops, and we will use them to define the Euclidean part of the physical Hamiltonian. This element of the construction plays a key role in ensuring that the operator possesses a satisfactory adjoint. When the Euclidean operator creates loops according to the special loop assignment, the case never arises where the operator would remove some structures from a node, causing the adjoint operator to re-introduce these structures in an infinite number of inequivalent ways. In fact, by making a suitable choice of the order of tangentiality between the special loop and the edges to which it is attached, one can guarantee that in any node containing an arbitrary number of special loops, there will always be only a single, uniquely determined loop, which will be removed by the action of the adjoint operator.

\begin{figure}[t]
	\centering
		\includegraphics[width=0.2\textwidth]{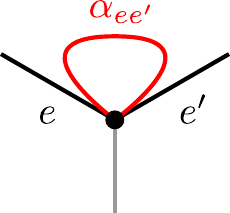}
		\caption{A special loop $\alpha_{ee'}$ attached to the edges $e$ and $e'$. The loop does not overlap with the two edges, but is tangent to them at a sufficiently high order, so that the loop is uniquely distinguished from any other edges present at the node.}
	\label{fig14-1}
\end{figure}

\subsection{The curvature operator}\label{sec:curvature}

Our quantization of the curvature term in \Eq{hphys-2} is based on the ideas of \cite{curvature}, where a loop quantum gravity operator corresponding to the integrated scalar curvature of the spatial manifold, 
\be\label{R-int}
\int d^3x\,\sqrt q\,{}^{(3)}\!R,
\ee
was constructed. The main point behind the construction is that the integral \eqref{R-int} happens to be the action integral for Euclidean general relativity in three dimensions. Using the well-known framework of Regge calculus \cite{regge}, the action integral (for arbitrary spacetime dimension $n$) can be related to the geometric quantities of a piecewise flat simplicial decomposition of the spacetime manifold, in which curvature is concentrated on the ''hinges'' of the decomposition, \ie the $(n-2)$-dimensional ''sub-simplices'' at which several $n$-dimensional simplices meet.

In the three-dimensional case, the precise relation is
\be\label{regge}
\int_{\Sigma_\Delta} d^3x\,\sqrt q\,{}^{(3)}\!R= 2\sum_h L_h\delta_h,
\ee
where the integral on the left-hand side is taken over the triangulated, piecewise flat manifold $\Sigma_\Delta$, which is regarded as an approximation of the smooth spatial manifold $\Sigma$. The sum on the right-hand side runs over the hinges, or the edges of the tetrahedra of the triangulation. Within the sum, $L_h$ is the length of the hinge $h$, and
\be
\delta_h = 2\pi - \sum_{\text{$\Delta$ at $h$}} \varphi_h(\Delta),
\ee
is the deficit angle associated to the hinge, with $\varphi_h(\Delta)$ the dihedral angle of the tetrahedron $\Delta$ at $h$. In Regge's original formulation, only simplicial decompositions of the spacetime manifold were considered; however, as argued in \cite{curvature}, \Eq{regge} is a valid approximation to the continuum integral \eqref{R-int} also for a certain class of arbitrary piecewise flat cellular decompositions of $\Sigma$.

The relevance of Regge's formula to loop quantum gravity arises from the fact that, as we have seen in Chapter \ref{ch:geometric}, well-defined operators describing angles and lengths are available in loop quantum gravity, and therefore \Eq{regge} indicates a very straightforward way in which the integral \eqref{R-int} can be promoted into an operator in the theory. To this end, it is convenient to split the sum in \Eq{regge} as 
\be\label{regge-sum}
\sum_h L_h\delta_h = \sum_{\Delta} \sum_{h\in\Delta} L_h\biggl(\frac{2\pi}{\alpha_h} - \varphi_h(\Delta)\biggr),
\ee
where the first sum runs over the cells of a suitable cellular decomposition of $\Sigma$, the inner sum runs over the hinges of the cell $\Delta$, and $\alpha_h$ denotes the number of cells sharing the hinge $h$. Moreover, the dihedral angle is given by
\be
\varphi_h(\Delta) = \pi - \theta_h(\Delta),
\ee
where $\theta_h(\Delta)$ is the angle between the two faces of $\Delta$ which intersect each other at $h$. 

With these preparations, the sum \eqref{regge-sum} has been expressed entirely in terms of quantities which are directly analogous to well-known operators already available in loop quantum gravity. Therefore \Eq{regge} could be naively quantized simply by placing hats over the quantities $L_h$ and $\theta_h(\Delta)$, which clearly correspond to the length and angle operators described in Chapter \ref{ch:geometric}. The sums $\sum_\Delta$ and $\sum_{h\in\Delta}$ in \Eq{regge-sum} are expected to be replaced with a sum over the nodes of a spin network state, and one over the pairs of edges at each node, since the geometric picture associated with a spin network state suggests that a pair of edges emerging from a node defines a hinge through the intersection of the elementary surfaces dual to the two edges.

The result of the naive quantization is confirmed, up to a certain numerical factor at each node, by the detailed construction of \cite{curvature}, in which the curvature operator is defined through a careful regularization of the relevant classical expressions. The resulting operator acts on a state based on a graph $\Gamma$ as
\be\label{R-op}
\int d^3x\,\widehat{\sqrt q\,{}^{(3)}\!R}\,\ket{\Psi_\Gamma} = \sum_{v\in\Gamma} R_v\ket{\Psi_\Gamma}
\ee
with
\be\label{R_v}
R_v\ket{\Psi_\Gamma} = \kappa_R(v)\sum_{(e,e')\in v}\biggl(\frac{2\pi}{\alpha_{ee'}}L_v^{(e,e')} - \frac{1}{2}\Bigl(L_v^{(e,e')}\varphi_v^{(e,e')} + \varphi_v^{(e,e')}L_v^{(e,e')}\Bigr)\biggr)\ket{\Psi_\Gamma}.
\ee
The sum in the above equation runs over all pairs of edges $(e,e')$ at the node, $L_v^{(e,e')}$ is the length operator of section \ref{sec:length}, and the dihedral angle operator is
\be
\varphi_v^{(e,e')} = \pi - \theta_v^{(e,e')},
\ee
with $\theta_v^{(e,e')}$ the angle operator of section \ref{sec:angle}. Furthermore, $\kappa_R(v)$ is a coefficient which results from an averaging over the cellular structures involved in the regularization, and whose value depends only on the valence of the node $v$. Essentially, $\kappa_R(v)$ arises to compensate for the fact that some pairs of edges in the sum of \Eq{R_v} generally do not correspond to any hinge in the sum \eqref{regge-sum} (see \cite{curvature} for further details).

In \Eq{R_v} a straightforward symmetric ordering has been chosen between the non-commuting operators $L_v^{(e,e')}$ and $\varphi_v^{(e,e')}$, following the choice originally made in \cite{curvature}. However, other symmetrizations of this term could certainly be considered. An alternative symmetrization can be obtained by recalling from \Eq{L_v} that the length operator itself is a product of several factors. By ordering these factors differently relative to the angle operator, the term $L_v^{(e,e')}\varphi_v^{(e,e')}$ can be symmetrized as
\be\label{curv-sym2}
\frac{1}{2}(8\pi\beta G)^2\sqrt{\displaystyle {\cal V}_v^{-1}}\bigl|Y_v^{(e,e')}\bigr|\varphi_v^{(e,e')}\sqrt{\displaystyle {\cal V}_v^{-1}},
\ee
where
\be
\bigl|Y_v^{(e,e')}\bigr| \equiv \sqrt{\displaystyle Y_i^{(v,e,e')}Y_{\phantom{i}}^{(v,e,e')i}},
\ee
and the operators $\bigl|Y_v^{(e,e')}\bigr|$ and $\varphi_v^{(e,e')}$ commute\footnote{This can be seen for example from \Eqs{angle-eq} and \eqref{Y^2-eq}, which show that the two operators share the same set of eigenstates.}, so there is no ambiguity in their relative ordering. Under such a symmetrization, the action of the curvature operator on a node of a spin network state takes the form
\be\label{R-sym}
R_v\ket{\Psi_\Gamma} = (8\pi\beta G)^2\kappa_R(v)\sum_{(e,e')\in v}\frac{1}{2}\sqrt{\displaystyle {\cal V}_v^{-1}}\bigl|Y_v^{(e,e')}\bigr|\biggl(\frac{2\pi}{\alpha_{ee'}} - \varphi_v^{(e,e')}\biggr)\sqrt{\displaystyle {\cal V}_v^{-1}}\ket{\Psi_\Gamma}.
\ee
In the author's opinion, \Eq{R-sym} seems to provide a somewhat simpler and perhaps more natural definition of the curvature operator. In particular, this definition guarantees that nodes carrying zero volume are annihilated by the curvature operator, which is a very reasonable property in light of the fact that the classical expression \eqref{R-int}, whose quantization the curvature operator is supposed to be, certainly vanishes when the integral is taken over a region of zero volume. 

\subsection{Regularization: Euclidean part}\label{sec:reg-E}

To construct the Euclidean part of the physical Hamiltonian according to the strategy outlined in section \ref{sec:strategy} and summarized by \Eq{int a+b reg}, we must consider how to quantize the integral
\be\label{Eucl-int}
\int d^3x\,\sqrt{\epsilon\updown{ij}{k}E^a_iE^b_jF_{ab}^k}.
\ee
As always, we should start by expressing the integral in terms of the elementary variables of loop quantum gravity. To illustrate the way in which this is done, let us introduce a cubic partition of the spatial manifold $\Sigma$, in which the edges of the cubes have coordinate length $\epsilon$. Approximating the integral \eqref{Eucl-int} by the Riemann sum associated to the partition, and choosing a point $x_\Box$ inside each cube $\Box$, we may write
\begin{align}
\int d^3x\,\sqrt{\epsilon\updown{ij}{k}E^a_iE^b_jF_{ab}^k} &\simeq \sum_\Box \epsilon^3 \sqrt{\epsilon\updown{ij}{k}E^a_i(x_\Box)E^b_j(x_\Box)F_{ab}^k(x_\Box)} \notag \\
&= \sum_\Box \sqrt{\epsilon\updown{ij}{k}\bigl(\epsilon^2E^a_i(x_\Box)\bigr)\bigl(\epsilon^2E^b_j(x_\Box)\bigr)\bigl(\epsilon^2F_{ab}^k(x_\Box)\bigr)}, \label{Eucl-sum}
\end{align}
where on the second line the factors of $\epsilon$ have been distributed in a suggestive way, showing how the triads and the curvature will be transformed into fluxes and holonomies.

For the factors involving the triads, we will use the parallel transported flux variable
\be\label{Eucl-flux}
E_i^{(x_\Box)}(S_\Box^a) = -2\Tr\biggl(\int_{S_\Box^a} d^2\sigma\,n_a(\sigma)\,h_{x_\Box\leftarrow x(\sigma)}E^a\bigl(x(\sigma)\bigr)h_{x(\sigma)\leftarrow x_\Box}\biggr),
\ee
in which $S_\Box^a$ is either one of the two sides of $\Box$ on which the coordinate $x^a = {\rm const.}$ Points on the surface are transported to the point $x_\Box$ along a fixed system of paths (for example, along straight lines in the background coordinates). The expansion of \eqref{Eucl-flux} in powers of $\epsilon$ reads
\be
E_i^{(x_\Box)}(S_\Box^a) = \epsilon^2 E^a_i + {\cal O}(\epsilon^3),
\ee
and at leading order in $\epsilon$, $E^a_i$ can be evaluated at any point inside $\Box$. Hence the factors $\epsilon^2E^a_i(x_\Box)$ in \Eq{Eucl-sum} can be regularized simply by replacing them with $E_i^{(x_\Box)}(S_\Box^a)$.

In order to regularize the factor involving the curvature, let us specify a set of loops $\{\alpha(S_\Box^a)\}$, one for each surface $S_\Box^a$, such that the loop $\alpha(S_\Box^a)$ lies within the corresponding surface $S_\Box^a$ and has coordinate area $\epsilon^2/2$. By \Eq{h-loop}, the holonomy around the loop $\alpha(S_\Box^a)$ is related to the curvature by
\be
h_{\alpha(S_\Box^a)} = 1 - \frac{\epsilon^2}{4}\epsilon^{abc}F_{bc}^i\tau_i + {\cal O}(\epsilon^3),
\ee
where again $F_{bc}^i$ can be evaluated at any point inside $\Box$. The factor involving the curvature in \Eq{Eucl-sum} can therefore be extracted from the holonomy $h_{\alpha(S_\Box^a)}$ as
\be
\epsilon^2F_{ab}^i = 2\epsilon_{abc}\Tr\bigl(\tau^ih_{\alpha(S_\Box^c)}\bigr) + {\cal O}(\epsilon^3).
\ee
More generally, we may consider regularizing the curvature in terms of a holonomy carrying an arbitrary spin $l$. Recalling the normalization of the $SU(2)$ generators in the spin-$l$ representation,
\be
\Tr\bigl(\tau_i^{(l)}\tau_{\phantom{i}}^{(l)j}\bigr) = -\frac{1}{3}l(l+1)(2l+1)\delta_i^j,
\ee
we have
\be\label{e^2F}
\epsilon^2F_{ab}^i = 2\epsilon_{abc}\bigl(h_{\alpha(S_\Box^c)}^i\bigr)^{(l)} + {\cal O}(\epsilon^3),
\ee
where we introduced the shorthand notation
\be\label{h_alpha^l}
\bigl(h_{\alpha(S_\Box^c)}^i\bigr)^{(l)} \equiv -\frac{3}{W_l^2}\Tr\bigl(\tau^{(l)i}D^{(l)}(h_{\alpha(S_\Box^c)})\bigr)
\ee
with $W_l = \sqrt{l(l+1)(2l+1)}$. Using \Eq{e^2F} to replace the factor $\epsilon^2F_{ab}^k$ in \Eq{Eucl-sum}, we have managed to express the integral \eqref{Eucl-int} in terms of holonomies and fluxes as
\be
\int d^3x\,\sqrt{\epsilon\updown{ij}{k}E^a_iE^b_jF_{ab}^k} \simeq \sum_\Box \sqrt{2\epsilon\updown{ij}{k}\epsilon_{abc}E_i^{(x_\Box)}(S_\Box^a)E_j^{(x_\Box)}(S_\Box^b)\bigl(h_{\alpha(S_\Box^c)}^k\bigr)_{\phantom{i}}^{(l)}},
\ee
the sum on the right-hand side converging to the integral in the limit of an arbitrarily fine partition.

In our article \cite{paper1}, the integral \eqref{Eucl-int} was ultimately regularized by using a more general partition, in which the spatial manifold $\Sigma$ is decomposed into a family of cells $\Delta$ of coordinate size $\epsilon$. The boundary of each cell $\Delta$ is subdivided into surfaces $S_\Delta^I$, so that $\cup_I S_\Delta^I = \partial\Delta$. To each cell there is also assigned a family of paths $p_\Delta(x)$, which are labeled by points $x\in\partial\Delta$, and along which parallel transports will be defined from the boundary of $\Delta$ to a fixed point $x_\Delta$ inside $\Delta$. We furthermore specify a family of loops $\alpha_\Delta^K$ and a family of coefficients $\kappa_{IJK}^{(\Delta)}$ such that the sum
\be\label{Eucl-sum2}
\sum_\Delta \sqrt{\epsilon\updown{ij}{k}\sum_{IJK}\kappa_{IJK}^{(\Delta)} E_i^{(x_\Delta)}(S_\Delta^I)E_j^{(x_\Delta)}(S_\Delta^J)\bigl(h^k_{\alpha_\Delta^K}\bigr)_{\phantom{i}}^{(l)}},
\ee
where the parallel transported flux variables are defined using the paths $p_\Delta(x)$, converges to the integral \eqref{Eucl-int} in the limit where the coordinate size of the cells is taken to zero.

While each term inside the square root in \eqref{Eucl-sum2} can be immediately promoted into an operator on the kinematical Hilbert space ${\cal H}_{\rm kin}$, the resulting overall operator is highly ambiguous due to the large freedom available in choosing the structures used to write down the regularized expression \eqref{Eucl-sum2}. On the other hand, the operator defined by \eqref{Eucl-sum2} carries a dependence on the regularization parameter $\epsilon$, and the ambiguity involved in the choice of partitions can be significantly reduced by imposing certain conditions, which guarantee that the dual of the operator, defined by \Eq{Astar}, converges to a well-defined operator on the diffeomorphism invariant Hilbert space in the limit $\epsilon\to 0$ \cite{LS14, paper1}. (As the size of the loops $\alpha_\Delta^K$ depends on $\epsilon$, the limit $\epsilon\to 0$ is certainly not well-defined in the kinematical Hilbert space.)

Further conditions on the partition can be derived from requiring that the general structure of the operaror defined by \eqref{Eucl-sum2} is consistent with what one would expect on grounds of a naive, formal quantization of the term $\epsilon\downup{i}{jk}F_{ab}^iE^a_jE^b_k$. Applying the quantization rule $E^a_i(x) \to -i\,\delta/\delta A_a^i(x)$, we see that the action of the formal operator on a cylindrical function has the form
\begin{align}
\epsilon\downup{i}{jk}&F_{ab}^i(x)\frac{\delta}{\delta A_a^j(x)}\frac{\delta}{\delta A_b^k(x)} \psi(h_{e_1},\dots,h_{e_N}) \notag \\
&= \sum_{e,e'}\int ds\int dt\,\delta\bigl(x,e(s)\bigr)\delta\bigl(x,e'(t)\bigr)\epsilon\downup{i}{jk}\dot e^a(s){\dot e}'{}^b(t)F_{ab}^i(x) \notag \\
&\quad\times \bigl(h_{e(1,s)}\tau_jh_{e(s,0)}\bigr)\updown{A}{B}\bigl(h_{e'(1,t)}\tau_kh_{e'(t,0)}\bigr)\updown{C}{D}\frac{\partial}{\partial {h_e}\updown{A}{B}}\frac{\partial}{\partial {h_{e'}}\updown{C}{D}}\psi(h_{e_1},\dots,h_{e_N}).\label{FEE*psi}
\end{align}
In general this expression is ill-defined due to the product of delta functions, but we can nevertheless read off from \Eq{FEE*psi} the conditions that must be met in order for the right-hand side to not be identically zero. If the edges $e$ and $e'$ do not intersect each other, the product of delta functions vanishes for every value of $x$. If $e$ and $e'$ have a point in common, the factor $\dot e^a(s){\dot e}'{}^b(t)F_{ab}^i$ will give zero if the tangent vectors $\dot e$ and $\dot e'$ are parallel to each other at the intersection point. From the formal calculation, we have therefore collected the following expectations:
\begin{itemize}
\item The action of the operator will receive non-zero contributions only from the nodes of a spin network state;
\item The operator acts separately on every pair of edges at the node;
\item The action on a pair of edge is non-zero only if the edges are not tangent to each other at the node.
\end{itemize}
We would also expect that in a proper, regularized quantization, the factor $\dot e^a{\dot e}'{}^bF_{ab}^i$ will be replaced with a holonomy around a loop lying in the plane spanned by the vectors $\dot e$ and $\dot e'$ at the node.

We are now ready to write down the requirements which we impose on the structures used in the regularized expression \eqref{Eucl-sum2}. The role of these requirements is twofold. On one hand, they guarantee that the resulting operator is diffeomorphism invariant, and that the regulator can be removed by taking the limit $\epsilon\to 0$ in the diffeomorphism invariant Hilbert space; on the other hand, they ensure that the resulting operator satisfies the expectations formulated above concerning the general structure of its action.

As is familiar by now, the partition will be adapted to the graph of a given spin network state. To begin with, we assume that the partition is sufficiently refined so that each cell $\Delta$ contains at most a single node of the graph, and each surface $S_\Delta^I\subset\partial\Delta$ is intersected by at most a single edge of the graph. If $\Delta$ contains a node, the point $x_\Delta$ is taken to coincide with the node, and if $S_\Delta^I$ is intersected by an edge, we assume that the intersection is transversal. The case where $\Delta$ does not contain a node but contains the segment of an edge is treated by splitting the edge into two by introducing an ''artificial'' node inside $\Delta$, thereby transforming it into the case where $\Delta$ contains a two-valent node.

To each (ordered) pair of non-tangential edges $(e_I,e_J)$ at a node, we then assign a loop $\alpha_\Delta^{IJ}$. The loop is oriented according to the orientation of the pair $(e_I,e_J)$, \ie to the pair $(e_J,e_I)$ there is assigned the loop $\alpha_\Delta^{JI} = (\alpha_\Delta^{IJ})^{-1}$. Since the loop is uniquely specified by the pair $(e_I,e_J)$, and each edge $e_I$ corresponds to a unique surface $S_\Delta^I$, we denote the coefficient $\kappa_{IJK}^{(\Delta)}$ more shortly as $\kappa_{IJ}^{(\Delta)}$. The loops $\alpha_\Delta^{IJ}$ and the coefficients $\kappa_{IJ}^{(\Delta)}$ are now required to satisfy the following conditions:
\begin{enumerate}[(i)]
\item For a pair of edges $(e_I,e_J)$ meeting transversally at a node, the value of $\kappa_{IJ}^{(\Delta)}$ is non-zero.
\item For a pair of edges $(e_I,e_J)$ meeting tangentially at a node, $\kappa_{IJ}^{(\Delta)} = 0$.
\item The non-vanishing value of $\kappa_{IJ}^{(\Delta)}$ depends only on the valence of the node and is independent of the labels $\Delta$ and $IJ$. Hence we denote the coefficient by $\kappa_E(v)$.
\item The shape of the loop $\alpha_\Delta^{IJ}$ depends on $\epsilon$ in a diffeomorphism covariant manner. In other words, if $\Delta$ and $\Delta'$ are the cells of two partitions defined by two different values of $\epsilon$, there exists a diffeomorphism $\phi$ which relates the corresponding loops to each other: $\alpha_\Delta^{IJ} = \phi(\alpha_{\Delta'}^{IJ})$.
\end{enumerate}
Requirements (i) and (ii) guarantee that the action of the operator is consistent with the requirements derived from the calculation \eqref{FEE*psi}. Requirement (iii) implies that the action of the operator has the same form on a pair of edges $(e,e')$ at a node $v$ as it does on the diffeomorphic image $\bigl(\phi(e),\phi(e')\bigr)$ at $\phi(v)$, and therefore makes it possible for the operator to be diffeomorphism invariant. From requirement (iv) it follows that the operators obtained from partitions defined by different values of $\epsilon$ are diffeomorphically equivalent to each other, ensuring that the limit $\epsilon\to 0$ results in a well-defined operator on ${\cal H}_{\rm diff}$.

Let us then discuss in detail the properties of the loops created by the action of the operator. First of all, the position of the loop relative to the graph should be defined in a diffeomorphism invariant way. This can be accomplished by using a scheme invented by Ashtekar and Lewandowski \cite{AL_unpub} and used by Thiemann in the construction of his Hamiltonian \cite{QSD}. For Thiemann's operator, it is essential that the family of edges created by the regularized operators characterized by different values of $\epsilon$ lie in a surface which does not intersect any other edges of the graph, and is defined in a suitably diffeomorphism-invariant manner. This requirement ensures that the edge can be moved arbitrarily close to the node by a diffeomorphism without it intersecting any other edges along the way. In the case at hand, we make use of the surface defined in this way, and require that the loops $\alpha_\Delta^{IJ}(\epsilon)$, corresponding to different values of $\epsilon$, all lie within this surface. We refer to \cite{QSD} for the details of the construction.

The loop created when the operator acts on the pair of edges $(e,e')$ -- denoted from now on by $\alpha_{ee'}$ -- is attached to $e$ and $e'$ according to the ''special loops'' prescription of section \ref{sec:special}. That is, the loop does not overlap with the edges $e$ and $e'$, but is tangent to them at the node at a sufficiently high order. As explained in section \ref{sec:special}, this prescription ensures that the edges belonging to the loop can be distinguished from any other edges that might be tangent to the edges $e$ and $e'$. In particular, consequtive loops attached to the same pair of edges by successive actions of the Hamiltonian will be perfectly distinguishable from each other.

In the article \cite{paper1}, the precise orders of tangentiality between the loops created by the Hamiltonian and the edges to which they are attached were specified by the following assumption:
\begin{quote}
The special loop $\alpha_{ee'}$ is tangent to $e$ and $e'$ respectively at orders $T(e)+1$ and $T(e')+1$, where $T(e)$ is the highest order of tangentiality between $e$ and the other edges at the node before the loop was introduced (and $T(e')$ is defined similarly for $e'$).
\end{quote}
This choice is clearly sufficient to make the edges of the loop distinguishable from any other edges tangent to $e$ and $e'$ on the basis of their orders of tangentiality. However, it could be advantageous to consider alternative prescriptions for the order of tangentiality between the loop $\alpha_{ee'}$ and the edges $e$ and $e'$.

\begin{figure}[t]
	\centering
		\includegraphics[width=0.7\textwidth]{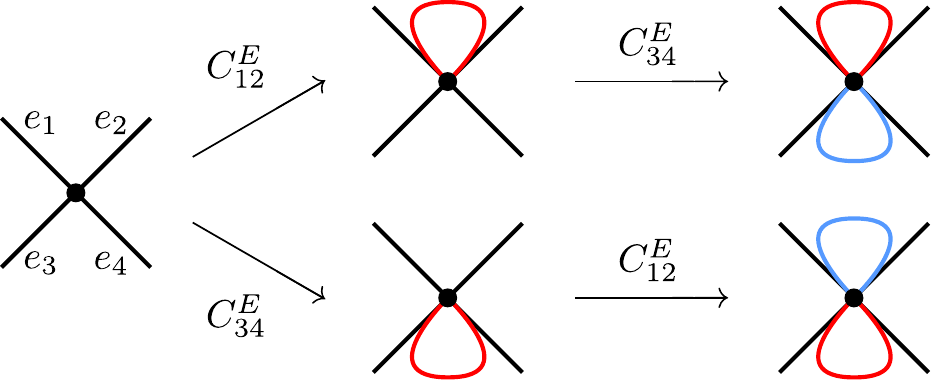}
		\caption{According to the original prescription for the order of tangentiality of the special loops, the two rightmost graphs, generated by two successive actions of the Hamiltonian, are identical. Under the modified prescription, the two graphs are distinguishable from each other.}
	\label{fig14-2}
\end{figure}

As motivation for a particular alternative prescription, consider acting repeatedly with the Hamiltonian on a state based on a fixed initial graph. The result will be a sum of terms based on graphs in which special loops are attached to the initial graph in various ways. A feature of the prescription defined above is that the graph corresponding to a typical term will generally be generated in many different ways, from many different sequences of actions of the operator. For example, if one acts with the Hamiltonian twice, the term where the first Hamiltonian acted on a pair $(e_I,e_J)$ and the second one acted on an independent pair $(e_K,e_L)$ will be based on the same graph as the term where the first Hamiltonian acted on the pair $(e_K,e_L)$ and the second on the pair $(e_I,e_J)$. If no edges in the initial state were tangent to each other, then in both terms the special loops will be tangent to the edges $(e_I,e_J)$ and $(e_K,e_L)$ at order 1. See \Fig{fig14-2} for an illustration.

When making calculations with the Hamiltonian and having to keep track of the resulting graphs, one might be inclined to consider the above feature a nuisance (for instance, because several different graphs may be produced by the action of the adjoint operator on a state based on a single, fixed graph). If one wishes to eliminate the nuisance, one can do so by adopting the following, slightly modified prescription: 
\begin{quote}
The special loop $\alpha_{ee'}$ is tangent to both $e$ and $e'$ at order $T(v)+1$, where $T(v)$ is the highest order of tangentiality between any two edges at the node $v$ before the loop was attached.
\end{quote}
Under this prescription, $\alpha_{ee'}$ is tangent to $e$ and $e'$ at a higher order than any other special loop already present in the state is to its corresponding pair of edges. Therefore the order in which the special loops were created can always be reconstructed from the orders of tangentiality; the terms in which special loops were attached to the same pairs of edges in different sequences will always be based on different graphs. In this work, we will nevertheless follow the choice originally made in \cite{paper1} for the order of tangentiality between the loop $\alpha_{ee'}$ and the edges to which it is attached.

As a result of the regularization described in this section, the function inside the square root in \eqref{Eucl-sum2} gives rise to an $\epsilon$-dependent operator on the kinematical Hilbert space. Its action on a state based on a graph $\Gamma$ is given by
\be\label{C_E(e)}
C^E_v(\epsilon)\ket{\Psi_\Gamma} = (8\pi\beta G)^2\kappa_E(v)\sum_{(e,e')\in v} \kappa(e,e')\epsilon\updown{ij}{k}\bigl(h_{\alpha_{ee'}(\epsilon)}^k\bigr)^{(l)}J_i^{(v,e)}J_j^{(v,e')}\ket{\Psi_\Gamma},
\ee
where $v$ is the node of $\Gamma$ contained inside the cell $\Delta$, and
\be
\kappa(e,e') = \begin{cases} 0 & \text{if $e$ and $e'$ are tangent at $v$,} \\ 1 & \text{if they are not.} \end{cases}
\ee
In \Eq{C_E(e)} we have chosen what seems to be the simplest possible ordering of the operators, the holonomy being ordered to the left so that it will not be acted on by the angular momentum operators.

As a gauge invariant operator, $C^E_v(\epsilon)$ is also a well-defined operator on the gauge invariant Hilbert space. While the limit $\epsilon\to 0$ cannot be taken at the kinematical level, the situation is different for the dual operator $C^E_v(\epsilon)^\star$, defined on the diffeomorphism invariant Hilbert space by
\be\label{C^star}
\bigl(C^E_v(\epsilon)^\star P_{\rm diff}[\Psi]\bigr)\bigl(\ket\Phi\bigr) \equiv P_{\rm diff}[\Psi]\bigl(C^E_v(\epsilon)\ket\Phi\bigr),
\ee
where $P_{\rm diff}[\Psi] \in {\cal H}_{\rm diff}$ and $\ket\Phi\in{\cal H}_{\rm kin}$. Since the loops $\alpha_{ee'}(\epsilon)$ are shrank in a diffeomorphism covariant manner as $\epsilon\to 0$, the operator $C^E_v(\epsilon)^\star$ is in fact independent of $\epsilon$, so ''taking the limit'' $\epsilon\to 0$ simply amounts to dropping the label $\epsilon$. We have therefore managed to promote the function \eqref{Eucl-int} into a well-defined operator on ${\cal H}_{\rm diff}$. It acts on a state $\ket{\Psi_\Gamma}\in{\cal H}_{\rm diff}$ as
\be
\int d^3x\,\widehat{\sqrt{\epsilon\updown{ij}{k}E^a_iE^b_jF_{ab}^k}}\,\ket{\Psi_\Gamma} = \sum_{v\in\Gamma}\sqrt{\displaystyle(C^E_v)^\star}\,\ket{\Psi_\Gamma},
\ee
with
\be
C^E_v = (8\pi\beta G)^2\kappa_E(v)\sum_{(e,e')\in v} \kappa(e,e')\epsilon\updown{ij}{k}\bigl(h_{\alpha_{ee'}}^k\bigr)^{(l)}J_i^{(v,e)}J_j^{(v,e')}.
\ee
This completes the first half of the quantization program indicated by \Eq{int a+b reg}.

\subsection{Regularization: Curvature part}\label{sec:reg-R}

The second half of the construction of the physical Hamiltonian consists of quantizing the integral
\be\label{curv-int}
\int d^3x\,\sqrt{\displaystyle q\,{}^{(3)}\!R}.
\ee
We introduce again a cellular decomposition of the spatial manifold, in which the size of the cells is controlled by a parameter $\epsilon$. In order to approximate the integral \eqref{curv-int} by a sum over the cells, we split the determinant under the square root into two factors of $\sqrt q$, creating two objects of density weight 1 that can be integrated over a cell. Then we can write
\be\label{curv-sum}
\int d^3x\,\sqrt{\displaystyle q\,{}^{(3)}\!R} = \int d^3x\,\sqrt{\displaystyle \sqrt q\,\sqrt q\,{}^{(3)}\!R} \simeq \sum_\Delta \sqrt{\displaystyle \biggl(\int_\Delta d^3x\,\sqrt q\biggr)\biggl(\int_\Delta d^3x\,\sqrt q\,{}^{(3)}\!R\biggr)}.
\ee
It seems like the quantization could be immediately completed by quantizing the two integrals inside the square root as the volume operator and the curvature operator associated to the cell. However, it proves to be more advantageous to keep manipulating the regularized classical expression before quantizing it, since in this way one can bring about a substantial cancellation of factors, which leads to the volume operator being eliminated entirely from the resulting quantization of \eqref{curv-int}.

We therefore proceed by approximating the integrals inside the square root in \Eq{curv-sum} with suitable functions of flux variables, writing
\be\label{curv-sum2}
\int d^3x\,\sqrt{\displaystyle q\,{}^{(3)}\!R} \simeq \sum_\Delta \sqrt{\sqrt{q_\Delta(E)}R_{\Delta}(E)}.
\ee
Here $\sqrt{q_\Delta(E)}$ is a regularization of the integral $\int_\Delta d^3x\,\sqrt q$ in terms of fluxes, and $R_{\Delta}(E)$ is a regularized Regge action for the cell $\Delta$, which was used in the construction of the curvature operator in \cite{curvature}. For concreteness, we choose
\be
q_\Delta(E) = \kappa_q(\Delta)\sum_{IJK}\biggl|\frac{1}{3!}\epsilon_{IJK}\epsilon^{ijk}E_i^{(x_\Delta)}(S_\Delta^I)E_j^{(x_\Delta)}(S_\Delta^J)E_k^{(x_\Delta)}(S_\Delta^K)\biggr|,
\ee
where, as in the previous section, the boundary of the cell $\Delta$ is subdivided into the surfaces $S_I^{(\Delta)}$, and parallel transports are defined from these surfaces to a fixed point $x_\Delta$ inside $\Delta$. Morevoer, $\kappa_q(\Delta)$ is a coefficient whose value depends on the shape of the cell $\Delta$. 

The function $R_{\Delta}(E)$ in \Eq{curv-sum2} has the form
\be
R_{\Delta}(E) = \sum_{h\in\partial\Delta} L_h^{(\Delta)}(E)\delta^{(\Delta)}_h(E),
\ee
where the sum runs over the ''hinges'' of the cell $\Delta$, \ie the one-dimensional curves at which the surfaces $S_\Delta^I$ meet each other. Denoting by $S_h$ and $S_h'$ the two surfaces which meet at the hinge $h$, the functions $L_h^{(\Delta)}(E)$ and $\delta^{(\Delta)}_h(E)$ are given by
\be\label{L(Delta)}
L_h^{(\Delta)}(E) = \sqrt{\delta_{il}\frac{\epsilon^{ijk}E_j^{(x_\Delta)}(S_h)E_k^{(x_\Delta)}(S'_h)}{2\sqrt{q_\Delta(E)}}\frac{\epsilon^{lmn}E_m^{(x_\Delta)}(S_h)E_n^{(x_\Delta)}(S'_h)}{2\sqrt{q_\Delta(E)}}}
\ee
and
\be
\delta_h^{(\Delta)}(E) = \frac{2\pi}{\alpha_h} - \pi + \cos^{-1}\biggl(\frac{\delta^{ij}E_i^{(x_\Delta)}(S_h)E_j^{(x_\Delta)}(S'_h)}{\bigl|E^{(x_\Delta)}(S_h)\bigr|\bigl|E^{(x_\Delta)}(S'_h)\bigr|}\biggr),
\ee
where $\alpha_h$ is the number of cells sharing the hinge $h$, and $|E(S)| = \sqrt{E_i(S)E^i(S)}$. 

Now we see that when $R_\Delta(E)$ is multiplied by $\sqrt{q_\Delta(E)}$ in \Eq{curv-sum2}, the $\sqrt{q_\Delta(E)}$ is cancelled against the factors of $\sqrt{q_\Delta(E)}$ in the expression \eqref{L(Delta)}. We have
\be\label{curv-sum3}
\sqrt{q_\Delta(E)}R_\Delta(E) = \sum_{h\in\partial\Delta} \frac{1}{2}\sqrt{\delta_{ij}Y_h^{(\Delta)i}(E)Y_h^{(\Delta)j}(E)}\,\delta_h^{(\Delta)}(E),
\ee
where
\be
Y_h^{(\Delta)i}(E) = \epsilon^{ijk}E_j^{(x_\Delta)}(S_h)E_k^{(x_\Delta)}(S'_h).
\ee
Therefore, as advertized, the factors of $\sqrt{q_\Delta(E)}$ have disappeared from the regularized classical expression. Consequently, the volume operator will be absent from the resulting quantum operator, which is a significant practical advantage.

The function $Y_h^{(\Delta)i}(E)$ approximates the unregularized function $\epsilon^{ijk}\epsilon_{abc}\dot h^aE^b_jE^c_k$ in the sense that
\be
\epsilon^{ijk}E_j^{(x_\Delta)}(S_h)E_k^{(x_\Delta)}(S'_h) \simeq \epsilon^2\epsilon'{}^2\epsilon^{ijk}\epsilon_{abc}\dot h^aE^b_jE^c_k,
\ee
where $\epsilon^2$ and $\epsilon'{}^2$ are the coordinate areas of the surfaces $S_h$ and $S_h'$. Before promoting the expression \eqref{curv-sum3} into an operator, we wish to derive a requirement on the form of the resulting operator by studying the straightforward formal quantization of $\epsilon^{ijk}\epsilon_{abc}\dot h^aE^b_jE^c_k$, as we did in the construction of the Euclidean part. Making the replacement $E^a_i \to -i\,\delta/\delta A_a^i$, a calculation similar to \Eq{FEE*psi} produces the factor
\be
\delta\bigl(x,e(s)\bigr)\delta\bigl(x,e(t)\bigr)\epsilon_{abc}\dot e^b(s){\dot e'}^c(t),
\ee
which vanishes if the edges $e$ and $e'$ do not intersect each other, or if they share a common node but are tangent to each other at the node.

In order to pass this property to the operator obtained from \Eq{curv-sum3}, we introduce suitable coefficients in the sum over hinges, writing
\be\label{curv-sum-kappa}
\sqrt{q_\Delta(E)}R_\Delta(E) = \sum_{h\in\partial\Delta} \kappa_h(\Delta) \frac{1}{2}\sqrt{\delta_{ij}Y_h^{(\Delta)i}(E)Y_h^{(\Delta)j}(E)}\,\delta_h^{(\Delta)}(E).
\ee
As in the previous section, this expression can be promoted into a well-defined operator only if the cellular decomposition is adapted to the graph of a given spin network state through the requirements which we repeat here:
\begin{itemize}
\item Each cell $\Delta$ contains at most a single node of the graph.
\item Each surface $S_\Delta^I\subset\partial\Delta$ is intersected by at most a single edge of the graph. 
\item If $\Delta$ contains a node, the point $x_\Delta$ is taken to coincide with the node.
\item If $S_\Delta^I$ is intersected by an edge, we assume that the intersection is transversal. 
\item The case where $\Delta$ does not contain a node but contains the segment of an edge is treated by splitting the edge into two by introducing an ''artificial'' node inside $\Delta$, thereby transforming it into the case where $\Delta$ contains a two-valent node.
\end{itemize}
In \Eq{curv-sum-kappa}, we further demand that the coefficient $\kappa_h(\Delta)$ vanishes if the two edges defining the hinge $h$ are tangential to each other at the node inside $\Delta$. Then, as in the case of the Euclidean part, diffeomorphism invariance requires that the non-vanishing value of $\kappa_h(\Delta) \equiv \kappa_R(v)$ depends only on the valence of the node, and is independent of $\Delta$ and $h$. The actual value of $\kappa_R(v)$ is determined through an averaging procedure which removes the dependence of the operator on the cellular structures used to construct it, as explained in detail in \cite{curvature}.

The way in which the integral \eqref{curv-int} will be quantized is now indicated by \Eq{curv-sum-kappa}, in which the factors on the right-hand side can be readily promoted into operators, together with the approximation
\be
\int d^3x\,\sqrt{\displaystyle q\,{}^{(3)}\!R} = \sum_\Delta \int_\Delta d^3x\,\sqrt{\displaystyle q\,{}^{(3)}\!R} \simeq \sum_\Delta\sqrt{\displaystyle \sqrt{q_\Delta(E)}R_\Delta(E)}.
\ee
The resulting operator acts on a state $\ket{\Psi_\Gamma}$ as
\be\label{sum C^R_v}
\int d^3x\,\widehat{\sqrt{\displaystyle q\,{}^{(3)}\!R}}\,\ket{\Psi_\Gamma} = \sum_{v\in\Gamma}\sqrt{C^R_v}\,\ket{\Psi_\Gamma},
\ee
where
\be\label{C^R_v}
C^R_v = (8\pi\beta G)^2\kappa_R(v)\sum_{(e,e')\in v}\kappa(e,e')\frac{1}{2}\bigl|Y_v^{(e,e')}\bigr|\biggl(\frac{2\pi}{\alpha_{ee'}} - \pi + \theta_v^{(e,e')}\biggr)
\ee
Here the operators appearing on the right-hand side are given by
\be\label{curv-Y}
\bigl|Y_v^{(e,e')}\bigr| = \sqrt{\displaystyle Y_i^{(v,e,e')}Y_{\phantom{i}}^{(v,e,e')i}} \qquad \text{with} \qquad Y_v^{(e,e')i} = \epsilon^{ijk}J_j^{(v,e)}J_k^{(v,e')},
\ee
and
\be\label{curv-theta}
\theta_v^{(e,e')} = \cos^{-1}\biggl(\frac{J_i^{(v,e_1)}J_{\phantom{i}}^{(v,e_2)i}}{\bigl|J^{(v,e_1)}\bigr|\bigl|J^{(v,e_2)}\bigr|}\biggr). 
\ee
Since the operators $\bigl|Y_v^{(e,e')}\bigr|$ and $\theta_v^{(e,e')}$ commute with each other, there is no ordering ambiguity in the operator \eqref{C^R_v}.

The operator defined by \Eqs{sum C^R_v} and \eqref{C^R_v} is gauge invariant, and therefore it is well-defined not only as an operator on ${\cal H}_{\rm kin}$, but also on the gauge-invariant Hilbert space ${\cal H}_G$. Moreover, since the operator $C^R_v$ is composed entirely of the operators $J_i^{(v,e)}$, its action affects only the intertwiners at the nodes of a spin network state, and leaves the graph of the state unchanged. Hence the dual operator $(C^R_v)^\star$ is immediately a well-defined operator on ${\cal H}_{\rm diff}$; in contrast to the Euclidean part, no discussion concerning the removal of the regulator is needed. In a sense, the regulator is removed already at the level of the kinematical Hilbert space.

\subsection{Summary of the construction}\label{sec:Hphys-summ}

With both steps of the construction completed, we may now summarize our findings. By quantizing the classical function defined by \Eqs{hphys-1} and \eqref{hphys-2}, we obtained the physical Hamiltonian operator
\be\label{Hphys-fin}
H_{\rm phys}\ket{\Psi_\Gamma} = \frac{1}{\sqrt{8\pi G\beta^2}}\sum_{v\in\Gamma} \sqrt{\displaystyle {\rm Sym}\bigl(C_v^E + (1+\beta^2)C_v^R\bigr)}\,\ket{\Psi_\Gamma},
\ee
where ${\rm Sym}$ indicates that the operator inside the square root should be symmetrized. Since the adjoint ${C_v^E}^\dagger$ exists as a densely defined operator, there is no obstruction against carrying out the symmetrization. The obvious choice is
\be\label{Sym}
{\rm Sym}\bigl(C_v^E + (1+\beta^2)C_v^R\bigr) = \frac{1}{2}\bigl(C_v^E + {C_v^E}^\dagger\bigr) + (1+\beta^2)C_v^R;
\ee
however, one could also consider other symmetrizations, based on symmetrizing a non-symmetric operator $C$ as $\sqrt{C^\dagger C}$, as opposed to $\half(C+C^\dagger)$. Such symmetrizations would have the advantage that the operator under the square root in \Eq{Hphys-fin} could be made automatically positive definite, whereas the positive-definiteness of the operator \eqref{Sym} is not clear at this stage.

The operators $C^E_v$ and $C^R_v$ in \Eq{Sym} have the form
\be\label{CE-fin}
C^E_v = (8\pi\beta G)^2\kappa_E(v)\sum_{(e,e')\in v} \kappa(e,e')\biggl(-\frac{3}{W_l^2}\biggr)\epsilon^{ijk}\Tr\bigl(\tau_k^{(l)}D^{(l)}(h_{\alpha_{ee'}})\bigr)J_i^{(v,e)}J_j^{(v,e')},
\ee
and
\be\label{CR-fin}
C^R_v = (8\pi\beta G)^2\kappa_R(v)\sum_{(e,e')\in v}\kappa(e,e')\frac{1}{2}\bigl|Y_v^{(e,e')}\bigr|\biggl(\frac{2\pi}{\alpha_{ee'}} - \pi + \theta_v^{(e,e')}\biggr),
\ee
with the operators $\bigl|Y_v^{(e,e')}\bigr|$ and $\theta_v^{(e,e')}$ defined by \Eqs{curv-Y} and \eqref{curv-theta}. The loop $h_{\alpha_{ee'}}$ is attached to the edges $e$ and $e'$ according to the special loop prescription, and the coefficient $\kappa(e,e')$ restricts the action of the operators to those pairs of edges which meet each other transversally at the node $v$. The task of computing matrix elements of the operators \eqref{CE-fin} and \eqref{CR-fin} in the spin network basis will be considered in Chapter \ref{ch:elements}.

\Eqs{Hphys-fin}--\eqref{CR-fin} can be interpreted equally well as equations in ${\cal H}_{\rm kin}$ or in ${\cal H}_{\rm diff}$. In the former case, the Euclidean operator $C^E_v$ carries an implicit dependence on the regularization parameter $\epsilon$. At the diffeomorphism invariant level, the regularization parameter is absent, and each of the operators is defined by duality in the way shown by \Eq{C^star} (and therefore carries a suppressed $\star$ -superscript). Note that it follows from the definition \eqref{C^star} that the dual operator $(C^E_v)^\star$ acts by removing loops; in ${\cal H}_{\rm diff}$ loops are created by the adjoint operator $(C^E_v{}^\dagger)^\star$.

In principle, the physical Hamiltonian provided by \Eq{Hphys-fin} completes the definition of a concrete and mathematically consistent model of quantum gravity. However, from a practical perspective \Eq{Hphys-fin} is not fully concrete, since the square root on the right-hand side cannot be expressed in an explicit form unless a spectral decomposition of the operator under the square root is available. While no progress has been made towards deriving an exact, analytic expression for the spectrum of the operator \eqref{Sym} (or any of its differently symmetrized variants), under certain conditions an approximate, partial solution to the eigenvalue problem can be obtained using perturbation theory, as we will show in section \ref{sec:perturbation}.

\newpage

\addtocontents{toc}{\protect\newpage}
\section{A new Hamiltonian constraint operator}\label{ch:constraint}

The technical ideas introduced in the previous chapter can naturally also be applied to the quantization of the Hamiltonian constraint, even though they were primarily developed with a view towards constructing a satisfactory physical Hamiltonian for loop quantum gravity deparametrized with respect to a free scalar field. In this way one would obtain a Hamiltonian constraint operator whose Euclidean part acts by creating ''special loops'', and whose remaining part is quantized as the curvature operator of section \ref{sec:curvature}. Most importantly, the properties of the special loops guarantee that the resulting operator can be symmetrized. Even if one prefers the point of view that the Hamiltonian constraint operator should not be symmetric, a quantization of the constraint as a symmetric operator is nevertheless needed in order to provide a physical Hamiltonian for deparametrized dust models.

A concrete construction of a Hamiltonian constraint operator of this type was given by the author together with Mehdi Assanioussi and Jerzy Lewandowski in the article \cite{paper2}. In this chapter we will display two slightly modified versions of the construction presented in \cite{paper2}. The first version differs from the operator defined in \cite{paper2} essentially only by the way in which we choose to symmetrize the curvature operator, while in the second version the quantization of the Euclidean part of the constraint is also modified in comparison to \cite{paper2}.

\subsection{The vertex Hilbert space}\label{sec:vertex}

When one considers the quantization of the Hamiltonian constraint, there arises an additional complication which was not encountered in the case of the physical Hamiltonian discussed in the previous chapter. Namely, due to the presence of the lapse function in the functional $C(N)$, it is not possible to quantize the Hamiltonian constraint as a diffeomorphism invariant operator (\ie as an operator on ${\cal H}_{\rm diff}$ which would map ${\cal H}_{\rm diff}$ into itself). A way to deal with this difficulty was proposed by Lewandowski and Sahlmann, who in \cite{LS14} introduced a suitable intermediate Hilbert space, denoted ${\cal H}_{\rm vtx}$, of partially diffeomorphism-invariant states. As it turns out, the Hamiltonian constraint can be promoted into a well-defined operator which preserves the space ${\cal H}_{\rm vtx}$, and after solutions of the Hamiltonian constraint on ${\cal H}_{\rm vtx}$ are found, they can be turned into solutions of the complete set of constraints by averaging them over the remaining diffeomorphisms, which were not accounted for in the construction of the space ${\cal H}_{\rm vtx}$.

The general idea for constructing the space ${\cal H}_{\rm vtx}$ is to start with the space ${\cal H}_G^\Gamma$ of gauge-invariant states\footnote{Here we are following the construction presented in \cite{paper1} and \cite{paper2}, rather than the original construction of \cite{LS14}. Thus, the space ${\cal H}_{\rm vtx}$ defined in this section corresponds to the space denoted by ${\cal H}^G_{\rm vtx}$ in the notation of \cite{paper1} and \cite{paper2}, and not to the space denoted by ${\cal H}_{\rm vtx}$ in \cite{LS14}.} based on a given graph $\Gamma$, and average the elements of ${\cal H}^G_{\rm vtx}$ (in the sense discussed in Chapter \ref{ch:diff}) with respect to diffeomorphisms which act trivially on the nodes of the graph $\Gamma$. More precisely, let us denote by ${\cal D}_{N(\Gamma)}$ the set of diffeomorphisms $\phi$ for which $\phi(v)=v$ for every node $v$ of $\Gamma$ and by ${\cal TD}_\Gamma$ the set of diffeomorphisms acting trivially on elements of ${\cal H}_G^\Gamma$, and consider the averaging of states in ${\cal H}_G^\Gamma$ with respect to the quotient group
\be
\Delta_\Gamma = \frac{{\cal D}_{N(\Gamma)}}{{\cal TD}_\Gamma}.
\ee
The averaging is again carried out in the dual space $\Cyl^*$. Given a state $\ket{\Psi_\Gamma}\in{\cal H}_G^\Gamma$, we define
\be\label{eta_vtx}
\eta[\Psi_\Gamma] = \frac{1}{N_{{\cal S}_\Gamma}}\sum_{\phi\in\Delta_\Gamma}\bbra{U(\phi)\Psi_\Gamma},
\ee
where, as in Chapter \ref{ch:diff}, $N_{{\cal S}_\Gamma}$ is the number of symmetries of the graph $\Gamma$. \Eq{eta_vtx} defines a linear functional $\eta:{\cal H}_G^\Gamma\to\Cyl^*$ which is invariant under the action of ${\cal D}_{N(\Gamma)}$. The state $\eta[\Psi_\Gamma]$ is gauge invariant and invariant under diffeomorphisms contained in ${\cal D}_{N(\Gamma)}$, but it is not invariant under arbitrary diffeomorphisms $\phi\in{\rm Diff}$. Therefore, in order to turn the states in $\eta({\cal H}_G^\Gamma)$ into full solutions of the Gauss and diffeomorphism constraint, they must be averaged with the remaining diffeomorphisms, which act non-trivially on the nodes of $\Gamma$.

By linearity, the action of $\eta$ can be extended into the entire gauge-invariant Hilbert space, resulting in a functional $\eta:{\cal H}_G\to\Cyl^*$.  The space ${\cal H}_{\rm vtx}$ is then defined as
\be
{\cal H}_{\rm vtx} = \overline{\eta\bigl({\cal H}_G\cap\Cyl\bigr)},
\ee
the completion being taken with respect to the norm induced by the natural scalar product
\be
\bbraket{\eta[\Psi]}{\eta[\Psi']}_{\rm vtx} \equiv \eta[\Psi]\bigl(\ket{\Psi'}\bigr).
\ee
Any gauge invariant operator $A:{\cal H}_G\to{\cal H}_G$ gives rise by duality to an operator $A^\star$, which has a well-defined action on ${\cal H}_{\rm vtx}$. However, the image $A^\star({\cal H}_{\rm vtx})$ is generally not contained in ${\cal H}_{\rm vtx}$, but is only some subspace of $\Cyl^*$. Concerning the conditions for $A^\star$ to be an operator on ${\cal H}_{\rm vtx}$, the following theorem was established in \cite{LS14}: Suppose an operator $A(f):{\cal H}_G\to{\cal H}_G$ has the form
\be\label{vtx-cond1}
A(f) = \sum_{x\in\Sigma} f(x)A_x,
\ee
where the operators $A_x$ satisfy
\be\label{vtx-cond2}
U(\phi_x)A_x = A_xU(\phi_x)
\ee
for every diffeomorphism $\phi_x$ which preserves the point $x$, and
\be\label{vtx-cond3}
A_x\ket{\Psi_\Gamma}=0
\ee
whenever $\ket{\Psi_\Gamma}\in{\cal H}_G^\Gamma$ and $x$ is not a node of $\Gamma$. Then $A^\star(f)$ is an operator on ${\cal H}_{\rm vtx}$, \ie $A^\star(f): {\cal H}_{\rm vtx}\to {\cal H}_{\rm vtx}$. The conditions \eqref{vtx-cond1}--\eqref{vtx-cond3} are met, in particular, by the Hamiltonian constraint operators constructed in sections \ref{sec:C(N)-1} and \ref{sec:C(N)-2}, and so the above result guarantees that these operators are indeed well-defined as operators on ${\cal H}_{\rm vtx}$.

\subsection{A Thiemann-like construction}\label{sec:C(N)-1}

The starting point for the quantization of the Hamiltonian constraint performed in \cite{paper2} is given by the classical expression \eqref{C-AE-2}, in which the Lorentzian part of the constraint has been traded for the integral of the Ricci scalar over the spatial manifold. That is,
\be
C(N) = C_E(N) + C_R(N),
\ee
where the Euclidean part of the constraint is
\be\label{C_E(N)-2}
C_E(N) = \frac{1}{16\pi G}\frac{1}{\beta^2}\int d^3x\, N\frac{\epsilon\updown{ij}{k}E^a_iE^b_jF_{ab}^k}{\sqrt{\det E}},
\ee
and the remaining term
\be\label{C_R(N)}
C_R(N) = \frac{1}{16\pi G}\frac{1+\beta^2}{\beta^2}\int d^3x\,N\sqrt{\det E}\,{}^{(3)}\! R
\ee
will be referred to as the curvature part.

To express the Euclidean part in a form suitable for quantization, we adopt a suitably modified version of the regularization used by Thiemann in his construction of the constraint operator. Using the notation of section \ref{sec:Thiemann-reg}, and allowing the holonomies involved in the regularization to carry an arbitrary spin $l$, we have instead of \eqref{C_E-reg} the regularized expression\footnote{The numerical prefactor in \Eq{C_E-reg-l} is adjusted to account for the normalization of the $SU(2)$ generators, which in the spin-$l$ representation is given by
\[
{\rm Tr}\,\bigl(\tau_i^{(l)}\tau_j^{(l)}\bigr) = -\frac{W_l^2}{3}\delta_{ij}.
\]}
\begin{align}
C_E(N) &\simeq -\frac{1}{(8\pi G)^2\beta^3}\frac{1}{W_l^2}\sum_\Delta N(v_\Delta)\epsilon^{IJK} \notag \\
&\qquad\times\Tr\Bigl(D^{(l)}\bigl(h_{\alpha_{IJ}(\Delta)}\bigr)D^{(l)}\bigl(h_{s_K(\Delta)}^{-1}\bigr)\bigl\{D^{(l)}\bigl(h_{s_K(\Delta)}\bigr),V(\Delta)\bigr\}\Bigr).\label{C_E-reg-l}
\end{align}
Here the symbols have the same meaning as in section \ref{sec:Thiemann-reg}, except for the loop $\alpha_{IJ}(\Delta)$, which is now chosen according to the special loop prescription introduced in section \ref{sec:special}. That is, the loop $\alpha_{IJ}(\Delta)$ does not overlap with the edges $e_I$ and $e_J$, but is tangent to them at a sufficiently high order, allowing it to be distinguished from any other loops which may have been present before the loop $\alpha_{IJ}(\Delta)$ was introduced. However, at the level of the classical regularized expression \eqref{C_E-reg-l}, we still assume that the shape of the loop $\alpha_{IJ}(\Delta)$ follows arbitrarily closely the shape of the triangular loop spanned by the segments $s_I(\Delta)$ and $s_J(\Delta)$. We also require the family of loops loops $\alpha_{IJ}(\Delta)$ to satisfy the conditions spelled out in section \ref{sec:reg-E}. In particular, we assume that the loop $\alpha_{IJ}(\Delta)$ depends on the regularization parameter $\epsilon$ in a diffeomorphism covariant manner: If $\Delta$ and $\Delta'$ are the cells of two triangulations characterized by two different values of $\epsilon$, there exists a diffeomorphism $\phi$ such that $\alpha_{IJ}(\Delta) = \phi\bigl(\alpha_{IJ}({\Delta'})\bigr)$.

By quantizing the classical expression \eqref{C_E-reg-l}, we obtain a triangulation-dependent constraint operator, whose action on a state based on a graph $\Gamma$ is given by
\begin{align}
C_E^{(\Delta)}(N)&\ket{\Psi_\Gamma} = \frac{i}{(8\pi G)^2\beta^3}\frac{1}{W_l^2}\sum_{v\in\Gamma} N(v)\frac{8}{E(v)} \notag \\
&\times\sum_{\Delta(v)} \epsilon^{IJK}\Tr\Bigl(D^{(l)}\bigl(h_{\alpha_{IJ}(\Delta)}\bigr)D^{(l)}\bigl(h_{s_K(\Delta)}^{-1}\bigr)\bigl[D^{(l)}\bigl(h_{s_K(\Delta)}\bigr),V_v\bigr]\Bigr)\ket{\Psi_\Gamma},\label{C_E^Delta-l}
\end{align}
where, as in \Eq{C_E^Delta}, $E(v)$ denotes the number of distinct triples of edges at the node $v$. In order to remove the dependence on the triangulation, we must consider the limit $\epsilon\to 0$. As discussed in the preceding chapters, the operator \eqref{C_E^Delta-l} does not converge into a well-defined operator on ${\cal H}_{\rm kin}$ as $\epsilon\to 0$, and therefore the limit cannot be taken at the level of the kinematical Hilbert space.

On the other hand, the dual operator $C_E^{(\Delta)}(N)^\star$ is a well-defined operator on the vertex Hilbert space, and the diffeomorphism covariance of the loops $\alpha_{IJ}(\Delta)$ ensures that the limit $\epsilon\to 0$ is trivial on ${\cal H}_{\rm vtx}$, and can be performed simply by dropping the (implicit) label $\epsilon$ in \Eq{C_E^Delta-l}. Thus, the Euclidean part of the constraint can be defined as an operator on ${\cal H}_{\rm vtx}$ as
\be
C_E(N) \equiv \lim_{\epsilon\to 0} C_E^{(\Delta)}(N)^\star.
\ee
The action of the operator on a state based on a graph $\Gamma$ takes the form
\be
C_E(N)\ket{\Psi_\Gamma} = \sum_{v\in\Gamma} N(v){\cal C}^E_v\ket{\Psi_\Gamma},
\ee
where the explicit form of the operator ${\cal C}^E_v$ can be read off from \Eq{C_E^Delta-l}.

The quantization of the curvature part \eqref{C_R(N)} is given directly by the construction outlined in section \ref{sec:curvature}; the presence of the lapse function in the integral \eqref{C_R(N)} does not alter the construction in any essential way. The operator defined by the appropriate modification of \Eq{R-op} acts in a graph-preserving way, so it can be immediately passed to an operator on ${\cal H}_{\rm vtx}$ The resulting operator acts on ${\cal H}_{\rm vtx}$ as
\be
C_R(N)\ket{\Psi_\Gamma} = \sum_{v\in\Gamma} N(v){\cal C}^R_v\ket{\Psi_\Gamma},
\ee
where, choosing the symmetrization suggested in \Eq{curv-sym2}, we have\footnote{Since ${\cal C}^R_v$ is an operator on ${\cal H}_{\rm vtx}$, all the operators on the right-hand side of \Eq{cal C^R_v} should strictly speaking carry a $\star$ -superscript. However, since each of these operators is symmetric and acts on the node $v$ merely by modifying the intertwiner, there is essentially no difference between the action of the operator on ${\cal H}_G$ and that of the corresponding dual operator on ${\cal H}_{\rm vtx}$. For such operators, we omit the $\star$ -superscript, using the same notation to denote the operator on ${\cal H}_G$ and its dual on ${\cal H}_{\rm vtx}$. The distinction is more significant in the case of the Euclidean part, since the dual of the loop-creating operator \eqref{C_E^Delta-l} acts in ${\cal H}_{\rm vtx}$ by removing loops.}
\be\label{cal C^R_v}
{\cal C}^R_v\ket{\Psi_\Gamma} = 4\pi G(1+\beta^2)\kappa_R(v)\sum_{(e,e')\in v}\frac{1}{2}\sqrt{\displaystyle {\cal V}_v^{-1}}\bigl|Y_v^{(e,e')}\bigr|\biggl(\frac{2\pi}{\alpha_{ee'}} - \pi + \theta_v^{(e,e')}\biggr)\sqrt{\displaystyle {\cal V}_v^{-1}}\ket{\Psi_\Gamma}.
\ee
All in all, we have therefore managed to quantize the Hamiltonian constraint as a well-defined operator on the space ${\cal H}_{\rm vtx}$. The action of the operator on states in ${\cal H}_{\rm vtx}$ has the form
\be\label{C(N)-v1}
C(N)\ket{\Psi_\Gamma} = \sum_{v\in\Gamma} N(v)\bigl({\cal C}_v^E + {\cal C}_v^R\bigr)\ket{\Psi_\Gamma}.
\ee
Some general properties of the operator will be discussed in section \ref{C(N)-prop}. However, we will first present an alternative way of quantizing the Euclidean part of the constraint.

\subsection{An alternative construction}\label{sec:C(N)-2}

In the previous section, the Euclidean part of the constraint was quantized using Thiemann's trick, which is designed to deal with the problematic factor of $1/\sqrt{\det E}$ in the classical expression \eqref{C_E(N)-2}. On the other hand, in the construction of the curvature operator the same factor of $1/\sqrt{\det E}$ (which, as we recall, enters the construction when the lengths of hinges in Regge's formula are expressed in terms of fluxes) was essentially quantized as the regularized ''inverse volume'' operator ${\cal V}_v^{-1}$. Therefore, once we have decided to quantize the curvature part of the constraint in this way, there seems to be little reason to avoid using the operator ${\cal V}_v^{-1}$ in the quantization of the Euclidean part as well.

In order to derive a suitably regularized classical expression, from which the Euclidean part can be quantized by means of the operator ${\cal V}_v^{-1}$, let us consider a generic cellular decomposition of the spatial manifold $\Sigma$ into cells denoted by $\Delta$. We may then trivially rewrite the integral \eqref{C_E(N)-2} as
\be
\int d^3x\, N\frac{\epsilon\updown{ij}{k}E^a_iE^b_jF_{ab}^k}{\sqrt{\det E}} = \sum_\Delta \int_\Delta d^3x\, N\frac{\epsilon\updown{ij}{k}E^a_iE^b_jF_{ab}^k}{\sqrt{\det E}}.
\ee
Here the integral over a cell can be approximated as
\be
\int_\Delta d^3x\, N\frac{\epsilon\updown{ij}{k}E^a_iE^b_jF_{ab}^k}{\sqrt{\det E}} \simeq \dfrac{\biggl(\displaystyle\int_\Delta d^3x\,\sqrt{N\epsilon\updown{ij}{k}E^a_iE^b_jF_{ab}^k}\biggr)^2}{{\displaystyle\int_\Delta} d^3x\,\sqrt{\det E}},\label{C_E-newreg}
\ee
since at leading order in the regularization parameter, both sides of the above equation reduce to $\epsilon^3N\epsilon\updown{ij}{k}E^a_iE^b_jF_{ab}^k/\sqrt{\det E}$, where $\epsilon^3$ is the volume of the cell $\Delta$.

In the denominator on the right-hand side of \Eq{C_E-newreg}, we have the integral which was already considered in section \ref{sec:reg-E}. Assuming that the cell $\Delta$ contains a single spin network node $v$, the denominator of \Eq{C_E-newreg} gives rise upon quantization to the operator $C^E_v(\epsilon)$ defined by \Eq{C_E(e)}, while the factor $\bigl(\int_\Delta d^3x\,\sqrt{\det E}\bigr)^{-1}$ can be promoted into the operator ${\cal V}_v^{-1}$. The relative ordering between the two operators may in principle be chosen freely, and we fix it by ordering a square root of ${\cal V}_v^{-1}$ to either side of $C^E_v(\epsilon)$ (mimicking the way in which we have chosen to symmetrize the curvature operator).

After passing to the space ${\cal H}_{\rm vtx}$, in which it is trivial to remove the regulator still present in the operator $C^E_v(\epsilon)$, we obtain the Euclidean part of the constraint operator in the form
\be
C_E(N)\ket{\Psi_\Gamma} = \frac{1}{16\pi G\beta^2} \sum_{v\in\Gamma} N(v)\sqrt{\displaystyle {\cal V}_v^{-1}}C^E_v\sqrt{\displaystyle {\cal V}_v^{-1}}\ket{\Psi_\Gamma},
\ee
where we have denoted
\be
C^E_v \equiv \lim_{\epsilon\to 0} C^E_v(\epsilon)^\star.
\ee
The quantization of the curvature part can be carried over without modification from the previous section. Thus we have quantized the entire constraint as 
\be\label{C(N)-v2}
C(N)\ket{\Psi_\Gamma} = \sum_{v\in\Gamma} N(v)C_v\ket{\Psi_\Gamma},
\ee
where
\be\label{C_v}
C_v = \frac{1}{16\pi G\beta^2}\sqrt{\displaystyle {\cal V}_v^{-1}} \Bigl(C^E_v + (1+\beta^2)C^R_v\Bigr) \sqrt{\displaystyle {\cal V}_v^{-1}},
\ee
and $C^E_v$ and $C^R_v$ are (the duals of) the operators defined by \Eqs{CE-fin} and \eqref{CR-fin}.

Besides the operator $C^E_v$, the adjoint operator $(C^E_v{})^\dagger$ can be considered as an equally good quantization of the classical functional \eqref{Eucl-int}. Therefore the operator $C^E_v$ in \Eq{C_v} could be replaced with the adjoint operator $(C^E_v)^\dagger$, or with any combination ''of total weight 1'' between $C^E_v$ and its adjoint. If a symmetric combination of $C^E_v$ and the adjoint is chosen, the operator between the factors of ${\cal V}_v^{-1/2}$ in \Eq{C_v} is the same as the operator under the square root in the physical Hamiltonian defined by \Eq{Hphys-fin}.

\subsection{Properties of the constraint operator}\label{C(N)-prop}

The Hamiltonian constraint operators constructed in this chapter are operators on the space ${\cal H}_{\rm vtx}$ of partially diffeomorphism-invariant states. Notably, the space ${\cal H}_{\rm vtx}$ is preserved by the action of these operators. The operators can be expressed in a general form, without reference to any graph, as
\be\label{C(N)-gen}
C(N) = C_E(N) + C_R(N) = \sum_{x\in\Sigma} N(x)C_x,
\ee
where $C_x$ denotes either the operator ${\cal C}^E_v + {\cal C}^R_v$ in \Eq{C(N)-v1} or the operator $C_v$ of \Eq{C_v}, and it is understood that the action of $C_x$ on a state $\ket{\Psi_\Gamma}$ gives zero whenever the point $x$ does not coincide with any node of the graph $\Gamma$. Since the adjoint operator $C^\dagger(N)$ also exists as a well-defined operator on ${\cal H}_{\rm vtx}$, we are free to define a symmetric constraint operator, such as
\be
C_{\rm sym}(N) = \frac{1}{2}\bigl(C(N) + C^\dagger(N)\bigr),
\ee
which will again be of the general form \eqref{C(N)-gen}. As shown by the discussion in Chapter \ref{ch:deparametrized}, a quantization of the Hamiltonian constraint as a symmetric operator (and evaluated at unit lapse function) is needed as the physical Hamiltonian for deparametrized models in which a dust field is used as the relational time variable, even if one disapproves of the proper constraint operator -- that is, the operator whose kernel determines the physical Hilbert space of loop quantum gravity seen as a fully constrained theory -- being symmetric.

Concerning the commutator algebra of the constraint operator \eqref{C(N)-gen}, we may begin by noting that the operator is gauge invariant, and therefore commutes with the Gauss constraint. It is also not difficult to see that the operator behaves under diffeomorphisms in the expected way, namely
\be
U^{-1}(\phi)C(N)U(\phi) = C(\phi^* N).
\ee
The commutator between two Hamiltonian constraints can be evaluated with a straightforward calculation. Using the expression \eqref{C(N)-gen}, we get
\be
[C(M),C(N)] = \sum_{x\neq y}M(x)N(y)[C_x,C_y],
\ee
where the summation can be restricted to the points $x\neq y$, since when $x=y$, the commutator $[C_x,C_x]$ evidently vanishes\footnote{However, we should note that this is the case only because $C_x$ is a well-defined operator on ${\cal H}_{\rm vtx}$ as such, without having an implicit dependence on some regularization parameter. When calculating the commutator of two regularized constraint operators at the kinematical level, one would encounter a commutator of the form $[C_x(\epsilon),C_x(\epsilon')]$, which does not generally vanish, because the two operators $C_x$ are characterized by two different values of the regularization parameter.}. When the commutator $[C_x,C_y]$ (with $x\neq y$) acts on a state based on a graph $\Gamma$, the only case which does not immediately give zero is when the points $x$ and $y$ coincide with two nodes $v$ and $v'$ of the graph. However, even in this case we have
\be
[C_v,C_{v'}]\ket{\Psi_\Gamma} = 0,
\ee
since the action of the operator $C_v$ is local at the node $v$, in the sense that the ''quantum numbers'' at other nodes of the state $\ket{\Psi_\Gamma}$ are not changed by the action of $C_v$, and neither does the action on the node $v$ depend on the values of any quantum numbers outside of the node $v$. Hence we have shown that
\be
[C(M),C(N)] = 0,
\ee
so we may conclude that the constraint algebra contains no anomalies at the diffeomorphism invariant level. However, the above calculations are strictly speaking not sufficient to determine whether the operator \eqref{C(N)-gen} is truly anomaly-free, since they are carried out in the space ${\cal H}_{\rm vtx}$, and not in the kinematical Hilbert space, where all the constraint operators have a non-trivial action\footnote{On the other hand, in the present work the primary purpose of the operator \eqref{C(N)-gen} is to serve as a physical Hamiltonian for the non-rotational dust model, and in the deparametrized framework the question of anomalies does not seem to have the same fundamental importance as it does in the case of the fully constrained theory.}.

The different versions of the constraint operator \eqref{C(N)-gen} possess a large number of trivial solutions. Any state whose edges are coplanar (\ie the tangent vectors of the edges are coplanar) at every node of the graph belongs to the kernel of the operator constructed in section \ref{sec:C(N)-1}, while any eigenstate of the volume operator with eigenvalue zero is also annihilated by the operator of section \ref{sec:C(N)-2}

The structure of the non-trivial solutions can be described in a general, qualitative manner, even though we have not been able to derive any concrete examples of such solutions. A generic solution of the constraint \eqref{C(N)-gen} will be an infinite linear combination of states based on different graphs. This linear combination starts with any graph which does not contain any special loops of the type created by the Euclidean operator; in the present context, such a graph could be called a ''seed graph''. The linear combination also includes the graphs where the nodes of the seed graph are decorated with special loops in all the possible ways which would be generated by repeated actions of the Euclidean operator on the seed graph. Since the constraint acts locally at the nodes of the graph, the problem of deriving solutions boils down to solving the constraint separately node-by-node. The solution at each node will be an infinite linear combination having the following schematic structure:
\begin{align}
&c_0\RealSymb{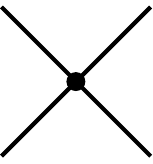}{0.6} \quad + \quad c_1\RealSymb{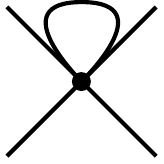}{0.6}\quad + \quad c_2\RealSymb{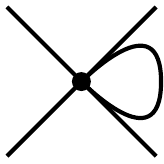}{0.6} \quad + \;\dots \notag \\
&c_{11}\RealSymb{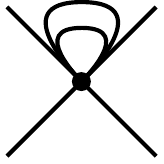}{0.6} \quad + \quad c_{12}\RealSymb{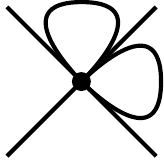}{0.6}\quad + \quad c_{13}\RealSymb{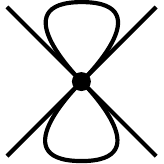}{0.6} \quad + \;\dots \notag \\
&c_{111}\RealSymb{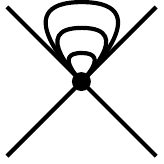}{0.6} \quad + \quad c_{112}\RealSymb{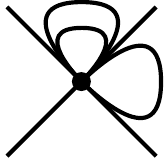}{0.6}\quad + \quad c_{123}\RealSymb{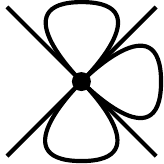}{0.6} \quad + \;\dots \label{superposition}
\end{align}
If solutions of this kind can be found, they will be states in ${\cal H}_{\rm vtx}$. In order to promote them into solutions of the full set of constraints, and hence to elements of the physical Hilbert space, each such solution would still have to be averaged with respect to the diffeomorphisms which act non-trivially on the nodes of the seed graph from which the solution was constructed.

\section{Matrix elements of the Hamiltonian}\label{ch:elements}

In this chapter we take up the task of computing matrix elements of the Hamiltonian operators introduced in the preceding chapters. The calculations are performed using the graphical formalism which is introduced in the Appendix, and which is generally an invaluable tool for analyzing the action of operators in loop quantum gravity in the spin network basis.

The material in this chapter is divided into three sections. In the first two sections we consider the Euclidean operator $C^E_v$ and the curvature operator $C^R_v$, defined respectively by \Eqs{CE-fin} and \eqref{CR-fin}. In the third section we examine the volume operator, which enters the physical Hamiltonian of the non-rotational dust model. We will not attempt to calculate the action of any of these operators on a general spin network state. Rather, we will only derive those matrix elements of the operators which are needed for the numerical applications presented in the next chapter. In any case, once the methods of calculation are known, they can be easily generalized to more complicated states, should the need ever arise to evaluate the action of the Hamiltonian on such states.

The calculations presented below have not been previously published anywhere. They were performed by the author during the course of the work that eventually lead to the article \cite{paper5}, and they formed the basis for the numerical computations which were presented in that article.

\subsection{Euclidean part}\label{sec:CE-el}

Let us start by considering the Euclidean operator $C^E_v$ defined by \Eq{CE-fin}. The ''elementary'' Euclidean operator, out of which the operator $C^E_v$ is composed, is given by\footnote{From now on, we let $s$ denote the spin of the loop created by the Euclidean operator. The letter $l$ is reserved to denote an internal spin of the intertwiner in a state on which the operator $C^E_{(ee')}$ acts.}
\be\label{CE_ee'}
C^E_{v,(e,e')} = -\frac{3}{W_s^2}\epsilon^{ijk}\Tr\bigl(\tau_k^{(s)}D^{(s)}(h_{\alpha_{ee'}})\bigr)J_i^{(v,e)}J_j^{(v,e')},
\ee
and acts on a pair of edges $(e,e')$ at the node $v$.

We will first calculate the action of the operator $C^E_{v,(e_1,e_2)} \equiv C^E_{12}$ on a state of the form
\be\label{E12-state}
\RealSymb{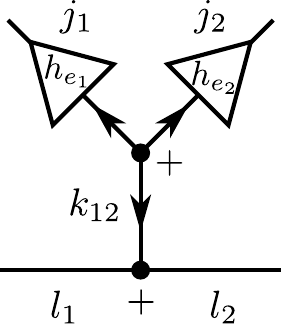}{0.6}
\ee
Since some orientation of the edges must be chosen in order to conveniently carry out the calculation, we will assume that the edges $e_1$ and $e_2$ are oriented outwards from the node $v$. However, it is not difficult to verify that the matrix elements of the operator are independent of the orientation of the edges on which it acts, and we will do so at the end of the calculation.

The action of the angular momentum operators in \Eq{CE_ee'} on the corresponding holonomies is given by \Eqs{J*h source} and \eqref{J*h target}. In the case that the edge is oriented away from the node, we have \vspace{-28pt}
\be\label{J source}
J_i^{(v,e)}\;\RealSymb{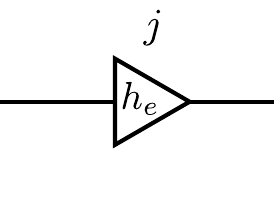}{0.6} \quad = \quad -W_j\;\RealSymb{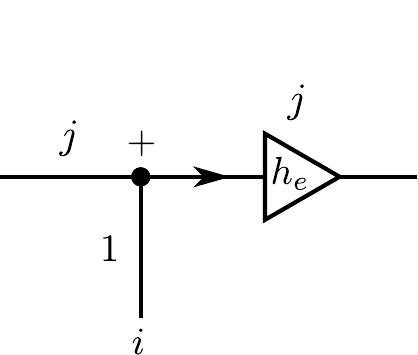}{0.6}
\ee
The remaining part of the operator \eqref{CE_ee'} may also be expressed in graphical form. Making use of the graphical representations of $\epsilon^{ijk}$ and $\tau_k^{(s)}$ from \Eqs{epsuvw g} and \eqref{tau=C g}, we find
\be\label{epsTr g}
\epsilon^{ijk}\Tr\bigl(\tau_k^{(s)}D^{(s)}(h_{\alpha_{ee'}})\bigr) \quad = \quad (i\sqrt 6)(iW_s)\;\RealSymb{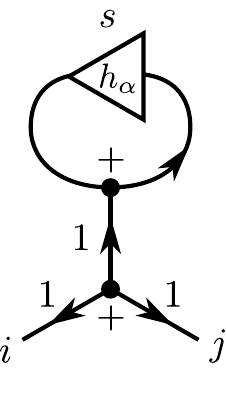}{0.6}
\ee
where the free indices $i$ and $j$ are to be contracted against the indices of the angular momentum operators.

With the help of \Eqs{J source} and \eqref{epsTr g}, we can now write down the action of the operator $C^E_{12}$ on the state \eqref{E12-state}:
\be\label{CE*state12}
C^E_{12}\;\RealSymb{fig16-state12_E.pdf}{0.6} \quad = \quad \frac{3\sqrt 6}{W_s}W_{j_1}W_{j_2}\RealSymb{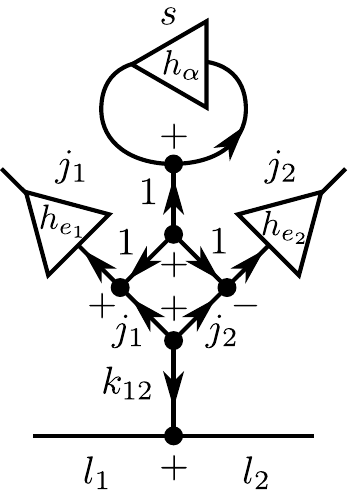}{0.6}
\ee
In order to bring out the matrix elements of the operator with respect to a particular basis, the intertwiner on the right-hand side must be expanded in the desired basis. Focusing on the five-valent intertwiner carrying spins $j_1$, $j_2$, $l_1$, $l_2$ and 1, we choose to expand it as
\newpage
\be\label{E12-intw-expanded}
\RealSymb{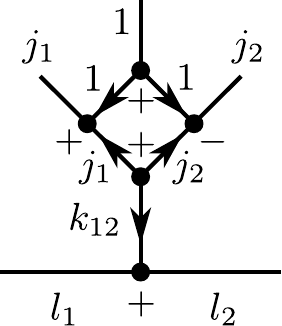}{0.6} \quad = \quad \sum_{x_{12}y} d_{x_{12}}d_y\,E_{12}(k_{12},x_{12},y)\;\RealSymb{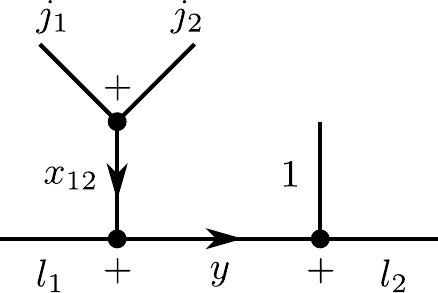}{0.6}
\ee
At the end of the calculation, the loop created by the Euclidean operator,
\be\label{E12-loop}
\RealSymb{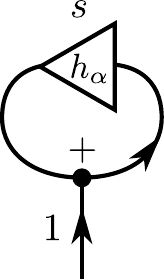}{0.6}
\ee
will be attached to the line carrying spin 1 on the right-hand side of \Eq{E12-intw-expanded}.

The coefficients $E_{12}(k_{12},x_{12},y)$ in \Eq{E12-intw-expanded} are obtained by contracting the intertwiner on the left-hand side of the equation with that on the right-hand side:\footnote{For simplicity of notation, we indicate explicitly only the dependence of $E_{12}(k_{12},x_{12},y)$ on the variable internal spins of the intertwiner; however, $E_{12}(k_{12},x_{12},y)$ is certainly also a function of the external spins $j_1$, $j_2$, $l_1$ and $l_2$. The same remark applies to other similar functions defined later in this chapter, such as the function $E_{13}$ of \Eq{E_13}.}
\be\label{E_12}
E_{12}(k_{12},x_{12},y) \quad = \quad \RealSymb{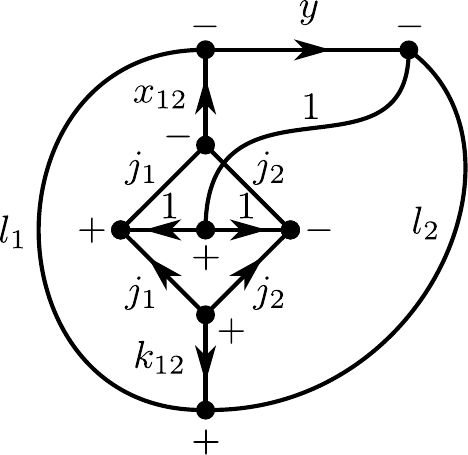}{0.6}
\ee
To deal with this diagram, we appeal to the ''fundamental theorem of graphical calculus'' in the form \eqref{thm3'}. The diagram can be separated into two disconnected pieces by cutting three lines according to
\be\label{E12-cut}
\RealSymb{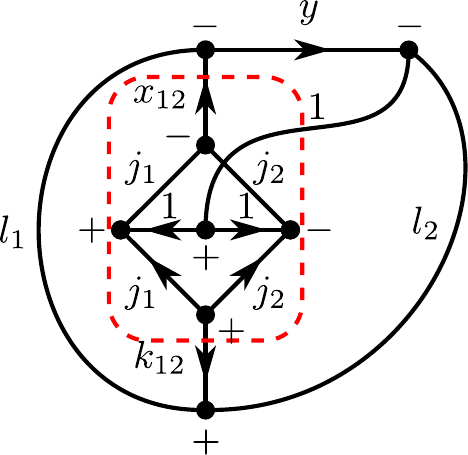}{0.6}
\ee
Therefore the diagram is equal to the product of the two pieces resulting from the cutting. The simpler of the two pieces is \vspace{-10pt}
\be
\RealSymb{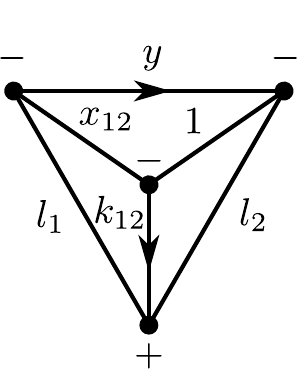}{0.6} \quad = \quad (-1)^{k_{12}+l_1-l_2}\sixj{k_{12}}{x_{12}}{1}{y}{l_2}{l_1},
\ee
where the factor $(-1)^{k_{12}+l_1-l_2}$ arises when \Eqs{invarrow}, \eqref{3jminus g} and \eqref{3jarrows g} are used to adjust the arrows and signs in the diagram so that they agree with \Eq{6j g}. Similarly, by comparing the remaining piece of the diagram \eqref{E12-cut} with \Eq{9j g}, we find
\be
\RealSymb{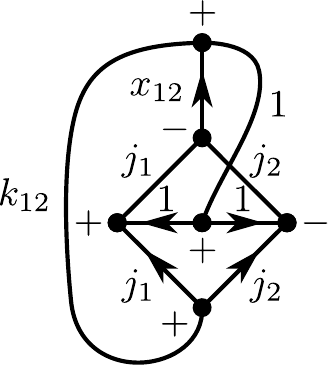}{0.6} \quad = \quad \RealSymb{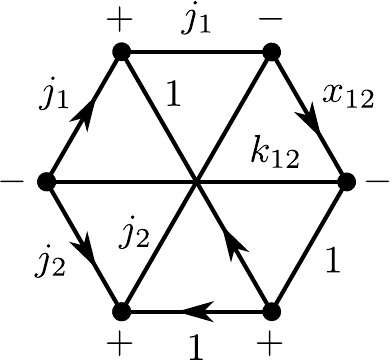}{0.6} \quad = \quad (-1)^{k_{12}-x_{12}}\ninej{j_1}{j_1}{1}{j_2}{j_2}{1}{k_{12}}{x_{12}}{1}.
\ee
Hence we have shown that
\be\label{E_12-result}
E_{12}(k_{12},x_{12},y) = (-1)^{l_1-l_2+x_{12}}\sixj{k_{12}}{x_{12}}{1}{y}{l_2}{l_1}\ninej{j_1}{j_1}{1}{j_2}{j_2}{1}{k_{12}}{x_{12}}{1}.
\ee
We may now read off the result of our calculation from \Eqs{CE*state12} and \eqref{E12-intw-expanded}. We see that the action of the Euclidean operator on the state \eqref{E12-state} is given by
\begin{align}
C^E_{12}\;&\RealSymb{fig16-state12_E.pdf}{0.6} \quad = \quad 3\sqrt 6\frac{W_{j_1}W_{j_2}}{W_s} \sum_{x_{12}y}d_{x_{12}}d_y(-1)^{l_1-l_2+x_{12}} \notag \\
&\times\sixj{k_{12}}{x_{12}}{1}{y}{l_2}{l_1}\ninej{j_1}{j_1}{1}{j_2}{j_2}{1}{k_{12}}{x_{12}}{1}\;\RealSymb{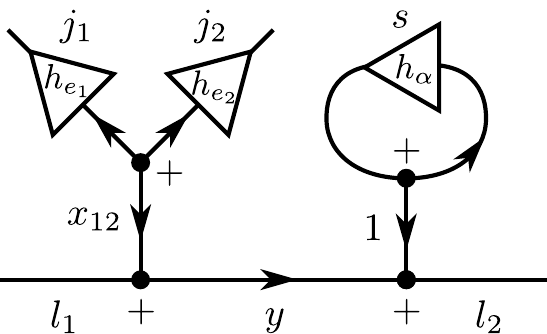}{0.6}\label{CE12-result}
\end{align}

\newpage

The calculation having been completed, let us then address the question of whether the result depends on the orientation of the edges $e_1$ and $e_2$ in the state \eqref{E12-state}. If one of the edges, say $e_1$, is oriented into the node, we must consider the action of the operator $C^E_{12}$ on the state
\be\label{E12-state_inv}
\RealSymb{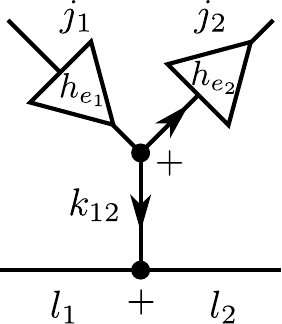}{0.6}
\ee
The action of the angular momentum operator on the first holonomy is now given by \eqref{J*h target}, namely \vspace{-28pt}
\be\label{J target}
J_i^{(v,e)}\;\RealSymb{fig16-h_e-source.pdf}{0.6} \quad = \quad W_j\;\RealSymb{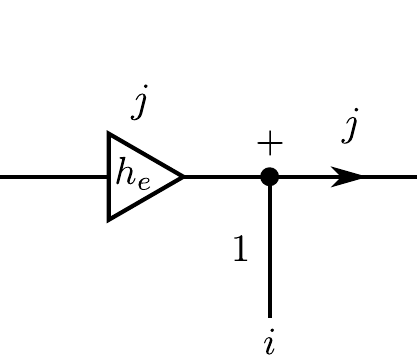}{0.6}
\ee
We then obtain, instead of \Eq{CE*state12},
\be\label{CE*state12_inv}
C^E_{12}\;\RealSymb{fig16-state12_inv.pdf}{0.6} \quad = \quad -\frac{3\sqrt 6}{W_s}W_{j_1}W_{j_2}\RealSymb{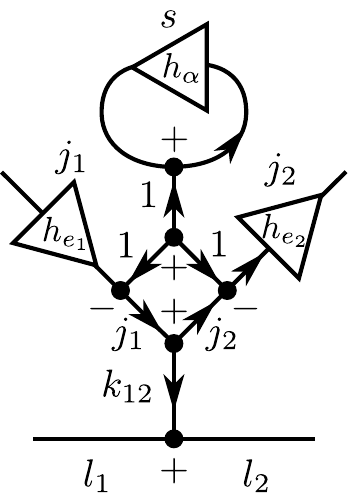}{0.6}
\ee
Compared against \Eq{CE*state12}, there are three differences on the right-hand side:
\begin{itemize}
\item The arrow on the line carrying spin $j_1$ is reversed;
\item The node with spins $j_1$, $j_1$ and $1$ has the opposite sign;
\item There is an overall factor of $(-1)$, due to the absence of $(-1)$ in \Eq{J target}.
\end{itemize}
By \Eqs{invarrow} and \eqref{3jminus g}, the first two differences contribute respectively the factors $(-1)^{2j_1}$ and $(-1)^{2j_1+1}$. Thus the overall factor in \Eq{CE*state12_inv} relative to \Eq{CE*state12} is $(-1)^{4j_1+2} = +1$. This shows that the matrix elements of not only the Euclidean operator \eqref{CE_ee'}, but those of any operator which acts on holonomies only through the angular momentum operator, are not sensitive to the orientation of the edges in the state on which the operator is acting.

\newpage

The other case which we need to consider for the Euclidean operator is the action of the operator $C^E_{v,(e_1,e_3)}\equiv C^E_{13}$ on a state of the form
\be\label{E13-state}
\RealSymb{fig16-state13}{0.6}
\ee
where we again assume that the edges $e_1$ and $e_3$ are oriented outwards from the node $v$. The calculation naturally proceeds along the same lines as the calculation that lead from \Eq{CE*state12} to \eqref{CE12-result}. Applying the graphical representation of the Euclidean operator in the same way as in \Eq{CE*state12}, we find
\begin{align}
&C^E_{13}\;\RealSymb{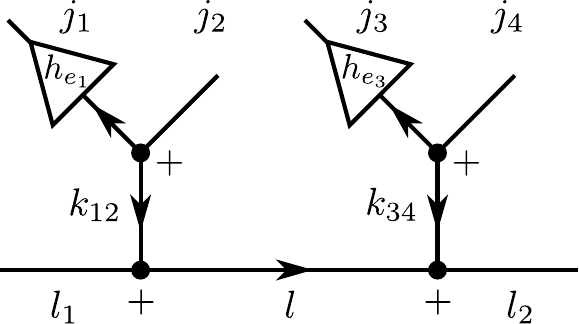}{0.6} \notag \\
&\hspace{3cm} =\quad 3\sqrt 6\frac{W_{j_1}W_{j_3}}{W_s}\RealSymb{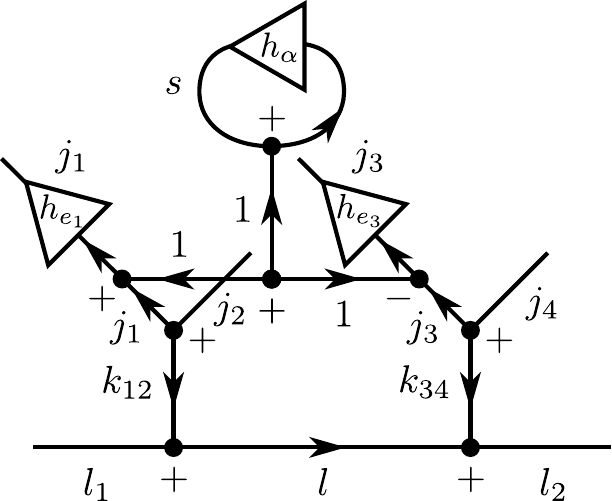}{0.6}\label{CE*state13}
\end{align}
We now expand the intertwiner on the right-hand side as
\begin{align}
&\RealSymb{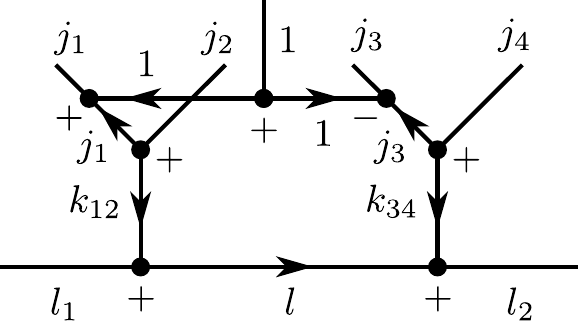}{0.6} \quad = \quad \sum_{x_{12}x_{34}yz} d_{x_{12}}d_{x_{34}}d_yd_z \notag \\
&\hspace{1cm}\times E_{13}(x_{12},x_{34},y,z|k_{12},k_{34},l)\;\RealSymb{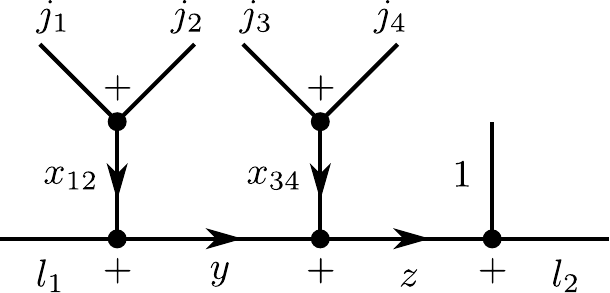}{0.6}\label{E13-intw-expanded}
\end{align}
with the coefficients of the expansion given by
\be\label{E_13}
E_{13}(x_{12},x_{34},y,z|k_{12},k_{34},l) \quad = \quad \RealSymb{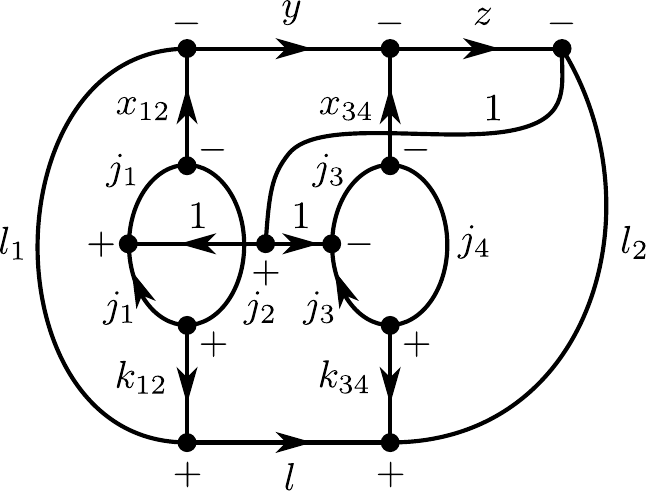}{0.6}
\ee
While this diagram is more complicated than the analogous diagram \eqref{E_12}, it can still be reduced into its ''elementary constituents'' by repeated cuts of no more than three lines. Making the cuts indicated by the dashed red lines in
\be
\RealSymb{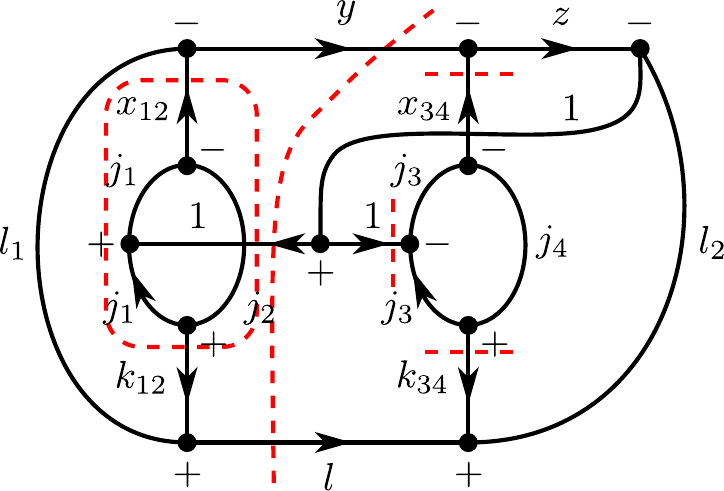}{0.6}
\ee
we find that the diagram splits into a product of four pieces. Three of the pieces are proportional to 6$j$-symbols:
\be
\RealSymb{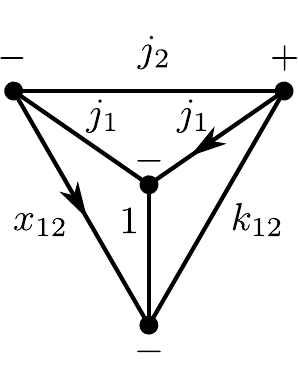}{0.6} \quad = \quad (-1)^{j_1+j_2+k_{12}}\sixj{j_1}{j_1}{1}{k_{12}}{x_{12}}{j_2},
\ee
\vspace{-10pt}
\be
\RealSymb{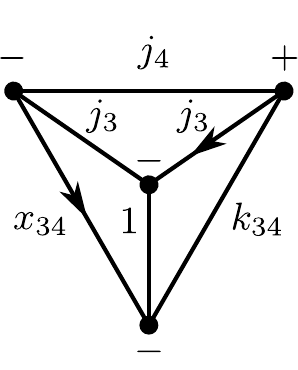}{0.6} \quad = \quad (-1)^{j_3+j_4+k_{34}}\sixj{j_3}{j_3}{1}{k_{34}}{x_{34}}{j_4},
\ee
\be
\RealSymb{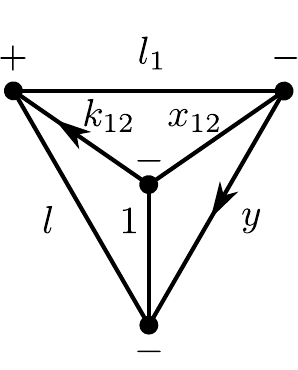}{0.6} \quad = \quad (-1)^{k_{12}+l_1-l}\sixj{k_{12}}{x_{12}}{1}{y}{l}{l_1}.
\ee
The remaining piece gives a 9$j$-symbol:
\begin{align}
&\RealSymb{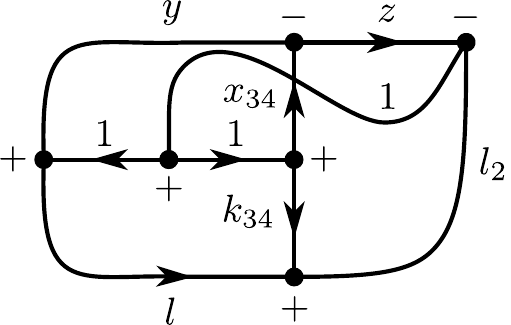}{0.6} \notag \\
&\hspace{1cm} =\quad \RealSymb{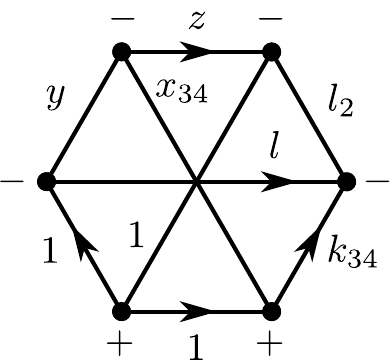}{0.6} \quad = \quad (-1)^{l+l_2+y+z+1}\ninej{l}{y}{1}{l_2}{z}{1}{k_{34}}{x_{34}}{1}.
\end{align}
Putting all the pieces together in \Eq{E13-intw-expanded}, we obtain the result
\begin{align}
&\quad C^E_{13}\;\RealSymb{fig16-state13.pdf}{0.6} \notag \\
&\notag \\
&=\;3\sqrt 6\frac{W_{j_1}W_{j_2}}{W_s}\sum_{x_{12}x_{34}yz}d_{x_{12}}d_{x_{34}}d_yd_z(-1)^{j_1+j_2+j_3+j_4+k_{34}-l_1-l_2-y-z+1} \notag \\
&\quad\times\;\sixj{j_1}{j_1}{1}{k_{12}}{x_{12}}{j_2}\sixj{j_3}{j_3}{1}{k_{34}}{x_{34}}{j_4}\sixj{k_{12}}{x_{12}}{1}{y}{l}{l_1}\ninej{l}{y}{1}{l_2}{z}{1}{k_{34}}{x_{34}}{1} \notag \\
&\hspace{3cm}\times\;\RealSymb{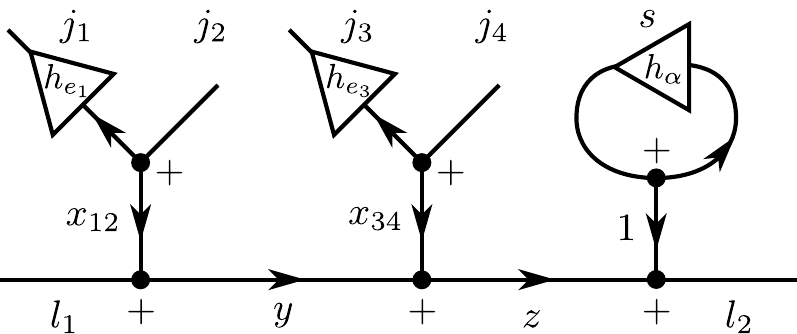}{0.6}\label{CE13-result}
\end{align}

\subsection{Curvature part}\label{sec:CR-el}

We then move on to consider the curvature operator\footnote{We will often refer to the operator of \Eq{CR-fin} simply as the ''curvature operator'', even though strictly speaking it is not the operator corresponding to the scalar curvature of the spatial surfaces.} $C^R_v$ of \Eq{CR-fin}. The operator $C^R_v$ is composed of ''elementary'' curvature operators of the form
\be\label{CR_ee'}
C^R_{v,(e,e')} = \frac{1}{2}\bigl|Y_v^{(e,e')}\bigr|\biggl(\frac{2\pi}{\alpha_{ee'}} - \pi + \theta_v^{(e,e')}\biggr),
\ee
where the operators $\bigl|Y_v^{(e,e')}\bigr|$ and $\theta_v^{(e,e')}$ are defined by \Eqs{curv-Y} and \eqref{curv-theta}. From the point of view of computing the action of the operator on a spin network state, the most important difference between the curvature operator \eqref{CR_ee'} and the Euclidean operator \eqref{CE_ee'} is that the former is a non-polynomial combination of the angular momentum operators $J_i^{(v,e)}$. For this reason, the graphical representation of the action of the angular momentum operator is of little direct use in calculating matrix elements of the curvature operator, and we must therefore approach this problem by different, largely non-graphical means.

The key observation, on which the calculation of the action of the curvature operator on any spin network state is based, is that the state
\be\label{R12-state}
\RealSymb{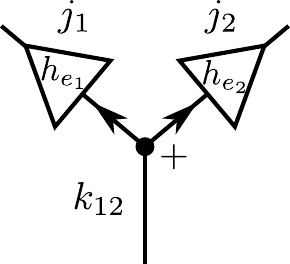}{0.6}
\ee
is an eigenstate of the operator $C^R_{v,(e_1,e_2)}\equiv C^R_{12}$. In order to see this, we recall that we have already discussed the operators $\bigl|Y_v^{(e,e')}\bigr|^2$ and $\theta_v^{(e,e')}$ in sections \ref{sec:angle} and \ref{sec:length}. In particular, we found that the operators act diagonally on the state \eqref{R12-state} -- see \Eqs{angle-eq} and \eqref{Y^2-eq}. Regarding $\bigl|Y_v^{(e,e')}\bigr|^2$ and $\theta_v^{(e,e')}$ as operators acting on the intertwiner space of the node, we have the eigenvalue equations
\begin{align}
\bigl|Y_v^{(e_1,e_2)}\bigr|^2\;\RealSymb{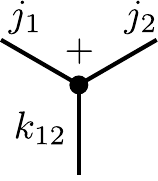}{0.6} \quad &= \quad \omega(j_1,j_2,k_{12})\;\RealSymb{fig16-iota12.pdf}{0.6} \\
\intertext{and}
\theta_v^{(e_1,e_2)}\;\RealSymb{fig16-iota12.pdf}{0.6} \quad &= \quad \theta(j_1,j_2,k_{12})\;\RealSymb{fig16-iota12.pdf}{0.6}
\end{align}
The eigenvalues are given by \Eqs{angle-eig} and \eqref{Y^2-eig} as
\begin{align}
\omega(j_1,j_2,k_{12}) = j_1(j_1+1)j_2(j_2+1) &- \biggl(\frac{k_{12}(k_{12}+1)-j_1(j_1+1)-j_2(j_2+1)}{2}\biggr)^2 \notag \\
&- \frac{k_{12}(k_{12}+1)-j_1(j_1+1)-j_2(j_2+1)}{2}
\end{align}
and \vspace{8pt}
\be
\theta(j_1,j_2,k_{12}) = \cos^{-1}\biggl(\frac{k_{12}(k_{12}+1) - j_1(j_1+1) - j_2(j_2+1)}{2\sqrt{j_1(j_1+1)}\sqrt{j_2(j_2+1)}}\biggr).
\ee
Since the operators $\bigl|Y_v^{(e_1,e_2)}\bigr|^2$ and $\theta_v^{(e_1,e_2)}$ act diagonally on the state \eqref{R12-state}, the same is true of the operator $C^R_{12}$:
\be\label{CR12-result}
C^R_{12}\;\RealSymb{fig16-iota12.pdf}{0.6} \quad = \quad \Lambda(j_1,j_2,k_{12})\;\RealSymb{fig16-iota12.pdf}{0.6}
\ee
From \Eq{CR_ee'}, we see that the eigenvalue is
\be
\Lambda(j_1,j_2,k_{12}) = \frac{1}{2}\sqrt{\omega(j_1,j_2,k_{12})}\biggl(\frac{2\pi}{\alpha_{12}} - \pi + \theta(j_1,j_2,k_{12})\biggr).
\ee
Now the strategy for computing the action of the operator $C^R_{v,(e,e')}$ on any spin network state will be as follows. If the edges $e$ and $e'$ are coupled together by the intertwiner at the node $v$, the action is immediately given by \Eq{CR12-result}. If the edges are not coupled together by the intertwiner, the action is calculated in three steps:
\begin{itemize}
\item We perform a suitable change of basis in the intertwiner space of the node, expressing the intertwiner at the node in a basis in which the edges $e$ and $e'$ are coupled together;
\item We apply the operator $C^R_{v,(ee')}$, which now acts diagonally on the new intertwiner basis;
\item We reverse the change of basis performed in the first step, transforming the intertwiner back into the original basis.
\end{itemize}
For the purposes of this work, it suffices to consider the action of the operator $C^R_{v,(e_1,e_3)} \equiv C^R_{13}$ on the state
\be\label{R13-state}
\RealSymb{fig16-state13.pdf}{0.6}
\ee
The required change of basis, which brings the intertwiner of \Eq{R13-state} into a basis in which the spins $j_1$ and $j_3$ are coupled to a definite total spin, is derived in section \ref{sec:ftgc}. Applying the transformation given by \Eq{iota6change}, we find
\begin{align}
&\RealSymb{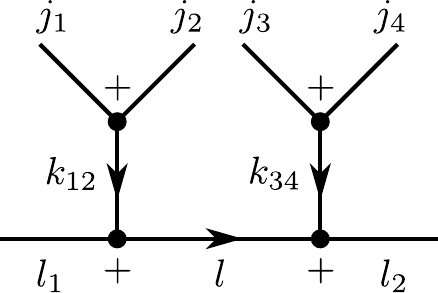}{0.6} \quad = \quad \sum_{a_{13}a_{24}b} d_{a_{13}}d_{a_{24}}d_b(-1)^{k_{12}+k_{34}-a_{13}-a_{24}} \notag \\
&\times\sum_c d_c\sixj{k_{12}}{k_{34}}{c}{l_2}{l_1}{l}\sixj{a_{13}}{a_{24}}{c}{l_2}{l_1}{b}\ninej{j_1}{j_2}{k_{12}}{j_3}{j_4}{k_{34}}{a_{13}}{a_{24}}{c}\quad\RealSymb{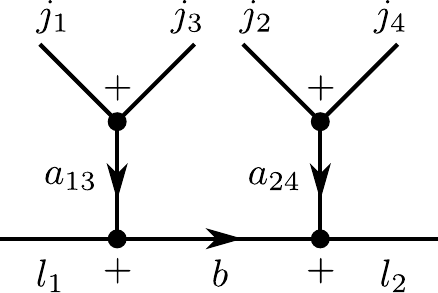}{0.6}
\end{align}
where the intertwiners on the right-hand side are of the type on which the operator $C^R_{13}$ acts diagonally. Hence, acting with the operator and then using \Eq{iota6change} in reverse to restore the intertwiners in the resulting state back into the basis of \Eq{R13-state}, we obtain
\begin{align}
&C^R_{13}\;\RealSymb{fig16-iota13.pdf}{0.6} \quad = \quad \sum_{a_{13}a_{24}b} d_{a_{13}}d_{a_{24}}d_b(-1)^{k_{12}+k_{34}-a_{13}-a_{24}} \notag \\
&\times\sum_c d_c\sixj{k_{12}}{k_{34}}{c}{l_2}{l_1}{l}\sixj{a_{13}}{a_{24}}{c}{l_2}{l_1}{b}\ninej{j_1}{j_2}{k_{12}}{j_3}{j_4}{k_{34}}{a_{13}}{a_{24}}{c}\Lambda(j_1,j_3,a_{13})\notag \\
&\times\sum_{x_{12}x_{34}y}d_{x_{12}}d_{x_{34}}d_y(-1)^{a_{13}+a_{24}-x_{12}-x_{34}}\sum_z d_z\sixj{a_{13}}{a_{24}}{z}{l_2}{l_1}{b}\sixj{x_{12}}{x_{34}}{z}{l_2}{l_1}{y} \notag \\
&\hspace{4cm}\times\ninej{j_1}{j_3}{a_{13}}{j_2}{j_4}{a_{24}}{x_{12}}{x_{34}}{z}\RealSymb{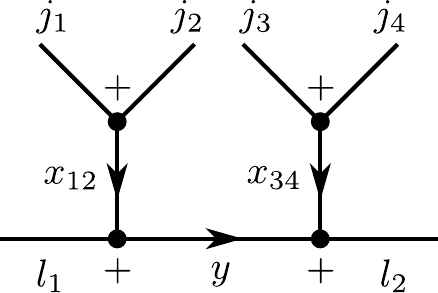}{0.6}
\end{align}
The sum over $b$ can now be performed using the orthogonality relation of 6$j$-symbols, \Eq{6j-orth}:
\be
\sum_b d_b\sixj{a_{13}}{a_{24}}{c}{l_2}{l_1}{b}\sixj{a_{13}}{a_{24}}{z}{l_2}{l_1}{b} \; = \; \frac{1}{d_c}\delta_{cz}.
\ee
This leads to
\begin{align}
&C^R_{13}\;\RealSymb{fig16-iota13.pdf}{0.6} \quad = \quad \sum_{\substack{a_{13}a_{24}c \\ x_{12}x_{34}y}} d_{a_{13}}d_{a_{24}}d_cd_{x_{12}}d_{x_{34}}d_y(-1)^{k_{12}+k_{34}-x_{12}-x_{34}} \notag \\
&\times\sixj{k_{12}}{k_{34}}{c}{l_2}{l_1}{l}\sixj{x_{12}}{x_{34}}{c}{l_2}{l_1}{y} \ninej{j_1}{j_2}{k_{12}}{j_3}{j_4}{k_{34}}{a_{13}}{a_{24}}{c}\ninej{j_1}{j_3}{a_{13}}{j_2}{j_4}{a_{24}}{x_{12}}{x_{34}}{c}\Lambda(j_1,j_3,a_{13}) \notag \\
&\hspace{7.5cm}\times\RealSymb{fig16-iota13_xy.pdf}{0.6}\label{CR13-result1}
\end{align}
In principle, \Eq{CR13-result1} completes the task of calculating the action of the curvature operator on the state \eqref{R13-state}. However, we can still manipulate the result by performing the sum over the spin $a_{24}$, which appears only in the two 9$j$-symbols -- for one thing, by doing so we will cast the matrix element into a form which is more suitable for numerical computations.

In order to compute the sum over $a_{24}$ in \Eq{CR13-result1}, we use the identity \cite{Varshalovich}
\be\label{9j=6j^3}
\ninej{j_1}{j_2}{j_3}{k_1}{k_2}{k_3}{l_1}{l_2}{l_3} = \sum_x d_x(-1)^{2x}\sixj{j_1}{j_2}{j_3}{k_3}{l_3}{x}\sixj{k_1}{k_2}{k_3}{j_2}{x}{l_2}\sixj{l_1}{l_2}{l_3}{x}{j_1}{k_1}.
\ee
Before applying this identity in \Eq{CR13-result1}, we shuffle the spins in the 9$j$-symbols so that in each symbol the position of the spin $a_{24}$ matches the position of $l_1$ on the left-hand side of \Eq{9j=6j^3}, and hence will appear in only one 6$j$-symbol on the right-hand side. In this way, the sum over $a_{24}$ becomes
\begin{align}
\sum_{a_{24}} d_{a_{24}}&\ninej{j_1}{j_2}{k_{12}}{j_3}{j_4}{k_{34}}{a_{13}}{a_{24}}{c}\ninej{j_1}{j_3}{a_{13}}{j_2}{j_4}{a_{24}}{x_{12}}{x_{34}}{c} \notag \\
=\sum_{a_{24}} d_{a_{24}}&\ninej{j_4}{j_3}{k_{34}}{j_2}{j_1}{k_{12}}{a_{24}}{a_{13}}{c}\ninej{j_4}{j_3}{x_{34}}{j_2}{j_1}{x_{12}}{a_{24}}{a_{13}}{c} \notag \\
=\sum_{a_{24}} d_{a_{24}}&\sum_x d_x(-1)^{2x}\sixj{j_4}{j_3}{k_{34}}{k_{12}}{c}{x}\sixj{j_2}{j_1}{k_{12}}{j_3}{x}{a_{13}}\sixj{a_{24}}{a_{13}}{c}{x}{j_4}{j_2} \notag \\
\times&\sum_{x'} d_{x'}(-1)^{2x'}\sixj{j_4}{j_3}{x_{34}}{x_{12}}{c}{x'}\sixj{j_2}{j_1}{x_{12}}{j_3}{x'}{a_{13}}\sixj{a_{24}}{a_{13}}{c}{x'}{j_4}{j_2}. \label{sum_a24}
\end{align}
By virtue of the orthogonality relation
\be
\sum_{a_{24}} d_{a_{24}} \sixj{a_{24}}{a_{13}}{c}{x}{j_4}{j_2}\sixj{a_{24}}{a_{13}}{c}{x'}{j_4}{j_2} = \frac{1}{d_x}\delta_{xx'}
\ee
the expression \eqref{sum_a24} now reduces to
\be
\sum_x d_x\sixj{j_4}{j_3}{k_{34}}{k_{12}}{c}{x}\sixj{j_2}{j_1}{k_{12}}{j_3}{x}{a_{13}}\sixj{j_4}{j_3}{x_{34}}{x_{12}}{c}{x}\sixj{j_2}{j_1}{x_{12}}{j_3}{x}{a_{13}}.
\ee
Going with this back to \Eq{CR13-result1}, and relabeling some of the dummy spins, we obtain the action of the curvature operator on the state \eqref{R13-state} in the form
\begin{align}
&C^R_{13}\;\RealSymb{fig16-iota13.pdf}{0.6} \quad = \quad \sum_{\substack{x_{12}x_{34}y \\ abc}} d_{x_{12}}d_{x_{34}}d_yd_ad_bd_c(-1)^{k_{12}+k_{34}-x_{12}-x_{34}} \notag \\
&\notag \\
&\quad\times\sixj{j_1}{j_2}{k_{12}}{b}{j_3}{a}\sixj{j_3}{j_4}{k_{34}}{c}{k_{12}}{b}\sixj{j_1}{j_2}{x_{12}}{b}{j_3}{a}\sixj{j_3}{j_4}{x_{34}}{c}{x_{12}}{b} \notag \\
&\quad\times\sixj{k_{12}}{k_{34}}{c}{l_2}{l_1}{l}\sixj{x_{12}}{x_{34}}{c}{l_2}{l_1}{y}\Lambda(j_1,j_3,a)\;\RealSymb{fig16-iota13_xy.pdf}{0.6}\label{CR13-result}
\end{align}

\subsection{Volume operator}\label{sec:V-el}

We then come to the volume operator, which, as we have seen, enters the Hamiltonian constraint operator (and thus the physical Hamiltonian of dust models) in addition to the operators $C^E_v$ and $C^R_v$. We will calculate matrix elements of the operator
\be\label{q_IJK}
q_v^{(IJK)} = \epsilon^{ijk}J_i^{(v,e_I)}J_j^{(v,e_J)}J_k^{(v,e_K)},
\ee
which is related to the actual volume operator in the way indicated by \Eqs{V(R)} and \eqref{q_v}. In particular, extracting the volume operator from the operator \eqref{q_IJK} involves taking a square root, which must be done numerically in all but the very simplest cases.

Let us first calculate the action of the operator $q_v^{(123)}$ on the state
\be\label{q123-state}
\RealSymb{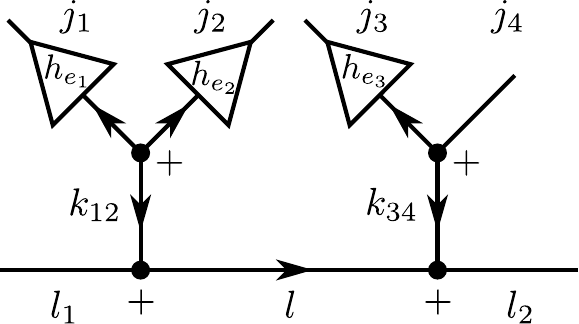}{0.6}
\ee
where we are again assuming that all the edges are oriented outwards from the node $v$, while recalling that the matrix elements of the operator do not depend on the orientation of the edges. When each of the angular momentum operators in \Eq{q_IJK} acts on the corresponding holonomy according to \Eq{J source}, and $\epsilon^{ijk}$ is converted into graphical form while picking up a factor of $i\sqrt 6$, the operator $q_v^{(123)}$ effectively acts on the intertwiner of the state \eqref{q123-state} as
\vspace{-12pt}
\be
q_v^{(123)}\;\RealSymb{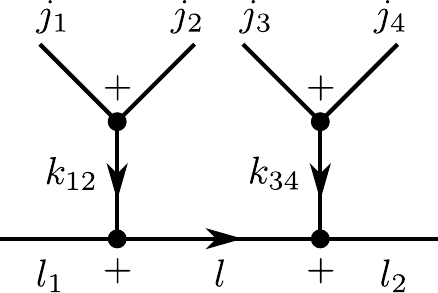}{0.6} \quad = \quad -i\sqrt 6 W_{j_1}W_{j_2}W_{j_3}\;\RealSymb{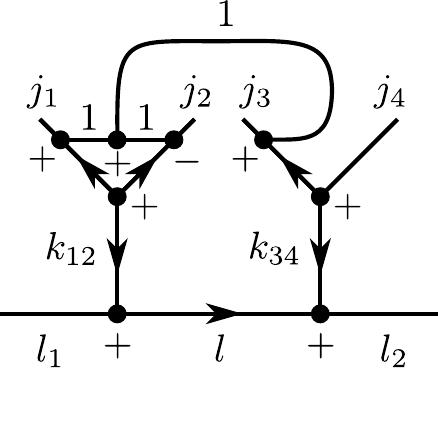}{0.6}
\ee
Now we must expand the intertwiner on the right-hand side in the basis of intertwiners used in the state \eqref{q123-state}. We have
\begin{align}
&\RealSymb{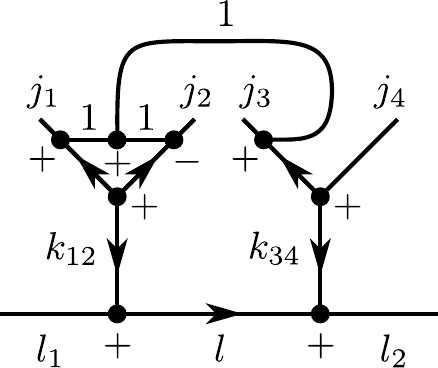}{0.6} \notag \\
&=\sum_{x_{12}x_{34}y} d_{x_{12}}d_{x_{34}}d_y\,q_{123}(x_{12},x_{34},y|k_{12},k_{34},l)\;\RealSymb{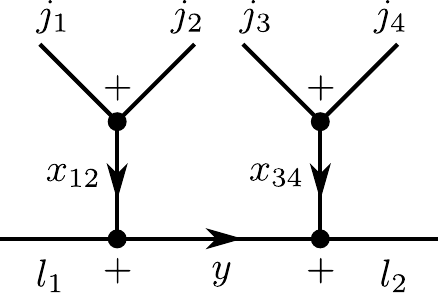}{0.6}
\end{align}
where the coefficients are given by
\be\label{q123}
q_{123}(x_{12},x_{34},y|k_{12},k_{34},l) \quad = \quad \RealSymb{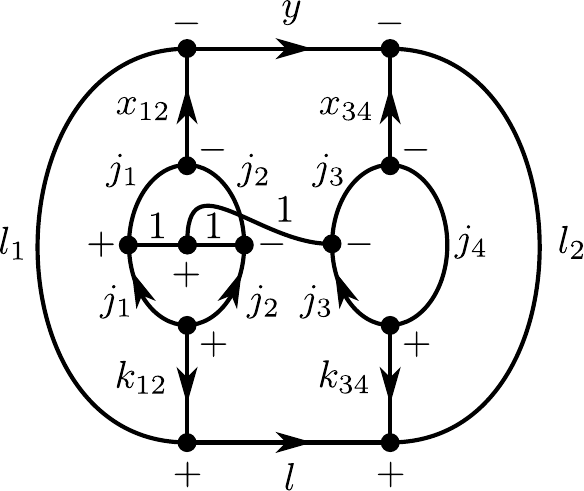}{0.6}
\ee
\newpage
\noindent
The diagram \eqref{q123} can be reduced into 6$j$- and 9$j$-symbols by making three cuts of three lines as follows:
\be
\RealSymb{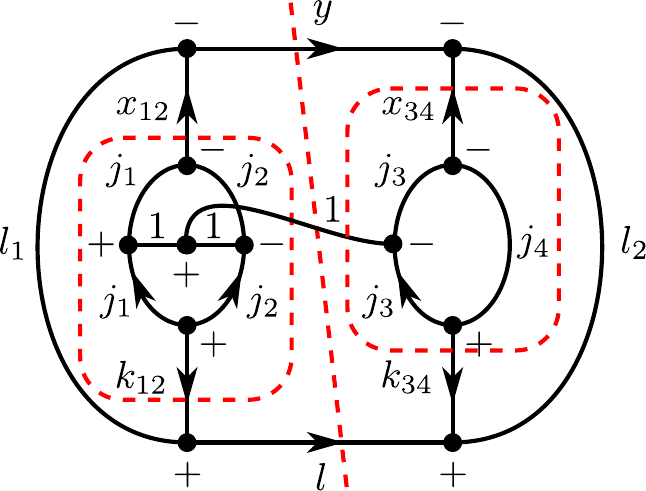}{0.6}
\ee
The piece on the inner left side of the diagram is a 9$j$-symbol,
\be
\RealSymb{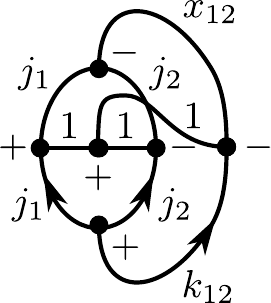}{0.6} \quad = \quad \RealSymb{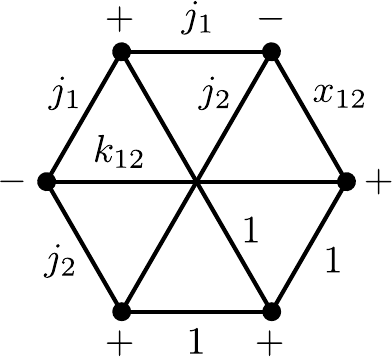}{0.6} \quad = \quad (-1)\ninej{j_1}{j_1}{1}{j_2}{j_2}{1}{k_{12}}{x_{12}}{1},
\ee
while each of the remaining pieces gives a 6$j$-symbol: \vspace{-8pt}
\begin{align}
\RealSymb{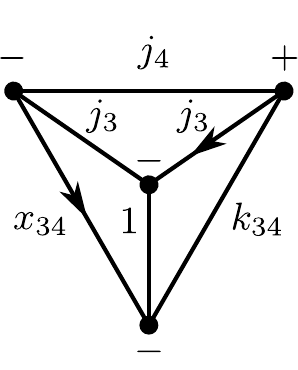}{0.6} \quad &= \quad (-1)^{j_3+j_4+k_{34}}\sixj{j_3}{j_3}{1}{k_{34}}{x_{34}}{j_4}, \\
\RealSymb{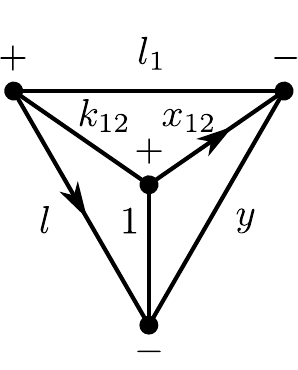}{0.6} \quad &= \quad (-1)^{l_1+l+x_{12}+1}\sixj{k_{12}}{x_{12}}{1}{y}{l}{l_1}, \\
\RealSymb{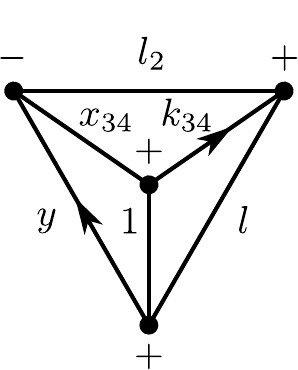}{0.6} \quad &= \quad (-1)^{l_2+y+x_{34}}\sixj{k_{34}}{x_{34}}{1}{y}{l}{l_2}.
\end{align}
Therefore we conclude that the action of the operator $q_v^{(123)}$ on the state \eqref{q123-state} reads
\begin{align}
&q_v^{(123)}\;\RealSymb{fig16-iota123.pdf}{0.6} \notag \\
&{}\notag \\
&=-i\sqrt 6 W_{j_1}W_{j_2}W_{j_3}\sum_{x_{12}x_{34}y} d_{x_{12}}d_{x_{34}}d_y\,(-1)^{j_3+j_4+k_{34}+l_1+l+l_2+x_{12}+x_{34}+y} \notag \\
&\quad\times \sixj{j_3}{j_3}{1}{k_{34}}{x_{34}}{j_4}\sixj{k_{12}}{x_{12}}{1}{y}{l}{l_1}\sixj{k_{34}}{x_{34}}{1}{y}{l}{l_2}\ninej{j_1}{j_1}{1}{j_2}{j_2}{1}{k_{12}}{x_{12}}{1}\notag \\
&\hspace{6.5cm}\times\;\RealSymb{fig16-iota123_xy.pdf}{0.6}\label{q123-result}
\end{align}
We must also consider the case of the operator $q_v^{(135)}$ acting on the state
\be\label{q135-state}
\RealSymb{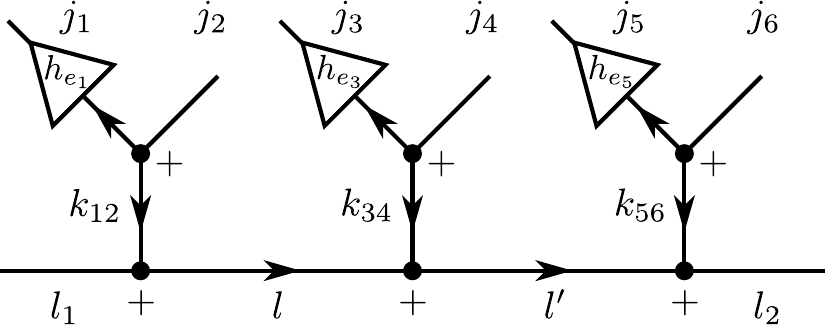}{0.6}
\ee
Letting again the angular momentum operators act on the holonomies in \Eq{q135-state}, and interpreting the resulting expression as an action on the intertwiner, we have
\begin{align}
&q_v^{(135)}\;\RealSymb{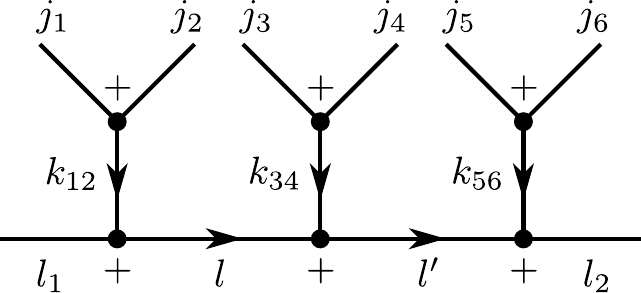}{0.6} \notag \\
&\hspace{2.5cm}=-i\sqrt 6 W_{j_1}W_{j_3}W_{j_5}\;\RealSymb{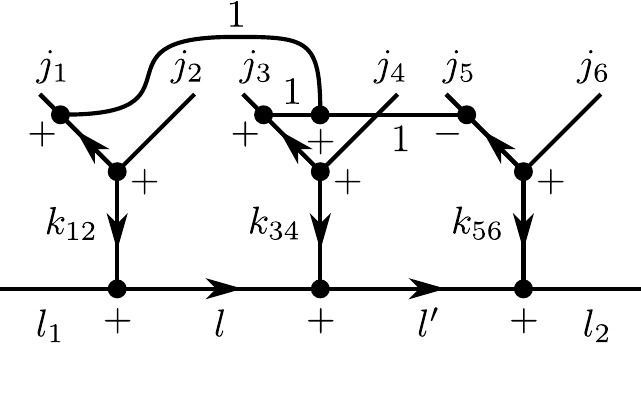}{0.6}
\end{align}
As before, the intertwiner on the right-hand side should be expanded with respect to the appropriate basis. This gives
\begin{align}
&\RealSymb{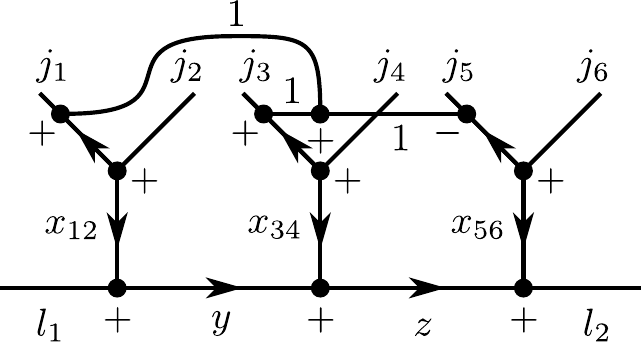}{0.6} \notag \\
&\notag \\
&=\sum_{x_{12}x_{34}x_{56}yz} d_{x_{12}}d_{x_{34}}d_{x_{56}}d_yd_z\,q_{135}(x_{12},x_{34},x_{56},y,z|k_{12},k_{34},k_{56},l,l')\notag \\
&\hspace{4.5cm}\times\;\RealSymb{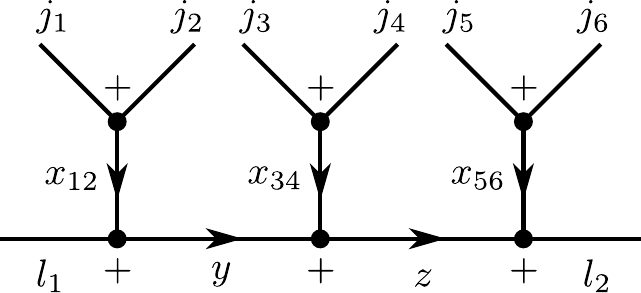}{0.6}
\end{align}
where
\begin{align}
q_{135}(x_{12},x_{34},x_{56}&,y,z|k_{12},k_{34},k_{56},l,l') \notag \\
&=\quad\RealSymb{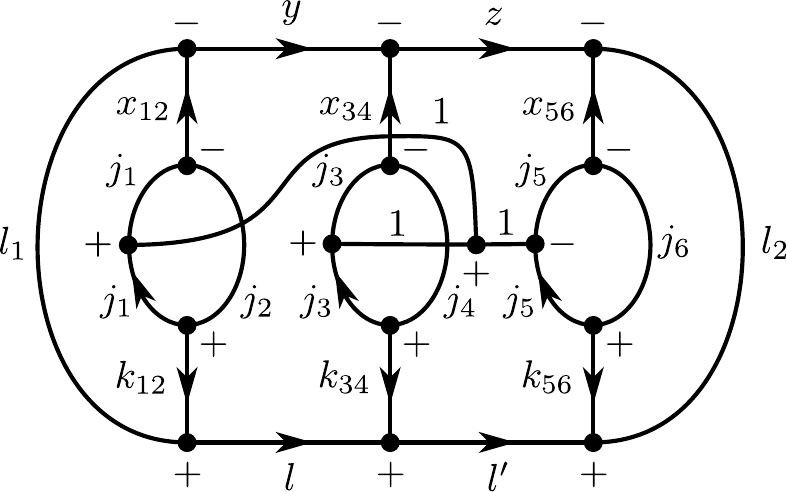}{0.6}\label{q135}
\end{align}
This diagram as well can be beaten down into a product of 6$j$- and 9$j$-symbols by repeatedly splitting off pieces by cutting three lines. We start by making the cuts shown below:
\be\label{q135-cut}
\RealSymb{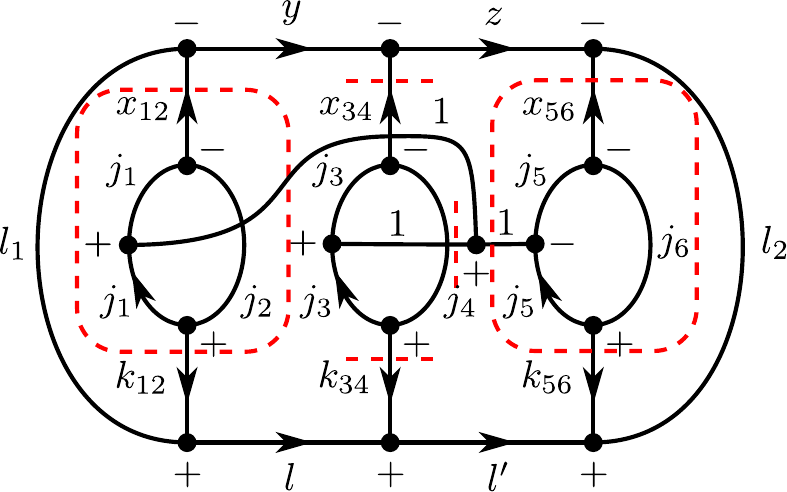}{0.6}
\ee
In this way we cut off three pieces of identical form, each of them being a 6$j$-symbol:
\be
\RealSymb{fig16-q135_cut1.pdf}{0.6} \quad = \quad (-1)^{j_1+j_2+k_{12}}\sixj{j_1}{j_1}{1}{k_{12}}{x_{12}}{j_2},
\ee
\vspace{-10pt}
\be
\RealSymb{fig16-q135_cut2.pdf}{0.6} \quad = \quad (-1)^{j_3+j_4+k_{34}}\sixj{j_3}{j_3}{1}{k_{34}}{x_{34}}{j_4},
\ee
and \vspace{-10pt}
\be
\RealSymb{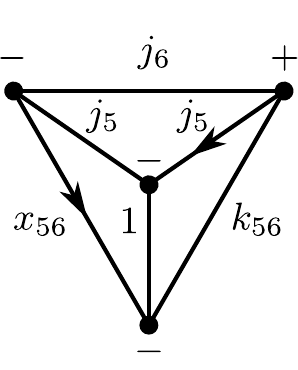}{0.6} \quad = \quad (-1)^{j_5+j_6+k_{56}}\sixj{j_5}{j_5}{1}{k_{56}}{x_{56}}{j_6}.
\ee
What remains of the diagram \eqref{q135-cut} can then be cut as follows:
\be
\RealSymb{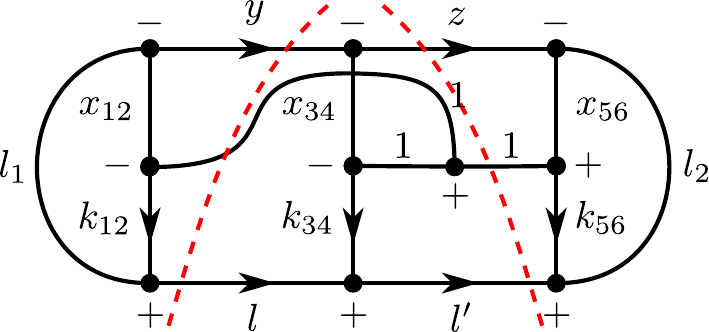}{0.6}
\ee
Now the left and right pieces each give a 6$j$-symbol,
\be
\RealSymb{fig16-q135_cut4.pdf}{0.6} \quad = \quad (-1)^{k_{12}+l_1-l}\sixj{k_{12}}{x_{12}}{1}{y}{l}{l_1}
\ee
\newpage
and \vspace{-10pt}
\be
\RealSymb{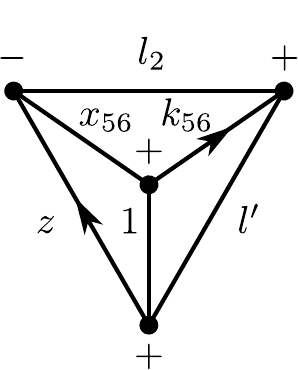}{0.6} \quad = \quad (-1)^{l_2+x_{56}+z}\sixj{k_{56}}{x_{56}}{1}{z}{l'}{l_2},
\ee
while the piece in the middle is a 9$j$-symbol,
\begin{align}
\RealSymb{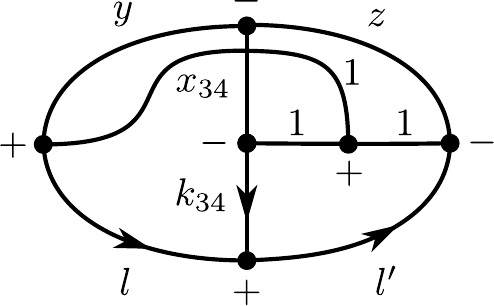}{0.6} \quad &= \quad (-1)^{2l'}\RealSymb{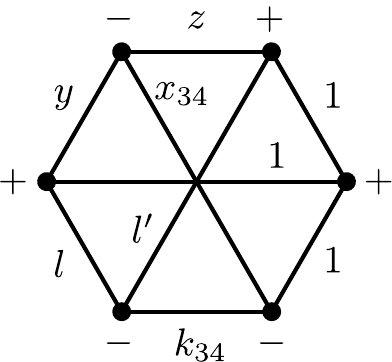}{0.6} \notag \\
&= \quad (-1)^{l-l'+y+z+1}\ninej{l}{y}{1}{l'}{z}{1}{k_{34}}{x_{34}}{1}.
\end{align}
Collecting all the pieces together, and carefully simplifying the powers of $(-1)$, we find the result
\begin{align}
&q_v^{(135)}\;\RealSymb{fig16-iota135.pdf}{0.6} \quad = \quad -i\sqrt 6 W_{j_1}W_{j_3}W_{j_5}\notag \\
&\notag \\
&\times\sum_{x_{12}x_{34}x_{56}yz} d_{x_{12}}d_{x_{34}}d_{x_{56}}d_yd_z\,(-1)^{j_1+j_2+j_3+j_4+j_5+j_6-l_1-l_2+k_{34}-l'-y+k_{56}-x_{56}+1} \notag \\
&\times \sixj{j_1}{j_1}{1}{k_{12}}{x_{12}}{j_2}\sixj{j_3}{j_3}{1}{k_{34}}{x_{34}}{j_4}\sixj{j_5}{j_5}{1}{k_{56}}{x_{56}}{j_6}\sixj{k_{12}}{x_{12}}{1}{y}{l}{l_1}\sixj{k_{56}}{x_{56}}{1}{z}{l'}{l_2} \notag \\
&\hspace{3cm}\times \ninej{l}{y}{1}{l'}{z}{1}{k_{34}}{x_{34}}{1}\;\RealSymb{fig16-iota135_xy.pdf}{0.6}
\end{align}

\newpage

\section[Approximation methods for deparametrized models]{Approximation methods for time evolution \\ in deparametrized models}\label{ch:approximation}

In deparametrized models of loop quantum gravity, where the role of a physical time variable is assigned to a scalar field contained in the model, the dynamics of diffeomorphism invariant states of the gravitational field is governed by the Schr\"odinger equation
\be\label{SE-17}
i\frac{d}{dt}\ket\Psi = H_{\rm phys}\ket\Psi,
\ee
with the physical Hamiltonian $H_{\rm phys}$ generating time evolution with respect to the time defined by the scalar field. For models in which the physical time variable is taken to be a free Klein--Gordon field or a non-rotational dust field, the mathematical structure of the model is completed by the constructions which are presented in Chapters \ref{ch:Hphys} and \ref{ch:constraint}, and which provide a concrete proposal for the form of the physical Hamiltonian.

In the case of the free scalar field model, the physical Hamiltonian is given by
\be\label{Hphys^S}
H_{\rm phys}^S\ket{\Psi_\Gamma} = \sqrt{\frac{1+\beta^2}{8\pi G\beta^2}}\sum_{v\in\Gamma}\sqrt{\displaystyle{\rm Sym}\,\biggl(C^R_v + \frac{1}{1+\beta^2}C^E_v\biggr)\bigg|_{\R_+}}\ket{\Psi_\Gamma},
\ee
where $C^R_v$ and $C^E_v$ are the operators defined by \Eqs{CE-fin} and \eqref{CR-fin} (here naturally considered as operators on ${\cal H}_{\rm diff}$). Inside the square root, ${\rm Sym}$ denotes a symmetric ordering of the operator. In this chapter, we choose the straightforward symmetrization
\be\label{Sym'}
{\rm Sym}\,\biggl(C^R_v + \frac{1}{1+\beta^2}C^E_v\biggr) = C^R_v + \frac{1}{1+\beta^2}\frac{1}{2}\Bigl(C^E_v + (C^E_v)^\dagger\Bigr).
\ee
Furthermore, we are forced to introduce a restriction of the operator within the square root to the positive part of its spectrum, defined as ${\cal O}\big|_{\R_+} \equiv \half\bigl({\cal O} + |{\cal O}|\bigr)$, in order to ensure that \Eq{Hphys^S} defines a symmetric physical Hamiltonian. 

For the non-rotational dust model, the physical Hamiltonian is obtained by evaluating the Hamiltonian constraint operator of Chapter \ref{ch:constraint} at unit lapse function. In this chapter we consider only the physical Hamiltonian obtained by symmetrizing the operator defined by \Eqs{C(N)-v2} and \eqref{C_v}. Thus,
\be\label{Hphys^D}
\hspace{0.5cm}H_{\rm phys}^D\ket{\Psi_\Gamma} = \frac{1}{16\pi G}\frac{1+\beta^2}{\beta^2}\sum_{v\in\Gamma}\sqrt{{\cal V}_v^{-1}}\biggl(C^R_v + \frac{1}{1+\beta^2}\frac{1}{2}\Bigl(C^E_v + (C^E_v)^\dagger\Bigr)\biggr)
\sqrt{{\cal V}_v^{-1}}\ket{\Psi_\Gamma},\hspace{0.5cm}
\ee
where $C^R_v$ and $C^E_v$ are the same operators as in \Eq{Hphys^S}, and ${\cal V}_v^{-1}$ is (the dual of) the operator \eqref{V^-1}. We emphasize that the operator \eqref{Hphys^D} is a well-defined operator on ${\cal H}_{\rm diff}$, even though the constraint operator \eqref{C(N)-v2} (for a general lapse function) is only an operator on ${\cal H}_{\rm vtx}$.

In order to study the dynamics generated by a given physical Hamiltonian through concrete calculations, a knowledge of the spectrum of the Hamiltonian is a necessary prerequisite. The time evolution operator $U(t) = e^{-iH_{\rm phys}t}$ must generally be defined through its spectral decomposition; moreover, in the case of the free scalar field model, even the action of the physical Hamiltonian itself can be made explicit only if the spectral decomposition of the operator under the square root in \Eq{Hphys^S} is available. However, so far no progress has been made towards obtaining an exact solution of the eigenvalue problem of the operator \eqref{Hphys^S} or \eqref{Hphys^D}, even within some restricted subspace of the entire diffeomorphism invariant Hilbert space.

This being the case, it becomes imperative to develop suitable approximation methods for dealing with the Hamiltonians \eqref{Hphys^S} or \eqref{Hphys^D}, so that the dynamics contained in them could be understood at the level of explicit calculations. A first step in this direction was taken by the author in collaboration with Mehdi Assanioussi and Jerzy Lewandowski in the article \cite{paper5}, in which we used ideas from elementary quantum mechanics to introduce methods which make it possible to approximate the time evolution of a given initial state under the Hamiltonian \eqref{Hphys^S} or \eqref{Hphys^D}. We also gave examples of using our methods to compute numerically the time evolution of expectation values of the volume and curvature operators in certain simple initial states. In this chapter we will review the work carried out in \cite{paper5}.

\subsection{Perturbation theory}\label{sec:perturbation}

As we have seen, the operator $C^R_v$ in \Eqs{Hphys^S} or \eqref{Hphys^D} is a graph-preserving operator, and acts only on the intertwiner space of the node $v$. The matrix elements of $C^R_v$ within a given intertwiner space can be calculated in the way discussed in section \ref{sec:CR-el}, and the resulting matrix can be diagonalized numerically, at least if the dimension of the intertwiner space is not unreasonably high. (The same remarks apply to the operator ${\cal V}_v^{-1/2}$ in the Hamiltonian of the dust field model.) The complexity in diagonalizing the Hamiltonian \eqref{Hphys^S} or \eqref{Hphys^D} arises from the Euclidean part, which has non-vanishing matrix elements between states carrying a different number of special loops. Consequently, even the simplest subspaces which are preserved by the action of the Hamiltonian, and within which the Hamiltonian could be diagonalized, are infinite-dimensional.

A possible way to deal with this situation is suggested by the way we have written \Eqs{Hphys^S} and \eqref{Hphys^D}, where a factor of $(1+\beta^2)$ has been pulled out, so that \mbox{$1/(1+\beta^2)$} appears as a prefactor for the Euclidean part relative to the curvature part. We see that for sufficiently large values of $\beta$, the factor $1/(1+\beta^2)$ plays the role of a small parameter multiplying the problematic part of the Hamiltonian. In such a situation, the familiar time-independent perturbation theory of quantum mechanics can be used to find approximate expressions for the eigenvalues and eigenstates of the full Hamiltonian in terms of the known eigenvalues and eigenstates of the ''unperturbed'' part of the Hamiltonian.

In our case we must consider an operator of the form
\be\label{H0+V}
H = H_R + \epsilon H_E,
\ee
which can be either the operator under the square root in the physical Hamiltonian of the free scalar field model, or the physical Hamiltonian itself for the non-rotational dust model. In either case the curvature part $H_R$ acts in a graph-preserving way, while the Euclidean part $H_E$ acts by creating and removing special loops. The perturbation parameter has the value $\epsilon = 1/(1+\beta^2)$. The solution of the perturbative problem of the operator \eqref{H0+V} requires some care due to degeneracies which are present in the spectrum of the unperturbed operator $H_R$, and which seem not to be removed by the perturbation $H_E$. We will therefore present the solution in full (up to second order in perturbation theory) in the following section; here we will summarize the results of the calculation.

Let us denote the eigenvalues and eigenstates of the unperturbed operator $H_R$ by $\lambda_n^{(0)}$ and $\bket{\lambda_n^{(0)}}$, and those of the entire operator $H$ by $\lambda_n$ and $\bket{\lambda_n}$. Then the two operators can be expressed in terms of their spectral decompositions as
\be\label{HR-spectral}
H_R = \sum_n \lambda_n^{(0)}\bket{\lambda_n^{(0)}}\bbra{\lambda_n^{(0)}}
\ee
and
\be\label{H-spectral}
H = \sum_n \lambda_n\bket{\lambda_n}\bbra{\lambda_n}.
\ee
To second order in perturbation theory, the eigenvalues $\lambda_n$ are given by
\be\label{lambda-approx}
\lambda_n = \lambda_n^{(0)} + \epsilon^2 \sum_k \frac{\bigl|\bbra{\lambda_n^{(0)}}H_E\bket{\lambda_k^{(0)}}\bigr|^2}{\lambda_n^{(0)}-\lambda_k^{(0)}} + {\cal O}(\epsilon^3),
\ee
where the first-order correction vanishes because $H_E$ changes the number of special loops, while all the eigenstates corresponding to a given unperturbed eigenvalue $\lambda_n^{(0)}$ have the same number of special loops. For the eigenstates, we have
\begin{align}
\bket{\lambda_n} = \bket{\lambda_n^{(0)}} &+ \epsilon\sum_k \frac{\bbra{\lambda_k^{(0)}}H_E\bket{\lambda_n^{(0)}}}{\lambda_n^{(0)}-\lambda_k^{(0)}}\bket{\lambda_k^{(0)}} \notag \\
&+\epsilon^2\hspace{-8pt}\sum_{\substack{k \\ \lambda_k^{(0)}\neq \lambda_n^{(0)}}}\hspace{-8pt}\biggl(\sum_l \frac{\bbra{\lambda_k^{(0)}}H_E\bket{\lambda_l^{(0)}}\bbra{\lambda_l^{(0)}}H_E\bket{\lambda_n^{(0)}}}{\bigl(\lambda_n^{(0)}-\lambda_k^{(0)}\bigr)\bigl(\lambda_n^{(0)}-\lambda_l^{(0)}\bigr)}\biggr)\bket{\lambda_k^{(0)}} \notag \\
&+\epsilon^2\biggl(-\frac{1}{2}\sum_k \frac{\bigl|\bbra{\lambda_n^{(0)}}H_E\bket{\lambda_k^{(0)}}\bigr|^2}{\bigl(\lambda_n^{(0)}-\lambda_k^{(0)}\bigr)^2}\biggr)\bket{\lambda_n^{(0)}} + {\cal O}(\epsilon^3).\label{state-approx}
\end{align}
In general, if the unperturbed eigenstate $\bket{\lambda_n^{(0)}}$ has $L$ special loops, the second-order approximation to the eigenstate $\bket{\lambda_n}$ will contain terms in which the number of special loops ranges from $L-2$ to $L+2$. In this chapter we will restrict ourselves to considering the second-order approximations to the eigenvalues and eigenstates, even though in principle there is nothing to prevent one from including terms of higher order in \Eqs{lambda-approx} and \eqref{state-approx}.

The Hamiltonian $H$ can now be approximated by inserting the expressions \eqref{lambda-approx} and \eqref{state-approx} into the spectral decomposition \eqref{H-spectral} (and discarding terms of higher than second order in $\epsilon$). The time evolution operator $U(t) = e^{-iH_{\rm phys}t}$ can be approximated in the same way. In particular, for the scalar field model we can write
\be\label{U-spectral}
U_S(t) = e^{-iH_{\rm phys}^S t} = \sum_{\substack{n \\ \lambda_n\geq 0}} \exp\biggl(-i\sqrt{\frac{1+\beta^2}{8\pi G\beta^2}}\sqrt{\lambda_n}\,t\biggr)\bket{\lambda_n}\bbra{\lambda_n},
\ee
where $\lambda_n$ and $\bket{\lambda_n}$ are the (approximate) eigenvalues and eigenstates of the operator \eqref{Sym'}, and the restriction to the positive part of the spectrum is accounted for by excluding the negative eigenvalues from the sum. Naturally a similar expression, but without the square root and the restriction on the eigenvalues, holds in the case of the dust model.

In actual calculations it would certainly be impractical to try to use \Eq{H-spectral} as an approximation for the Hamiltonian on the entire Hilbert space ${\cal H}_{\rm diff}$, by letting the sum run over the whole set of eigenstates of the unperturbed operator. Instead, the expressions \eqref{H-spectral} and \eqref{U-spectral} should be viewed as restrictions of the Hamiltonian and the time evolution operator to a suitable subspace of ${\cal H}_{\rm diff}$ which is sufficiently large for the calculation at hand. For example, consider using \Eq{U-spectral} to compute time evolution of the expectation value of a graph-preserving operator starting from an initial state which is based on a single graph containing no special loops. Under the second-order approximation given by \Eqs{lambda-approx} and \eqref{state-approx}, the Hamiltonian (and hence the evolution operator) has no non-vanishing matrix elements between the initial state and states in which three or more special loops have been attached to the graph of the initial state. Then the summation in \Eq{U-spectral} can be restricted to those states in which the graph of the unperturbed eigenstate is the graph of the initial state decorated by at most two special loops. (This example is discussed in more detail in section \ref{sec:analysis}.)

\subsection{Complete solution of the perturbative problem}

In this section we will go through the derivation of the eigenvalues and eigenstates of the operator \eqref{H0+V} up to second order in perturbation theory. The eigenvalues and eigenstates of the operator $H_R$ are assumed to be known, and are denoted by $\lambda_n^{(0)}$ and $\bket{\lambda_n^{(0)}}$. We look for approximate expressions for the eigenvalues and eigenstates of the full operator $H$ in the form
\be\label{lambda-ansatz}
\lambda_n = \lambda_n^{(0)} + \epsilon\lambda_n^{(1)} + \epsilon^2\lambda_n^{(2)} + \dots
\ee
and
\be\label{state-ansatz}
\bket{\lambda_n} = \bket{\lambda_n^{(0)}} + \epsilon\bket{\lambda_n^{(1)}} + \epsilon^2\bket{\lambda_n^{(2)}} + \dots
\ee
Inserting \Eqs{lambda-ansatz} and \eqref{state-ansatz} into \Eq{H0+V}, we obtain the equation
\begin{align}
\bigl(H_R+\epsilon H_E\bigr)&\Bigl(\bket{\lambda_n^{(0)}} + \epsilon\bket{\lambda_n^{(1)}} + \epsilon^2\bket{\lambda_n^{(2)}} + \dots\Bigr) \notag \\
&= \bigl(\lambda_n^{(0)} + \epsilon\lambda_n^{(1)} + \epsilon^2\lambda_n^{(2)} + \dots\bigr)\Bigl(\bket{\lambda_n^{(0)}} + \epsilon\bket{\lambda_n^{(1)}} + \epsilon^2\bket{\lambda_n^{(2)}} + \dots\Bigr),\label{master-eq}
\end{align}
from which the eigenvalues $\lambda_n$ and eigenstates $\ket{\lambda_n}$ will be determined iteratively, order-by-order in $\epsilon$.

As is well known, whenever some of the unperturbed eigenvalues $\lambda_n^{(0)}$ are degenerate, the corresponding eigenstates $\bket{\lambda_n^{(0)}}$ form a correct starting point for the perturbative expansion only if they are chosen such that the perturbation $H_E$ is diagonal in the subspace corresponding to the degenerate eigenvalue. In our case this requirement places no restriction on the states $\bket{\lambda_n^{(0)}}$, since all of these states have the same number of special loops, and the action of the Euclidean operator $H_E$ changes the number of special loops by one. In all the examples we have studied numerically, we have not found any instances of an eigenvalue of the curvature part being shared between states having a different number of special loops\footnote{With the exception that the eigenvalue zero occurs for every graph in the dust field model. However, in this case the eigenstates of zero eigenvalue are always annihilated also by the Euclidean part of the Hamiltonian, so the matrix elements of the perturbation in the subspace of the unperturbed eigenvalue zero are still all zeros.}.

The derivation of the corrections to the unperturbed eigenvalues then proceeds according to the textbook treatment, and presents no special problems. The first-order correction to an eigenvalue is given by the expectation value of the perturbation in the unperturbed eigenstate, or by the eigenvalues or the matrix of the perturbation in the degenerate subspace, depending on whether the unperturbed eigenvalue is degenerate. In either case, we have
\be\label{lambda(1)}
\lambda_n^{(1)} = 0.
\ee
To determine the second-order corrections, we look at the terms of order $\epsilon^2$ in \Eq{master-eq}, and find
\be\label{lambda(2)}
\lambda_n^{(2)} = \bbra{\lambda_n^{(0)}}H_E\bket{\lambda_n^{(1)}} = \sum_k \frac{\bigl|\bbra{\lambda_n^{(0)}}H_E\bket{\lambda_k^{(0)}}\bigr|^2}{\lambda_n^{(0)}-\lambda_k^{(0)}},
\ee
where the second equality follows from \Eq{psi(1)} together with the observation that the state $H_E\bket{\lambda_n^{(0)}}$ has no non-vanishing components on unperturbed eigenstates with eigenvalue $\lambda_n^{(0)}$.

Moving now on to the eigenstates, the projections of the first- and second-order corrections $\bket{\lambda_n^{(1)}}$ and $\bket{\lambda_n^{(2)}}$ onto unperturbed eigenstates outside of the degenerate subspace of the eigenvalue $\lambda_n^{(0)}$ are found easily in the standard way. Considering the terms of first and second order in \Eq{master-eq}, we see that
\be\label{psi(1)}
\bbraket{\lambda_k^{(0)}}{\lambda_n^{(1)}} = \frac{\bbra{\lambda_k^{(0)}}H_E\bket{\lambda_n^{(0)}}}{\lambda_n^{(0)}-\lambda_k^{(0)}} \qquad (\lambda_n^{(0)}\neq\lambda_k^{(0)}),
\ee
and, recalling that the first-order correction to the eigenvalue vanishes,
\be\label{psi(2)-v0}
\bbraket{\lambda_k^{(0)}}{\lambda_n^{(2)}} = \frac{\bbra{\lambda_k^{(0)}}H_E\bket{\lambda_n^{(1)}}}{\lambda_n^{(0)}-\lambda_k^{(0)}} \qquad (\lambda_n^{(0)}\neq\lambda_k^{(0)}).
\ee
\Eq{psi(1)'}, derived below, shows that the correction $\bket{\lambda_n^{(1)}}$ has no non-vanishing components on unperturbed eigenstates having eigenvalue $\lambda_n^{(0)}$. We may therefore use \Eq{psi(1)} in \Eq{psi(2)-v0}, leading to
\be
\bbraket{\lambda_k^{(0)}}{\lambda_n^{(2)}} = \sum_l \frac{\bbra{\lambda_k^{(0)}}H_E\bket{\lambda_l^{(0)}}\bbra{\lambda_l^{(0)}}H_E\bket{\lambda_n^{(0)}}}{\bigl(\lambda_n^{(0)}-\lambda_k^{(0)}\bigr)\bigl(\lambda_n^{(0)}-\lambda_l^{(0)}\bigr)}\qquad (\lambda_n^{(0)}\neq\lambda_k^{(0)}). \label{psi(2)}
\ee

Finding the components of the corrections $\bket{\lambda_n^{(1)}}$ and $\bket{\lambda_n^{(2)}}$ within the degenerate subspace turns out to be somewhat more involved. The projection $\bbraket{\lambda_n^{(0)}}{\lambda_n^{(1)}}$ is not determined by \Eq{master-eq}, but the requirement that the state \eqref{state-ansatz} should be normalized to 1 up to first order in $\epsilon$ shows that we can set
\be
\bbraket{\lambda_n^{(0)}}{\lambda_n^{(1)}} = 0.
\ee
The terms of first order in \Eq{H0+V} do not contain any information on the projection $\bbraket{\lambda_{n'}^{(0)}}{\lambda_n^{(1)}}$, where $\bket{\lambda_{n'}^{(0)}}$ is another unperturbed eigenstate with eigenvalue $\lambda_n^{(0)}$. They merely reproduce
\be
\bbra{\lambda_{n'}^{(0)}}H_E\bket{\lambda_n^{(0)}} = 0
\ee
as a consistency condition for the perturbative expansion. As we have argued above, in our case this condition is satisfied regardless of the choice of basis in the degenerate subspace.

Because the first-order correction $\lambda_n^{(1)}$ vanishes, the projection $\bbraket{\lambda_{n'}^{(0)}}{\lambda_n^{(1)}}$ is also not determined by the second-order terms of \Eq{master-eq}. Instead, we obtain another consistency condition,
\be
\bbra{\lambda_{n'}^{(0)}} H_E\bket{\lambda_n^{(1)}} = \sum_k \frac{\bbra{\lambda_{n'}^{(0)}} H_E\bket{\lambda_k^{(0)}}\bbra{\lambda_{k}^{(0)}} H_E\bket{\lambda_n^{(0)}}}{\lambda_n^{(0)}-\lambda_k^{(0)}} = 0.
\ee
A numerical evaluation of the sum shows that this condition also seems to be satisfied for any choice of basis in the degenerate subspace. (See section \ref{sec:analysis} for the details and the extent of our numerical calculations.)

We therefore move on to the terms of third order in \Eq{H0+V}. We now find
\be
\lambda_n^{(2)}\bbraket{\lambda_{n'}^{(0)}}{\lambda_n^{(1)}} = \bbra{\lambda_{n'}^{(0)}}H_E\bket{\lambda_n^{(2)}}.
\ee
In all the examples we have considered, the second-order corrections $\lambda_n^{(2)}$ given by \Eq{lambda(2)} are all non-vanishing, so the projections $\bbraket{\lambda_{n'}^{(0)}}{\lambda_n^{(1)}}$ are indeed determined by the above equation. Furthermore, by expanding the matrix element on the right-hand side as
\be
\bbra{\lambda_{n'}^{(0)}}H_E\bket{\lambda_n^{(2)}} = \sum_k \bbra{\lambda_{n'}^{(0)}}H_E\bket{\lambda_k^{(0)}}\bbraket{\lambda_k^{(0)}}{\lambda_n^{(2)}},
\ee
and recalling that each action of the Euclidean operator $H_E$ changes the number of special loops by one, we see that this matrix element actually vanishes. (The projections $\bbraket{\lambda_k^{(0)}}{\lambda_n^{(2)}}$ are given by \Eq{psi(2)}, which contains two actions of $H_E$, while the states $\bket{\lambda_n^{(0)}}$ and $\bket{\lambda_{n'}^{(0)}}$ have the same number of special loops.) Hence we conclude that
\be\label{psi(1)'}
\bbraket{\lambda_{n'}^{(0)}}{\lambda_n^{(1)}} = 0.
\ee

As to the second-order correction $\bket{\lambda_n^{(2)}}$, its projection on the unperturbed eigenstate $\bket{\lambda_n^{(0)}}$ can be found by requiring that the state \eqref{state-ansatz} is normalized up to second order in $\epsilon$. In this way we obtain
\be\label{psi(2)'}
\bbraket{\lambda_n^{(0)}}{\lambda_n^{(2)}} = -\frac{1}{2}\bbraket{\lambda_n^{(1)}}{\lambda_n^{(1)}} = -\frac{1}{2}\sum_k \frac{\bigl|\bbra{\lambda_n^{(0)}}H_E\bket{\lambda_k^{(0)}}\bigr|^2}{\bigl(\lambda_n^{(0)}-\lambda_k^{(0)}\bigr)^2}.
\ee
In an attempt to determine the projection $\bbraket{\lambda_{n'}^{(0)}}{\lambda_n^{(2)}}$, we look at the terms of fourth order in \Eq{master-eq}, which give
\be\label{psi(2)''-attempt}
\lambda_n^{(2)}\bbraket{\lambda_{n'}^{(0)}}{\lambda_n^{(2)}} = \bbra{\lambda_{n'}^{(0)}}H_E\bket{\lambda_n^{(3)}}.
\ee
At a first sight, it seems that we have managed to find $\bbraket{\lambda_{n'}^{(0)}}{\lambda_n^{(2)}}$, but we should realize that this projection is also involved in the matrix element on the right-hand side. By repeatedly inserting resolutions of identity in terms of the unperturbed eigenstates, we can rewrite the matrix element as
\begin{align}
\bbra{\lambda_{n'}^{(0)}}H_E\bket{\lambda_n^{(3)}} &= \sum_k \bbra{\lambda_{n'}^{(0)}}H_E\bket{\lambda_k^{(0)}}\bbraket{\lambda_k^{(0)}}{\lambda_n^{(3)}} \notag \\
&= \sum_k  \bbra{\lambda_{n'}^{(0)}}H_E\bket{\lambda_k^{(0)}}\Biggl[\frac{\bbra{\lambda_k^{(0)}}H_E\bket{\lambda_n^{(2)}}}{\lambda_n^{(0)}-\lambda_k^{(0)}} - \lambda_n^{(2)}\frac{\bbra{\lambda_k^{(0)}}H_E\bket{\lambda_n^{(0)}}}{\bigl(\lambda_n^{(0)}-\lambda_k^{(0)}\bigr)^2}\Biggr] \notag \\
&= \sum_{kl} \frac{\bbra{\lambda_{n'}^{(0)}}H_E\bket{\lambda_k^{(0)}}\bbra{\lambda_k^{(0)}}H_E\bket{\lambda_l^{(0)}}}{\lambda_n^{(0)}-\lambda_k^{(0)}}\bbraket{\lambda_l^{(0)}}{\lambda_n^{(2)}}, \label{attempt-2}
\end{align}
where the expression for $\bbraket{\lambda_k^{(0)}}{\lambda_n^{(3)}}$ has been derived straightforwardly from the third-order terms in \Eq{master-eq}, and we again resorted to numerics to discover that in all the examples we have considered,
\be
\sum_k  \frac{\bbra{\lambda_{n'}^{(0)}}H_E\bket{\lambda_k^{(0)}}\bbra{\lambda_k^{(0)}}H_E\bket{\lambda_n^{(0)}}}{\bigl(\lambda_n^{(0)}-\lambda_k^{(0)}\bigr)^2} = \bbraket{\lambda_{n'}^{(1)}}{\lambda_n^{(1)}} = 0.
\ee
Going now back to \Eq{attempt-2}, we see that the projection $\bbraket{\lambda_{n'}^{(0)}}{\lambda_n^{(2)}}$ indeed appears in the sum over $l$, the $l=n'$ term of the sum being equal to
\be
\sum_k \frac{\bbra{\lambda_{n'}^{(0)}}H_E\bket{\lambda_k^{(0)}}\bbra{\lambda_k^{(0)}}H_E\bket{\lambda_{n'}^{(0)}}}{\lambda_n^{(0)}-\lambda_k^{(0)}}\bbraket{\lambda_{n'}^{(0)}}{\lambda_n^{(2)}} = \lambda_{n'}^{(2)}\bbraket{\lambda_{n'}^{(0)}}{\lambda_n^{(2)}}.
\ee
In our application, the second-order corrections $\lambda_n^{(2)}$ and $\lambda_{n'}^{(2)}$ to a degenerate eigenvalue $\lambda_n^{(0)}$ are always equal to each other, at least in all the cases within the scope of our numerical analysis. Therefore we see that the projection $\bbraket{\lambda_{n'}^{(0)}}{\lambda_n^{(2)}}$ cancels out in \Eq{psi(2)''-attempt} and so is not determined by \Eq{master-eq}, apparently due to the fact that the degeneracy in the eigenvalues of the curvature part is not removed by the Euclidean part, at least up to second order in perturbation theory. It also does not seem to be possible to extract the projection $\bbraket{\lambda_{n'}^{(0)}}{\lambda_n^{(2)}}$ from terms of even higher order in \Eq{master-eq}, nor from any normalization or orthogonality conditions of the corrected eigenstates. In order to resolve this situation in some way, we simply set
\be\label{psi(2)''}
\bbraket{\lambda_{n'}^{(0)}}{\lambda_n^{(2)}} = 0,
\ee
since this is the simplest possible choice, and it is consistent with the normalization and orthogonality of the corrected eigenstates up to second order in $\epsilon$.

\subsection{Expansion in powers of time}

The perturbative treatment introduced in the preceding sections can be expected to give an accurate approximation for the spectrum of the physical Hamiltonian for sufficiently large values of $\beta$, but we cannot hope for such an approach to be valid if the value of $\beta$ is not particularly large. We are therefore not able to work with the physical Hamiltonian of the scalar field model with an arbitrary value of $\beta$, since the spectral decomposition of the Hamiltonian is required in order to resolve the square root in \Eq{Hphys^S}.

The situation is different for the dust model, where an alternative (even if rather primitive) approach is possible for approximating the time evolution of a given initial state over a short interval of time, irrespectively of the value of $\beta$. For small values of the time variable $t$, we can obtain a good approximation for the evolution operator $U(t)$ by expanding it in powers of $t$ and truncating the expansion at some finite power of $t$. Then each term in the resulting expansion can be evaluated by repeatedly applying the Hamiltonian to the initial state, and there is no fundamental obstruction against performing the required calculations, since the Hamiltonian of the dust model is available as an explicitly defined operator.

If one is interested in computing the time dependence of the expectation value of an operator $A$ starting from a given initial state $\ket{\psi_0}$, the most convenient way to proceed is to evaluate the coefficients in the (truncated) power series expansion of the expectation value,
\be\label{A(t)-exp}
\langle A(t)\rangle_{\psi_0} = \sum_n a_nt^n.
\ee
Here the coefficients are given by expectation values of repeated commutators of the operator $A$ with the Hamiltonian in the initial state:
\be\label{a_n time}
a_n = \frac{(-i)^n}{n!}\bbra{\psi_0}\bigl[\underbrace{H,\dots,\bigl[H,[H,A]\bigr]\dots}_{\text{$n$ commutators}}\bigr]\bket{\psi_0}.
\ee
The advantage of considering directly the expansion of the expectation value, as opposed to computing the evolved state vector $\ket{\Psi(t)}$ truncated at some order $t^N$, is that in many situations the expectation value can be determined up to order $t^N$ without having to derive all the components of the truncated state vector. Let us consider again the example of computing the expectation value of a graph-preserving operator $A$, when the initial state is based on a single graph having no special loops. If we want to find the expectation value $\langle A(t)\rangle$ up to order $t^N$, then states in which more than $N/2$ special loops have been attached to the initial state do not enter the calculation of the numbers $a_1,\dots,a_N$, even though the state vector truncated at order $t^N$ has components containing up to $N$ special loops.

\subsection{The setup for numerical analysis}\label{sec:analysis}

The approximation methods described in this chapter were used in our article \cite{paper5} to study numerically the time evolution of expectation values of the volume and curvature operators in certain simple spin network states. The initial states which we consider are based on a graph consisting of a single four-valent node. The spins on the edges of the graph are fixed to a common value $j$, and the intertwiner at the node is chosen to be an eigenstate of the volume operator. The node is assumed to have a non-degenerate tangential structure, in which none of the four edges are tangent to each other at the node (\ie the tangent vectors of the edges are diffeomorphic to the normal vectors of a regular tetrahedron). Furthermore, we choose $s=1/2$ for the spin of the special loops created by the Euclidean part of the physical Hamiltonian.

For our choice of the initial graph, there are six different graphs resulting from the action of the Euclidean operator on the initial state, corresponding to the six different pairs of edges of a four-valent node. After a careful counting, and taking into account that in some cases the same graph is obtained from two different sequences of acting with the Euclidean operator, one finds that there are 63 different graphs in which two special loops have been attached to the initial four-valent node by the Euclidean operator. We have not counted the exact number of graphs resulting from three actions of the Euclidean operator on the initial state, but this number seems to be of the order of one thousand, and in any case is so large that it would be infeasible to try to include the graphs containing three special loops into our numerical calculations. We must therefore truncate the perturbative expansion at second order in the parameter $\epsilon = 1/(1+\beta^2)$, and the expansion in powers of the time variable at the fourth power of $t$.

With these preparations, the main part of our numerical calculations consists of computing the matrices $C^E_v$ and $C^R_v$, as well as the matrix of the volume operator, in the finite-dimensional space spanned by the four-valent initial state together with all the states in which no more than two special loops have been attached to the initial state. Having the matrix of the volume operator, we can construct the matrix ${\cal V}_v^{-1/2}$, and thereafter the matrix of the physical Hamiltonian for the dust model. The matrix of the operator under the square root in the physical Hamiltonian of the scalar field model is given directly in terms of the matrices $C^E_v$ and $C^R_v$.

The matrix elements of the operators are computed from the formulas derived in Chapter \ref{ch:elements}, noting in particular that the intertwiners appearing in these formulas are not normalized, so the formulas must be suitably modified in order to obtain the matrix elements with respect to a basis of normalized states. The cases in which the Euclidean operator or the curvature operator acts on a pair of edges different from $(e_1,e_2)$ or $(e_1,e_3)$, or the squared volume operator acts on a triple of edges different from $(e_1,e_2,e_3)$ or $(e_1,e_3,e_5)$, can be reduced to the cases given in Chapter \ref{ch:elements} by appropriately using the symmetry relation \eqref{3jminus g}.

The matrix elements of the Euclidean operator $C^E_v$ and the squared volume operator are expressed entirely in terms of 6$j$-symbols in which one spin is equal to 1, and 9$j$-symbols in which a triple of spins are all equal to 1. In the computation of the matrix elements, the values of these symbols are calculated using their explicit algebraic expressions, which can be found \eg in \cite{Varshalovich}. The matrix elements of the operator $C^R_v$ contain generic 6$j$-symbols, in which none of the spins is equal to a given, fixed value. The values of such symbols are precalculated and stored in tables in order to speed up the computation of the matrix elements of the curvature part.

After the matrices required to form the physical Hamiltonian (or the operator \eqref{Sym'} in the case of the scalar field model) have been computed, the eigenvalues and eigenvectors of the Hamiltonian can be approximated using \Eqs{lambda-approx} and \eqref{state-approx}. Since the curvature part of the Hamiltonian is a graph-preserving operator, the unperturbed eigenvalues and eigenstates entering \Eqs{lambda-approx} and \eqref{state-approx} can be computed separately within the subspace of each graph. To evaluate the time dependence of an expectation value under our perturbative approximation, we then use the (approximate) spectral decomposition of the time evolution operator,
\be
U(t) = \sum_n e^{-i\lambda_n t}\ket{\lambda_n}\bra{\lambda_n},
\ee
to write the expectation value of an operator $A$ in the state $\ket{\Psi(t)} = U(t)\ket{\psi_0}$ as
\be\label{A(t)-pert}
\langle A(t)\rangle = \sum_{mn} e^{-i(\lambda_n-\lambda_m)t}\braket{\psi_0}{\lambda_m}\bra{\lambda_m}A\ket{\lambda_n}\braket{\lambda_n}{\psi_0}.
\ee
For the scalar field model, one should keep in mind that perturbation theory is applied to approximate the spectrum of the operator \eqref{Sym'}, from which the evolution operator will be constructed according to \Eq{U-spectral}.

In the examples we are considering, the operator $A$ is graph-preserving, while the initial state $\ket{\psi_0}$ is based on a single graph $\Gamma_0$ containing no special loops. Under these assumptions, some parts of the corrected state vectors \eqref{state-approx} can be discarded, since they do not contribute to the expectation value \eqref{A(t)-pert} at second order in the perturbative parameter $\epsilon$. For unperturbed eigenstates based on the initial graph $\Gamma_0$, one has to take the first-order correction, and the part of the second-order correction which is based on $\Gamma_0$. For unperturbed eigenstates whose graph is $\Gamma_0$ decorated with a single special loop, it is enough to take the part of the first-order correction which is based on $\Gamma_0$; the second-order correction can be neglected entirely. Unperturbed eigenstates in which more than one special loop has been attached to the initial graph do not contribute to the calculation at second order in $\epsilon$.

\subsection{Numerical results}

\subsubsection*{Perturbation theory}

\begin{figure}[h]
	\centering
		\includegraphics[width=0.9\textwidth]{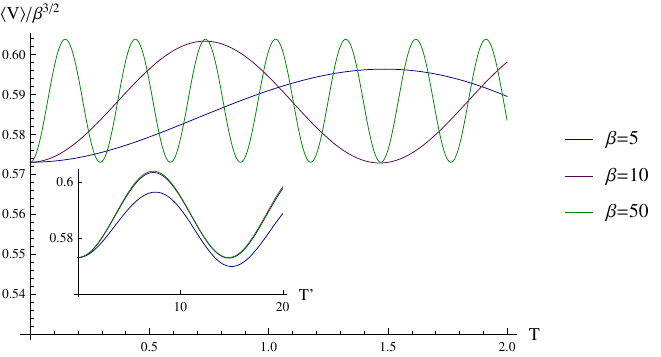}
		\caption{Expectation value of the volume operator in the scalar field model. \\ $j=2$, $v=0.5730$ }
	\label{VS2B}
\end{figure}

\begin{figure}[t]
	\centering
		\includegraphics[width=0.9\textwidth]{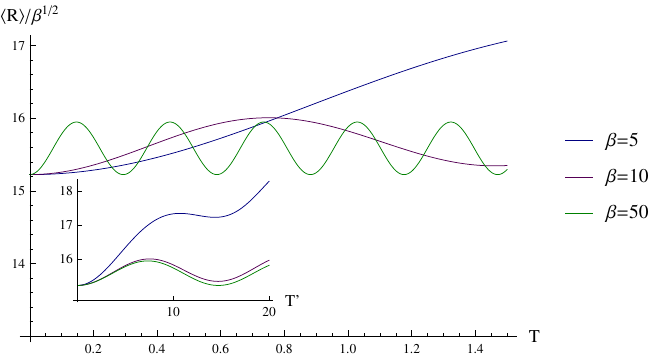}
		\caption{Expectation value of the curvature operator in the scalar field model. \\ $j=2$, $v=0.8725$ }
	\label{RS2B}
\end{figure}

\begin{figure}[b]
	\centering
		\includegraphics[width=0.9\textwidth]{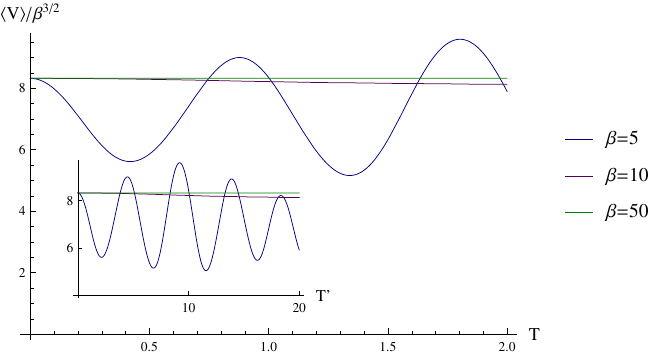}
		\caption{Expectation value of the volume operator in the scalar field model. \\ $j=10$, $v=8.3177$}
	\label{V1S10B1}
\end{figure}

\begin{figure}[t]
	\centering
		\includegraphics[width=0.9\textwidth]{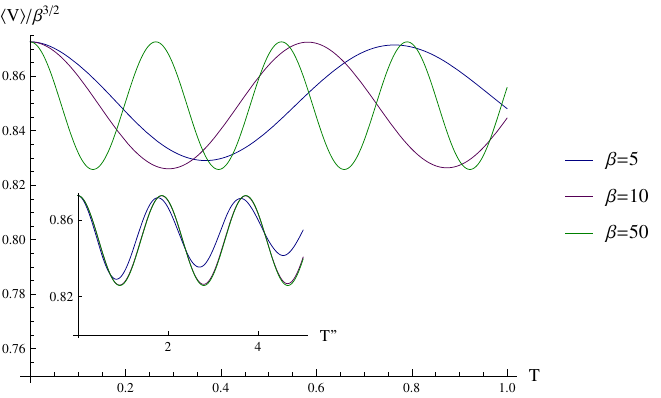}
		\caption{Expectation value of the volume operator in the dust model. \\ $j=2$, $v=0.8725$}
	\label{VD2B}
\end{figure}

\begin{figure}[b]
	\centering
		\includegraphics[width=0.9\textwidth]{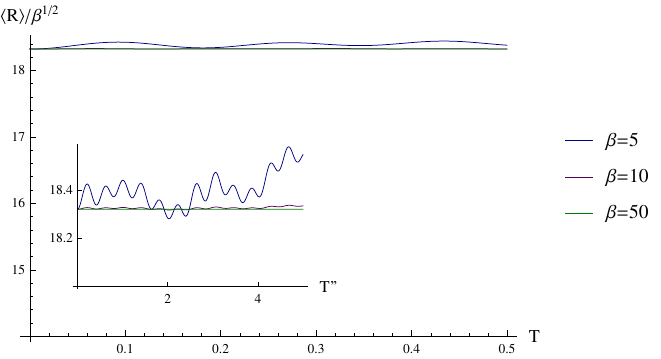}
		\caption{Expectation value of the curvature operator in the dust model. \\ $j=2$, $v=0.5730$}
	\label{RD2B}
\end{figure}

\begin{figure}[t]
	\centering
		\includegraphics[width=0.9\textwidth]{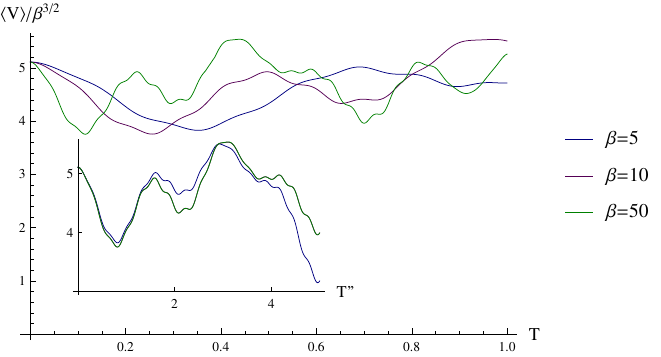}
		\caption{Expectation value of the volume operator in the dust model. \\ $j=10$, $v=5.1078$}
	\label{V1D10B}
\end{figure}

\begin{figure}[b]
	\centering
		\includegraphics[width=0.9\textwidth]{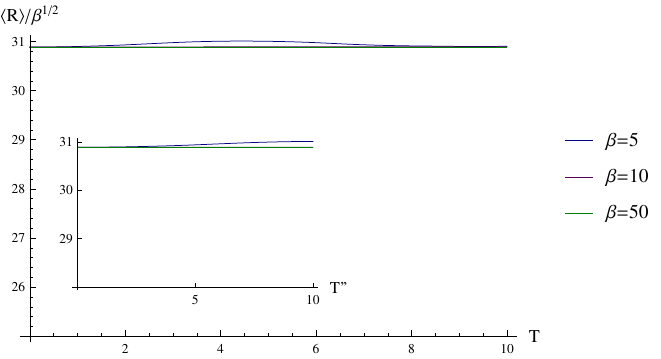}
		\caption{Expectation value of the curvature operator in the dust model. \\ $j=10$, $v=8.3177$}
	\label{R1D10B}
\end{figure}

Some sample results of expectation values calculated from \Eq{A(t)-pert} are shown in \Figs{VS2B}--\ref{V1S10B1} for the scalar field model, and in \Figs{VD2B}--\ref{R1D10B} for the dust model\footnote{The plots displayed in this chapter are used here with the permission of Mehdi Assanioussi, who created them for the article \cite{paper5} on the basis of our joint calculations.}. We consider expectation values of the volume and curvature operators, starting from an initial state which is an eigenstate of the volume operator. The spin $j$ and the volume eigenvalue $v$ in the initial state are reported in the caption under each figure. The horizontal axis in each plot is labeled by the physical time variable $t$ which appears in the Schr\"odinger equation \eqref{SE-17}, and which is given by the value of the scalar field or the dust field (in units where $16\pi G = 1$). The vertical axis is scaled by an appropriate factor of $\beta$ in order to factor out the $\beta$-dependence of the operator whose expectation value is shown in the plot. Furthermore, the small embedded plots show the evolution of the expectation values with respect to a rescaled time variable, which takes into account the fact that as the value of $\beta$ changes, the overall scale of time evolution is affected by the $\beta$-dependent multiplicative factor present in the physical Hamiltonian. Thus, the rescaled time variable is given by
\be
t' = \sqrt{1+\beta^2}\,t
\ee
for the scalar field model, and
\be
t'' = \frac{1+\beta^2}{\beta^{3/2}}t
\ee
in the case of the dust model.

Since the initial state is based on a single graph, and the second-order approximation to the evolution operator has no non-vanishing matrix elements between the initial state and states in which more than two special loops have been attached to the initial state, the expectation values of volume and curvature are bounded throughout the time evolution of the initial state. (However, we certainly do not expect the perturbative treatment to give an accurate approximation for the evolution of the initial state over arbitrarily long time intervals.) Given that the expectation values must remain bounded, it is not surprising to see that they exhibit an oscillatory behaviour as a function of time. It might be possible to obtain unbounded expectation values under the perturbative approximation by considering initial states which take the form of an infinite superposition of states based on different graphs.

Our numerical calculations show that the degeneracies present in the spectrum of the volume operator are preserved under time evolution, in the sense that the function $\langle V(t)\rangle$ obtained from an initial state corresponding to a degenerate eigenvalue remains the same regardless of which state in the degenerate subspace is chosen as the initial state. The same statement applies to the curvature operator if one considers the expectation value $\langle R(t)\rangle$, taking a degenerate eigenstate of curvature as the initial state. This observation seems to indicate that whichever symmetry is responsible for the degeneracies in the eigenvalues of volume and curvature is likely to be a symmetry of the physical Hamiltonian as well.

Looking at \Figs{RD2B} and \ref{R1D10B} for the expectation value of the curvature operator in the dust model, we see that for larger values of $\beta$ the expectation value of curvature remains nearly constant in time. This is as it should be, since in the limit of large $\beta$ the physical Hamiltonian of the dust model reduces to just the curvature operator.

\subsubsection*{Time expansion}

\begin{figure}[t]
	\centering
		\includegraphics[width=0.9\textwidth]{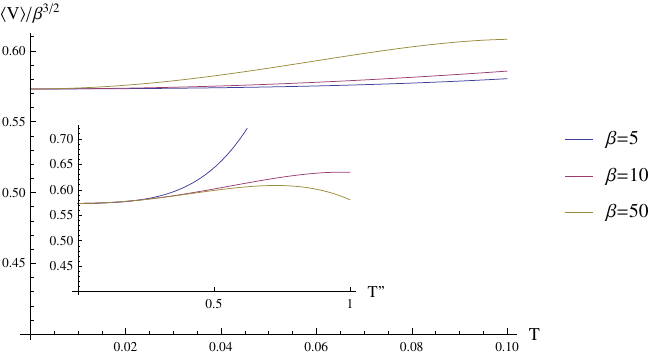}
		\caption{Time expansion for the expectation value of the volume operator. \\ $j=2$, $v=0.5730$}
	\label{EVS2B1}
\end{figure}

\begin{figure}[b]
	\centering
		\includegraphics[width=0.9\textwidth]{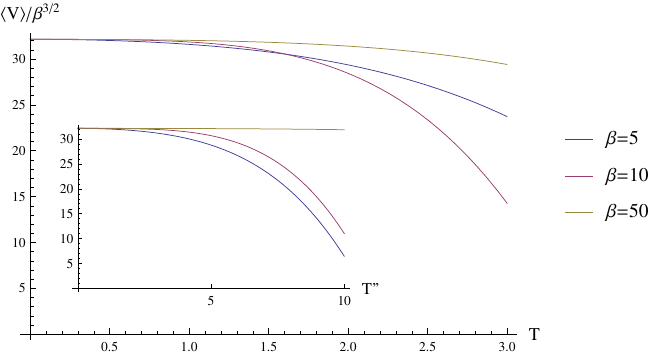}
		\caption{Time expansion for the expectation value of the volume operator. \\ $j=25$, $v=32.140$}
	\label{EVS25B}
\end{figure}

\begin{figure}[t]
	\centering
		\includegraphics[width=0.9\textwidth]{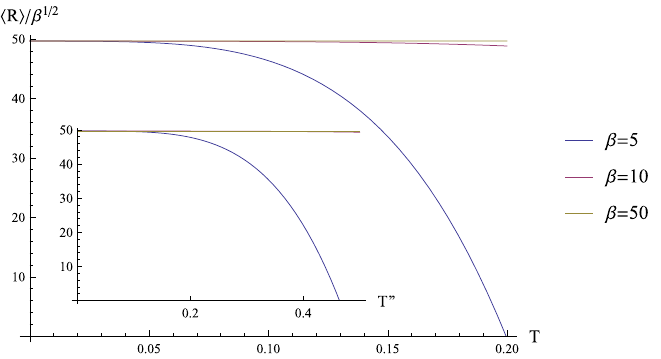}
		\caption{Time expansion for the expectation value of the curvature operator. \\ $j=25$, $v=20.176$}
	\label{ERS25B1}
\end{figure}

In the case of the dust model, we have also investigated the expansion of expectation values in powers of time, as given by \Eqs{A(t)-exp} and \eqref{a_n time}. The expansion is taken up to fourth order in time, and a small selection of the results is shown in \Figs{EVS2B1}--\ref{ERS25B1}. The computation of the coefficients of the expansion from \Eq{a_n time} reveals that only even powers of time enter the expansion of the expectation values of volume and curvature. The coefficients of the odd powers ($t$ and $t^3$) vanish up to numerical rounding error. This seems to suggest the invariance of the operators involved in the calculation under an appropriate notion of time reversal. We also find that the degeneracy in the eigenvalues of volume and curvature is again preserved under time evolution: The function $\langle V(t)\rangle$ (or $\langle R(t)\rangle$) does not depend on which eigenstate corresponding to a degenerate eigenvalue is selected as the initial state. 

The range of validity of the time expansion can be determined by estimating the value of $t$ at which the first term discarded from the expansion (in our examples, the term of order $t^6$) starts being comparable in magnitude to the terms included in the approximation. This criterion can be tested in a toy example in which the Hamiltonian consists only of the curvature part, and the dynamics can be evaluated exactly. In this case we find that the criterion correctly predicts the order of magnitude of the time at which an expectation value computed from the fourth-order time expansion begins to deviate significantly from the exact expectation value.

\begin{figure}[b]
	\centering
		\includegraphics[width=\textwidth]{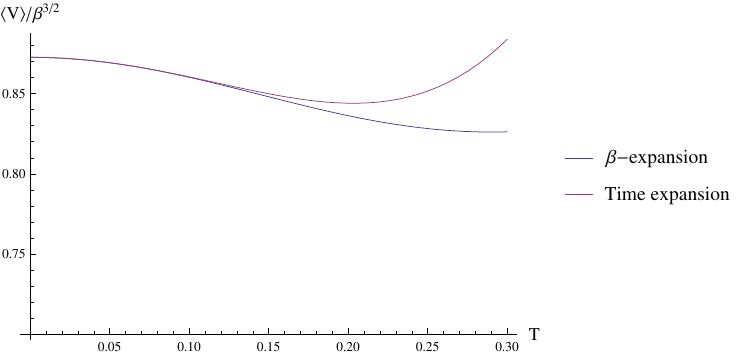}
		\caption{Comparison between the time expansion and the perturbative expansion. Expectation value of the volume operator in the dust model ($j=2$, $v=0.8725$, $\beta=10$)}
	\label{Compar}
\end{figure}

In \Fig{Compar} we show the time expansion of an expectation value of volume compared against the same expectation value computed within the perturbative approach when $\beta=10$. We see that at first the two approximations agree with each other, but around a certain time the time expansion starts to diverge from the result of the perturbative calculation. For this value of $\beta$, we expect that perturbation theory provides an accurate description of the dynamics over a longer time interval than the time expansion does. Hence, at the time when the time expansion begins to diverge from the perturbative approximation, the latter presumably still gives a very close approximation to the exact expectation value, whereas the former has reached the end of its range of validity.

\newpage

\subsection{Discussion}

The central result of this chapter is the observation that for large values of the Barbero--Immirzi parameter, standard time-independent perturbation theory can be used to approximately evaluate the dynamics in deparametrized models, even if an exact spectral decomposition of the physical Hamiltonian cannot be accomplished. As a sort of supplementary approximation, we also introduced the expansion in powers of time, which can be used in the dust model to approximate time evolution over a short interval of time, even if the value of $\beta$ is too small for perturbation theory to be applicable.

The results of our numerical analysis contain several hints of the existence of a symmetry which is shared by the volume and curvature operators and the Euclidean part of the Hamiltonian. The degeneracies present in the spectra of volume and curvature are preserved under time evolution, the degeneracy in the eigenvalues of the curvature operator is not removed by the perturbation consisting of the Euclidean part (at least up to second order in perturbation theory), and only even powers of time appear in the expansion of expectation values of volume and curvature when the initial state is an eigenstate of one of the operators. If this symmetry could be identified and its action on spin network states understood in detail, this knowledge could immediately be applied to make the numerical computations more efficient, by separating the states involved in the computations into subspaces which are characterized by different eigenvalues of the symmetry operator, and which are not mixed with each other by the Hamiltonian. A thorough understanding of the symmetry could even turn out to be useful for the problem of trying to derive exact eigenvalues and eigenstates of the Hamiltonian.

In the perturbative treatment of the dynamics, the unperturbed Hamiltonian is essentially the curvature operator, so it is tempting to speculate that the perturbative approach might be particularly suitable for obtaining an accurate approximation for the dynamics in physical situations where the spatial curvature is large compared to the value of the observable measured by the Euclidean part of the Hamiltonian. Such a situation could arise, for instance, in the interior of a collapsing star at late stages of the collapse, when the value of the curvature is large enough that quantum effects are expected to become relevant for the dynamics of the collapse.

Before the approximation methods introduced in this chapter -- or any other approximation schemes possibly established in the future -- can be used to truly investigate the physical content of deparametrized models of loop quantum gravity, the crucial problem which must be dealt with is the development of coherent states which have a clear physical interpretation while being compatible with the structure of the model under investigation. While the complexifier coherent states discussed in Chapter \ref{ch:CS} have good semiclassical properties with respect to a large class of operators in loop quantum gravity, they are based on a single graph, and for this reason are unlikely to be suitable coherent states for a model in which the dynamics is governed by a graph-changing Hamiltonian.

Consider a state $\ket{\psi_{(q_0,p_0)}}$ peaked on a point $(q_0,p_0)$ of a classical phase space. A reasonable requirement for $\ket{\psi_{(q_0,p_0)}}$ to be regarded as a ''good'' coherent state is that it behaves under time evolution according to
\be\label{U(t)cs}
U(t)\ket{\psi_{(q_0,p_0)}} \simeq \ket{\psi_{(q_t,p_t)}},
\ee
where $(q_t,p_t)$ is the classical time evolution of the phase space point $(q_0,p_0)$. In general we should certainly not expect \Eq{U(t)cs} to hold as an exact equation, but for macroscopic values of $q_0$ and $p_0$, it should be valid up to a small quantum correction over a macroscopically long interval of time. 

However, in the case at hand it seems that if the state $\ket{\psi_{(q_0,p_0)}}$ is based on a fixed graph, and the Hamiltonian acts by creating and removing special loops, then \Eq{U(t)cs} cannot hold even approximately, since the state $U(t)\ket{\psi_{(q_0,p_0)}}$ will not be based on a fixed graph, but has the form of a superposition of graphs in which arbitrarily high numbers of special loops have been attached to the graph of the initial state. It also seems clear, aside from any dynamical considerations, that a state based on a fixed graph cannot be a good coherent state even at the kinematical level for a model in which the Euclidean part of the Hamiltonian is a graph-changing operator, as the expectation value of the Euclidean part in the prospective coherent state will be identically zero, independently of the phase space point on which the state is peaked.

On these grounds we are lead to believe that satisfactory coherent states for the models considered in this work must not be based on a fixed graph; rather, they must have the structure of an infinite superposition of graphs. Possibly such a superposition should start with a loopless ''seed graph'', and include all the graphs generated by repeated actions of the Hamiltonian on the seed graph, as illustrated by \Eq{superposition}. A construction which leads to states having the desired graph structure has recently been proposed in \cite{Mehdi}, but it has not yet been established whether the resulting states have any peakedness properties with respect to the standard operators of loop quantum gravity.

\vspace{72pt}

\captionsetup{justification=centering,singlelinecheck=true}

\clearpage

\clearpage

\clearpage

\clearpage

\clearpage

\clearpage

\section[A new representation for intertwiners from spin coherent states]{A new representation for intertwiners from angular \\ momentum coherent states}\label{ch:coherent3j}

\captionsetup{justification=justified,singlelinecheck=true}

When matrix elements of operators in loop quantum gravity are calculated in the usual way, using the techniques described in the Appendix to compute the action of the operator on a spin network state, the results will be expressed in terms of objects of $SU(2)$ recoupling theory, such as the Wigner $nj$-symbols. While such results provide exact expressions for the action of operators in the spin network basis, and are well suited for studying the operators through numerics, it is difficult to develop any intuitive feel for equations such as \eqref{CR13-result}, which gives the matrix elements of the curvature operator in terms of a rather complicated combination of 6$j$-symbols.

Motivated by thoughts along these lines, an alternative method for dealing with calculations in loop quantum gravity was proposed in the article \cite{paper4} by the author in collaboration with Emanuele Alesci and Jerzy Lewandowski. This method is based on a new representation for intertwiners, constructed by expressing the standard spin network states \eqref{spinnetwork} in terms of representation matrices and intertwiners projected onto angular momentum coherent states -- which have been used extensively as a computational tool in spin foam models, but have so far found little use in the context of canonical loop quantum gravity.

The new representation provides an alternative approach for computing matrix elements of operators in loop quantum gravity, in which the heavy use of $SU(2)$ recoupling theory can be avoided. Instead, calculations are formulated in terms of complex variables, which encode (via the stereographic projection) the polar angles of the unit vectors labeling the angular momentum coherent states, and the results of the calculations can be expressed in a geometrical language, in terms of the unit vectors themselves. However, as we will see, the work performed in \cite{paper4} and reviewed in this chapter merely amounts to laying down the foundations of a new and intriguing framework which can be used for computations in loop quantum gravity, but which still has to be studied in much more detail in order to bring out its full potential as a useful practical tool.

We begin this chapter with a review of the angular momentum coherent states, which are the essential ingredient for everything that follows. After these preliminaries, we look at intertwiners projected onto the basis of coherent states, using them to write down a ''coherent representation'' of a spin network state. We also derive explicit expressions for the components of three- and four-valent intertwiners with respect to the coherent state basis. We then show how a representation of an $N$-valent intertwiner as a polynomial of $N$ complex variables can be extracted from the expression of the intertwiner in the basis of coherent states, and how angular momentum operators acting on the intertwiner can be formulated as differential operators acting on these complex variables. Finally, we close our presentation by discussing the possibility of applying the techniques developed in this chapter to concrete calculations in loop quantum gravity.

\subsection{Angular momentum coherent states}

The well-known coherent states of angular momentum \cite{Radcliffe, Perelomov1972} are the central tool for the work presented in this chapter. In this section we will therefore give a thorough review of these states and their most important properties. Following the conventions of \cite{Perelomov}, we construct the coherent state $\ket{j\vec n}$ by starting with the state of lowest magnetic number, $\ket{j,-{}j}$, and applying a rotation which rotates the vector $\hat e_z=(0,0,1)$ into the vector $\vec n = (\sin\theta\cos\phi, \sin\theta\sin\phi, \cos\theta)$. Thus, denoting the corresponding $SU(2)$ group element by $g(\vec n)$ and introducing the abbreviation $D^{(j)}\bigl(g(\vec n)\bigr) \equiv D^{(j)}(\vec n)$, we have
\be\label{jn}
\ket{j\vec n} = D^{(j)}(\vec n)\ket{j,-{}j}.
\ee
Recalling that the state $\ket{j,-{}j}$ is a $2j$-fold tensor product of the states $\ket - \equiv \ket{\half,-{}\half}$, we see that the coherent state \eqref{jn} can also be expressed in the form
\be\label{jn-product}
\ket{j\vec n} = \underbrace{\ket{\vec n}\otimes\cdots\otimes\ket{\vec n}}_{\text{2$j$ times}},
\ee
where $\ket{\vec n} \equiv g(\vec n)\ket -$ is the coherent state in the spin-1/2 representation.

By construction, the state $\ket{j\vec n}$ is an eigenstate of the operator $\vec n\cdot\vec J$ with eigenvalue $-j$. Moreover, the states \eqref{jn} form a basis of the Hilbert space ${\cal H}_j$, as indicated by the resolution of identity 
\be\label{cs-1}
\Id = d_j\int \frac{d^2n}{4\pi}\,\ket{j\vec n}\bra{j\vec n},
\ee
where $d^2n = \sin\theta\,d\theta\,d\phi$ is the standard measure on the unit sphere. The expectation value of angular momentum in the state $\ket{j\vec n}$ is given by $\bra{j\vec n}\vec J\ket{j\vec n} = -j\vec n$, while the relative uncertainty $\Delta J/\langle J\rangle = {\cal O}(j^{-1/2})$ is small in the limit of large $j$. Therefore, for large values of $j$, the state $\ket{j\vec n}$ has a semiclassical interpretation, describing a nearly classical angular momentum having magnitude $j$ and pointing in the direction $-\vec n$.

The form of the group element $g(\vec n)$ in \Eq{jn} is not uniquely determined by the requirement of rotating the vector $\hat e_z$ into the vector $\vec n$. In order to fix the ambiguity, we specify that the axis of rotation lies in the $xy$-plane. With this choice, the group element effecting the rotation is uniquely determined as
\be\label{gn}
g(\vec n) = e^{-i\theta\vec m\cdot\vec J} = \begin{pmatrix} \cos\theta/2 & -e^{-i\phi}\sin\theta/2 \\ e^{i\phi}\sin\theta/2 & \cos\theta/2 \end{pmatrix},
\ee
where the vector $\vec m = (-\sin\phi,\cos\phi,0)$ lies in the $xy$-plane and is orthogonal to the vector $\vec n$.

For practical purposes, it is convenient to introduce the complex parameter
\be\label{xi}
\xi = -e^{-i\phi}\tan\frac{\theta}{2},
\ee
which can be used instead of the vector $\vec n$ to label the coherent states \eqref{jn} (and which is constructed geometrically by projecting the point $\vec n(\theta,\phi)$ into the complex plane through the stereographic projection). When expressed in terms of the parameter $\xi$, the group element \eqref{gn} becomes
\be\label{gxi}
g(\xi) = \frac{1}{\sqrt{1+|\xi|^2}}\begin{pmatrix} 1&\xi \\ -\bar\xi&1 \end{pmatrix}.
\ee
Furthermore, the resolution of identity \eqref{cs-1} takes the form
\be\label{xi-1}
\Id = \int d\mu_j(\xi)\,\ket{j\xi}\bra{j\xi},
\ee
where the integration measure is
\be\label{dmu_j}
d\mu_j(\xi) = \frac{d_j}{\pi}\frac{d^2\xi}{(1+|\xi|^2)^2}.
\ee
For later use, let us note some properties of the matrix $g(\xi)$. Firstly, we see that the inverse matrix is obtained simply by reversing the sign of the parameter, \ie
\be\label{g-1}
g^{-1}(\xi) = g(-\xi).
\ee
Furthermore, the matrix associated to the vector $-\vec n$ is given by
\be\label{g(-n)}
g(-\vec n) = g(-1/\bar\xi),
\ee
where $\xi$ is the parameter corresponding to the vector $\vec n$. Finally, the composition law of the group elements $g(\xi)$ (derived straightforwardly by carrying out the matrix multiplication) reads
\be\label{g1g2}
g(\xi_1)g(\xi_2) = g(\xi_{12})e^{i\alpha_{12}\sigma_z},
\ee
where
\be
\xi_{12} = \frac{\xi_1+\xi_2}{1-\bar\xi_1\xi_2} \qquad \text{and} \qquad e^{i\alpha_{12}} = \frac{1-\xi_1\bar\xi_2}{|1-\xi_1\bar\xi_2|}.
\ee
Incidentally, this result shows that the action of the matrix $g(\xi)$ on an angular momentum coherent state produces another coherent state, but generally multiplied with a phase factor.

We can derive an alternative expression for the group element \eqref{gxi} by using the fact that a general element of $SU(2)$ can be decomposed as\footnote{\Eq{+0-dec} is easily verified by matrix multiplication in the fundamental representation, in which the components of the angular momentum operator are represented by
\[
J_+ = J_x+iJ_y = \begin{pmatrix} 0&1 \\ 0&0 \end{pmatrix}, \qquad
J_0 = \frac{1}{2}\begin{pmatrix} 1&0 \\ 0&-1 \end{pmatrix}, \qquad
J_- = J_x - iJ_y = \begin{pmatrix} 0&0 \\ 1&0 \end{pmatrix}.
\]
Note that in general, none of the matrices on the right-hand side of \Eq{+0-dec} belong to $SU(2)$, though they are elements of $SL(2,\C)$.}
\be\label{+0-dec}
g = \begin{pmatrix} \alpha & \beta \\ -\bar\beta & \bar\alpha \end{pmatrix} = e^{\xi J_+}e^{\eta J_0}e^{\zeta J_-},
\ee
where the parameters are given by
\be
e^{-\eta/2} = \bar\alpha, \qquad \xi = \frac{\beta}{\bar\alpha}, \qquad\zeta = -\frac{\bar\beta}{\bar\alpha}.
\ee
Applying \Eq{+0-dec} to \Eq{gxi}, we obtain
\be\label{g+0-}
g(\xi) = e^{\xi J_+}e^{\ln(1+|\xi|^2)J_0}e^{-\bar\xi J_-}.
\ee
Consequently, since the operator $J_-$ annihilates the state $\ket{j,-{}j}$, we find the expression
\be\label{jxi}
\ket{j\xi} = \frac{1}{(1+|\xi|^2)^j}e^{\xi J_+}\ket{j,-{}j}
\ee
for the coherent state $\ket{j\xi} = g(\xi)\ket{j,-{}j}$.

\Eqs{jn-product} and \eqref{jxi} are very useful in practical calculations involving angular momentum coherent states. For example, using the tensor product property \eqref{jn-product}, the scalar product between two coherent states is immediately found to be
\be\label{xieta}
\braket{j\xi}{j\eta} = \bra - g^\dagger(\xi)g(\eta)\ket -^{2j} = \frac{(1+\bar\xi\eta)^{2j}}{(1+|\xi|^2)^j(1+|\eta|^2)^j}.
\ee
In terms of the vectors $\vec n$, we have
\be\label{mn}
\braket{j\vec n_1}{j\vec n_2} = e^{2ij\alpha(\vec n_1,\vec n_2)}\biggl(\frac{1+\vec n_1\cdot\vec n_2}{2}\biggr)^j,
\ee
where
\be\label{sph-area}
\alpha(\vec n_1,\vec n_2) = -\frac{i}{2}\ln\frac{1+\overline{\xi_1}\xi_2}{1+\xi_1\overline{\xi_2}}.
\ee
According to \cite{Perelomov}, $\alpha(\vec n_1,\vec n_2)$ has a geometrical interpretation as one-half of the (oriented) area of the spherical triangle spanned by the vectors $\hat e_z$, $\vec n_1$ and $\vec n_2$.

With the help of \Eq{jxi}, we may also derive the matrix elements of the angular momentum operator between the states $\ket{j\xi}$. For the calculation, it is convenient to use the unnormalized states
\be\label{jxi-unnorm}
|j\xi) = (1+|\xi|^2)^j\ket{j\xi} = e^{\xi J_+}\ket{j,-{}j},
\ee
for which we have
\be
(j\xi|j\eta) = (1+\bar\xi\eta)^{2j}.
\ee
To find the matrix element of $J_+$, we consider
\be
(j\xi|e^{\epsilon J_+}|j\eta) = (j\xi|e^{\epsilon J_+}e^{\eta J_+}\ket{j,-{}j} = (j\xi|j,\eta+\epsilon).
\ee
Expanding both sides to first order in $\epsilon$, we see that
\be
(j\xi|J_+|j\eta) = \frac{\partial}{\partial\eta}(j\xi|j\eta) = 2j\bar\xi(1+\bar\xi\eta)^{2j-1}.
\ee
In the same way, we find
\be
(j\xi|J_-|j\eta) = \frac{\partial}{\partial\bar\xi}(j\xi|j\eta) = 2j\eta(1+\bar\xi\eta)^{2j-1}.
\ee
The derivation of the matrix element of $J_0$ is less straightforward, but can be accomplished by considering the identity
\be
e^{\eta J_+}J_-e^{-\eta J_+}|j\eta) = 0.
\ee
According to the Baker--Campbell--Hausdorff formula, we have
\be
e^{\eta J_+}J_-e^{-\eta J_+} = J_- + \eta[J_+,J_-] + \frac{\eta^2}{2}\bigl[J_+,[J_+,J_-]\bigr] + \dots
\ee
where $[J_+,J_-] = 2J_0, \bigl[J_+,[J_+,J_-]\bigr] = -2J_+,$ and the terms represented by the dots contain the commutator $[J_+,J_+]$ and therefore vanish identically. It follows that
\be
\bigl(J_- + 2\eta J_0 - \eta^2 J_+\bigr)|j\eta) = 0,
\ee
from which we deduce
\be
(\xi|J_0|\eta) = \frac{\eta}{2}(j\xi|J_+|j\eta) - \frac{1}{2\eta}(j\xi|J_-|j\eta) = j(\bar\xi\eta-1)(1+\bar\xi\eta)^{2j-1}.
\ee
Summarizing our findings, and expressing them in terms of the normalized states $\ket{j\xi}$, we have shown that
\begin{subequations}\label{Jxy}
\begin{align}
\bra{j\xi}J_+\ket{j\eta} &= 2j\bar\xi\frac{(1+\bar\xi\eta)^{2j-1}}{(1+|\xi|^2)^j(1+|\eta|^2)^j}, \label{J+xy}\\
\bra{j\xi}J_0\ket{j\eta} &= j(\bar\xi\eta-1)\frac{(1+\bar\xi\eta)^{2j-1}}{(1+|\xi|^2)^j(1+|\eta|^2)^j}, \label{J0xy} \\
\bra{j\xi}J_-\ket{j\eta} &= 2j\eta\frac{(1+\bar\xi\eta)^{2j-1}}{(1+|\xi|^2)^j(1+|\eta|^2)^j}. \label{J-xy}
\end{align}
\end{subequations}

\subsection{Intertwiners in the basis of coherent states}\label{sec:IinCSbasis}

For the remainder of this chapter, our interest will be focused on the object
\begin{align}
\iota_{\xi_1\cdots\xi_N} &\equiv \braket{\iota}{j_1\xi_1\otimes\cdots\otimes j_N\xi_N} \notag \\
&= \iota_{m_1\cdots m_N} \D{j_1}{m_1}{-j_1}{\xi_1}\cdots\D{j_N}{m_N}{-j_N}{\xi_N}. \label{iotaxi}
\end{align}
By construction, the complex conjugates of the numbers $\iota_{\xi_1\cdots\xi_N}$ give the projections of the intertwiner $\ket\iota$ onto the angular momentum coherent state basis of ${\cal H}_{j_1}\otimes\cdots\otimes{\cal H}_{j_N}$. These numbers appear when a spin network state $\Psi_{\Gamma,\{j_e\},\{\iota_n\}}(h_{e_1},\dots,h_{e_{N}})$ of the familiar form \eqref{spinnetwork} is transformed into what could be called a coherent representation of the spin network, by using the resolution of identity $\Id = \int d\mu_j(\xi)\,\ket{j\xi}\bra{j\xi}$ on ${\cal H}_j$ instead of $\Id = \sum_m \ket{jm}\bra{jm}$ to trade each sum over a magnetic index for an integration over a corresponding complex variable. In this way, the spin network state becomes expressed in the form
\begin{align}
\Psi_{\Gamma,\{j_e\},\{\iota_n\}}(h_{e_1},\dots,h_{e_{N}}) = \int &d\mu_{j_1}(\xi_1)\,d\mu_{j_1}(\eta_1)\cdots d\mu_{j_{N}}(\xi_{N})\,d\mu_{j_{N}}(\eta_{N}) \notag \\
&{}\times\biggl(\prod_{e\in\Gamma} \sqrt{d_{j_e}}\D{j_e}{\xi_e}{\eta_e}{h_e}\biggr)\biggl(\prod_{n\in\Gamma} \iota_n \biggr)_{\xi_1\cdots\xi_{N}}^{\eta_1\cdots\eta_{N}},\label{csSN}
\end{align}
where $\D{j}{\xi}{\eta}{g} = \bra{j\xi}D^{(j)}(g)\ket{j\eta}$ are the matrix elements of the Wigner matrix with respect to the coherent state basis of ${\cal H}_j$, and the intertwiner $\ket{\iota_n}$ at each node of the graph is likewise projected onto the coherent state basis of the space $\bigl(\bigotimes_{\text{$e$ in}} {\cal H}_{j_e}\bigr)\bigl(\bigotimes_{\text{$e$ out}} {\cal H}^*_{j_e}\bigr)$.

A secondary interpretation of the object \eqref{iotaxi} can be brought out by considering the Livine--Speziale coherent intertwiner
\be\label{LS-18}
\bket{{\iota({\vec n_1,\dots,\vec n_N})}} = \int dg\,D^{(j_1)}(g)\ket{j_1\vec n_1}\otimes\cdots\otimes D^{(j_N)}(g)\ket{j_N\vec n_N}.
\ee
The coherent intertwiner can be expanded in some standard basis $\{\ket\iota\}$ of the $N$-valent intertwiner space as
\be\label{LS-expanded}
\bket{{\iota({\vec n_1,\dots,\vec n_N})}} = \sum_\iota c_\iota({\vec n_1,\dots,\vec n_N})\ket\iota.
\ee
Keeping in mind the identity \eqref{int D...D}, a short calculation shows that the coefficients of the expansion are given by
\be\label{c_iota}
c_\iota({\vec n_1,\dots,\vec n_N}) = \iota_{m_1\cdots m_N} \D{j_1}{m_1}{-j_1}{\vec n_1}\cdots\D{j_N}{m_N}{-j_N}{\vec n_N}.
\ee
Hence $\iota_{\xi_1\cdots\xi_N}$ can also be interpreted as the projection of the coherent intertwiner $\ket{\iota(\vec n_{\xi_1},\dots,\vec n_{\xi_N})}$ onto the intertwiner $\ket\iota$. A yet another interpretation of $\iota_{\xi_1\cdots\xi_N}$ will be discussed in section \ref{sec:iasP}.

\subsubsection*{The three-valent intertwiner}

We will now derive explicit expressions for the object \eqref{iotaxi} in the case of three- and four-valent intertwiners. Let us start with the three-valent case, where we must consider the object
\be\label{iDDD}
\iota_{\xi_1\xi_2\xi_3} \equiv \begin{pmatrix} j_1&j_2&j_3 \\ \xi_1&\xi_2&\xi_3 \end{pmatrix} = \begin{pmatrix} j_1&j_2&j_3 \\ m_1&m_2&m_3 \end{pmatrix}
\D{j_1}{m_1}{-j_1}{\xi_1}\D{j_2}{m_2}{-j_2}{\xi_2}\D{j_3}{m_3}{-j_3}{\xi_3}.
\ee
In order to tackle the calculation indicated by \Eq{iDDD}, it is convenient to pass to the realization of the spin-$j$ representation of $SU(2)$ as a completely symmetrized tensor product of $2j$ copies of the fundamental representation (see sections \ref{sec:spin-j} and \ref{sec:int-stp}). In this realization, a magnetic index in the spin-$j$ representation corresponds to a symmetrized group of $2j$ spinor indices, $(A_1\cdots A_{2j})$. The representation matrix $\D{j}{m}{n}{g}$ has the form \eqref{g...g}, namely
\be
\D{j}{m}{n}{g} \equiv \D{j}{(A_1\cdots A_{2j})}{(B_1\cdots B_{2j})}{g} = g\updown{A_1}{(B_1} \cdots g\updown{A_{2j}}{B_{2j})}.
\ee
In particular, when the second index takes its minimal value,
\be\label{DA-}
\D{j}{m}{-j}{g} = \D{j}{(A_1\cdots A_{2j})}{(-...-)}{g}= g\updown{A_1}{-} \cdots g\updown{A_{2j}}{-}.
\ee
The components of the three-valent intertwiner with respect to the magnetic basis are given in explicit form by \Eq{iota_ABC}. We have
\be\label{iABC}
\iota_{(A_1\cdots A_{2j_1})(B_1\cdots B_{2j_2})(C_1\cdots C_{2j_3})} = N_{j_1j_2j_3}I_{(A_1\cdots A_{2j_1})(B_1\cdots B_{2j_2})(C_1\cdots C_{2j_3})}
\ee
with
\be
I_{A_1\cdots A_{2j_1}B_1\cdots B_{2j_2}C_1\cdots C_{2j_3}} = \epsilon_{A_1B_1}\cdots\epsilon_{A_aB_a}\epsilon_{B_{a+1}C_1}\cdots\epsilon_{B_{a+b}C_b}\epsilon_{C_{b+1}A_{a+1}}\cdots\epsilon_{C_{b+c}A_{a+c}},
\ee
where $a = j_1+j_2-j_3$, $b = j_2+j_3-j_1$ and $c = j_3+j_1-j_2$, and the normalization factor $N_{j_1j_2j_3}$ has the value \eqref{Njjj}.

Using now \Eqs{DA-} and \eqref{iABC} in \Eq{iDDD}, we immediately obtain
\begin{align}
\begin{pmatrix} j_1&j_2&j_3 \\ \xi_1&\xi_2&\xi_3 \end{pmatrix} = N_{j_1j_2j_3}(\epsilon_{\xi_1\xi_2})^{j_1+j_2-j_3}(\epsilon_{\xi_2\xi_3})^{j_2+j_3-j_1}(\epsilon_{\xi_3\xi_1})^{j_3+j_1-j_2},
\end{align}
where we have introduced the notation
\be\label{eps_xy}
\epsilon_{\xi\eta} = \epsilon_{AB}g\updown{A}{-}(\xi)g\updown{B}{-}(\eta).
\ee
Recalling the form of the matrix $g(\xi)$ from \Eq{gxi}, we see that
\be
\epsilon_{\xi\eta} = \frac{\xi-\eta}{\sqrt{1+|\xi|^2}\sqrt{1+|\eta|^2}}.
\ee
Hence we have found the expression
\be\label{xi3j}
\begin{pmatrix} j_1&j_2&j_3 \\ \xi_1&\xi_2&\xi_3 \end{pmatrix} = N_{j_1j_2j_3}\frac{(\xi_1-\xi_2)^{j_1+j_2-j_3}(\xi_2-\xi_3)^{j_2+j_3-j_1}(\xi_3-\xi_1)^{j_3+j_1-j_2}}{(1+|\xi_1|^2)^{j_1}(1+|\xi_2|^2)^{j_2}(1+|\xi_3|^2)^{j_3}}
\ee
for the components of the three-valent intertwiner with respect to the coherent-state basis.

An important property of the object \eqref{xi3j} can be inferred from the work of Livine and Speziale concerning the behaviour of the coherent intertwiner \eqref{LS-18} in the limit of large spins. As shown in \cite{LivineSpeziale}, for large values of $j$, the norm of the intertwiner \eqref{LS-18} is exponentially small in $j$, unless the vectors $\vec n_a$ satisfy the closure condition $\sum_a j_a\vec n_a=0$ (in which case the norm of the coherent intertwiner depends on $j$ through a power law). On the other hand, if we consider \eqref{xi3j} as a function of the vectors $\vec n_a$ corresponding to the parameters $\xi_a$, then \Eqs{LS-expanded} and \eqref{c_iota} show that the absolute value of \eqref{xi3j} gives the norm of the three-valent coherent intertwiner $\ket{\iota(\vec n_1,\vec n_2,\vec n_3)}$. Hence it follows that for large spins, the magnitude of \eqref{xi3j} is peaked on configurations satisfying $j_1\vec n_1+j_2\vec n_2+j_3\vec n_3=0$, and is exponentially suppressed when the closure condition is not fulfilled.

\subsubsection*{Four-valent intertwiners}

In the case of the four-valent intertwiner, we again start by recalling the expression of the intertwiner in the symmetric tensor product representation. From \Eq{iota_ABCD}, 
\be\label{i4fig}
\iota^{(k)}_{(A_1\cdots A_{2j_1})\cdots(D_1\cdots D_{2j_4})} = \sqrt{d_k}N_{j_1j_2k}N_{kj_3j_4}\;\makeSymbol{\includegraphics[scale=0.75]{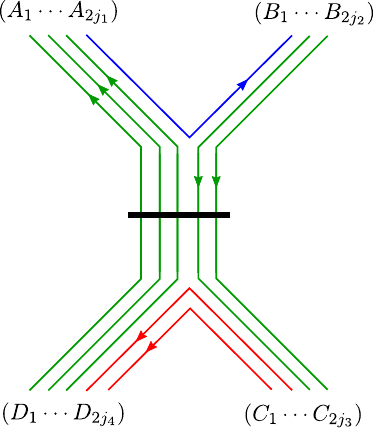}}
\ee
where the horizontal bar denotes a complete symmetrization of the internal group of lines. When the above expression is contracted with $\D{j_1}{m_1}{-j_1}{\xi_1}\cdots\D{j_4}{m_4}{-j_4}{\xi_4}$ according to \Eq{iotaxi}, each of the $j_1+j_2-k$ blue lines contributes a factor of $\epsilon_{\xi_1\xi_2}$, while each of the $j_3+j_4-k$ red lines contributes a $\epsilon_{\xi_3\xi_4}$. Keeping in mind the symmetrization, the contribution of the green lines has the form
\be\label{i4-sum}
{\cal N}\sum_{n_{ab}} N(n_{ab})(\epsilon_{\xi_3\xi_1})^{n_{13}}(\epsilon_{\xi_4\xi_1})^{n_{14}}(\epsilon_{\xi_2\xi_3})^{n_{23}}(\epsilon_{\xi_2\xi_4})^{n_{24}},
\ee
where $n_{ab}$ stands for the number of lines running from external spin $j_a$ to $j_b$, and $N(n_{ab})$ is the number of different ways of arranging the green lines for fixed values of the $n_{ab}$. The sum is taken over those values of the $n_{ab}$ that satisfy the constraints
\begin{subequations}\label{i4-conditions}
\begin{align}
n_{13} + n_{14} &= j_1-j_2+k, \\
n_{23} + n_{24} &= j_2-j_1+k, \\
n_{13} + n_{23} &= j_3-j_4+k, \\
n_{14} + n_{24} &= j_4-j_3+k,
\end{align}
\end{subequations}
which reduce the summation in \Eq{i4-sum} to one independent sum. Finally, the overall normalization factor ${\cal N}$ should be adjusted so that the symmetrization over the internal group of lines is performed with total weight 1. The values of $N(n_{ab})$ and ${\cal N}$ can be determined by a straightforward combinatorial counting, eventually yielding the result
\begin{align}
\iota^{(k)}_{\xi_1\cdots\xi_4} = \sqrt{d_k}N_{j_1j_2k}N_{kj_3j_4}\frac{(\xi_1-\xi_2)^{j_1+j_2-k}(\xi_3-\xi_4)^{j_3+j_4-k}Q_k(\xi_1,\xi_2,\xi_3,\xi_4)}{(1+|\xi_1|^2)^{j_1}(1+|\xi_2|^2)^{j_2}(1+|\xi_3|^2)^{j_3}(1+|\xi_4|^2)^{j_4}},\label{xi-int4}
\end{align}
where
\begin{align}
Q_k(\xi_1&,\xi_2,\xi_3,\xi_4) = \dfrac{1}{\begin{pmatrix} 2k \\ k+j_1-j_2 \end{pmatrix}} \sum_t \begin{pmatrix} j_3-j_4+k \\ t \end{pmatrix} \begin{pmatrix} j_4-j_3+k \\ j_1-j_2+k-t \end{pmatrix} \notag \\
&\times (\xi_3-\xi_1)^t (\xi_2-\xi_4)^{j_2-j_1-j_3+j_4+t} (\xi_4-\xi_1)^{j_1-j_2+k-t} (\xi_2-\xi_3)^{j_3-j_4+k-t},
\end{align}
the sum running over all values of $t$ for which all the factorials in the binomial coefficients have non-negative arguments.

In principle, the method described above could be used to derive the components of an intertwiner of any valence with respect to the coherent state basis, though the calculations would rapidly become more and more tedious with increasing valence. However, considering that an appropriate generalization of \Eq{i4fig} holds for intertwiners of higher valence as well, it is clear even without any detailed calculations that the components of the $N$-valent intertwiner will always have the general structure
\be\label{iN_xi}
\iota_{\xi_1\cdots\xi_N} = \frac{P_\iota(\xi_1,\dots,\xi_N)}{(1+|\xi_1|^2)^{j_1}\cdots(1+|\xi_N|^2)^{j_N}},
\ee
where the function $P_\iota(\xi_1,\dots,\xi_N)$ is a polynomial of order $2j_a$ in each variable $\xi_a$, and depends on its arguments only through the differences $\xi_a-\xi_b$.

\subsubsection*{Raising and lowering indices}

The calculations which have lead to \Eqs{xi3j}, \eqref{xi-int4} and \eqref{iN_xi} apply as such only to intertwiners having only lower indices. To complete our discussion of intertwiners in the basis of coherent states, we must therefore consider how the ''indices'' $\xi_a$ can be raised and lowered in this basis. This is necessary, for example, for a consistent contraction of indices in equations such as \eqref{csSN}.

To derive the rule for raising an index in the coherent state basis, we may compute the projection of
\be
\sum_{m_1\cdots m_N} \iota\,\updown{m_1}{m_2\cdots m_N}\,\bra{j_1m_1}\otimes\ket{j_2m_2}\otimes\cdots\otimes\ket{j_Nm_N}
\ee
(with $\iota\updown{m_1}{m_2\cdots m_N} = \epsilon^{(j_1)m_1n_1}\iota_{n_1m_2\cdots m_N}$) onto the state $\bra{j_1\xi_1}\otimes\ket{j_2\xi_2}\otimes\cdots\otimes\ket{j_N\xi_N}$. In this way we find the expected result,
\be\label{raising-xi}
\iota\,\updown{\xi_1}{\xi_2\cdots\xi_N} = \int d\mu_{j_1}(\eta_1)\,\epsilon^{(j_1)\xi_1\eta_1}\iota_{\eta_1\xi_2\cdots\xi_N},
\ee
where $\epsilon^{(j)\xi\eta}$ are the components of the epsilon tensor with respect to the coherent state basis, \ie 
\be
\epsilon^{(j)\xi\eta} \equiv \bra{j\xi}\bra{j\eta}\sum_{mn} \epsilon^{(j)mn}\ket{jm}\ket{jn} = \frac{(\bar\xi - \bar\eta)^{2j}}{(1+|\xi|^2)^j(1+|\eta|^2)^j}.
\ee
Inserting the explicit expression \eqref{iN_xi} for the intertwiner $\iota_{\eta_1\xi_2\cdots\xi_N}$, the integral \eqref{raising-xi} can now be evaluated using the identity \eqref{integration}. This yields the result
\be\label{xi-raised}
\iota\,\updown{\xi_1}{\xi_2\cdots\xi_N} = \frac{\bar\xi_1^{2j_1}P_\iota(-\bar\xi_1^{-1},\xi_2,\dots,\xi_N)}{(1+|\xi_1|^2)^{j_1}\cdots(1+|\xi_N|^2)^{j_N}},
\ee
which can be alternatively be written in the form
\be
\iota\,\updown{\xi_1}{\xi_2\cdots\xi_N} = \biggl(\frac{\bar\xi_1}{|\xi_1|}\biggr)^{2j_1}\frac{P_\iota(-\bar\xi_1^{-1},\xi_2,\dots,\xi_N)}{(1+|{-}\bar\xi_1^{-1}|^2)^{j_1}\cdots(1+|\xi_N|^2)^{j_N}}.
\ee
Comparing with \Eq{g(-n)}, we see that if we consider the object \eqref{iN_xi} as a function of the vectors $\vec n(\xi)$, then raising an index $\xi_a$ amounts to replacing the corresponding vector $\vec n_a$ with $-\vec n_a$ and adjusting the phase by $(\bar\xi_a/|\xi_a|)^{2j_a}$.

\subsection{The complex polynomial representation of $SU(2)$}\label{sec:SU2pol}

Our next goal in this chapter will be to explain how the polynomial $P_\iota(\xi_1,\dots,\xi_N)$ in \Eq{iN_xi} can itself be regarded as a representation of the intertwiner $\ket\iota$, and how angular momentum operators acting on the intertwiner can be formulated as differential operators acting on the variables $\xi_a$. The key point underlying this interpretation is the fact that the spin-$j$ representation of $SU(2)$ can be realized on the space of polynomials of degree $2j$ in a single complex variable (see \eg \cite{Perelomov}). Below we will recall the main properties of the polynomial representation of $SU(2)$.

Let ${\cal P}_j$ denote the space of polynomials of degree $2j$ in a complex variable $z$. Thus, a general element of ${\cal P}_j$ has the form $f^{(j)}(z) = \sum_{k=0}^{2j} c_kz^k$. A scalar product on ${\cal P}_j$ can be defined as
\be\label{Pj prod}
\braket{f^{(j)}}{g^{(j)}} = \int d\nu_j(z)\,\overline{\displaystyle f^{(j)}(z)}g^{(j)}(z),
\ee
where the integration measure is
\be\label{dnu_j}
d\nu_j(z) = \frac{d_j}{\pi}\frac{d^2z}{(1+|z|^2)^{2j+2}} = \frac{d\mu_j(z)}{(1+|z|^2)^{2j}},
\ee
with $d\mu_j(z)$ the measure defined by \Eq{dmu_j}. The monomials
\be\label{f_jm}
f^{(j)}_m(z) = \sqrt{\frac{(2j)!}{(j+m)!(j-m)!}}z^{j+m},
\ee
where the label $m$ takes the values $m=-j,-j+1,\dots,j$, form a basis on ${\cal P}_j$, orthonormal under the scalar product \eqref{Pj prod}. As an immediate consequence of the orthonormality of the functions \eqref{f_jm}, we have the useful identity
\be\label{integration}
\int d\nu_j(\xi)\,(1+\bar\xi z)^{2j}f^{(j)}(\xi) = f^{(j)}(z),
\ee
which holds for all functions $f^{(j)}\in{\cal P}_j$.

A representation of $SU(2)$ on the space ${\cal P}_j$ can now be defined by declaring the action of the group element $g = \bigl(\begin{smallmatrix} \alpha & \beta \\ -\bar\beta & \bar\alpha \end{smallmatrix}\bigr) \in SU(2)$ on a function $f^{(j)}(z)\in{\cal P}_j$ to be
\be\label{su2pol}
D^{(j)}(g)f^{(j)}(z) = (\beta z+\bar\alpha)^{2j}f^{(j)}\biggl(\frac{\alpha z-\bar\beta}{\beta z+\bar\alpha}\biggr).
\ee
In fact, this definition can be extended to the group $SL(2,\C)$. For $h = \bigl(\begin{smallmatrix} a&b \\ c&d \end{smallmatrix}\bigr) \in SL(2,\C)$, the assignment
\be\label{sl2cpol}
D^{(j)}(h)f^{(j)}(z) = (bz+d)^{2j}f^{(j)}\biggl(\frac{az+c}{bz+d}\biggr)
\ee
defines a representation of $SL(2,\C)$ on ${\cal P}_j$.

To find the form of the angular momentum operator on the space ${\cal P}_j$, we can use \Eq{sl2cpol} to calculate the action of the $SL(2,\C)$ matrices $e^{\epsilon J_i}$ on a function $f^{(j)}(z)\in{\cal P}_j$. Considering for example the operator $J_+$, we have
\be
e^{\epsilon J_+}f^{(j)}(z) = (1+\epsilon z)^{2j}f^{(j)}\biggl(\frac{z}{1+\epsilon z}\biggr).
\ee
Expanding both sides to first order in $\epsilon$ then reveals that
\be\label{J+}
J_+ = -z^2\frac{d}{dz} + 2jz.
\ee
By similar calculations one finds
\be\label{J0}
J_0 = z\frac{d}{dz}-j
\ee
and
\be\label{J-}
J_- = \frac{d}{dz}.
\ee
We note that the function $f^{(j)}_m$ of \Eq{f_jm} is an eigenstate of the operator $J_0$ with eigenvalue $m$, and hence corresponds to the state $\ket{jm}\in{\cal H}_j$ in the realization of the spin-$j$ representation normally used in physics.

Before concluding our review of the polynomial representation of $SU(2)$, let us write down the angular momentum coherent state $\ket{j\xi}$ as an element of the space ${\cal P}_j$. Using \Eqs{jxi} and \eqref{sl2cpol}, and observing that the state $\ket{j,-{}j}$ corresponds to the function $f^{(j)}_{-j}(z)=1$ in ${\cal P}_j$, we see that
\be\label{cspoly}
f_\xi^{(j)}(z) = \frac{1}{(1+|\xi|^2)^j}(1+\xi z)^{2j}.
\ee
Hence the polynomial representing the unnormalized coherent state $|j\xi)$ of \Eq{jxi-unnorm} is simply $(1+\xi z)^{2j}$. Furthermore, recalling that the scalar product between the states $|j\xi)$ is given by $(j\xi|j\eta) = (1+\bar\xi\eta)^{2j}$, the integration formula \eqref{integration} may be written as
\be\label{integration2}
\int d\nu_j(\xi)\,(j\xi|jz)f^{(j)}(\xi) = f^{(j)}(z),
\ee
showing that the scalar product $(j\xi|jz)$ acts as a so-called ''reproducing kernel'' (see \eg \cite{Gazeau-book}). Finally, by inspecting the matrix elements of the angular momentum operator in the basis of angular momentum coherent states, summarized in \Eqs{Jxy}, we see that the matrix elements between the unnormalized states $|j\xi)$ can be expressed in the form
\be\label{Jmu-unnorm}
(j\xi|J_i|j\eta) = J_i(\bar\xi)(j\xi|j\eta),
\ee
where $J_i(\bar\xi)$ (with $i = +,0,-$) denotes any of the differential operators \eqref{J+}--\eqref{J-} acting on the variable $\bar\xi$ in the function $(j\xi|j\eta) = (1+\bar\xi\eta)^{2j}$.

\subsubsection*{Example: A polynomial calculation of the coherent 3$j$-symbol}

As an example of working with the polynomial representation of $SU(2)$, we will use it to compute the ''coherent 3$j$-symbol'' defined by \Eq{iDDD}, following a similar calculation made in \cite{BasuKar} for the Clebsch--Gordan coefficient. Besides serving as an example of using the polynomial formalism, the calculation given below also provides an independent derivation of \Eq{xi3j}, which is one of the main results of this chapter.

Using the $SU(2)$-invariance of the standard 3$j$-symbol, \Eq{iDDD} can be rewritten as
\be\label{coh3j-2}
\begin{pmatrix} j_1&j_2&j_3 \\ \xi_1&\xi_2&\xi_3 \end{pmatrix} = \sum_{m_1m_2} \begin{pmatrix} j_1&j_2&j_3 \\ m_1&m_2&-j_3 \end{pmatrix}\D{j_1}{m_1}{-{}j_1}{\xi_3^{-1}\xi_1}\D{j_2}{m_2}{-{}j_2}{\xi_3^{-1}\xi_2},
\ee
where $D^{(j)}(\xi_a^{-1}\xi_b)$ is shorthand for $D^{(j)}\bigl(g^{-1}(\xi_a)g(\xi_b)\bigr)$. The products of coherent state rotations in \Eq{coh3j-2} can be computed by means of \Eqs{g-1} and \eqref{g1g2}. If we then introduce the state
\begin{align}
\ket{\Phi_0} &= \sum_{m_1m_2} \begin{pmatrix} j_1&j_2&j_3 \\ m_1&m_2&-j_3 \end{pmatrix} \ket{j_1m_1}\ket{j_2m_2} \notag \\
&= \frac{(-1)^{j_1-j_2+j_3}}{\sqrt{d_{j_3}}}\sum_{m_1m_2} \CGi{j_1j_2}{j_3}{m_1m_2}{j_3}\ket{j_1m_1}\ket{j_2m_2} \notag \\
&= \frac{(-1)^{j_1-j_2+j_3}}{\sqrt{d_{j_3}}}\ket{j_1j_2;j_3j_3}, \label{Phi_0}
\end{align}
which (up to a numerical factor) is the state of total angular momentum $j_3$ and maximal magnetic number $m=j_3$ constructed from the angular momenta $j_1$ and $j_2$, we can cast \Eq{coh3j-2} in the form
\be\label{prod2}
\begin{pmatrix} j_1&j_2&j_3 \\ \xi_1&\xi_2&\xi_3 \end{pmatrix} = e^{-2ij_1\beta_{13}-2ij_2\beta_{23}}\braket{\Phi_0}{j_1\xi_{13}\otimes j_2\xi_{23}},
\ee
where
\be\label{parameters}
\xi_{13} = \frac{\xi_1-\xi_3}{1+\xi_1\bar\xi_3}, \qquad e^{i\beta_{13}} = \frac{1+\bar\xi_1\xi_3}{|1+\bar\xi_1\xi_3|},
\ee
and $\xi_{23}$ and $\beta_{23}$ are defined similarly.

From here on, our strategy will be to express the states $\ket{\Phi_0}$ and $\ket{j_1\xi_{13}\otimes j_2\xi_{23}}$ as polynomials in the space ${\cal P}_{j_1}\otimes{\cal P}_{j_2}$, after which their scalar product in \Eq{prod2} can be computed using \Eq{Pj prod} for the scalar product in ${\cal P}_j$. The representation of the coherent states as polynomials is given by \Eq{cspoly}. To find the state $\ket{\Phi_0}$ in polynomial form, we write the defining equations of the state $\ket{j_1j_2;j_3j_3}$,
\be
J_+\ket{j_1j_2;j_3j_3}=0 \qquad \text{and} \qquad J_z\ket{j_1j_2;j_3j_3} = j_3\ket{j_1j_2;j_3j_3}
\ee
as differential equations for a function $f_{j_3}(z_1,z_2)$ in ${\cal P}_{j_1}\otimes{\cal P}_{j_2}$, using the representation of the angular momentum operators from \Eqs{J+}--\eqref{J-}. These equations are
\begin{align}
&\biggl(-z_1^2\frac{\partial}{\partial z_1} + 2j_1z_1 - z_2^2\frac{\partial}{\partial z_2} + 2j_2z_2\biggr)f_{j_3}(z_1,z_2) = 0, \label{J+eq}\\
&\biggl(z_1\frac{\partial}{\partial z_1} - j_1 + z_2\frac{\partial}{\partial z_2} - j_2\biggr)f_{j_3}(z_1,z_2) = j_3f_{j_3}(z_1,z_2). \label{Jzeq}
\end{align}
We look for the solution of these equations by introducing the ansatz
\be
f_{j_3}(z_1,z_2) = z_1^{2j_1}z_2^{2j_2}g_{j_3}\biggl(\frac{1}{z_1},\frac{1}{z_2}\biggr),
\ee
where $g_{j_3}$ is a polynomial of order $2j_3$ in $1/z_1$ and $1/z_2$. Inserting this into \Eq{J+eq}, and using the fact that the operator $J_+ = -z^2\,\partial/\partial z + 2jz$ annihilates the function $f_j(z) = z^{2j}$, we find that $g_{j_3}(1/z_1,1/z_2)$ satisfies the equation
\be
\biggl(-z_1^2\frac{\partial}{\partial z_1} - z_2^2\frac{\partial}{\partial z_2}\biggr)g_{j_3}\biggl(\frac{1}{z_1},\frac{1}{z_2}\biggr) = 0,
\ee
or, changing variables from $z$ to $\zeta = 1/z$,
\be
\biggl(\frac{d}{d\zeta_1} + \frac{d}{d\zeta_2}\biggr)g_{j_3}(\zeta_1,\zeta_2) = 0.
\ee
This shows that the function $g_{j_3}$ depends on its arguments only through their difference, \ie $g_{j_3}(\zeta_1,\zeta_2) = g_{j_3}(\zeta_1-\zeta_2)$. It follows that the function $f_{j_3}(z_1,z_2)$ has the form
\be
f_{j_3}(z_1,z_2) = z_1^{2j_1}z_2^{2j_2}g_{j_3}\biggl(\frac{1}{z_1}-\frac{1}{z_2}\biggr).
\ee
Inserting this into \Eq{Jzeq}, we obtain
\be
\biggl(\frac{1}{z_1}-\frac{1}{z_2}\biggr) g_{j_3}'\biggl(\frac{1}{z_1}-\frac{1}{z_2}\biggr)  = (j_1+j_2-j_3)g_{j_3}\biggl(\frac{1}{z_1}-\frac{1}{z_2}\biggr),
\ee
where $g_{j_3}'(z)$ denotes the derivative of $g_{j_3}(z)$ with respect to its single argument. From this we see that $g_{j_3} = {\rm const.}\times(1/z_1-1/z_2)^{j_1+j_2-j_3}$, and so
\begin{align}
f_{j_3}(z_1,z_2) &= {\rm const}.\times z_1^{2j_1}z_2^{2j_2}\biggl(\frac{1}{z_1}-\frac{1}{z_2}\biggr)^{j_1+j_2-j_3} \notag \\
&= {\rm const}.\times z_1^{j_1-j_2+j_3}z_2^{j_2-j_1+j_3}(z_1-z_2)^{j_1+j_2-j_3}. \label{f_j}
\end{align}
By evaluating the integral $\int d\nu_{j_1}(z_1)\,d\nu_{j_2}(z_2)\,|f_{j_3}(z_1,z_2)|^2 = 1$, the value of the normalization constant is determined to be $\sqrt{d_{j_3}}N_{j_1j_2j_3}$, with $N_{j_1j_2j_3}$ defined by \Eq{Njjj}. (The phase of the constant is fixed by the Condon--Shortley convention, according to which the Clebsch--Gordan coefficient $\CG{j_1j_2}{j_3}{j_1,j_3{-}j_1}{j_3}$ -- equivalently, the coefficient of $z_1^{2j_1}z_2^{j_2+j_3-j_1}$ in the expansion of the function $f_{j_3}(z_1,z_2)$ -- should be real and positive.)

As a result of the above calculation, we have found that the polynomial $\phi_0(z_1,z_2)$ representing the state $\ket{\Phi_0}$ of \Eq{Phi_0} as an element of ${\cal P}_{j_1}\otimes{\cal P}_{j_2}$ is given by
\be
\phi_0(z_1,z_2) = N_{j_1j_2j_3}(-z_1)^{j_1-j_2+j_3}z_2^{j_2-j_1+j_3}(z_1-z_2)^{j_1+j_2-j_3}.
\ee
The scalar product $\braket{\Phi_0}{j_1\xi_{13}\otimes j_2\xi_{23}}$ appearing in \Eq{prod2} can then be computed as
\begin{align}
&\braket{\Phi_0}{j_1\xi_{13}\otimes j_2\xi_{23}} = \int d\nu_{j_1}(z_1)\,d\nu_{j_2}(z_2)\,\overline{\phi_0(z_1,z_2)}f^{(j_1)}_{\xi_{13}}(z_1)f^{(j_2)}_{\xi_{23}}(z_2) \notag \\
&\hspace{0.5cm}= \frac{N_{j_1j_2j_3}}{(1+|\xi_{13}|^2)^{j_1}(1+|\xi_{23}|^2)^{j_2}}\int d\nu_{j_1}(z_1)\,d\nu_{j_2}(z_2)\,(1+\xi_{13}z_1)^{2j_1}(1+\xi_{23}z_2)^{2j_2} \notag \\
&\hspace{6cm}\times(-\bar z_1)^{j_1-j_2+j_3}\bar z_2^{j_2-j_1+j_3}(\bar z_1-\bar z_2)^{j_1+j_2-j_3}.
\end{align}
The factor multiplying $(1+\xi_{13}z_1)^{2j_1}(1+\xi_{23}z_2)^{2j_2}$ in the integrand is a polynomial of order $2j_1$ and $2j_2$ in $\bar z_1$ and $\bar z_2$. Therefore, after changing the variables of integration from $z_1$ and $z_2$ to $\bar z_1$ and $\bar z_2$, the integral can be evaluated immediately by means of \Eq{integration}, leading to
\be
\braket{\Phi_0}{j_1\xi_{13}\otimes j_2\xi_{23}} = \frac{N_{j_1j_2j_3}}{(1+|\xi_{13}|^2)^{j_1}(1+|\xi_{23}|^2)^{j_2}}(-\xi_{13})^{j_1-j_2+j_3}\xi_{23}^{j_2-j_1+j_3}(\xi_{13}-\xi_{23})^{j_1+j_2-j_3}.
\ee
When this is inserted into \Eq{prod2}, and the result expressed in terms of $\xi_1$, $\xi_2$ and $\xi_3$ using \Eq{parameters}, we reproduce \Eq{xi3j} for the coherent-state components of the three-valent intertwiner. 

\subsection{Intertwiners as complex polynomials}\label{sec:iasP}

With the preliminaries of the previous section, we are now ready to show how a polynomial representation of an intertwiner can be extracted from the explicit expression \eqref{iN_xi} for the components of the intertwiner with respect to the coherent state basis. We will demonstrate that the action of the angular momentum operator on the intertwiner can be formulated as an action on the polynomial $P_\iota(\xi_1,\dots,\xi_N)$ in \Eq{iN_xi} by the differential operators \Eqs{J+}--\eqref{J-}, which represent the angular momentum operator as an operator on ${\cal P}_j$. On grounds of this result, the polynomial $P_\iota(\xi_1,\dots,\xi_N)$ can indeed be interpreted as representing the intertwiner $\ket\iota\in{\rm Inv}\,\bigl({\cal H}_{j_1}\otimes\cdots\otimes{\cal H}_{j_N}\bigr)$ as an element of ${\cal P}_{j_1}\otimes\cdots\otimes{\cal P}_{j_N}$.

For clarity, let us start by considering a single angular momentum operator acting on the $a$-th spin of an intertwiner $\ket\iota$. The state $\ket\iota$ can be expanded in the basis of coherent states on ${\cal H}_{j_1}\otimes\cdots\otimes{\cal H}_{j_N}$ as
\be\label{iota-expansion}
\ket\iota = \int d\mu_{j_1}(\xi_1)\cdots d\mu_{j_N}(\xi_N)\,\overline{\iota_{\xi_1\cdots\xi_N}}\,\ket{j_1\xi_1\otimes\cdots\otimes j_N\xi_N}.
\ee
Letting the operator $J_i^{(a)}$ act on $\ket\iota$, and projecting the resulting state on the basis element $\bra{j_1\xi_1\otimes\cdots\otimes j_N\xi_N}$, we obtain
\begin{align}
\bra{j_1\xi_1\otimes\cdots\otimes j_N\xi_N}J_i^{(a)}\ket\iota = &\int d\mu_{j_1}(\eta_1)\cdots d\mu_{j_N}(\eta_N)\frac{P_\iota(\bar\eta_1,\dots,\bar\eta_N)}{(1+|\eta_1|^2)^{j_1}\cdots(1+|\eta_N|^2)^{j_N}} \notag \\
&\times\braket{j_1\xi_1}{j_1\eta_1}\cdots\bra{j_a\xi_a}J_i\ket{j_a\eta_a}\cdots\braket{j_N\xi_N}{j_N\eta_N}.
\end{align}
To perform the integral, we rewrite the integrand in terms of the unnormalized states $|j\xi)$, recalling from \Eq{Jmu-unnorm} that $(j\xi|J_i|j\eta) = J_i(\bar\xi)(j\xi|j\eta)$. We then write $(j\xi|j\eta) = (j\bar\eta|j\bar\xi)$, and change the variables of integration from $\eta_1,\dots,\eta_N$ to their complex conjugates, leading to
\begin{align}
&\bra{j_1\xi_1\otimes\cdots\otimes j_N\xi_N}J_i^{(a)}\ket\iota = \frac{1}{{(1+|\xi_1|^2)^{j_1}\cdots(1+|\xi_N|^2)^{j_N}}} \notag \\
&\times J_i(\bar\xi_a)\int d\nu_{j_1}(\eta_1)\cdots d\nu_{j_N}(\eta_N)\,P_\iota(\eta_1,\dots,\eta_N)(j_1\eta_1|j_1\bar\xi_1)\cdots(j_N\eta_N|j_N\bar\xi_N),
\end{align}
where the integration measure $d\nu_j$ of \Eq{dnu_j} has naturally appeared. Now, taking into account the ''reproducing kernel'' property \eqref{integration2}, we see immediately that the integral in the above equation is equal to $P_\iota(\bar\xi_1,\dots,\bar\xi_N)$. We have therefore found
\be
\bra{j_1\xi_1\otimes\cdots\otimes j_N\xi_N}J_i^{(a)}\ket\iota = \frac{J_i(\bar\xi_a)P_\iota(\bar\xi_1,\dots,\bar\xi_N)}{(1+|\xi_1|^2)^{j_1}\cdots(1+|\xi_N|^2)^{j_N}}.
\ee
A similar calculation can be made to show that for any operator ${\cal O}(J^{(1)},\cdots,J^{(N)})$, constructed from the angular momentum operators $J^{(1)},\cdots,J^{(N)}$ by a power series expansion, we have
\be\label{O(J)P}
\bra{j_1\xi_1\otimes\cdots\otimes j_N\xi_N}{\cal O}(J^{(1)},\cdots,J^{(N)})\ket\iota = \frac{{\cal O}\bigl(J(\bar\xi_1),\cdots,J(\bar\xi_N)\bigr)P_\iota(\bar\xi_1,\dots,\bar\xi_N)}{(1+|\xi_1|^2)^{j_1}\cdots(1+|\xi_N|^2)^{j_N}}.
\ee
Thus we have shown that the action of the angular momentum operator on the state $\ket\iota$ translates to an action on the polynomial $P_\iota(\xi_1,\dots,\xi_N)$ by the representation of the angular momentum operator as a differential operator on the space ${\cal P}_{j_1}\otimes\cdots\otimes{\cal P}_{j_N}$. This result guarantees that we can consistently interpret $P_\iota(\xi_1,\dots,\xi_N)$ as the polynomial representing the intertwiner $\ket\iota$ as an element of ${\cal P}_{j_1}\otimes\cdots\otimes{\cal P}_{j_N}$. We should note that under such an interpretation, the variable $\xi$ is to be regarded merely as an abstract variable, whose first $2j$ powers span the space ${\cal P}_j$, and it no longer plays a role as a parameter specifying the direction of a unit vector $\vec n$.

\subsection{Application to calculations in loop quantum gravity}

The results developed in this chapter can be used in loop quantum gravity to set up an alternative approach towards analyzing the matrix elements of operators (such as the Hamiltonian), in which spin network states are expressed in the ''coherent representation'' of \Eq{csSN}, and the action of angular momentum operators on intertwiners is formulated as the action of differential operators on the polynomials $P_\iota(\xi_1,\dots,\xi_N)$. This makes it possible to compute matrix elements of operators in the spin network basis in a way which largely bypasses the use of $SU(2)$ recoupling theory, and to describe the action of the operator geometrically, in terms of the complex parameters $\xi$ and the geometry of the associated unit vectors $\vec n$ -- as opposed to expressions which involve combinations of 6$j$- and 9$j$-symbols, and to which it is difficult to assign any immediate geometric or otherwise intuitive meaning.

As a concrete example of applying the techniques of this chapter to loop quantum gravity, let us consider the Euclidean operator
\be\label{CE-18}
C^E_{ab} = \epsilon^{ijk}\Tr\bigl(\tau_k^{(s)}D^{(s)}(h_{\alpha_{ab}})\bigr)J_i^{(v,e_a)}J_j^{(v,e_b)}.
\ee
The matrix elements of this operator were computed in section \ref{sec:CE-el} using the standard basis of magnetic indices, in which the action of the operator on an intertwiner $\iota_{m_1\cdots m_N}$ produces a new intertwiner of the form $(C^E_{ab}\iota)\downup{m_1\cdots m_Nn_1}{n_2}$. The two new indices of the intertwiner are contracted with the indices of the holonomy $\D{s}{n_1}{n_2}{h_{\alpha_{ab}}}$ created by the action of the operator.

Similarly, in the coherent state basis the action of $C^E_{ab}$ on an intertwiner $\iota_{\xi_1\cdots\xi_N}$ produces the intertwiner
\be\label{C_E*i-def}
(C^E_{ab}\iota){}\downup{\xi_1\cdots\xi_N\eta_1}{\eta_2} \equiv \bbraket{\epsilon^{ijk}\Tau{s}{k}{\eta_2}{\eta_1}J_i^{(a)}J_j^{(b)}\iota}{j_1\xi_1\otimes\cdots\otimes j_N\xi_N},
\ee
where the indices $\eta_1$ and $\eta_2$ are eventually contracted with the indices of the holonomy $\D{s}{\eta_1}{\eta_2}{h_{\alpha_{ab}}}$ by the integral $\int d\nu_s(\eta_1)\,d\nu_s(\eta_2)$. According to \Eq{iN_xi} -- and recalling from \Eq{raising-xi} that the variable corresponding to an upper index of the intertwiner enters the polynomial $P_\iota(\xi_1,\dots,\xi_N)$ as a complex conjugate -- the intertwiner \eqref{C_E*i-def} has the general form
\be\label{Ciota}
(C^E_{ab}\iota){}\downup{\xi_1\cdots\xi_N\eta_1}{\eta_2} = \frac{(C^E_{ab}P_\iota)(\xi_1,\dots,\xi_N,\eta_1,\bar\eta_2)}{(1+|\xi_1|^2)^{j_1}\cdots(1+|\xi_N|^2)^{j_N}(1+|\eta_1|^2)^s(1+|\eta_2|^2)^s},
\ee
where $(C^E_{ab}P_\iota)(\xi_1,\dots,\xi_N,\eta_1,\bar\eta_2)$ is the polynomial representing the intertwiner $C^E_{ab}\iota$ as an element of the space ${\cal P}_{j_1}\otimes\cdots\otimes{\cal P}_{j_N}\otimes{{\cal P}_l}\otimes{\cal P}^*_l$.

In order to derive an expression for the polynomial $C^E_{ab}P_\iota$, we use \Eqs{J+}--\eqref{J-} for the angular momentum operator, and \Eqs{Jxy} for the matrix elements $\Tau{s}{k}{\eta_2}{\eta_1} = -i\bra{s\eta_2}J_k\ket{s\eta_1}$, to translate the operator \eqref{CE-18} into a differential operator acting on the polynomial $P_\iota(\xi_1,\dots,\xi_N)$. Introducing the abbreviations $\partial_a = \partial/\partial\xi_a$ and $\partial^2_{ab} = \partial^2/\partial\xi_a\partial\xi_b$, we find
\begin{align}
(C^E_{ab}&P_\iota)(\xi_1,\cdots,\xi_N,\eta_1,\bar\eta_2) = 2s(1+\eta_1\bar\eta_2)^{2s-1}\biggl[\eta_1\Bigl((\xi_a-\xi_b)\partial^2_{ab} + j_b\partial_a - j_a\partial_b\Bigr) \notag \\
&+\bar\eta_2\Bigl(-\xi_a\xi_b(\xi_a-\xi_b)\partial^2_{ab} + j_b\xi_a(\xi_a-2\xi_b)\partial_a - j_a\xi_b(\xi_b-2\xi_a)\partial_b - 2j_aj_b(\xi_a-\xi_b)\Bigr) \notag \\
&+(\eta_1\bar\eta_2-1)\Bigl(\half(\xi_a^2-\xi_b^2)\partial^2_{ab} + j_b\xi_b\partial_a - j_a\xi_a\partial_b\Bigr)\biggr] P_\iota(\xi_1,\dots,\xi_N). \label{CP}
\end{align}
On a general level, we cannot make the result given by \Eqs{Ciota} and \eqref{CP} any more explicit, since we lack a general expression for the polynomial $P_\iota(\xi_1,\dots,\xi_N)$ in the $N$-valent case. Let us therefore specialize to consider the action of the operator $C^E_{12}$ on a three-valent intertwiner. The polynomial $P_\iota(\xi_1,\xi_2,\xi_3)$ representing the three-valent intertwiner can be read off from \Eq{xi3j} as
\be
P_\iota(\xi_1,\xi_2,\xi_3) = (\xi_1-\xi_2)^{j_1+j_2-j_3}(\xi_2-\xi_3)^{j_2+j_3-j_1}(\xi_3-\xi_1)^{j_3+j_1-j_2}.
\ee
Then, denoting $\xi_{ab} = \xi_a-\xi_b$ and $k_a = j_1+j_2+j_3-2j_a$, the derivatives involved in \Eq{CP} can be written as
\begin{align}
\partial_1P_\iota(\xi_1,\xi_2,\xi_3) &= \biggl(\frac{k_3}{\xi_{12}} - \frac{k_2}{\xi_{31}}\biggr)P_\iota(\xi_1,\xi_2,\xi_3), \label{d1P} \\
\partial_2P_\iota(\xi_1,\xi_2,\xi_3) &= \biggl(\frac{k_1}{\xi_{23}} - \frac{k_3}{\xi_{12}}\biggr)P_\iota(\xi_1,\xi_2,\xi_3),
\end{align}
and
\be\label{d12P}
\partial_{12}P_\iota(\xi_1,\xi_2,\xi_3) = \biggl(-\frac{k_3^2-k_3}{\xi_{12}^2} - \frac{k_1k_2}{\xi_{23}\xi_{31}} + \frac{k_2k_3}{\xi_{31}\xi_{12}} + \frac{k_3k_1}{\xi_{12}\xi_{23}} \biggr)P_\iota(\xi_1,\xi_2,\xi_3).
\ee
The form of these derivatives suggests to express the components of the intertwiner $C_{12}\iota$, given by \Eq{Ciota}, by factoring out the component of the original intertwiner, $\bigl(\begin{smallmatrix} j_1&j_2&j_3 \\ \xi_1&\xi_2&\xi_3 \end{smallmatrix}\bigr)$. We can additionally separate the factor $(1+\eta_1\bar\eta_2)^{2s}/(1+|\eta_1|^2)^s(1+|\eta_2|^2)^s = \braket{s\eta_2}{s\eta_1}$, which, up to a phase, is equal to $[\tfrac{1}{2}(1+\vec m_1\cdot\vec m_2)]^s$, where $\vec m_1$ and $\vec m_2$ are the unit vectors corresponding to the variables $\eta_1$ and $\eta_2$. After these factors have been pulled out, let us denote the remainder on the right-hand side of \Eq{Ciota} as $F(\xi_1,\xi_2,\xi_3,\eta_1,\bar\eta_2)$. In this way we have expressed the components of the new intertwiner in the form
\be\label{Ci3}
(C^E_{12}\iota){}\downup{\xi_1\xi_2\xi_3\eta_1}{\eta_2} = F(\xi_1,\xi_2,\xi_3,\eta_1,\bar\eta_2)\biggl(\frac{1+\vec m_1\cdot\vec m_2}{2}\biggr)^s\begin{pmatrix} j_1&j_2&j_3 \\ \xi_1 &\xi_2 &\xi_3 \end{pmatrix}.
\ee
The explicit expression of the function $F(\xi_1,\xi_2,\xi_3,\eta_1,\bar\eta_2)$ is not particularly enlightening, but in principle it could be determined by inserting \Eqs{d1P}--\eqref{d12P} into the right-hand side of \Eq{CP}.

In the limit of large spins, the expression \eqref{Ci3} has the following structure in terms of the unit vectors associated to the variables $\xi_a$ and $\eta_a$. As argued in section \ref{sec:IinCSbasis}, the coherent 3$j$-symbol in \Eq{Ci3} is exponentially suppressed unless the vectors $\vec n_a$ corresponding to the parameters $\xi_a$ satisfy the closure condition $j_1\vec n_1+j_2\vec n_2+j_3\vec n_3=0$. Similarly, if we also assume the spin $s$ belonging to the loop created by the Hamiltonian to be large, the factor $[\tfrac{1}{2}(1+\vec m_1\cdot\vec m_2)]^s$ forces the vectors $\vec m_1$ and $\vec m_2$ to be parallel to each other. Hence it follows that the magnitude of the intertwiner \eqref{Ci3} is peaked on configurations in which its five vectors satisfy the closure condition in the form $j_1\vec n_1+j_2\vec n_2+j_3\vec n_3+s\vec m_1-s\vec m_2=0$ (in general, each vector associated to an upper index of the intertwiner should be counted in the closure condition with a negative sign). 

If we assume that the vectors $\vec n_1$, $\vec n_2$ and $\vec n_3$ close, and that $\vec m_1=\vec m_2\equiv\vec m$, we can look at the function $F(\xi_1,\xi_2,\xi_3,\eta_1,\bar\eta_2)$ to determine the preferred orientation of the vector $\vec m$ relative to the vectors $\vec n_a$. Figuratively speaking, $F$ can be thought of as a sort of probability amplitude for the vector $\vec m$ to point in a given direction, for a fixed orientation of the vectors $\vec n_1$, $\vec n_2$ and $\vec n_3$.

As a first example, let us consider the case in which the spins $j_1$, $j_2$ and $j_3$ are all equal to a common value $j$. Choosing $\vec n_1=\tfrac{1}{2}(\sqrt 3,0,-1)$, $\vec n_2=\tfrac{1}{2}(-\sqrt 3,0,-1)$ and $\vec n_3=(0,0,1)$, we find that the absolute value of the function $F$ is given in terms of the polar angles of the vector $\vec m$ as
\be
|F(\theta,\phi)| = \sqrt{3}j^2s\Bigl(|\sin\theta\sin\phi| + {\cal O}(j^{-1})\Bigr).
\ee
We see that this function reaches its maximum value at the points $(\theta,\phi) = (\pi/2,\pi/2)$ and $(\theta,\phi) = (\pi/2,3\pi/2)$, corresponding to $\vec m = (0,1,0)$ and $\vec m = (0,-1,0)$. In other words, we have found that the preferred orientation of the vector $\vec m$ is orthogonal to the plane spanned by the vectors $\vec n_1$, $\vec n_2$ and $\vec n_3$. In the large-spin limit, this conclusion is independent of the values of the spins $j$ and $s$, since at leading order the magnitude of $F(\theta,\phi)$ depends on the spins only through an overall multiplicative factor.

\begin{figure}[tb]
	\centering
		\includegraphics[width=1\textwidth]{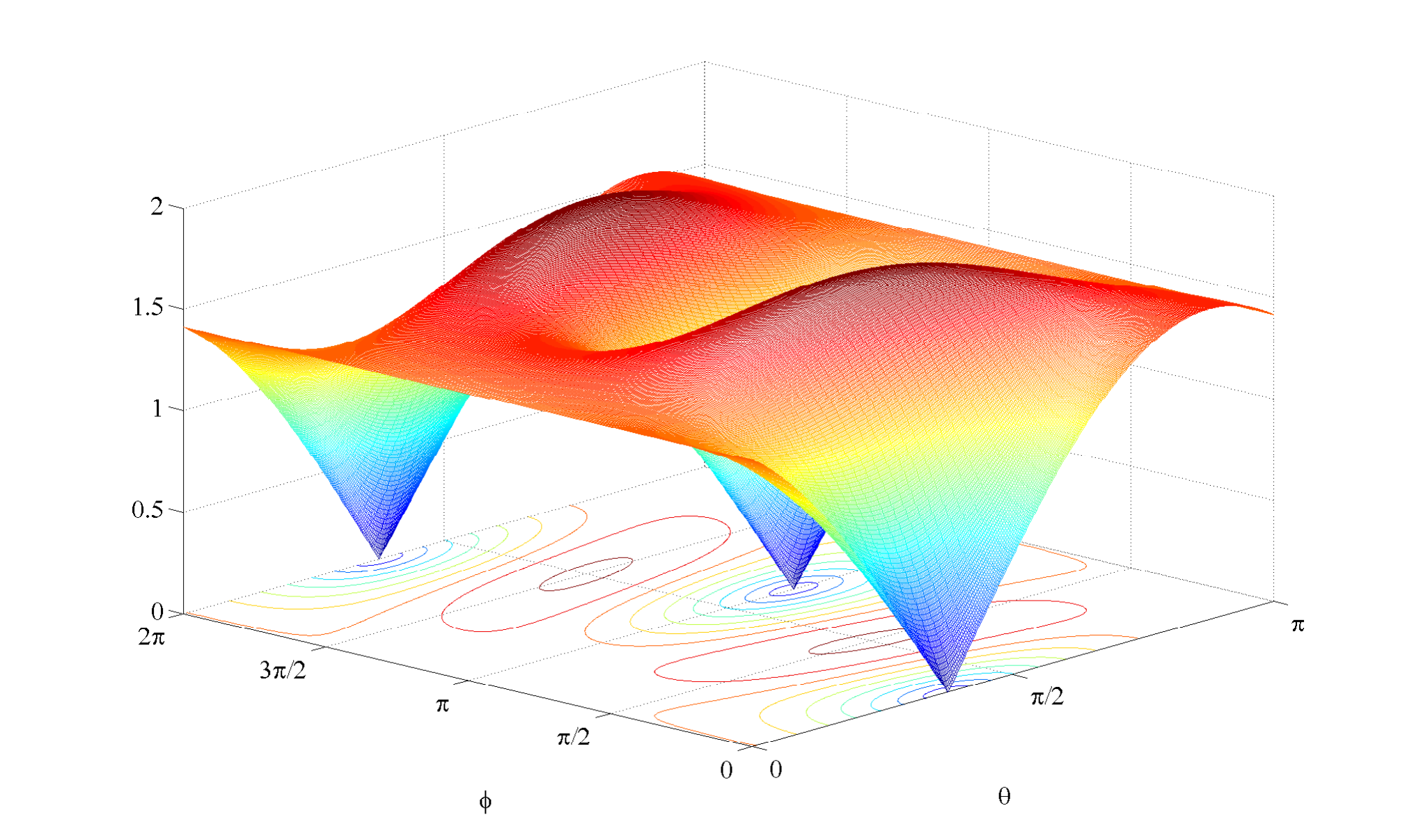}
		\caption{The function $|F(\theta,\phi)|/j^2s$ for the case where the spins of the three-valent intertwiner $\iota$ are given by $j_1=j_2=j$ and $j_3\simeq \sqrt 2j$. The maxima of $|F(\theta,\phi)|$ determine the preferred orientation of the vectors $\vec m_1 = \vec m_2 \equiv \vec m$ created by the action of the operator $C_{12}$ on the intertwiner $\iota$. The maxima are located at $(\theta,\phi) = (\pi/2,\pi/2)$ and $(\theta,\phi) = (\pi/2,3\pi/2)$, corresponding to the vector $\vec m$ being orthogonal to the plane spanned by the vectors $\vec n_1$, $\vec n_2$ and $\vec n_3$.}
	\label{Fplot}
\end{figure}

To give another example, we examine the case $j_1=j_2=j$ and $j_3\simeq \sqrt 2 j$. To satisfy the closure condition of the vectors $\vec n_a$, we now fix $\vec n_1=(0,0,1)$, $\vec n_2=(1,0,0)$ and $\vec n_3=\tfrac{1}{\sqrt 2}(-1,0,-1)$. The resulting expression for $|F(\theta,\phi)|$ is rather complicated and not very easy to study analytically, so we show in \Fig{Fplot} a numerical plot of $|F(\theta,\phi)|/j^2s$ (which is independent of $j^2$ and $s$ at leading order). Locating the maxima of $|F(\theta,\phi)|$ in the plot, we see that the preferred orientation of the vector $\vec m$ is again perpendicular to the plane of the vectors $\vec n_1$, $\vec n_2$ and $\vec n_3$.

When considering the integral
\be\label{CE12-int}
\int d\nu_s(\eta_1)\,d\nu_s(\eta_2)\,(C^E_{12}\iota){}\downup{\xi_1\xi_2\xi_3\eta_1}{\eta_2}\D{s}{\eta_1}{\eta_2}{h_{\alpha_{12}}},
\ee
which represents the complete result of acting with the operator \eqref{CE-18} on a three-valent intertwiner in the coherent representation, it is important to realize the following: While a distinguished orientation of the vector $\vec m$ has indeed emerged from the function $F$ in the above examples, the preference for the distinguished orientation is not very strong, in the sense that the peak of $F$ around the maximum is rather wide. Most importantly, the width of the peak is essentially independent of the spins involved in the calculation, and so the peak remains wide even in the limit of large spins -- in contrast to, say, the peak in the norm of the Livine--Speziale intertwiner, which becomes more and more narrow as the value of $j$ increases. Therefore, even if one of the integrals in \Eq{CE12-int} could be dealt with by a suitable saddle point approximation, based on the observation that the second factor on the right-hand side of \Eq{Ci3} is peaked on the parallel configuration of the vectors $\vec m_1$ and $\vec m_2$, it seems unlikely that the remaining integral could be approximated in a similar way.

It seems reasonable to hope that the situation would be better if one would use the coherent representation described in this chapter to study the action of operators on coherent intertwiners, rather than standard intertwiners which are merely expressed in the coherent representation. In this way one could potentially see some interesting and non-trivial interplay between the vectors labeling the coherent intertwiner, and the vectors associated to the complex variables $\xi$ of the coherent representation. Provided that an explicit and sufficiently manageable expression can be found for the coherent intertwiner in the form \eqref{iN_xi}, making a detailed investigation of coherent intertwiners in the coherent representation would be an interesting topic for future work.

In any case, the work presented in the article \cite{paper4} and reviewed in this chapter can be seen merely as a demonstration that there exists an alternative framework, based on projecting intertwiners onto the basis of angular momentum coherent states, which can be used for performing calculations with the Hamiltonian and other loop quantum gravity operators. More work must be done in order to understand the extent to which this formalism is a useful computational tool for loop quantum gravity, and whether it can be used to obtain a clear, geometrical understanding of the action of the Hamiltonian on spin network states, and consequently to perform a systematic analysis of the dynamics.

\begin{nopage}

$\phantom{x}$

\vspace{96pt}

\vspace{24pt}

\noindent\textbf{\huge Conclusions} 

\vspace{72pt}

\begin{quote}

Consider, then, the sum total of our accumulated knowledge as constituting an island, which I call the ''Island of Knowledge.'' (\ldots) A vast ocean surrounds the Island of Knowledge, the unexplored ocean of the unknown, hiding countless tantalizing mysteries. (\ldots) As the Island of Knowledge grows, so do the shores of our ignorance -- the boundary between the known and the unknown. Learning more about the world doesn't lead to a point closer to a final destination -- whose existence is nothing but a hopeful assumption anyways -- but to more questions and mysteries. The more we know, the more exposed we are to our ignorance, and the more we know to ask.

\begin{flushright}
-- Marcelo Gleiser
\end{flushright}
\end{quote}

\end{nopage}

\sis{Conclusions}

\subsection*{Summary of results} 

The central theme of this work has been the issue of finding a satisfactory formulation of the dynamics for canonical loop quantum gravity. This problem was considered particularly in the context of deparametrized models, in which a scalar field is used as a relational time variable for the dynamics of the quantized gravitational field. Accordingly, one of the main results presented was the introduction of a new Hamiltonian operator (more precisely, a class of Hamiltonian operators) for loop quantum gravity. These operators were constructed specifically for the purpose of serving as a physical Hamiltonian in the deparametrized framework, even though it can also be used as the Hamiltonian constraint operator in the fully constrained theory.

Our construction of the Hamiltonian enjoys two important advantages over proposals which have been available in the literature so far. Firstly, by regularizing the Euclidean part of the Hamiltonian in terms of the so-called special loops, we ensure that the adjoint of the Euclidean operator is a densely defined operator on the space of diffeomorphism invariant states, and hence the Hamiltonian can be consistently defined as a symmetric operator. While the question of the eventual self-adjointness of the Hamiltonian is still open, having a symmetric Hamiltonian is of course a necessary requirement for the operator to be a mathematically consistent candidate for the generator of unitary physical time evolution.

The other novel feature which characterizes our Hamiltonian is the use of the scalar curvature of the spatial surface in place of the traditional Lorentzian part of the Hamiltonian. This leads to a considerable simplification in the structure of the operator, and therefore offers a very substantial practical advantage. We recall that the Lorentzian part of Thiemann's Hamiltonian is composed of several factors of holonomies, volume operators and Euclidean operators. In contrast, the complexity of the curvature operator is not substantially higher than that of a single volume operator.

Having completed the definition of the Hamiltonian, we moved on to consider the problem of calculating with the operator. We began by showing how the matrix elements of the operators out of which the Hamiltonian is constructed can be computed in explicit form in the spin network basis, even if it is difficult to develop any intuitive grasp for the physical significance of the resulting expressions. The essential tool, which enables us to carry out the calculations with relatively little difficulty, is the powerful graphical formalism designed for computations involving $SU(2)$ recoupling theory.

As far as working with the Hamiltonian is concerned, an important element of our work is the development of elementary approximation methods, which can be used to study the time evolution of spin network states in deparametrized models even if an exact solution to the eigenvalue problem of the physical Hamiltonian is not available. In particular, due to the relatively simple form of the curvature operator, a viable strategy is to consider the more complicated Euclidean part of the Hamiltonian as a perturbation over the curvature part. Standard time-independent perturbation theory can then be used to approximate the spectrum of the complete Hamiltonian in terms of the eigenvalues and eigenstates of the curvature operator. (Even if the spectrum of the curvature operator cannot be accessed analytically, it is straightforward to compute the eigenvalues and eigenstates numerically within the intertwiner space of a given, not too complicated node.)

The perturbative treatment of the Hamiltonian seems to require that the Barbero--Immirzi parameter has a sufficiently large value, since it is only in this case that the Euclidean part of the Hamiltonian is multiplied with a small numerical parameter relative to the curvature part. It seems reasonable to expect that the perturbative approach could also be used, regardless of the value of $\beta$, in physical situations where the value of the spatial curvature is large in comparison to the value of the observable corresponding to the Euclidean operator. However, for now this remark is not a concrete proposition, but merely an interesting speculation, since we are not able to write down states which would be peaked on desired values of the curvature operator and the Euclidean operator. 

Even when perturbation theory is not applicable, there is still available the primitive approach of computing time evolution over a short time interval simply by expanding the quantity of interest in powers of the time variable and evaluating the coefficients of the expansion. However, this method can be used only for the non-rotational dust model, since the square root present in the physical Hamiltonian of the free scalar field model makes it impossible to compute the action of the Hamiltonian unless the spectrum of the operator under the square root is known.

The last one among the results presented in this thesis was the introduction of a new description of intertwiners in loop quantum gravity. The new representation is constructed by projecting intertwiners onto the basis of angular momentum coherent states, rather than the conventional basis of magnetic indices. This line of development leads naturally to a description in which intertwiners are represented as polynomials of certain complex variables (which originate from the parametrization of the unit vectors labeling the coherent states by means of the stereographic projection), and the usual operators of loop quantum gravity are formulated as differential operators acting on these complex variables.

The primary motivation behind this work is the possibility that the formalism could be used to describe the action of the Hamiltonian in a more transparent, geometric language, in terms of the unit vectors which parametrize the angular momentum coherent states, and which play the role of a sort of ''magnetic indices'' in the new representation. The example of the Euclidean operator acting on a three-valent intertwiner shows that there is indeed a preferred relation between the two vectors ''created'' by the action of the operator and the three vectors belonging to the original intertwiner. However, the matrix element is not very sharply peaked on the preferred configuration of the vectors, not even in the limit of large spins, so the result is not as useful as one could have hoped. The situation might be better if we considered the action of the Hamiltonian on coherent intertwiners, which themselves contain geometrical information encoded in spins and unit vectors. However, such a calculation would require a manageable expression for the coherent intertwiner as a polynomial, which we currently do not have.

In principle, our construction of the Hamiltonian completes the definition of a class of mathematically consistent models of loop quantum gravity, in which each fundamental element of the formalism is given in explicit form\footnote{Strictly speaking, one might object that the physical Hamiltonian of the free scalar field model involves a square root, and hence is not really explicitly defined unless the spectral decomposition of the operator under the square root is known. But in any case the claim is valid at least for the non-rotational dust model.}. While deparametrized models of loop quantum gravity have already been considered in the literature before, all previous treatments of the subject have been partially formal, with some essential ingredients of the theory having been left undefined. Most importantly, our work provides the first fully concrete proposal for a satisfactory physical Hamiltonian governing the dynamics in deparametrized loop quantum gravity.

To conclude, we have shown that deparametrization can be used to formulate the dynamics of canonical loop quantum gravity in a way which offers a number of clear advantages, and at least from a practical point of view is preferable over a fully constrained formulation. In a sense, this can be regarded as the central claim advocated in this thesis. The formidable practical difficulties involved in extracting solutions of the Hamiltonian constraint and uncovering their physical meaning, and understanding the structure of the physical Hilbert space and the scalar product thereon, are largely circumvented when a physical time variable is introduced and the constraint operator is traded for a physical Hamiltonian. The deparametrized models considered in this work are nevertheless constructed with sufficient mathematical rigor to ensure that they are free from any fundamental inconsistencies, and conform to the high mathematical standards traditionally expected in the context of canonical loop quantum gravity.

\subsection*{Some open issues}

While the deparametrized models whose definition is completed in this work by supplying an explicit proposal for the physical Hamiltonian are mathematically consistent and complete models of loop quantum gravity, virtually nothing is known yet about the physics contained in them. The results of the numerical computations presented in Chapter \ref{ch:approximation} do little to clarify the issue of the physical viability of these models, since the initial states used in the calculations do not correspond to any familiar physical situation, in which one would have a prior expectation of how the volume or the curvature is supposed to evolve.

Perhaps the main purpose of our numerical calculations is to serve as a sort of proof of concept, showing that it is possible to perform explicit calculations concerning the dynamics of deparametrized loop quantum gravity. Moreover, the results of the calculations display at least some decidedly interesting features: The degeneracies present in the spectra of volume and curvature are be preserved under time evolution, and the time-dependent expectation values of volume and curvature behave as even functions of time. These observations seem to point towards the existence of a certain symmetry shared by the Hamiltonian and the volume and curvature operators. The symmetry might well be related to some appropriate notion of time reversal in the theory. Identifying this symmetry and understanding its properties in detail would be a worthwhile challenge for future study, especially in order to see whether the symmetry could be used to simplify the analysis of the Hamiltonian in any way.

The most significant obstacle which prevents us from attempting physically interesting calculations within the models considered in this work seems to be the lack of states which would have a clear physical, semiclassical interpretation, and which would also be compatible with the structure of the model, particularly its dynamics. Thus, our inability to obtain any definite, physically significant results using the methods developed in this work should not be taken as evidence that the methods themselves are flawed; it should rather be attributed to our not having been able to identify any suitable states from which such results could be extracted. All coherent states available in loop quantum gravity so far are based on a fixed graph, and as we have argued earlier, such states cannot be good coherent states even kinematically, let alone at the dynamical level, for a model whose dynamics is defined by a graph-changing Hamiltonian. 

Before the physical content of our models can be tested through concrete calculations, it therefore seems that one more major challenge must still be overcome: To write down states describing a specific physical situation (for example, a homogeneous and isotropic semiclassical geometry), while also having a chance to behave nicely under the dynamics of the model, at least over a short interval of time. It is natural to expect that such states would have the form of an infinite superposition of graphs, in which the nodes of a given ''seed graph'' are decorated with special loops in all possible ways. The search for coherent states having an appropriate graph structure is still essentially fully open, though some preliminary steps have been taken recently in \cite{Mehdi}. If such states were available, my expectation is that the methods introduced in this work would be powerful enough to be able to produce meaningful physical conclusions when used to investigate the dynamics of these states.

That being said, I think there is a definite possibility that the computational technology currently available in loop quantum gravity may not be the optimal one for the problems to which it is being applied. Imagine an alternative history of physics, in which the formalism of quantum mechanics was discovered before the theory of differential equations had become very well developed. The Schr\"odinger or Heisenberg of this hypothetical universe might have tried to derive the energy levels of the harmonic oscillator by computing the matrix elements of the operator $H = \half(-d^2\!/dx^2 + x^2)$ with respect to some arbitrarily chosen basis of $L_2(\R)$ and attempting to diagonalize the resulting matrix. While it is not impossible that he could eventually have found the solution in this way, we know that it is certainly not the ideal way of attacking the problem. Sometimes I cannot help but think that this is what we are doing when we try to analyze a Hamiltonian in loop quantum gravity in terms of its matrix elements in the spin network basis. It is difficult to suggest how one should look for alternative methods of calculation, especially not having in mind any specific difficulty that the new technology is supposed to solve, but our formalism of intertwiners in the basis of angular momentum coherent states can be seen as the beginning of an attempt in this direction.

Our discussion here has been limited to a selection of unresolved issues directly related to the work presented in this thesis. Looking at loop quantum gravity as a whole, there are certainly other, even more substantial open questions, in particular the issue of the continuum limit, and the crucial challenge of deriving falsifiable physical predictions from the theory. It is clear that conclusive answers to these questions must be found before loop quantum gravity can be considered as a correct theory of physics, and not just a promising mathematical framework. The work at hand has little to say on these issues, but what we have accomplished is to have established a class of coherent, well-defined models of loop quantum gravity, within which answers to these questions (and others) can be sought.

\subsection*{Parting thoughts}

When compared to most other theories of physics, a distinctive feature of loop quantum gravity is the exceptionally high level of mathematical rigor which characterizes a large part of the research on the subject, and which has been present throughout the history of the theory, particularly in its canonical formulation. The mathematical groundwork on which the entire structure of loop quantum gravity is built is extremely solid, and this is undoubtedly a great strength of the theory.

On the other hand, it is clear that a certain price has been paid for this strength. Physical applications of canonical loop quantum gravity are a rarity even now, thirty years after the birth of the theory. In the development of the theory, the strict requirements of mathematical rigor have been taken as the foremost consideration so often that not much room has been left for exploring the theory by trying to use it to answer concrete questions about physics. However, thorough mathematical reasoning and heuristic physical exploration are two complementary lines of research, and it seems unlikely that either of them alone could lead to a complete theory of physics.

Pushing a tentative physical theory to make calculations about physics often plays an important role in the development of a consistent formalism, even if the calculations initially involve several speculative steps which can be justified only by heuristic physical arguments, and from the point of view of mathematical precision fall somewhere on the spectrum between ''dubious'' and ''plainly wrong''. (This is shown by the history of how virtually any now established theory of physics has grown from a disorganized collection of hypothetical ideas into a valuable constituent of our knowledge about the physical world.) A calculation which leads to a physically correct result clearly counts as evidence that one is on the right track. Even if the result is wrong, the calculation may still offer valuable insights on how the formalism can be improved. In the words of John Wheeler, ''We can afford many mistakes in the search. The main thing is to make them as fast as possible.''

I think that canonical loop quantum gravity is by now more than ready to be used to address concrete physical questions. In the recent years, the canonical theory has indeed seen the emergence of such physically motivated research programs as the model of quantum-reduced loop gravity \cite{qrlg1, qrlg2, qrlg2.5, qrlg3, qrlg4} and the derivation of effective Hamiltonians for cosmology \cite{effective1, effective2, effective3}. (For comparison, the covariant version of the theory has been used to study, for instance, the dynamics of cosmological spacetimes \cite{SFC1, SFC2, SFC3, SFC4, SFC5} and the hypothesized bounce of a collapsing star \cite{planck1, planck2, planck3, planck4, planck5}.) I have no doubt that such lines of research will eventually turn out to have a crucial importance in carrying the theory closer to its completion, by providing us with clues on how to establish the elements of the formalism which are still missing. I close these concluding thoughts by invoking the words of another giant among physicists to express my sentiment:
\begin{quote}
We have to take a viewpoint of how to deal with problems where no experiments are available. There are two choices. The first choice is that of mathematical rigor. (\ldots) However, one can do an enormous amount by various approximations which are non-rigorous and unproved mathematically, perhaps for the first few years. Historically, the rigorous analysis of whether what one says is true or not comes many years later after the discovery of what is true. And, the discovery of what is true is helped by experiments. The attempt at mathematical rigorous solutions without guiding experiments is exactly the reason the subject is difficult. 

The second choice of action is to ''play games'' by intuition. (\ldots) Make up your mind which way it is and calculate without rigor in an exploratory way. (\ldots) I think the best viewpoint is to pretend that there are experiments and calculate. In this field since we are not pushed by experiments we must be pulled by imagination.

The real challenge is not to find an elegant formalism, but to solve a series of problems whose results could be checked. This is a different point of view. Don't be so rigorous or you will not succeed.

\begin{flushright}
-- Richard P.\@\xspace Feynman
\end{flushright}
\end{quote}

\begin{nopage}

\sis{Appendix. $SU(2)$ recoupling theory and graphical techniques}

$\phantom{x}$

\vspace{96pt}

\noindent\textbf{\LARGE Appendix \\}

\vspace{24pt}

\noindent\textbf{\huge $SU(2)$ recoupling theory and \vspace{10pt} \\
graphical techniques} 

\vspace{72pt}

\begin{quote}

Ninety percent of most magic merely consists of knowing one extra fact.

\begin{flushright}
-- Terry Pratchett
\end{flushright}
\end{quote}

\end{nopage}

\appendix
\addtocounter{page}{1}

\section*{Introduction}

In this Appendix we give a concise but self-contained presentation of those aspects of $SU(2)$ representation and recoupling theory that form the indispensable basis for performing practical calculations in loop quantum gravity. The two central themes of our discussion are the theory of intertwiners, or invariant tensors of $SU(2)$, from which the spin network states of loop quantum gravity are constructed; and the graphical techniques for calculations in $SU(2)$ recoupling theory, which are presented in the last chapter of the Appendix, and which provide an invaluable computational tool in loop quantum gravity. Another objective of our presentation is to clearly set down our notation and conventions, which are probably not in complete agreement with any of the standard references on the subject.

\section{Representations of $SU(2)$}

The first chapter of the Appendix consists of a review of the necessary elements of the representation theory of $SU(2)$. While the material in this chapter is elementary and most of it can be found in any good textbook on quantum mechanics or Lie groups, the chapter nevertheless serves a number of purposes: to establish notation, to make our treatment fully self-contained, and to put forward the author's personal preference for the quantum-mechanical theory of angular momentum, as opposed to the formal machinery of mathematical group theory, as the natural language for discussing the theory of $SU(2)$ in the context of physics.

\subsection{Fundamental representation}

A general element of $SU(2)$, the group of unitary $2\times 2$ -matrices with determinant $+1$, has the form
\be\label{gAB}
g\updown{A}{B} = \begin{pmatrix} \alpha&\beta \\ -\bar\beta &\bar\alpha \end{pmatrix}\qquad \text{with} \qquad |\alpha|^2+|\beta|^2=1.
\ee
The fundamental representation of the group is realized by the action of the matrices $g\updown{A}{B}$ on two-component vectors of the form
\be
v^A = \begin{pmatrix} v^0 \\ v^1 \end{pmatrix}.
\ee
We denote the vector space consisting of such vectors as ${\cal H}_{1/2}$. The natural scalar product on ${\cal H}_{1/2}$, defined by
\be\label{Hprod}
\braket{u}{v} = \bar u^0 v^0 + \bar u^1 v^1,
\ee
is invariant under the action of $SU(2)$. The antisymmetric tensors
\be
\epsilon_{AB} = \begin{pmatrix} 0&1 \\ -1&0 \end{pmatrix}, \qquad \epsilon^{AB} = \begin{pmatrix} 0&1 \\ -1&0 \end{pmatrix}
\ee
are also $SU(2)$-invariant:
\be
\epsilon_{AB}g\updown{A}{C}g\updown{B}{D} = \epsilon_{CD}
\ee
and similarly for $\epsilon^{AB}$. By manipulating this relation, one can show that the matrix elements of the inverse matrix $g^{-1}$ are related to those of $g$ by
\be
(g^{-1})\updown{A}{B} = \epsilon^{AC}\epsilon_{BD}g\updown{D}{C}.
\ee
Since we define both $\epsilon_{AB}$ and $\epsilon^{AB}$ to have the same numerical value, their contraction gives
\be
\epsilon_{AB}\epsilon^{BC} = -\delta_A^C.
\ee
The epsilon tensor can be used for raising and lowering of $SU(2)$ indices. We adopt the convention
\be\label{raise-1/2}
v^A = \epsilon^{AB}v_B, \qquad v_A = v^B\epsilon_{BA}.
\ee
Using $\epsilon_{AB}$, one may also define the invariant antisymmetric product
\be\label{eprod}
(u|v) = \epsilon_{AB}u^Av^B = u^0v^1 - u^1v^0
\ee
between two vectors in ${\cal H}_{1/2}$.

By introducing the Pauli matrices
\be
\sigma_x = \begin{pmatrix} 0&1 \\ 1&0 \end{pmatrix}, \qquad \sigma_y = \begin{pmatrix} 0&-i \\ i&0 \end{pmatrix}, \qquad \sigma_z = \begin{pmatrix} 1&0 \\ 0&-1 \end{pmatrix},
\ee
a general $SU(2)$-element can be expressed in terms of an angle $\alpha$ and a unit vector $\vec n$ as
\be\label{g_n(a)}
g(\alpha,\vec n) = e^{-i\alpha\vec n\cdot\vec\sigma/2} = \cos\frac{\alpha}{2} - i\sin\frac{\alpha}{2}(\vec n\cdot\vec\sigma),
\ee
where the second equality follows from expanding the exponential and observing that $(\vec n\cdot\vec\sigma)^2 = 1$. The parametrization \eqref{g_n(a)} makes it particularly clear that elements of $SU(2)$ can be viewed as representing rotations in three-dimensional space. Let us give a geometrical argument to justify this interpretation. Consider the following sequence of infinitesimal rotations:
\begin{itemize}
\item Around the $x$-axis by an infinitesimal angle $\epsilon$;
\item Around the $y$-axis by another infinitesimal angle $\epsilon'$;
\item Around the $x$-axis by the angle $-\epsilon$;
\item Around the $y$-axis by the angle $-\epsilon'$.
\end{itemize}
By elementary geometry it is possible to convince oneself that, at lowest nontrivial order in the angles, the sequence is equivalent to a single rotation around the $z$-axis by the angle $-\epsilon\epsilon'$.

Suppose that rotations around the coordinate axes are generated by the operators $(J_x,J_y,J_z)$, so that the operator
\be
R_i(\alpha) = e^{-i\alpha J_i}
\ee
gives a rotation by an angle $\alpha$ around the $i$-axis. Then the rotations involved in the above sequence are given by
\begin{align}
R_x(\epsilon) &= 1 - i\epsilon J_x - \frac{\epsilon^2}{2}J_x^2 + {\cal O}(\epsilon^3), \\
R_y(\epsilon') &= 1 - i\epsilon' J_y - \frac{\epsilon'{}^2}{2}J_y^2 + {\cal O}(\epsilon'{}^3).
\end{align}
Now a direct calculation shows that the entire sequence is represented by the operator
\be
R_y(-\epsilon')R_x(-\epsilon)R_y(\epsilon')R_x(\epsilon) = 1 + \epsilon\epsilon'[J_x,J_y] + \dots
\ee
But this operator is supposed to be $R_z(-\epsilon\epsilon') = 1+i\epsilon\epsilon'J_z + \dots$, so we conclude that the generators must satisfy $[J_x,J_y] = iJ_z$. By considering cyclic permutations of the coordinate axes, we find the complete commutation relation
\be
[J_i,J_j] = i\epsilon\downup{ij}{k}J_k.
\ee
Hence we see that this commutation relation has a direct geometrical significance: it encodes the way in which successive rotations in three-dimensional space are combined.

The Pauli matrices satisfy the commutation relation
\be
[\sigma_i,\sigma_j] = 2i\epsilon\downup{ij}{k}\sigma_k,
\ee
which shows that they can be interpreted as generators of rotations by making the identification $J_i=\sigma_i/2$. Under this interpretation, $SU(2)$ elements of the form
\be
g_i(\alpha) = e^{-i\alpha\sigma_i/2}
\ee
describe rotations around the coordinate axes. The general element of \Eq{g_n(a)} corresponds to a rotation by the angle $\alpha$ around the direction given by the vector $\vec n$.

\subsection{The angular momentum operator}

In quantum mechanics, any (Hermitian) vector operator $\vec J$ whose components satisfy the commutation relation
\be
[J_i,J_j] = i\epsilon\downup{ij}{k}J_k
\ee
is called an angular momentum operator. All components of $\vec J$ commute with the squared angular momentum
\be
J^2 = J_x^2 + J_y^2 + J_z^2.
\ee
Therefore one can simultaneously diagonalize $J^2$ and one of the components, conventionally chosen as $J_z$. Let us write the eigenvalue equations as
\begin{align}
J^2\ket{\lambda,\mu} &= \lambda\ket{\lambda,\mu}, \label{J2ab} \\
J_z\ket{\lambda,\mu} &= \mu\ket{\lambda,\mu}. \label{Jzab}
\end{align}
To derive the solution to the eigenvalue problem, it is useful to define the raising and lowering operators
\be
J_\pm = J_x\pm iJ_y.
\ee
Their commutators with $J^2$ and $J_z$ are given by
\begin{align}
[J^2,J_\pm] = 0, \qquad [J_z,J_\pm] = \pm J_\pm.
\end{align}
These relations imply that the raising and lowering operators indeed raise and lower the eigenvalue of $J_z$ by one, while leaving the eigenvalue of $J^2$ unchanged: 
\begin{align}
J^2(J_\pm\ket{\lambda,\mu}) &= \lambda J_\pm\ket{\lambda,\mu}, \\
J_z(J_\pm\ket{\lambda,\mu}) &= (\mu\pm 1)J_\pm\ket{\lambda,\mu}.
\end{align}
In other words, the states $J_\pm\ket{\lambda,\mu}$ must be proportional to $\ket{\lambda,\mu\pm 1}$:
\be\label{Jpmab}
J_\pm\ket{\lambda,\mu} = A_\pm(\lambda,\mu)\ket{\lambda,\mu\pm 1}.
\ee
This shows that if any one of the eigenstates $\ket{\lambda,\mu}$ is known, then all the eigenstates for that value of $\lambda$ can be derived by repeatedly applying $J_+$ and $J_-$.

On the other hand, the inequality
\be
\bra{\lambda,\mu}J^2-J_z^2\ket{\lambda,\mu} = \bra{\lambda,\mu}J_x^2+J_y^2\ket{\lambda,\mu} \geq 0
\ee
implies that the eigenvalue $\mu$ is restricted by
\be
\mu^2\leq\lambda.
\ee
This condition can be satisfied only if there exists a maximal eigenvalue $\mu_{\rm max}$, such that the action of the raising operator on the state $\ket{\lambda,\mu_{\rm max}}$ does not give a new eigenstate, but instead
\be\label{bmax}
J_+\ket{\lambda,\mu_{\rm max}} = 0.
\ee
Similarly, there must exist a minimal eigenvalue $\mu_{\rm min}$, such that
\be\label{bmin}
J_-\ket{\lambda,\mu_{\rm min}} = 0.
\ee
Using the identities
\begin{align}
J_-J_+ &= J^2 - J_z^2 - J_z, \\
J_+J_- &= J^2 - J_z^2 + J_z
\end{align}
in \Eqs{bmax} and \eqref{bmin}, one finds the relation
\be
\lambda = \mu_{\rm max}(\mu_{\rm max}+1) = -\mu_{\rm min}(-\mu_{\rm min}+1),
\ee
showing that $\mu_{\rm max}$ and $\mu_{\rm min}$ are related to each other by
\be
\mu_{\rm min} = -\mu_{\rm max}.
\ee
Consider now acting repeatedly with $J_-$ on the state $\ket{\lambda,\mu_{\rm max}}$. After a certain number of actions, say $N$, one must arrive at the state $\ket{\lambda,\mu_{\rm min}}$. Since the eigenvalue $\mu$ is lowered by one at each step, the difference $\mu_{\rm max}-\mu_{\rm min}$ must be equal to $N$, giving
\be
\mu_{\rm max} = \frac{N}{2}.
\ee
The eigenvalue $\lambda$ then is
\be
\lambda = \mu_{\rm max}(\mu_{\rm max}+1) = \frac{N}{2}\biggl(\frac{N}{2}+1\biggr).
\ee
The possible eigenvalues of the operators $J^2$ and $J_z$ have therefore been determined. The eigenvalue equations \eqref{J2ab} and \eqref{Jzab} can be written as
\begin{align}
J^2\ket{jm} &= j(j+1)\ket{jm}, \label{J2jm} \\
J_z\ket{jm} &= m\ket{jm}, \label{Jzjm}
\end{align}
where $j$ may be any integer or half-integer, and $m$ ranges from $-j$ to $j$ in steps of $1$.

\newpage

To complete the solution of the eigenvalue problem, it remains to find the coefficients $A_\pm(\lambda,\mu) \equiv A_\pm(j,m)$ in \Eq{Jpmab}, since this will show how the angular momentum operator acts on the eigenstates $\ket{jm}$. By multiplying the equation with its adjoint, we find
\be
|A_\pm(j,m)|^2 = \bra{jm}J_\mp J_\pm\ket{jm} = \bra{jm}J^2-J_z^2\mp J_z\ket{jm} = j(j+1)-m(m\pm 1),
\ee
which determines $A_\pm(j,m)$ up to a phase. In this work we follow the nearly universally adopted Condon--Shortley phase convention, according to which $A_\pm(j,m)$ are taken to be real and positive. This leads to
\be\label{Jpm}
J_\pm\ket{jm} = \sqrt{j(j+1)-m(m\pm 1)}\ket{j,m\pm 1}.
\ee

\subsection{Spin-$j$ representation}\label{sec:spin-j}

The states $\ket{jm}$ with a fixed value of $j$ span the $(2j+1)$-dimensional vector space ${\cal H}_j$. A general element of ${\cal H}_j$ has the form
\be
\ket v = \sum_m v^m\ket{jm},
\ee
the index $m$ taking the values $-j$, $-j+1$, $\dots$, $j$. The natural definition
\be
\braket{u}{v} = \sum_m \bar u^m v^m
\ee
for a scalar product between elements of ${\cal H}_j$ promotes ${\cal H}_j$ into a Hilbert space. 

The relevance of the space ${\cal H}_j$ to $SU(2)$ representation theory is due to the fact that the matrices representing the operators $g = e^{-i\alpha\vec n\cdot\vec J}$ on ${\cal H}_j$ define an irreducible representation of $SU(2)$. These matrices, whose elements are given by
\be\label{WignerD}
\D{j}{m}{n}{g} = \bra{jm}e^{-i\alpha\vec n\cdot\vec J}\ket{jn},
\ee
are known as the Wigner matrices. The great orthogonality theorem of group theory implies that the Wigner matrices satisfy
\be\label{int DbarD}
\int dg\,\overline{\displaystyle \D{j}{m}{n}{g}} \D{j'}{m'}{n'}{g} = \frac{1}{d_j}\delta_{jj'}\delta_m^{m'}\delta^n_{n'},
\ee
where $dg$ is the normalized Haar measure\footnote{The Haar measure is determined uniquely by the conditions
\[
\int dg\,f(g_0g) = \int dg\,f(g), \qquad \int dg\,f(gg_0) = \int dg\,f(g), \qquad \int dg = 1,
\]
where $g_0$ is an arbitrary, fixed element of $SU(2)$. In the parametrization \eqref{gAB},
\[
dg = \frac{1}{\pi^2}d({\rm Re}\,\alpha)\,d({\rm Im}\,\alpha)\,d({\rm Re}\,\beta)\,d({\rm Im}\,\beta)\,\delta(1-|\alpha|^2-|\beta|^2),
\]
from which formulas for other parametrizations can be derived by making the appropriate change of variables and performing one of the integrals to remove the delta function. For more details, see \eg \cite{Tung}.
} of $SU(2)$, and
\be
d_j=2j+1
\ee
is a common shorthand for the dimension of ${\cal H}_j$.

\newpage

To derive the form of the invariant epsilon tensor in the spin-$j$ representation, we may consider the problem of constructing a state of zero total angular momentum on the tensor product space ${\cal H}_j\otimes{\cal H}_j$. Suppose that the state
\be
\ket{\Psi_0} = \sum_{mn} c_{mn}\ket{jm}\ket{jn}
\ee
satisfies
\begin{align}
\bigl(J^{(1)} + J^{(2)}\bigr)^2\ket{\Psi_0} &= 0, \\
\bigl(J^{(1)}_z + J^{(2)}_z\bigr)\ket{\Psi_0} &= 0,
\end{align}
where each angular momentum operator acts on the corresponding factor of the tensor product; for example, $J^{(1)}$ stands for $J^{(1)}\otimes\Id^{(2)}$. This is equivalent to the state being invariant under rotations generated by the total angular momentum $J = J^{(1)} + J^{(2)}$, which in turn implies that the tensor defined by the coefficients $c_{mn}$ is invariant under the action of $SU(2)$ by the matrices $\D{j}{m}{n}{g}$.

The eigenvalue equation for the $z$-component immediately shows that the coefficient $c_{mn} = 0$ unless $n = -m$. Then, requiring that the state
\be
\ket{\Psi_0} = \sum_m c_m\ket{jm}\ket{j,-m}
\ee
is annihilated by either one of $J_+$ or $J_-$ leads to the condition $c_{m+1} = -c_m$. Taken together, the two conditions imply that $c_{mn}$ is proportional to $(-1)^m\delta_{m,-n}$. Defining the tensor $\epsilon^{(j)}_{mn}$ so that $\epsilon^{(j)}_{j,-j} = +1$, we therefore have
\be
\epsilon^{(j)}_{mn} = (-1)^{j-m}\delta_{m,-n}, \qquad \epsilon_{\phantom{m}}^{(j)mn} = (-1)^{j-m}\delta_{m,-n}.
\ee
We see that $\epsilon^{(j)}_{mn}$ satisfies
\be\label{eps_nm}
\epsilon^{(j)}_{nm} = (-1)^{2j}\epsilon^{(j)}_{mn}
\ee
and
\be\label{epseps=delta}
\epsilon^{(j)}_{mm'}\epsilon_{\phantom{m'}}^{(j)m'n} = (-1)^{2j}\delta_m^n.
\ee
The remaining properties of the epsilon tensor are analogous to those in the fundamental representation. For the matrix elements of the inverse matrix, there holds the relation
\be\label{Dj-inv}
\D{j}{m}{n}{g^{-1}} = \epsilon_{\phantom{m'}}^{(j)mm'}\epsilon^{(j)}_{nn'}\D{j}{n'}{m'}{g},
\ee
and indices are raised and lowered according to the convention
\be
v^m = \epsilon_{\phantom{m}}^{(j)mn}v_n, \qquad v_m = v^n\epsilon^{(j)}_{nm}.
\ee
An antisymmetric, $SU(2)$-invariant product on ${\cal H}_j$ can be defined as
\be
(u|v) = \epsilon^{(j)}_{mn}u^mv^n = \sum_m (-1)^{j-m} u^m v^{-m}.
\ee

The way in which we introduced the space ${\cal H}_j$ does not make it particularly clear how the spin-$j$ representation of $SU(2)$ is related to the fundamental representation. We will now clarify this relation by showing how the states $\ket{jm}$ can be constructed from the states $\ket +$ and $\ket -$, which span the space ${\cal H}_{1/2}$, and are eigenstates of the angular momentum operator on ${\cal H}_{1/2}$ with eigenvalues $j=\half$ and $m=\pm\half$. To this end, let us consider the state
\be
\ket{\Psi_j} = \ket +\otimes\ket + \otimes \cdots \otimes\ket +,
\ee
which is an element of the $2j$-fold tensor product space ${\cal H}_{1/2}\otimes\cdots\otimes{\cal H}_{1/2}$. We would like to show that the state $\ket{\Psi_j}$ is an eigenstate of the total angular momentum operator
\be
J^{({\rm tot})} = J^{(1)} + J^{(2)} + \dots + J^{(2j)},
\ee
where each operator $J^{(i)}$ acts on the $i$-th factor of the tensor product $\otimes_{i=1}^{2j} {\cal H}_{1/2}^{(i)}$, \ie
\be
J^{(i)} \equiv \Id^{(1)} \otimes \cdots  \otimes \Id^{(i-1)} \otimes J^{(i)} \otimes \Id^{(i+1)} \otimes \cdots \otimes \Id^{(2j)}.
\ee
For the $z$-component of $J^{({\rm tot})}$, we immediately find
\be\label{JzPsij}
J^{({\rm tot})}_z\ket{\Psi_j} = j\ket{\Psi_j},
\ee
since each state $\ket +$ is an eigenstate of the corresponding operator $J_z$ with the eigenvalue $+\half$. For the square of the total angular momentum, we have
\be\label{Psij-calc}
\bigl(J^{({\rm tot})}\bigr)^2\ket{\Psi_j} = \biggl(\sum_i \bigl(J^{(i)}\bigr)^2 + \sum_{i\neq k} \vec J^{\,(i)}\cdot\vec J^{\,(k)}\biggr)\ket{\Psi_j}.
\ee
The first sum contains $2j$ terms, and in each of them the $J^2$ acts on the corresponding state $\ket +$, producing the eigenvalue $3/4$. To evaluate the cross terms, we note that each of the $2j(2j-1)$ terms in the sum cam be written as
\be
\vec J^{\,(i)}\cdot\vec J^{\,(k)} = J^{(i)}_z J^{(k)}_z + \frac{1}{2}\bigl(J^{(i)}_+ J^{(k)}_- + J^{(i)}_- J^{(k)}_+\bigr).
\ee
Here the action of the first term on the state $\ket{+}_{(i)}\ket{+}_{(k)}$ gives the eigenvalue $m_im_k = 1/4$. The action of the other two terms gives zero, because the state $\ket +$ is annihilated by the raising operator $J_+$. Hence, going back to \Eq{Psij-calc}, we conclude that
\be\label{J2Psij}
\bigl(J^{({\rm tot})}\bigr)^2\ket{\Psi_j} = j(j+1)\ket{\Psi_j}.
\ee
Together, \Eqs{JzPsij} and \eqref{J2Psij} show that $\ket{\Psi_j}$ can be identified with the state $\ket{jj}$. That is,
\be\label{jj}
\ket{jj} = \underbrace{\ket +\otimes\ket + \otimes \cdots \otimes\ket +}_{\text{$2j$ times}}.
\ee
The remaining states $\ket{jm}$ can now be constructed by repeatedly acting on \Eq{jj} with the lowering operator
\be
J_-^{({\rm tot})} = J_-^{(1)} + J_-^{(2)} + \dots + J_-^{(2j)}.
\ee
Recalling \Eq{Jpm}, we obtain
\be\label{jm}
\ket{jm} = \sqrt{\frac{(j+m)!(j-m)!}{(2j)!}}\Bigl(\underbrace{\ket +\otimes\cdots\otimes\ket +}_{\text{$j+m$ times}} \otimes \underbrace{\ket -\otimes \cdots \otimes\ket -}_{\text{$j-m$ times}}\;+\;\text{all permutations}\Bigr).
\ee
In this way we have established a direct relation between the spaces ${\cal H}_j$ and ${\cal H}_{1/2}$.

\newpage

In particular, an explicit expression for the matrix elements $\D{j}{m}{n}{g}$ can be derived from \Eq{jm} by using the known action of the matrix $g$ on the states $\ket +$ and $\ket -$ in ${\cal H}_{1/2}$. The derivation is valid not only for elements of $SU(2)$, but extends to elements of the group $SL(2,\C)$, which have the form
\be
h = \begin{pmatrix} a&b \\ c&d \end{pmatrix},
\ee
where $\det h = ad-bc=1$, but the matrix elements are otherwise unrestricted. By replacing the states $\ket +$ and $\ket -$ on the right-hand side of \Eq{jm} with
\begin{align}
h\ket + &= a\ket + + c\ket -, \\
h\ket - &= b\ket + + d\ket -,
\end{align}
one finds, after a somewhat tedious but in principle straightforward calculation,
\be\label{WignerD-expl}
\D{j}{m}{n}{h} = \sum_k \frac{\sqrt{(j+m)!(j-m)!(j+n)!(j-n)!}}{k!(j-m-k)!(j+n-k)!(m-n+k)!}a^{j+n-k}b^{m-n+k}c^kd^{j-m-k},
\ee
where the sum runs over all the values of $k$ for which the argument of every factorial is non-negative.

Note that the right-hand side of \Eq{jm} is a completely symmetric combination of the states $\ket +$ and $\ket -$. In other words, the states $\ket{jm}$ given by \Eq{jm} belong to the completely symmetric subspace of the tensor product space ${\cal H}_{1/2}\otimes\cdots\otimes{\cal H}_{1/2}$. Indeed, the spin-$j$ representation of $SU(2)$ is often introduced in the literature by defining the space ${\cal H}_j$ as the completely symmetrized part of ${\cal H}_{1/2}\otimes\cdots\otimes{\cal H}_{1/2}$, or equivalently as the space spanned by objects of the form
\be\label{v^A...A}
v^{(A_1\cdots A_{2j})},
\ee
where the ''index'' $(A_1\cdots A_{2j})$ is a completely symmetric combination of the spin-1/2 indices $A_i$. In other words, only the total number of $+$'s and $-$'s among the indices $A_1,\dots,A_{2j}$ is relevant to the value of $(A_1\cdots A_{2j})$. The relation between the symmetrized index $(A_1\cdots A_{2j})$ and the magnetic index $m$ used so far in this chapter is given by $m = \half\sum_i A_i$, where each $A_i$ takes the value $+$ or $-$. The representation matrix of an $SU(2)$ element $g$ in the space of the vectors \eqref{v^A...A} is defined by
\be\label{g...g}
\D{j}{(A_1\cdots A_{2j})}{(B_1\cdots B_{2j})}{g} = g\updown{A_1}{(B_1} \cdots g\updown{A_{2j}}{B_{2j})},
\ee
with $g\updown{A}{B}$ the matrix in the fundamental representation.

In practical $SU(2)$ calculations it is almost always more convenient to use the realization of the space ${\cal H}_j$ in terms of the magnetic indices $m$, rather than the symmetric tensor product indices $(A_1\cdots A_{2j})$. This is especially true when it comes to the graphical techniques of Chapter \ref{ch:graphical}, which are adapted to the magnetic index representation, the corresponding graphical calculus of the symmetric tensor product representation being much more primitive and cumbersome in comparison. Nevertheless, Chapter \ref{ch:coherent3j} contains one example of a calculation which is the most easily performed in the symmetric tensor product representation.

\section{Theory of intertwiners}

Intertwiners, or invariant tensors of $SU(2)$, lie at the core of the spin network states of loop quantum gravity. In this chapter we introduce the Clebsch--Gordan coefficient and the closely related Wigner 3$j$-symbol, show how the latter serves as the elementary building block from which intertwiners can be constructed, and derive the basic properties of the intertwiners obtained in this way. The material presented in this chapter provides the essential foundations underlying the powerful graphical formalism for calculations in $SU(2)$ recoupling theory, which will be introduced in the last chapter of the Appendix.

\subsection{Clebsch--Gordan coefficients}

Consider the tensor product space ${\cal H}_{j_1}\otimes{\cal H}_{j_2}$. An obvious basis on this space is provided by the tensor product states $\ket{j_1m_1}\ket{j_2m_2}$, which are eigenstates of the mutually commuting operators
\be
\bigl(J^{(1)}\bigr)^2, \qquad \bigl(J^{(2)}\bigr)^2, \qquad J^{(1)}_z, \qquad J^{(2)}_z.
\ee
Another complete set of commuting operators on ${\cal H}_{j_1}\otimes{\cal H}_{j_2}$ is formed by the operators
\be
\bigl(J^{(1)}\bigr)^2, \qquad \bigl(J^{(2)}\bigr)^2, \qquad \bigl(J^{(1)}+J^{(2)}\bigr)^2, \qquad J^{(1)}_z+J^{(2)}_z.
\ee
Let us denote their eigenstates by $\ket{j_1j_2;jm}$. Since both sets of states span the space ${\cal H}_{j_1}\otimes{\cal H}_{j_2}$, they must be related to each other by a unitary transformation of the form
\be\label{uncoupled}
\ket{j_1m_1}\ket{j_2m_2} = \sum_{jm} \CG{j_1j_2}{j}{m_1m_2}{m}\ket{j_1j_2;jm}
\ee
and
\be\label{coupled}
\ket{j_1j_2;jm} = \sum_{m_1m_2} \CGi{j_1j_2}{j}{m_1m_2}{m}\ket{j_1m_1}\ket{j_2m_2}.
\ee
The coefficients in these expansions are known as the Clebsch--Gordan coefficients. In the physics literature, a notation such as $\braket{j_1m_1,j_2m_2}{jm}$ is normally used for them. The notation adopted in this work is designed to display the tensorial structure of the Clebsch--Gordan coefficient when interpreted as an $SU(2)$ tensor; see \Eq{DDC=CD} below.

A number of properties satisfied by the Clebsch--Gordan coefficients follow immediately from their definition:
\begin{itemize}
\item The coefficient $\CG{j_1j_2}{j}{m_1m_2}{m}$ is trivially zero unless the conditions
\be\label{CGtriangle}
|j_1-j_2| \leq j \leq j_1+j_2 \qquad \text{and} \qquad j_1+j_2+j = {\rm integer}
\ee
are met. These conditions are referred to as the Clebsch--Gordan conditions, or the triangular conditions.
\item Moreover, $\CG{j_1j_2}{j}{m_1m_2}{m} = 0$ whenever $m\neq m_1+m_2$.
\item The orthogonality relations of the Clebsch--Gordan coefficients read
\be\label{CG-orth-jm}
\sum_{jm} \CG{j_1j_2}{j}{m_1m_2}{m}\CGi{j_1j_2}{j}{m_1'm_2'}{m} = \delta_{m_1}^{m_1'}\delta_{m_2}^{m_2'}
\ee
\newpage
\noindent and
\be\label{CG-orth-mm}
\sum_{m_1m_2} \CG{j_1j_2}{j}{m_1m_2}{m}\CGi{j_1j_2}{j'}{m_1m_2}{m'} = \delta_{jj'}\delta^m_{m'}.
\ee
\end{itemize}
The Condon--Shortley phase convention fixes the phases of the Clebsch--Gordan coefficients by requiring that all $\CG{j_1j_2}{j}{m_1m_2}{m}$ are real, and $\CG{j_1j_2}{j}{j_1,j-j_1}{j} > 0$; the relative phases between the coefficients for a fixed value of $j$ are determined by the choice already made in \Eq{Jpm}. Under this convention, the numerical value of the inverse coefficient $\CGi{j_1j_2}{j}{m_1m_2}{m}$ is equal to that of $\CG{j_1j_2}{j}{m_1m_2}{m}$, and for this reason, $\CG{j_1j_2}{j}{m_1m_2}{m}$ and $\CGi{j_1j_2}{j}{m_1m_2}{m}$ are usually not distinguished from each other in the physics literature.

Numerical values of the Clebsch--Gordan coefficients can be derived algebraically, using the properties of the raising and lowering operators $J_\pm$ (though in practice one would of course look up the coefficients using a tool such as Mathematica). For a given value of the total angular momentum $j$, one starts with the state of ''highest weight'' $\ket{j_1j_2;jj}$, in which the magnetic number is equal to its highest possible value. On grounds of the condition $m_1+m_2=m$ for the magnetic numbers in $\CG{j_1j_2}{j}{m_1m_2}{m}$, this state must have the form
\be\label{j1j2jj}
\ket{j_1j_2;jj} = \sum_m c_m \ket{j_1m}\ket{j_2,j-m}.
\ee
Applying the raising operator $J_+ = J^{(1)}_+ + J^{(2)}_+$ now gives
\be\label{sum c_m}
\sum_m c_m\Bigl(A_+(j_1,m)\ket{j_1,m+1}\ket{j_2,j-m} + A_+(j_2,j-m)\ket{j_1m}\ket{j_2,j-m+1}\Bigr) = 0,
\ee
where $A_+(j,m)$ is defined by \Eq{Jpm}. The information contained in \Eq{sum c_m} is sufficient to fully determine the state $\ket{j_1j_2;jj}$, since \Eq{sum c_m} gives $N-1$ conditions for the $N$ coefficients $c_m$, and one more condition is obtained by requiring that the state \eqref{j1j2jj} is normalized. The state $\ket{j_1j_2;jj}$ having been found, the remaining states $\ket{j_1j_2;jm}$ can then be derived by repeatedly applying the lowering operator.

Let us consider the effect of an $SU(2)$ rotation on the equation \eqref{uncoupled}. On the left-hand side, the rotation acts as $D^{(j_1)}(g)\otimes D^{(j_2)}(g)$. On the right-hand side, the terms having a given value of $j$ transform among themselves according to the matrix $D^{(j)}(g)$. Thus,
\be
D^{(j_1)}(g)\ket{j_1m_1}D^{(j_2)}(g)\ket{j_2m_2} = \sum_{jm} \CG{j_1j_2}{j}{m_1m_2}{m}D^{(j)}(g)\ket{j_1j_2;jm}.
\ee
Taking the product of this equation with the state $\bra{j_1n_1}\bra{j_2n_2}$ and using \Eq{uncoupled} on the right-hand side, we obtain the so-called Clebsch--Gordan series
\be\label{CG-ser}
\D{j_1}{m_1}{n_1}{g}\D{j_2}{m_2}{n_2}{g} = \sum_{jmn} \CGi{j_1j_2}{j}{m_1m_2}{m}\CG{j_1j_2}{j}{n_1n_2}{n}\D{j}{m}{n}{g},
\ee
which is, among other things, the basic rule for computing products of holonomies in loop quantum gravity. Contracting \Eq{CG-ser} with a Clebsch--Gordan coefficient and using the orthogonality relation \eqref{CG-orth-mm}, we further find
\be\label{DDC=CD}
\D{j_1}{n_1}{m_1}{g}\D{j_2}{n_2}{m_2}{g}\CG{j_1j_2}{j}{n_1n_2}{m} = \CG{j_1j_2}{j}{m_1m_2}{n}\D{j}{m}{n}{g},
\ee
showing how the Clebsch--Gordan coefficient itself behaves under $SU(2)$ transformations, and justifying the index structure used in the notation $\CG{j_1j_2}{j}{m_1m_2}{m}$.

\newpage

\subsection{The 3$j$-symbol}

The Wigner 3$j$-symbol is defined by lowering the upper index of the Clebsch--Gordan coefficient using the epsilon tensor, and multiplying with a numerical factor (which is inserted in order to optimize the symmetry properties of the resulting object):
\begin{align}
\threej{j_1}{j_2}{j_3}{m_1}{m_2}{m_3} &= \frac{1}{\sqrt{d_{j_3}}}(-1)^{j_1-j_2+j_3}\CG{j_1j_2}{j_3}{m_1m_2}{n}\epsilon^{(j_3)}_{nm_3} \notag \\
&= \frac{1}{\sqrt{d_{j_3}}}(-1)^{j_1-j_2-m_3}\CG{j_1j_2}{j_3}{m_1m_2}{-m_3}. \label{3j}
\end{align}
While the 3$j$-symbol is often introduced as a more symmetric version of the Clebsch--Gordan coefficient, its relevance to loop quantum gravity follows from its behaviour under $SU(2)$ transformations. Starting from \Eq{DDC=CD} and recalling \Eq{Dj-inv} for the elements of an inverse Wigner matrix, one can show that the 3$j$-symbol is invariant under the action of $SU(2)$:
\be\label{DDD3j}
\D{j_1}{m_1}{n_1}{g}\D{j_2}{m_2}{n_2}{g}\D{j_3}{m_3}{n_3}{g}\threej{j_1}{j_2}{j_3}{m_1}{m_2}{m_3} = \threej{j_1}{j_2}{j_3}{n_1}{n_2}{n_3}.
\ee
In the language of angular momentum, the Clebsch--Gordan coefficient couples two angular momenta $j_1$ and $j_2$ to a total angular momentum $j$, whereas the 3$j$-symbol couples the three angular momenta $j_1$, $j_2$ and $j_3$ to total angular momentum zero. The $SU(2)$ invariance of the 3$j$-symbol implies that the state
\be
\ket{\Psi_0} = \sum_{m_1m_2m_3} \threej{j_1}{j_2}{j_3}{m_1}{m_2}{m_3} \ket{j_1m_1}\ket{j_2m_2}\ket{j_3m_3}
\ee
is rotationally invariant, and hence is an eigenstate of the total angular momentum operator $J^{(1)} + J^{(2)} + J^{(3)}$ with eigenvalue zero.

The basic properties of the 3$j$-symbol follow from the corresponding properties of the Clebsch--Gordan coefficients:
\begin{itemize}
\item The value of the 3$j$-symbol can be non-zero only if the triangular conditions
\be
|j_1-j_2| \leq j \leq j_1+j_2 \qquad \text{and} \qquad j_1+j_2+j = {\rm integer}
\ee
as well as the condition
\be
m_1+m_2+m_3=0
\ee
are satisfied.
\item The 3$j$-symbol satisfies the orthogonality relations
\be\label{3j-orth-mm}
\sum_{m_1m_2} \threej{j_1}{j_2}{j}{m_1}{m_2}{m}\threej{j_1}{j_2}{j'}{m_1}{m_2}{m'} = \frac{1}{d_j}\delta_{jj'}\delta_m^{m'}
\ee
and
\be\label{3j-orth-jm}
\sum_{jm} d_j\threej{j_1}{j_2}{j}{m_1}{m_2}{m}\threej{j_1}{j_2}{j}{m_1'}{m_2'}{m} = \delta_{m_1}^{m_1'}\delta_{m_2}^{m_2'}.
\ee
\end{itemize}
When the Condon--Shortley phase convention is followed, the 3$j$-symbol is real-valued, and possesses several convenient symmetry properties. Interchanging any two columns in the symbol produces the factor $(-1)^{j_1+j_2+j_3}$; for example,
\be\label{3jperm}
\threej{j_2}{j_1}{j_3}{m_2}{m_1}{m_3} = (-1)^{j_1+j_2+j_3}\threej{j_1}{j_2}{j_3}{m_1}{m_2}{m_3}.
\ee
This implies in particular that the symbol is invariant under cyclic permutations of its columns. The same phase factor results from reversing the sign of all the magnetic numbers:
\be\label{3j-m}
\threej{j_1}{j_2}{j_3}{-m_1}{-m_2}{-m_3} = (-1)^{j_1+j_2+j_3}\threej{j_1}{j_2}{j_3}{m_1}{m_2}{m_3}.
\ee
The definition of the Clebsch--Gordan coefficient immediately implies that the coefficient $\CG{j0}{j}{m0}{n}$ is equal to $\delta_m^n$. Using this in \Eq{3j}, we find that when one of the angular momenta in the 3$j$-symbol is zero, the 3$j$-symbol reduces to the epsilon tensor:
\be\label{3jzero}
\threej{j}{j'}{0}{m}{n}{0} = \delta_{jj'}\frac{1}{\sqrt{d_j}}\epsilon^{(j)}_{mn}.
\ee
Further properties of the 3$j$-symbol, relations satisfied by it, and explicit expressions for its values in particular cases can be found in any of the standard references on angular momentum theory. The most comprehensive source of such information is the encyclopedic collection of formulas by Varshalovich, Moskalev and Khersonskii \cite{Varshalovich}.

\subsection{Three-valent intertwiners}\label{sec:intertwiners-3}

Intertwiners, or invariant tensors of $SU(2)$, play a crucial role in loop quantum gravity, entering the construction of the spin network states as solutions of the Gauss constraint. \Eq{DDD3j} shows that the 3$j$-symbol can be interpreted as such an invariant tensor. To emphasize the tensorial character of the 3$j$-symbol, we introduce the notation
\be\label{iota-3}
\iota_{m_1m_2m_3} = \threej{j_1}{j_2}{j_3}{m_1}{m_2}{m_3}.
\ee
Whenever we want to indicate explicitly the spins entering the 3$j$-symbol, the notation $\iota^{(j_1j_2j_3)}_{m_1m_2m_3}$ will be used for the tensor \eqref{iota-3}. The 3$j$-symbol is in fact (up to normalization) the only three-valent invariant tensor with indices in three given representations $j_1$, $j_2$ and $j_3$. Therefore the 3$j$-symbol alone spans the one-dimensional space of three-valent intertwiners, denoted by ${\rm Inv}\,\bigl({\cal H}_{j_1}\otimes{\cal H}_{j_2}\otimes{\cal H}_{j_3}\bigr)$.

By using epsilon to raise an index of the tensor \eqref{iota-3}, we obtain the tensor
\be
\iota\updown{m_1}{m_2m_3} = \epsilon_{\phantom{m}}^{(j_1)m_1m}\iota_{mm_2m_3},
\ee
which spans the intertwiner space ${\rm Inv}\,\bigl({\cal H}^*_{j_1}\otimes{\cal H}_{j_2}\otimes{\cal H}_{j_3}\bigr)$. Up to a numerical factor, the tensor $\iota\updown{m_1}{m_2m_3}$ is equal to the Clebcsh--Gordan coefficient $\CG{j_2j_3}{j_1}{m_2m_3}{m_1}$. Elements of a space such as ${\rm Inv}\,\bigl({\cal H}^*_{j_1}\otimes{\cal H}_{j_2}\otimes{\cal H}_{j_3}\bigr)$ are invariant under the action of $SU(2)$ in the sense of \Eq{DDC=CD}, \ie when a matrix $D^{(j)}(g)$ acts on each lower index of the tensor, while an inverse matrix $D^{(j)}(g^{-1})$ acts on each upper index.

The symmetry relation \Eq{3j-m} and the condition $m_1+m_2+m_3=0$ imply that the tensor
\be\label{iota3-upper}
\iota^{m_1m_2m_3} = \epsilon_{\phantom{m}}^{(j_1)m_1n_1}\epsilon_{\phantom{m}}^{(j_2)m_2n_2}\epsilon_{\phantom{m}}^{(j_3)m_3n_3}\iota_{n_1n_2n_3},
\ee
obtained by raising all the indices of $\iota_{m_1m_2m_3}$, is numerically equal to $\iota_{m_1m_2m_3}$. The orthogonality relation \eqref{3j-orth-mm} then shows that the three-valent intertwiner defined by \Eq{iota-3} is normalized:
\be
\iota^{m_1m_2m_3}\iota_{m_1m_2m_3} = 1.
\ee

The $SU(2)$ generator $\tau^{(j)}_i$, defined in the spin-$j$ representation as
\be\label{tau-j}
\Tau{j}{i}{m}{n} = -i\bra{jm}J_i\ket{jn},
\ee
can be interpreted as a three-valent intertwiner between the representations $j$, $1$ and $j$. This follows from the Wigner--Eckart theorem, which states that the matrix elements of a spherical tensor operator\footnote{A spherical tensor operator of rank $j$ is an operator whose $2j+1$ components $T^{(j)}_m$ ($m=-j,\dots,j$) transform under $SU(2)$ rotations in the same way as the states $\ket{jm}$. That is,
\[
U(g)T^{(j)}_mU^\dagger(g) = \sum_n \D{j}{n}{m}{g}T^{(j)}_n,
\]
where $U(g)$ is the unitary operator representing the rotation on the Hilbert space in which $T^{(j)}$ acts.} $T^{(j)}_m$ satisfy
\be
\bra{j_1m_1}T^{(j)}_m\ket{j_2m_2} = (j_1||T^{(j)}||j_2)\CG{j_2j}{j_1}{m_2m}{m_1},
\ee
where the so-called reduced matrix element $(j_1||T^{(j)}||j_2)$ is independent of the magnetic numbers; the dependence on $m_1$, $m_2$ and $m$ is given by the Clebsch--Gordan coefficient independently of the operator $T^{(j)}_m$. In \Eq{tau-j}, the vector operator $J_i$ is a spherical tensor operator of rank 1; therefore the matrix elements of the generators are proportional to the Clebsch--Gordan coefficient $\CG{j1}{j}{n1}{m}$. The coefficient of proportionality can be determined by contracting the equation with itself, using the orthogonality of the Clebsch--Gordan coefficient on one side, and
\be
\Tr\bigl(\tau_i^{(j)}\tau_{\phantom{i}}^{(j)i}\bigr) = -\sum_m \bra{jm}J_x^2+J_y^2+J_z^2\ket{jm} = -j(j+1)(2j+1)
\ee
on the other side. In this way one finds
\be\label{tau=C}
\Tau{j}{i}{m}{n} = -i\sqrt{j(j+1)}\CG{j1}{j}{ni}{m}.
\ee

Another familiar object which can be expressed in terms of a three-valent intertwiner is the antisymmetric symbol $\epsilon_{ijk}$. Since the 3$j$-symbol with $j_1=j_2=j_3=1$ is completely antisymmetric in its indices due to the symmetry relation \eqref{3jperm}, one might think that $\epsilon_{ijk}$ is directly proportional to the 3$j$-symbol:
\be\label{eps=3j}
\epsilon_{ijk} = -\sqrt 6\threej{1}{1}{1}{i}{j}{k}.
\ee
While this is a valid numerical relation, some care should be taken when using it, since the indices of $\epsilon_{ijk}$ usually refer to the Cartesian basis, in which an index takes the values $i=x,y,z$, whereas the indices of the 3$j$-symbol take the values $m=+1,0,-1$, and hence refer to the so-called spherical basis. (Strictly speaking, the index $i$ in \Eq{tau=C} should also be interpreted as a spherical index, not a Cartesian index.) The components of a vector with respect to the spherical basis are defined in terms of the Cartesian components by\footnote{Under a rotation descibed by a matrix $R\in SO(3)$, the Cartesian components of a vector $\vec v$ transform as $v^i \to R\updown{i}{j}v^j$. Under the same rotation, the spherical components defined by \Eq{v-spherical} transform according to $v^m \to \D{1}{m}{n}{g_R}v^n$, where $g_R$ is an $SU(2)$ element corresponding to the rotation $R$. Thus the spherical components of a vector are a special case of the definition of a spherical tensor given in the previous footnote.}
\be\label{v-spherical}
v^+ = -\frac{1}{\sqrt 2}(v^x - iv^y), \qquad v^0 = v^z, \qquad v^- = \frac{1}{\sqrt 2}(v^x+iv^y).
\ee
Now one can verify by direct calculation that a triple product of the form $\epsilon_{ijk}u^iv^jw^k$, where the indices $i$, $j$, $k$ refer to the Cartesian basis, is equal to $-i\epsilon_{\lambda\mu\nu}u^\lambda v^\mu w^\nu$, where $\lambda$, $\mu$, $\nu$ refer to the spherical basis, and the antisymmetric symbol in the spherical basis is defined so that $\epsilon_{10{}-1}=+1$. Therefore the correct way to express the triple product in terms of the 3$j$-symbol is given by
\be\label{epsuvw}
\epsilon_{ijk}u^iv^jw^k = i\sqrt 6\threej{1}{1}{1}{\lambda}{\mu}{\nu}u^\lambda v^\mu w^\nu.
\ee

\subsection{Intertwiners of higher valence}\label{sec:intertwiners-N}

The invariant tensors $\iota_{m_1m_2m_3}$ and $\epsilon^{(j)}_{mn}$ are the basic building blocks out of which intertwiners of higher valence can be constructed. For example, by using epsilon to contract two three-valent intertwiners on one index, we obtain the four-valent intertwiner
\begin{align}
\bigl(\iota_{12}^{(k)}\bigr)_{m_1m_2m_3m_4} &= \threej{j_1}{j_2}{k}{m_1}{m_2}{m}\epsilon_{\phantom{m}}^{(k)mn}\threej{k}{j_3}{j_4}{n}{m_3}{m_4} \notag \\
&= \sum_m (-1)^{k-m}\threej{j_1}{j_2}{k}{m_1}{m_2}{m}\threej{k}{j_3}{j_4}{-m}{m_3}{m_4}. \label{iota412}
\end{align}
The invariance of $\iota_{m_1m_2m_3}$ and $\epsilon^{(j)}_{mn}$ implies that the intertwiner \eqref{iota412} is invariant under the action of $SU(2)$ on its indices:
\be
\D{j_1}{m_1}{n_1}{g}\D{j_2}{m_2}{n_2}{g}\D{j_3}{m_3}{n_3}{g}\D{j_4}{m_4}{n_4}{g}\bigl(\iota_{12}^{(k)}\bigr)_{m_1m_2m_3m_4} = \bigl(\iota_{12}^{(k)}\bigr)_{n_1n_2n_3n_4}.
\ee
When the internal spin $k$ ranges over all the values allowed by the Clebsch--Gordan conditions, the tensors \eqref{iota412} span the intertwiner space ${\rm Inv}\,\bigl({\cal H}_{j_1}\otimes{\cal H}_{j_2}\otimes{\cal H}_{j_3}\otimes{\cal H}_{j_4}\bigr)$. Using the orthogonality relation of the 3$j$-symbols, one finds
\be\label{iota4-prod}
\bbraket{\iota_{12}^{(k)}}{\iota_{12}^{(k')}} = \bigl(\iota_{12}^{(k)}\bigr)^{m_1m_2m_3m_4}\bigl(\iota_{12}^{(k')}\bigr)_{m_1m_2m_3m_4} = \frac{1}{d_k}\delta_{kk'},
\ee
showing that the basis given by the intertwiners \eqref{iota412} is orthogonal but not normalized. To obtain normalized intertwiners, \Eq{iota412} should be multiplied by $\sqrt{d_k}$. 

Another basis on the four-valent intertwiner space is provided by the intertwiners
\be\label{iota413}
\bigl(\iota_{13}^{(l)}\bigr)_{m_1m_2m_3m_4} = \threej{j_1}{j_3}{l}{m_1}{m_3}{m}\epsilon_{\phantom{m}}^{(l)mn}\threej{l}{j_2}{j_4}{n}{m_2}{m_4},
\ee
in which the spins $j_1$ and $j_3$ have been coupled to the internal spin. The change of basis between the bases \eqref{iota412} and \eqref{iota413} will be discussed in section \ref{sec:6j9j}.

Intertwiners of arbitrarily high valence can evidently be derived by continuing to attach three-valent intertwiners to each other by contraction with epsilon. An $N$-valent intertwiner constructed according to this scheme has the form
\begin{align}
\iota^{(k_1\cdots k_{N-2})}_{m_1\cdots m_N} &= \threej{j_1}{j_2}{k_1}{m_1}{m_2}{\mu_1}\epsilon_{\phantom{m}}^{(k_1)\mu_1\nu_1} \threej{k_1}{j_3}{k_2}{\nu_1}{m_3}{\mu_2}\epsilon_{\phantom{m}}^{(k_2)\mu_2\nu_2} \cdots \notag \\
&\quad\cdots \threej{k_{N-3}}{j_{N-2}}{k_{N-2}}{\nu_{N-3}}{m_{N-2}}{\mu_{N-2}}\epsilon_{\phantom{m}}^{(k_{N-2})\mu_{N-2}\nu_{N-2}} \threej{k_{N-2}}{j_{N-1}}{j_N}{\nu_{N-2}}{m_{N-1}}{m_N}. \label{iotaN}
\end{align}
It is labeled by $N-2$ internal spins, which determine the eigenvalues of the operators $\bigl(J^{(1)} + J^{(2)}\bigr)^2$, $\bigl(J^{(1)} + J^{(2)} + J^{(3)})^2$, $\dots$, $\bigl(J^{(1)} + \dots + J^{(N-2)}\bigr)^2$. The intertwiner \eqref{iotaN} is not normalized; to normalize it, it should be multiplied by $\sqrt{d_{k_1}}\cdots\sqrt{d_{k_{N-2}}}$. Just as in the three-valent case, the intertwiner $\iota^{(k_1\cdots k_{N-2})m_1\cdots m_N}$, obtained by raising all the indices of the intertwiner \eqref{iotaN} by means of the epsilon tensor, is numerically equal to $\iota^{(k_1\cdots k_{N-2})}_{m_1\cdots m_N}$.

A useful fact of the $N$-valent intertwiner space concerns the integral
\be
I^{(j_1\cdots j_N)} = \int dg\,D^{(j_1)}(g)\cdots D^{(j_N)}(g).
\ee
The invariance and normalization of the Haar measure imply that the integral is invariant under the action of $SU(2)$, and satisfies $(I^{(j_1\cdots j_N)})^2 = I^{(j_1\cdots j_N)}$. Therefore $I^{(j_1\cdots j_N)}$, viewed as an operator on ${\cal H}_{j_1}\otimes\cdots\otimes{\cal H}_{j_N}$, must be the projection operator onto the $SU(2)$ invariant subspace of ${\cal H}_{j_1}\otimes\cdots\otimes{\cal H}_{j_N}$, \ie onto the intertwiner space \mbox{${\rm Inv}\,\bigl({\cal H}_{j_1}\otimes\cdots\otimes{\cal H}_{j_N}\bigr)$. Hence} the integral can be expressed as $I^{(j_1\cdots j_N)} = \sum_\iota \ket{\iota}\bra{\iota}$, or
\be\label{int D...D}
\int dg\,\D{j_1}{m_1}{n_1}{g}\cdots\D{j_N}{m_N}{n_N}{g} = \sum_\iota \overline{\displaystyle \iota^{m_1\cdots m_N}}\iota_{n_1\cdots n_N},
\ee
where the sum runs over any orthonormal basis of ${\rm Inv}\,\bigl({\cal H}_{j_1}\otimes\cdots\otimes{\cal H}_{j_N}\bigr)$; in general the basis may be complex-valued, although the basis given by the intertwiners \eqref{iotaN} is always real.

\subsection{6$j$- and 9$j$-symbols}\label{sec:6j9j}

The intertwiners \eqref{iota412} and \eqref{iota413} provide two inequivalent bases of the four-valent intertwiner space. One basis is expressed in terms of the other by the relation
\be\label{6j-def}
\bket{\iota_{13}^{(l)}} = \sum_k d_k(-1)^{j_2+j_3+k+l}\sixj{j_1}{j_2}{k}{j_4}{j_3}{l}\bket{\iota_{12}^{(k)}},
\ee
where the object with curly brackets is the Wigner 6$j$-symbol. From \Eq{6j-def} one can derive the expression
\be\label{6j=iiii}
\sixj{j_1}{j_2}{j_3}{k_1}{k_2}{k_3} = \bigl(\iota^{(j_1j_2j_3)}\bigr)^{m_1m_2m_3}\bigl(\iota^{(j_1k_2k_3)}\bigr)\downup{m_1}{n_2}{\vphantom{\bigl(\iota^{(j_1k_2k_3)}\bigr)}}_{n_3}\bigl(\iota^{(k_1j_2k_3)}\bigr)\downup{n_1m_2}{n_3}\bigl(\iota^{(k_1k_2j_3)}\bigr)\updown{n_1}{n_2m_3}
\ee
\newpage
\noindent for the 6$j$-symbol as a contraction of four three-valent intertwiners. This shows that the 6$j$-symbol vanishes unless the triples of spins indicated by 
\be
\sixj{\circ}{\circ}{\circ}{}{}{} \qquad \sixj{\circ}{}{}{}{\circ}{\circ} \qquad \sixj{}{\circ}{}{\circ}{}{\circ} \qquad \sixj{}{}{\circ}{\circ}{\circ}{}{}
\ee
satisfy the Clebsch--Gordan conditions. Furthermore, orthogonality of the states \eqref{6j-def} for different values of $l$ implies that the 6$j$-symbol satisfies the orthogonality relation
\be\label{6j-orth}
\sum_x d_x\sixj{j_1}{j_2}{x}{j_3}{j_4}{k}\sixj{j_1}{j_2}{x}{j_3}{j_4}{l} = \frac{1}{d_k}\delta_{kl}.
\ee
Symmetry properties of the 6$j$-symbol can be derived from \Eq{6j=iiii}, though they are more easily seen using the graphical representation of the symbol, which will be introduced in the next chapter. The value of the 6$j$-symbol is unchanged by any permutation of its columns:
\be
\sixj{j_1}{j_2}{j_3}{k_1}{k_2}{k_3} = \sixj{j_1}{j_3}{j_2}{k_1}{k_3}{k_2} = \sixj{j_2}{j_3}{j_1}{k_2}{k_3}{k_1}, \quad \text{etc.}
\ee
and by interchanging the upper and lower spins simultaneously in any two columns:
\be
\sixj{j_1}{j_2}{j_3}{k_1}{k_2}{k_3} = \sixj{j_1}{k_2}{k_3}{k_1}{j_2}{j_3}, \quad \text{etc.}
\ee

The Wigner 9$j$-symbol arises when changes of basis between five-valent intertwiners are performed. Two different bases in the intertwiner space ${\rm Inv}\,\bigl({\cal H}_{j_1}\otimes\cdots\otimes{\cal H}_{j_5}\bigr)$ are provided by the intertwiners
\be\label{iota512}
\bigl(\iota_{12}^{(k_{12}k_{34})}\bigr)_{m_1\cdots m_5} = \bigl(\iota^{(j_1j_2k_{12})}\bigr)\downup{m_1m_2}{n_{12}}\bigl(\iota^{(j_3j_4k_{34})}\bigr)\downup{m_3m_4}{n_{34}}\bigl(\iota^{(k_{12}k_{34}j_5)}\bigr)_{n_{12}n_{34}m_5}
\ee
and
\be\label{iota513}
\bigl(\iota_{13}^{(k_{13}k_{24})}\bigr)_{m_1\cdots m_5} = \bigl(\iota^{(j_1j_3k_{13})}\bigr)\downup{m_1m_3}{n_{13}}\bigl(\iota^{(j_2j_4k_{24})}\bigr)\downup{m_2m_4}{n_{24}}\bigl(\iota^{(k_{13}k_{24}j_5)}\bigr)_{n_{13}n_{24}m_5}.
\ee
Elements of one basis are expanded in the other basis as
\be
\bket{\iota_{13}^{(k_{13}k_{24})}} = \sum_{k_{12}k_{34}} d_{k_{12}}d_{k_{34}}\ninej{j_1}{j_2}{k_{12}}{j_3}{j_4}{k_{34}}{k_{13}}{k_{24}}{j_5}\bket{\iota_{12}^{(k_{12}k_{34})}},
\ee
where the 9$j$-symbol appears on the right-hand side. From this definition it follows that the 6$j$-symbol is given by a contraction of six three-valent intertwiners as
\begin{align}
\ninej{j_1}{j_2}{j_3}{k_1}{k_2}{k_3}{l_1}{l_2}{l_3} &= \bigl(\iota^{(j_1j_2j_3)}\bigr)_{m_1m_2m_3}\bigl(\iota^{(k_1k_2k_3)}\bigr)_{n_1n_2n_3}\bigl(\iota^{(l_1l_2l_3)}\bigr)_{\mu_1\mu_2\mu_3} \notag \\
&\quad\times \bigl(\iota^{(j_1k_1l_1)}\bigr)^{m_1n_1\mu_1}\bigl(\iota^{(j_2k_2l_2)}\bigr)^{m_2n_2\mu_2}\bigl(\iota^{(j_3k_3l_3)}\bigr)^{m_3n_3\mu_3}.\label{9j=i^6}
\end{align}
Hence the 9$j$-symbol can have a non-zero value only if the Clebsch--Gordan conditions are satisfied by the spins in each row and each column.

The 9$j$-symbol possesses a high degree of symmetry. Transposing the array of spins in the symbol preserves its value:
\be
\ninej{j_1}{k_1}{l_1}{j_2}{k_2}{l_2}{j_3}{k_3}{l_3} = \ninej{j_1}{j_2}{j_3}{k_1}{k_2}{k_3}{l_1}{l_2}{l_3}.
\ee
Moreover, interchanging any rows or any two columns produces a sign factor:
\be
\ninej{j_1}{j_2}{j_3}{k_1}{k_2}{k_3}{l_1}{l_2}{l_3} = (-1)^S\ninej{j_2}{j_1}{j_3}{k_2}{k_1}{k_3}{l_2}{l_1}{l_3} = (-1)^S\ninej{k_1}{k_2}{k_3}{j_1}{j_2}{j_3}{l_1}{l_2}{l_3}, \quad \text{etc.} \\
\ee
where
\be
S = j_1+j_2+j_3+k_1+k_2+k_3+l_1+l_2+l_3.
\ee
These symmetry relations again become the most transparent when the 9$j$-symbol is expressed in graphical form. A large number of further properties satisfied by the 6$j$- and 9$j$-symbols can be found in any of the standard references, such as \cite{Varshalovich}.

\subsection{Intertwiners in the symmetric tensor product representation}\label{sec:int-stp}

When the representation of the space ${\cal H}_j$ as a completely symmetrized tensor product of the spaces ${\cal H}_{1/2}$ is used, intertwiners can be expressed directly in terms of the fundamental invariant tensor $\epsilon_{AB}$, which is the only invariant tensor available in the space ${\cal H}_{1/2}$. The basic three-valent intertwiner
\be\label{qwerty}
\iota_{(A_1\cdots A_{2j_1})(B_1\cdots B_{2j_2})(C_1\cdots C_{2j_3})},
\ee
out of which higher intertwiners can be derived, must be constructed from the object
\be\label{eps..-eps-converted-to.pdf}
\epsilon_{A_1B_1}\cdots\epsilon_{A_aB_a}\epsilon_{B_{a+1}C_1}\cdots\epsilon_{B_{a+b}C_b}\epsilon_{C_{b+1}A_{a+1}}\cdots\epsilon_{C_{b+c}A_{a+c}},
\ee
which is the only possible combination of epsilons with the correct index structure. The requirement that the total number of $A$, $B$ and $C$ indices is respectively $2j_1$, $2j_2$ and $2j_3$ determines the numbers $a$, $b$ and $c$ as
\be
a = j_1+j_2-j_3, \qquad b = j_2+j_3-j_1, \qquad c = j_3+j_1-j_2.
\ee
The three-valent intertwiner \eqref{qwerty} is then obtained from the expression \eqref{eps..-eps-converted-to.pdf} by symmetrizing each group of indices. This gives
\be\label{iota_ABC}
\iota_{(A_1\cdots A_{2j_1})(B_1\cdots B_{2j_2})(C_1\cdots C_{2j_3})} = N_{j_1j_2j_3}I_{(A_1\cdots A_{2j_1})(B_1\cdots B_{2j_2})(C_1\cdots C_{2j_3})},
\ee
where
\be
I_{A_1\cdots A_{2j_1}B_1\cdots B_{2j_2}C_1\cdots C_{2j_3}} = \epsilon_{A_1B_1}\cdots\epsilon_{A_aB_a}\epsilon_{B_{a+1}C_1}\cdots\epsilon_{B_{a+b}C_b}\epsilon_{C_{b+1}A_{a+1}}\cdots\epsilon_{C_{b+c}A_{a+c}},
\ee
and the normalization factor is
\be\label{Njjj}
N_{j_1j_2j_3} = \sqrt{\frac{(2j_1)!(2j_2)!(2j_3)!}{(j_1+j_2+j_3+1)!(j_1+j_2-j_3)!(j_2+j_3-j_1)!(j_3+j_1-j_2)!}}.
\ee
The three-valent intertwiner having been found, intertwiners of higher valence can be constructed by contracting several three-valent intertwiners, in the same way as in \Eqs{iota412} and \eqref{iotaN}. For example, a four-valent intertwiner is given by
\be\label{iota_ABCD}
\iota^{(k)}_{(A_1\cdots A_{2j_1})\cdots(D_1\cdots D_{2j_4})} = \sqrt{d_k}\iota_{(A_1\cdots A_{2j_1})(B_1\cdots B_{2j_2})(E_1\cdots E_{2k})}\iota\updown{(E_1\cdots E_{2k})}{(C_1\cdots C_{2j_3})(D_1\cdots D_{2j_4})},
\ee
where the spin-1/2 indices are raised with $\epsilon^{AB}$ according to the convention \eqref{raise-1/2}, and the factor $\sqrt{d_k}$ is inserted for normalization.

Intertwiners in the symmetric tensor product representation can be expressed graphically by introducing a primitive graphical notation, in which
\begin{itemize}
\item $\epsilon_{AB}$ or $\epsilon^{AB}$ is represented by a line with an arrow pointing from index $A$ to $B$;
\item $\delta^A_B = \epsilon^{AC}\epsilon_{BC}$ is represented by a line with no arrow;
\item Contraction of an index is represented by connecting two lines at the contracted index.
\end{itemize}
Then we can write the intertwiner \eqref{iota_ABC} in the form 
\be\label{i3graphic}
\iota_{(A_1\cdots A_{2j_1})(B_1\cdots B_{2j_2})(C_1\cdots C_{2j_3})} = N_{j_1j_2j_3}\;\RealSymb{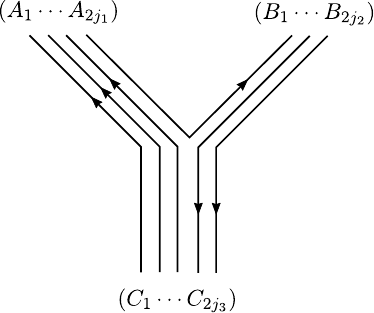}{0.8}
\ee
where each external group of lines is understood to be completely symmetrized. Similarly, the four-valent intertwiner \eqref{iota_ABCD} is given by
\be\label{i4graphic}
\iota^{(k)}_{(A_1\cdots A_{2j_1})\cdots(D_1\cdots D_{2j_4})} = \sqrt{d_k}N_{j_1j_2k}N_{kj_3j_4}\;\RealSymb{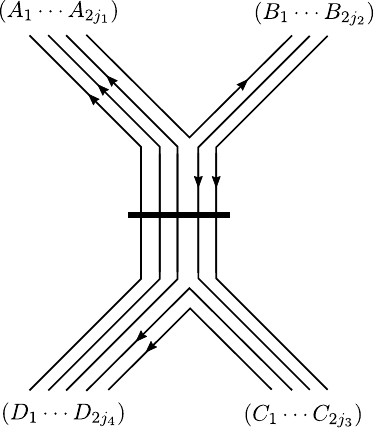}{0.8}
\ee
where we draw the horizontal bar to indicate a complete symmetrization of the internal group of lines.

\newpage

\section{The graphical method}\label{ch:graphical}

We now come to the last and in a sense the most important chapter of the Appendix, in which we introduce the powerful graphical techniques for calculations in $SU(2)$ recoupling theory. On the surface the graphical framework may simply seem to consist of adopting a diagrammatic notation for $SU(2)$ calculations. However, in reality the graphical approach serves a highly efficient computational device, since the visual graphical diagrams are invariably easier to comprehend and simpler to work with than the corresponding non-graphical expressions. In loop quantum gravity, the graphical techniques can be generally used in any calculation involving intertwiners. In particular, in this work they have an indispensable role as the technical tools by which we calculate the matrix elements of our Hamiltonian in the spin network basis (see Chapter \ref{ch:elements}). 

Several closely related but slightly different versions of the graphical formalism can be found in the literature of quantum angular momentum \cite{BrinkSatchler, Varshalovich, YLV}, and in articles in which graphical methods are used for calculations in loop quantum gravity \cite{paper1, paper3, AlesciThiemannZipfel, AlesciLiegenerZipfel, YangMa1, YangMa2}. In these circumstances it seems preferable (and it certainly requires less effort) to develop a consistent set of conventions for the graphical formalism on one's own, using the available references as a guide, rather than trying to ensure that one is in full agreement with the conventions chosen in some previous treatment of the subject. When compared to conventions available in the literature, the conventions adopted in this work match the most closely with those of Brink and Satchler \cite{BrinkSatchler}, which was the main reference used by the author to learn the graphical techniques in the early stages of his PhD studies.

\subsection{Elements of the graphical formalism}

The basic elements of the graphical formalism are provided by the graphical representations of the elementary objects of $SU(2)$ representation theory: the delta and epsilon tensors, the 3$j$-symbol, and $SU(2)$ group elements. The properties satisfied by these objects lead to graphical rules according to which diagrams appearing in a graphical calculation can be manipulated. The relation of the formalism to loop quantum gravity is indicated by \Eqs{DD g}--\eqref{J*h target}, which show how the action of the elementary operators of the theory can be expressed in graphical form.

\subsubsection*{Delta and epsilon tensors}

The unit tensor $\delta_m^n$ is represented graphically by a line which carries the indices $m$ and $n$ at its two ends:
\be\label{delta g}
\delta_m^n\; = \;\;\RealSymb{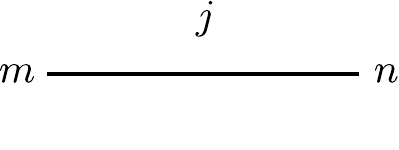}{0.6}\vspace{-12pt}
\ee
If necessary, the line can be labeled with a spin $j$ to indicate the representation which the indices of $\delta_m^n$ refer to.

The invariant tensor $\epsilon^{(j)}_{mn}$ is represented by a line with an arrow pointing from index $m$ to index $n$:
\be\label{epsilon g}
\epsilon^{(j)}_{mn}\; = \;\;\RealSymb{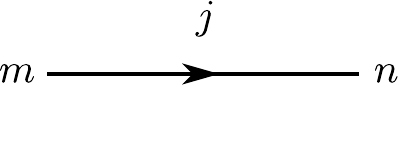}{0.6}\vspace{-12pt}
\ee
Since tensor $\epsilon^{(j)mn}_{\phantom{m}}$ is numerically equal to $\epsilon^{(j)}_{mn}$, it is represented by the same diagram:
\be\label{epsilon2 g}
\epsilon^{(j)mn}_{\phantom{m}}\; = \;\;\RealSymb{figA-epsilon-indices.pdf}{0.6}\vspace{-12pt}
\ee
The magnetic indices in graphical diagrams such as \eqref{delta g}--\eqref{epsilon2 g} are usually not shown explicitly, if leaving them out is not likely to cause any confusion. For example, the indices of the epsilon tensor can be safely omitted, since the direction of the arrow shows which end of the line corresponds to which index of $\epsilon^{(j)}_{mn}$. From now on we will follow this practice and not write out the magnetic indices unless there is a particular reason for doing so.

Contraction of indices is carried out in the graphical formalism by connecting the ends of the two lines corresponding to the contracted index. The relations \eqref{eps_nm} and \eqref{epseps=delta} satisfied by the epsilon tensor then imply that the arrow behaves according to the rules
\be\label{invarrow}
\RealSymb{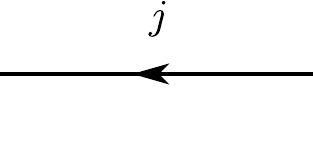}{0.6}\quad = \quad(-1)^{2j}\;\RealSymb{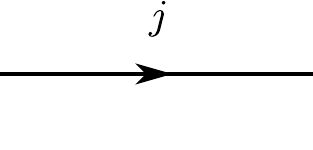}{0.6}\vspace{-12pt}
\ee
and
\be\label{2arrows-same}
\RealSymb{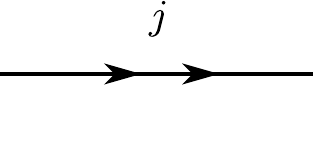}{0.6}\quad = \quad(-1)^{2j}\;\RealSymb{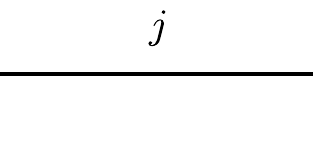}{0.6}\vspace{-8pt}
\ee
Using the first relation, the second relation can be equivalently written as
\be\label{2arrows-opp}
\RealSymb{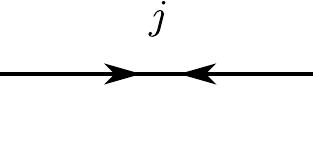}{0.6}\quad = \quad\RealSymb{figA-delta.pdf}{0.6}\vspace{-12pt}
\ee

\subsubsection*{The 3$j$-symbol}

The graphical representation of the 3$j$-symbol is given by three lines connected at a node:
\be\label{3j g}
\threej{j_1}{j_2}{j_3}{m_1}{m_2}{m_3}\; = \;\RealSymb{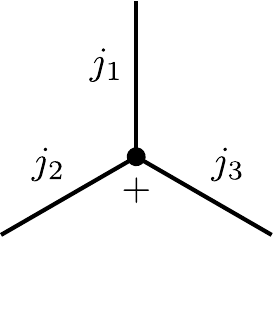}{0.6}\vspace{-20pt}
\ee
The sign at the node encodes the cyclic order of the spins in the symbol, a $+$ sign corresponding to the anticlockwise order of \Eq{3j g}, while a $-$ sign corresponds to a clockwise order. Thus,
\be\label{3j- g}
\threej{j_1}{j_3}{j_2}{m_1}{m_3}{m_2}\; = \quad\RealSymb{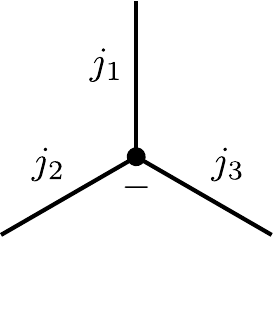}{0.6}\quad = \quad \RealSymb{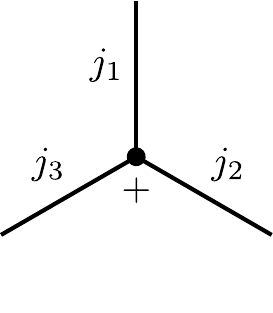}{0.6}\vspace{-20pt}
\ee
\Eq{3jperm} then implies that reversing the sign produces the factor $(-1)^{j_1+j_2+j_3}$:
\be\label{3jminus g}
\RealSymb{figA-3j-minus.pdf}{0.6} \quad = \quad (-1)^{j_1+j_2+j_3}\RealSymb{figA-3j.pdf}{0.6}\vspace{-20pt}
\ee
From the other symmetry relation \eqref{3j-m} it follows that
\be\label{3jarrows g}
\RealSymb{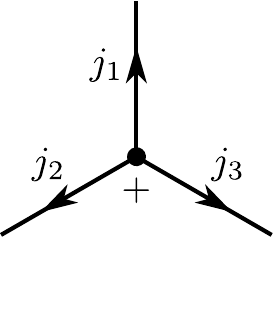}{0.6} \quad = \quad \RealSymb{figA-3j.pdf}{0.6}\vspace{-20pt}
\ee
which is a graphical representation of the fact that $\iota^{m_1m_2m_3}$ is numerically equal to $\iota_{m_1m_2m_3}$. Furthermore, the orthogonality relations \eqref{3j-orth-mm} and \eqref{3j-orth-jm} read
\be\label{3j-orth-mm g}
\RealSymb{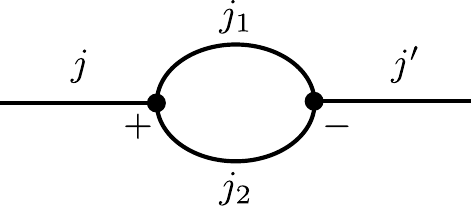}{0.6} \quad = \quad\delta_{jj'}\frac{1}{d_j}\;\RealSymb{figA-delta.pdf}{0.6}
\ee
and
\be\label{3j-orth-jm g}
\sum_j d_j\RealSymb{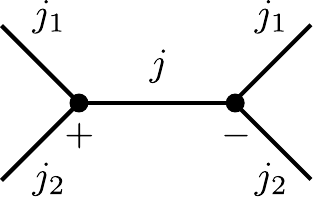}{0.6} \quad = \quad \RealSymb{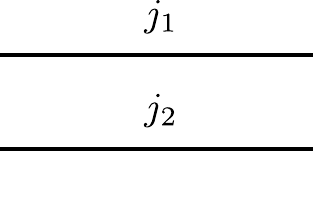}{0.6}
\ee
The normalization of the 3$j$-symbol, implied by the first orthogonality relation, gives
\be\label{3j-theta g}
\RealSymb{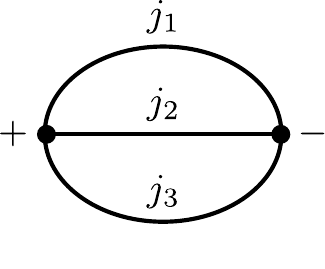}{0.6}\quad =\quad 1.\vspace{-12pt}
\ee
\Eq{3jzero}, which shows how the 3$j$-symbol reduces to the epsilon tensor when one of the spins is zero, is written graphically as
\be\label{3j-zero g}
\RealSymb{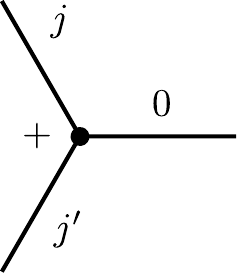}{0.6} \quad = \quad\delta_{jj'}\frac{1}{\sqrt{d_j}}\quad\RealSymb{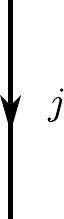}{0.6}
\ee
In equations such as \eqref{3j-orth-jm g} and \eqref{3j-zero g}, where the relation between the indices on the two sides of the equation is in principle ambiguous unless one writes them out explicitly, we follow the convention that the relative position of the indices is the same on both sides. For example, if we consider the index at the top left corner on the left-hand side of \Eq{3j-orth-jm g} as the index $m_1$ of the corresponding non-graphical equation \eqref{3j-orth-jm}, then $m_1$ is at the top left corner also on the right-hand side of \Eq{3j-orth-jm g}.

From \Eq{3j}, which defines the 3$j$-symbol in terms of the Clebsch--Gordan coefficient, we find that the graphical representation of the Clebsch--Gordan coefficient is
\be\label{clebsch g}
\CG{j_1j_2}{j}{m_1m_2}{m} \; = \; (-1)^{j_1-j_2-j}\sqrt{d_j}\;\RealSymb{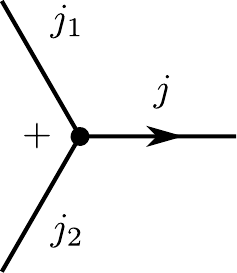}{0.6}
\ee
\Eq{tau=C} then shows that the matrix elements of the $SU(2)$ generators are given by \vspace{-20pt}
\be\label{tau=C g}
\Tau{j}{i}{m}{n} \; = \; iW_j\;\RealSymb{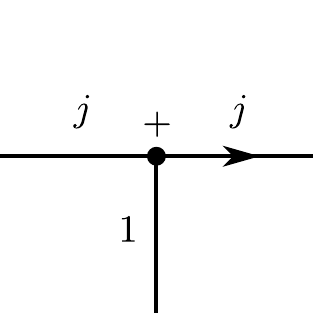}{0.6}
\ee
where we used the shorthand notation
\be
W_j = \sqrt{j(j+1)(2j+1)}.
\ee
To write the triple product $\epsilon_{ijk}u^iv^jw^j$ in graphical form, let us introduce the diagram
\be
v^m \; = \; \RealSymb{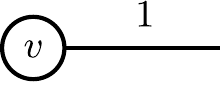}{0.6}
\ee
to represent a vector with an index in the $j=1$ representation. Then, according to \Eq{epsuvw}, the triple product can be expressed graphically as
\be\label{epsuvw g}
\epsilon_{ijk}u^iv^jw^j \; = \;i\sqrt{6}\;\RealSymb{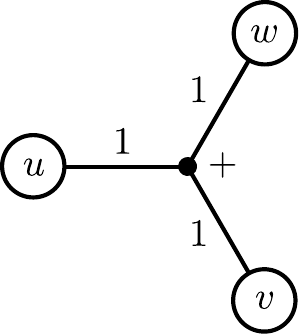}{0.6}
\ee

\subsubsection*{Group elements}

For the matrix elements of the Wigner matrices $D^{(j)}(g)$, we adopt the graphical representation
\be\label{WignerD g}
\D{j}{m}{n}{g} \; = \; \RealSymb{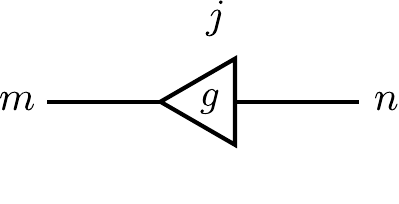}{0.6}\vspace{-12pt}
\ee
If the group element $g$ is the holonomy of an edge in a spin network state, then the transformation law
\be
\D{j}{m}{n}{h_e} \to \Db{j}{m}{\mu}{g(t_e)}\D{j}{\mu}{\nu}{h_e}\Db{j}{\nu}{n}{g^{-1}(s_e)}
\ee
of the holonomy under $SU(2)$ gauge transformations suggests that the indices $m$ and $n$ in $\D{j}{m}{n}{h_e}$ are associated respectively to the endpoint and the beginning point of the edge $e$. Therefore the direction of the triangle in \Eq{WignerD g} is consistent with the orientation of the edge.

From \Eq{Dj-inv}, we see that that the inverse matrix is given by
\be\label{D^-1 g}
\RealSymb{figA-Dginv.pdf}{0.6} \quad = \quad \RealSymb{figA-epsDeps.pdf}{0.6}
\ee
The orthogonality theorem for the Wigner matrices can be translated into graphical form, if we denote the complex conjugate of the matrix elements by a bar over the group element $g$ inside the triangle in \Eq{WignerD g}. We then have
\be
\int dg\quad \RealSymb{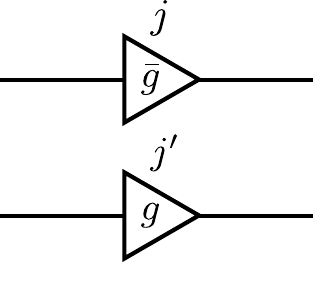}{0.6} \quad\; = \quad \delta_{jj'}\frac{1}{d_j}\quad\RealSymb{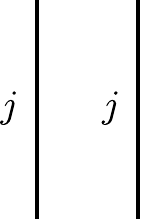}{0.6}
\ee
Alternatively, the relation $D^{(j)}(g^\dagger) = D^{(j)}(g^{-1})$ can be used to eliminate the complex conjugate. This leads to
\begin{align}
\int dg\,&\D{j}{m}{n}{g}\D{j'}{m'}{n'}{g} \notag \\
&= \epsilon^{(j)m\mu}_{\phantom{m}}\epsilon^{(j)}_{n\nu}\int dg\,\D{j}{\nu}{\mu}{g^{-1}}\D{j'}{m'}{n'}{g} = \delta_{jj'}\frac{1}{d_j}\epsilon^{(j)mm'}_{\phantom{m}}\epsilon^{(j)}_{nn'}
\end{align}
The corresponding graphical equation is
\be
\int dg\quad \RealSymb{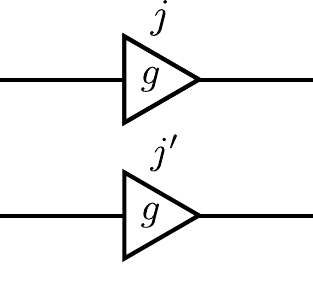}{0.6} \quad\; = \quad \delta_{jj'}\frac{1}{d_j}\quad\RealSymb{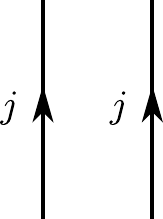}{0.6}
\ee
The Clebsch--Gordan series \eqref{CG-ser} is expressed graphically as
\begin{align}
\RealSymb{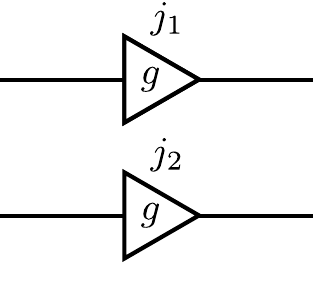}{0.6} \quad = \quad \sum_j d_j\;\RealSymb{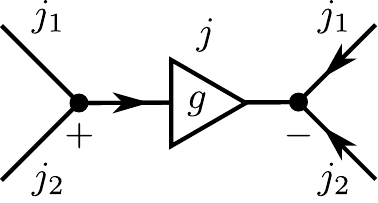}{0.6} \label{DD g} \\
\intertext{Using \Eq{D^-1 g}, we can also derive the coupling of a group element with its inverse:} 
\RealSymb{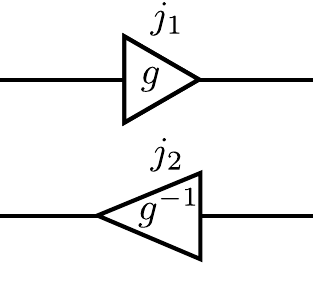}{0.6} \quad = \quad \sum_j d_j\;\RealSymb{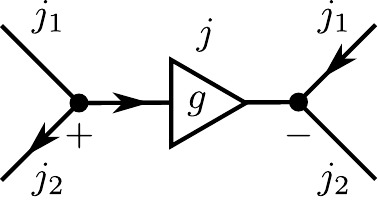}{0.6}\label{D^-1D g}
\end{align}
\Eqs{DD g} and \eqref{D^-1D g} are the basic graphical rules for computing the action of the holonomy operator on an edge of a spin network state. The action of the elementary ''momentum operator'' $J_i^{(v,e)}$ on a holonomy, given by \Eq{J*h_e}, takes the graphical form 
\vspace{-32pt}
\begin{align}
J_i^{(v,e)}\,\RealSymb{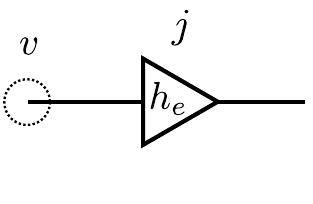}{0.6} \quad &= \quad -W_j\;\RealSymb{figA-tauh_e.pdf}{0.6} \label{J*h source}\\
\intertext{when the operator acts at the beginning point of the edge $e$, and \vspace{-24pt}}
J_i^{(v,e)}\;\RealSymb{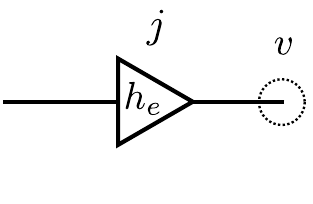}{0.6} \quad &= \quad W_j\;\RealSymb{figA-h_etau.pdf}{0.6} \label{J*h target}
\end{align}
when the operator acts at the endpoint of $e$.

\subsection{Intertwiners in graphical form}

The graphical representation of the basic three-valent intertwiner $\iota_{m_1m_2m_3}$ is given by \Eq{3j g}. Intertwiners of higher valence can be constructed by using the epsilon tensor to attach several three-valent intertwiners to each other, as discussed in section \ref{sec:intertwiners-N}. Hence the four-valent intertwiner of \Eq{iota412} is represented graphically as
\be\label{iota412 g}
\bigl(\iota_{12}^{(k)}\bigr)_{m_1m_2m_3m_4} \; = \;\RealSymb{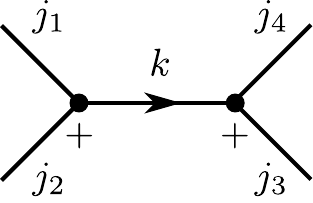}{0.6}
\ee
where the epsilon tensor appears as an arrow on the internal line. By graphical means it is easy to show that the scalar product between two intertwiners of this form is indeed given by \Eq{iota4-prod}. Using \Eqs{3j-orth-mm g} and \eqref{2arrows-opp}, we find
\be
\bbraket{\iota_{12}^{(k)}}{\iota_{12}^{(k')}} \; = \quad \RealSymb{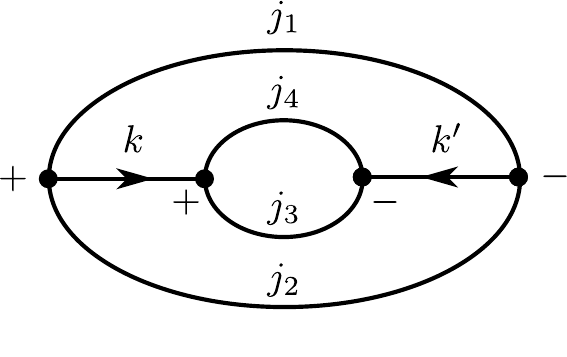}{0.6} \quad = \; \frac{1}{d_k}\delta_{kk'}.\vspace{-12pt}
\ee
The intertwiner \eqref{iota413}, in which the spins $j_1$ and $j_3$ have been coupled to the internal spin, has the graphical form
\be\label{iota413 g}
\bigl(\iota_{13}^{(l)}\bigr)_{m_1m_2m_3m_4} \; = \;\RealSymb{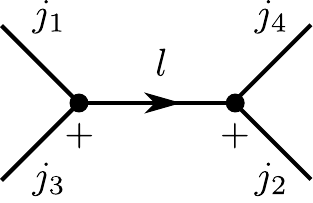}{0.6}
\ee
The change of basis between the bases \eqref{iota412 g} and \eqref{iota413 g} is given by
\be\label{6j-def g}
\RealSymb{figA-iota413.pdf}{0.6} \quad = \quad \sum_k d_k(-1)^{j_2+j_3+k+l}\sixj{j_1}{j_2}{k}{j_4}{j_3}{l}\;\RealSymb{figA-iota412.pdf}{0.6}
\ee
By contracting both sides of this equation with the intertwiner \eqref{iota412 g}, one can derive the graphical expression \vspace{-12pt}
\be\label{6j g}
\sixj{j_1}{j_2}{j_3}{k_1}{k_2}{k_3} \quad = \; \RealSymb{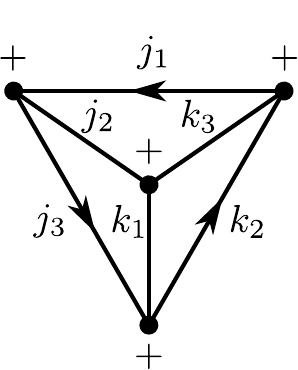}{0.6}
\ee
for the 6$j$-symbol. \Eq{6j g} is of course the graphical equivalent of \Eq{6j=iiii}.

The five-valent intertwiners \eqref{iota512} and \eqref{iota513} are written graphically as
\begin{align}
\bigl(\iota_{12}^{(k_{12}k_{34})}\bigr)_{m_1\cdots m_5} \; &= \; \RealSymb{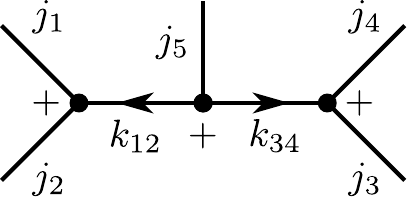}{0.6} \label{iota512 g} \\
\intertext{and}
\bigl(\iota_{13}^{(k_{13}k_{24})}\bigr)_{m_1\cdots m_5} \; &= \; \RealSymb{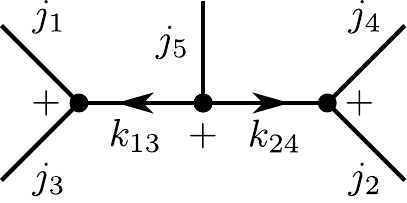}{0.6} \label{iota513 g}
\end{align}
The intertwiner \eqref{iota513 g} is expressed in the basis \eqref{iota512 g} by the relation
\begin{align}
&\RealSymb{figA-iota513.pdf}{0.6} \notag \\
&\qquad\qquad=\quad\sum_{k_{12}k_{34}} d_{k_{12}}d_{k_{34}}\ninej{j_1}{j_2}{k_{12}}{j_3}{j_4}{k_{34}}{k_{13}}{k_{24}}{j_5}\;\RealSymb{figA-iota512.pdf}{0.6}\label{9j-def g}
\end{align}
Contracting both sides with the intertwiner \eqref{iota512 g}, one finds that the 9$j$-symbol has the graphical representation \vspace{-12pt}
\be\label{9j g}
\ninej{j_1}{j_2}{j_3}{k_1}{k_2}{k_3}{l_1}{l_2}{l_3} \quad = \quad \RealSymb{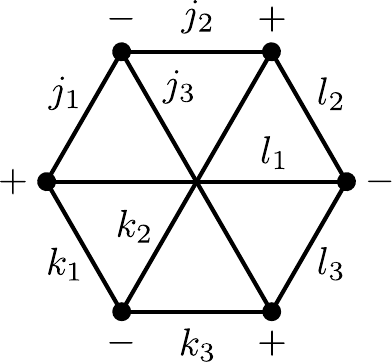}{0.6}
\ee
which is just \Eq{9j=i^6} translated into graphical form.

$N$-valent intertwiners can be derived by continuing to contract three-valent intertwiners in the way indicated by \Eqs{iota412 g} and \eqref{iota512 g}, as shown by the non-graphical equation \eqref{iotaN}. However, it can sometimes be convenient to use a different pattern of contractions to construct a basis of the $N$-valent intertwiner space. When calculating the matrix elements of the Hamiltonian, we use a basis of the form
\be\label{iotaN-loops g}
\RealSymb{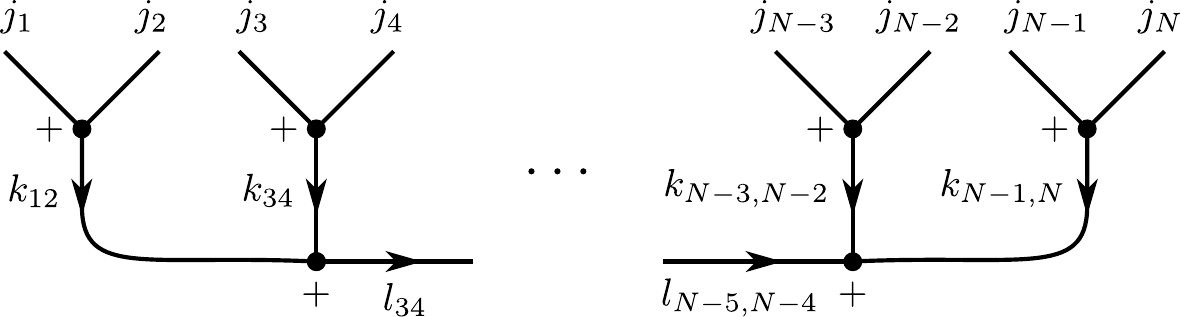}{0.6}
\ee
in which the spins of the intertwiner are coupled pairwise to internal spins. Just as the intertwiners given by \Eq{iotaN}, the intertwiners \eqref{iotaN-loops g} are not normalized, but intertwiners carrying different internal spins are orthogonal to each other. To normalize the intertwiner \eqref{iotaN-loops g}, it should be multiplied by the factor
\be
\sqrt{\displaystyle d_{k_{12}}\cdots d_{k_{N-1,N}}d_{l_{34}}\cdots d_{l_{N-5,N-4}}}.
\ee
Note that there is no loss of generality in using intertwiners of the form \eqref{iotaN-loops g}, even though at a first sight it may seem that one is assuming the valence $N$ to be even. However, to obtain an intertwiner of odd valence from \eqref{iotaN-loops g}, it suffices to set one of the spins, say $j_N$, equal to zero.

\subsection{The fundamental theorem of graphical calculus}\label{sec:ftgc}

The basic rules for working with graphical diagrams are given by \Eqs{invarrow}--\eqref{2arrows-opp} and \eqref{3j- g}--\eqref{3jarrows g}, which show how the arrows and signs in a graphical expression can be manipulated. However, the most essential tool for performing graphical calculations with intertwiners arises from the seemingly simple observation that an invariant tensor having $N$ indices can be expanded in a basis of the corresponding space of $N$-valent intertwiners. To express the implications of this observation in graphical form, we will represent a $N$-valent invariant tensor graphically as a rectangular block with $N$ lines attached to it.

A tensor carrying a single index cannot be invariant, unless the index belongs to the trivial representation. Thus,
\be\label{thm1}
\RealSymb{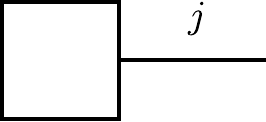}{0.6} \quad = \quad \delta_{j,0}\quad\RealSymb{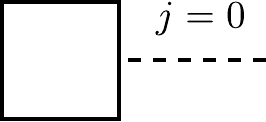}{0.6}
\ee
A two-valent tensor $T_{mn}$ can be invariant only if both of its indices belong to the same representation. In this case, $T_{mn}$ must be proportional to $\epsilon^{(j)}_{mn}$, which is the only invariant tensor having two lower indices. The coefficient of proportionality can be determined by contracting both sides of the equation with the epsilon tensor. Since the contraction of $\epsilon^{(j)}_{mn}$ with itself gives $d_j$, we have the graphical rule
\be\label{thm2}
\RealSymb{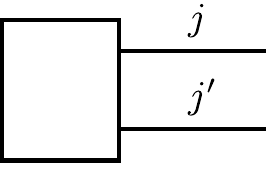}{0.6} \quad = \quad \delta_{jj'}\frac{1}{d_j}\quad\RealSymb{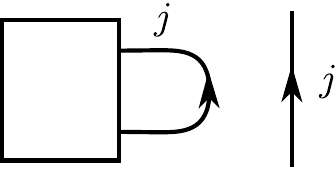}{0.6}
\ee
A three-valent invariant tensor $T_{m_1m_2m_3}$ belongs to the space ${\rm Inv}\,\bigl({\cal H}_{j_1}\otimes{\cal H}_{j_2}\otimes{\cal H}_{j_3}\bigr)$, which is again one-dimensional (provided that the spins $j_1$, $j_2$ and $j_3$ satisfy the Clebsch--Gordan conditions). Therefore $T_{m_1m_2m_3}$ must be proportional to the 3$j$-symbol, the coefficient of proportionality again being determined by contracting the equation with the 3$j$-symbol. That is,
\be\label{thm3}
\RealSymb{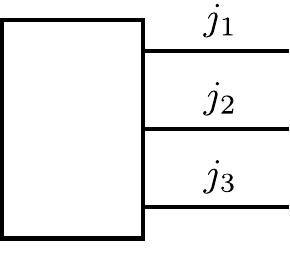}{0.6} \quad = \quad \RealSymb{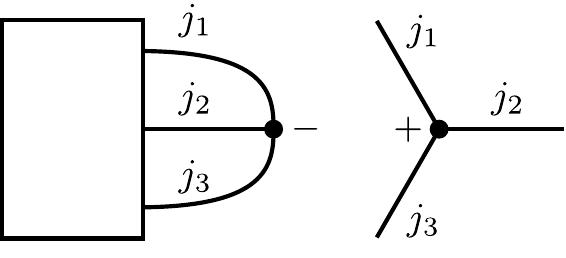}{0.6}
\ee
For invariant tensors of valence four or higher, there generally no longer is a unique intertwiner to which the tensor would have to be proportional. Nevertheless, an invariant tensor with $N$ indices can be expanded in any basis of the corresponding $N$-valent intertwiner space. For example, expanding a four-valent invariant tensor in the basis given by the intertwiners \eqref{iota412 g}, and recalling that the norm of the intertwiner \eqref{iota412 g} is $1/\sqrt{d_k}$, we obtain
\be\label{thm4}
\RealSymb{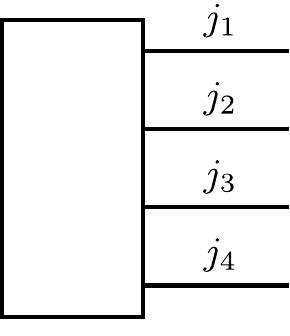}{0.6} \quad = \quad \sum_x d_x\quad\RealSymb{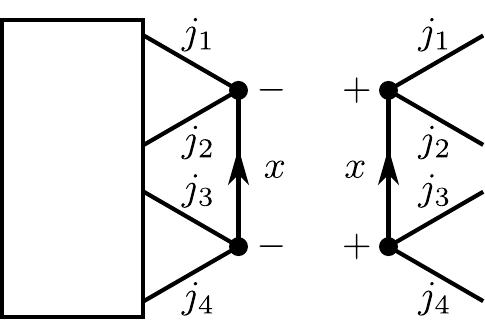}{0.6}
\ee

The relations \eqref{thm1}--\eqref{thm4} play such a central role in graphical calculations that they would fully deserve to be known as the fundamental theorem of graphical calculus. The last part of the theorem, given by \Eq{thm4}, generalizes in a straightforward way to tensors carrying more than four indices. Moreover, even though we wrote \Eqs{thm1}--\eqref{thm4} for tensors having only lower indices, similar relations are naturally valid for tensors having a different index structure. \Eqs{thm1}--\eqref{thm4} can be extended to such tensors simply by using the epsilon tensor to raise indices, corresponding graphically to attaching arrows to some lines on both sides of the equation.

An important way in which the fundamental theorem can be used is to break down a complicated graphical diagram into simpler constituents. Consider a graphical diagram representing an arbitrary invariant contraction of 3$j$-symbols or other invariant tensors. Suppose that the diagram contains a subdiagram with $N$ external lines, such that the subdiagram itself is an $N$-valent invariant tensor, and the graph of the entire diagram can be separated into two disconnected pieces by cutting the $N$ lines of the subdiagram. In this case the diagram can potentially be simplified by using the fundamental theorem to expand the subdiagram in a basis of the $N$-valent intertwiner space.

If a diagram can be divided into two pieces by cutting one, two or three lines, then \Eqs{thm1}--\eqref{thm3} imply that the diagram simply splits into a product of two factors according to the rules
\begin{align}
\RealSymb{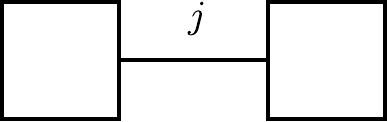}{0.6} \quad &= \quad \delta_{j,0}\quad\RealSymb{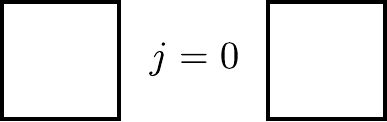}{0.6} \label{thm1'} \\[8pt]
\RealSymb{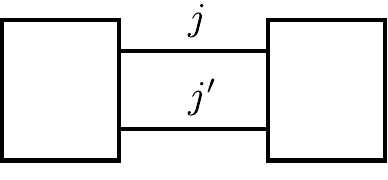}{0.6} \quad &= \quad \delta_{jj'}\frac{1}{d_j}\quad\RealSymb{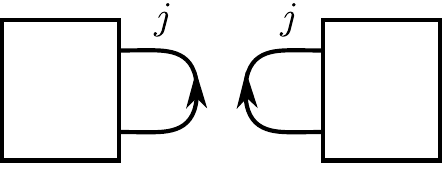}{0.6} \label{thm2'} \\[8pt]
\RealSymb{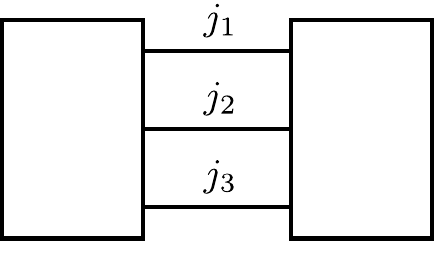}{0.6} \quad &= \quad \RealSymb{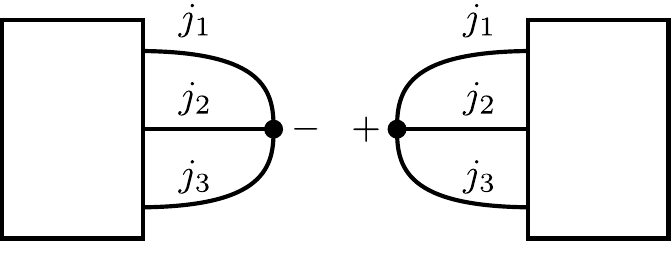}{0.6} \label{thm3'} \\
\intertext{In the case of cutting four lines to break a diagram into two, \Eq{thm4} shows that one does not obtain simply a product of two factors, but rather a sum in which each term is a product of two disconnected diagrams:}
\RealSymb{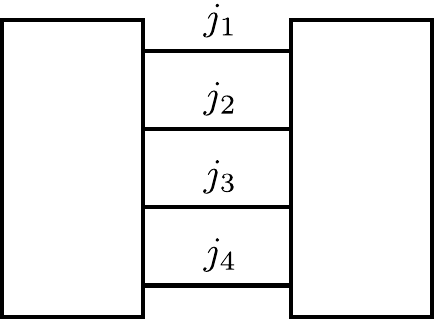}{0.6} \quad &= \quad \sum_x d_x\quad\RealSymb{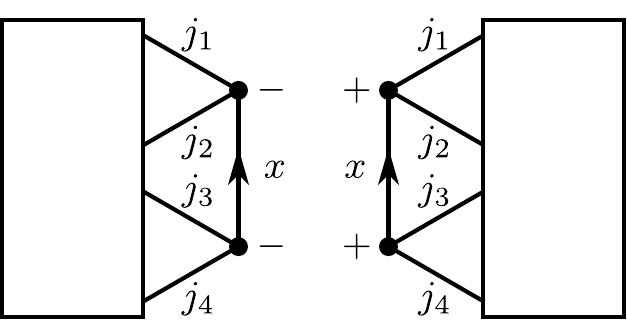}{0.6} \label{thm4'}
\end{align}
It is again straightforward to generalize \Eq{thm4'} to the case where a diagram is cut on more than four lines (though it seems that such generalizations are very rarely needed in practice).

When using the fundamental theorem in the form \eqref{thm1'}--\eqref{thm4'}, it is important to make sure that each of the two blocks actually represents a proper invariant tensor. If there is doubt as to whether this is the case, the invariance of the block can be checked by using the following criterion: A diagram representing a tensor constructed by contracting 3$j$-symbols corresponds to an invariant tensor if and only if it is possible to use \Eqs{2arrows-opp} and \eqref{3jarrows g} to bring the diagram into a form in which each internal line of the diagram carries exactly one arrow.

Let us mention at this point a useful theorem concerning diagrams which represent invariant scalar contractions of 3$j$-symbols, and hence carry no external, uncontracted lines. The theorem states that the value of such a diagram remains unchanged if one simultaneously reverses the direction of every arrow and the sign at every node in the diagram. This theorem implies, for example, that the 6$j$- and 9$j$-symbols of \Eqs{6j g} and \eqref{9j g} are equivalently represented by the diagrams
\vspace{-8pt}
\be
\sixj{j_1}{j_2}{j_3}{k_1}{k_2}{k_3} \quad = \quad \RealSymb{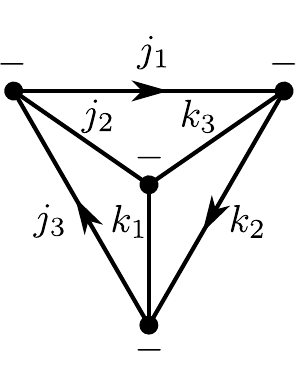}{0.6} \label{6jminus g}
\ee
and
\be
\ninej{j_1}{j_2}{j_3}{k_1}{k_2}{k_3}{l_1}{l_2}{l_3} \quad = \quad \RealSymb{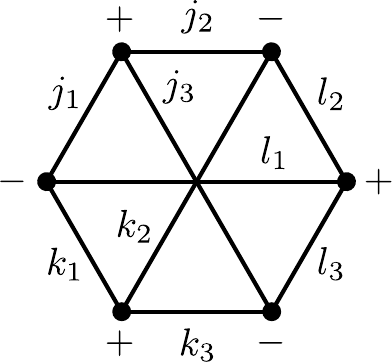}{0.6} \vspace{8pt} \label{9jminus g}
\ee
To prove the theorem, suppose that we have brought a given invariant diagram into a form where each line carries precisely one arrow. Since reversing an arrow on a line carrying spin $j$ produces the factor $(-1)^{2j}$, we see that reversing all the arrows in the diagram multiplies the diagram by the factor $(-1)^{2J}$, where $J$ is the sum of all the spins appearing in the diagram. Similarly, reversing the sign at a node where spins $j_1$, $j_2$ and $j_3$ meet produces the factor $(-1)^{j_1+j_2+j_3}$, so reversing all the signs in the diagram also multiplies the diagram by $(-1)^{2J}$, because every line in the diagram is connected to two nodes. In total, the diagram is therefore multiplied by $(-1)^{4J} = +1$, since $2J$ is an integer, and so $4J$ is even.

While the principal use of \Eqs{thm1'}--\eqref{thm4'} is to break large diagrams into smaller pieces, in some cases \Eq{thm4'} can be used in the reverse direction in order to evaluate a sum over an internal spin in a graphical expression. A very simple example of using the fundamental theorem in this way is provided by a graphical proof of the orthogonality relation \eqref{6j-orth} for the 6$j$-symbol. Using the graphical representation \eqref{6j g}, we can write the sum as
\be
\sum_x d_x\sixj{j_1}{j_2}{x}{j_3}{j_4}{k}\sixj{j_1}{j_2}{x}{j_3}{j_4}{l} \;\; = \;\; \sum_x d_x\;\;\RealSymb{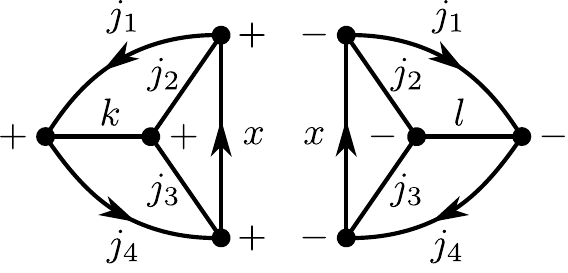}{0.6} \vspace{12pt}
\ee
\newpage
\noindent We see that the form of this expression matches the right-hand side of \Eq{thm4'}, the two blocks with four lines being
\be
\RealSymb{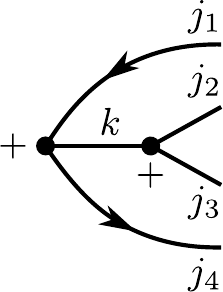}{0.6} \qquad\qquad \text{and} \qquad\qquad \RealSymb{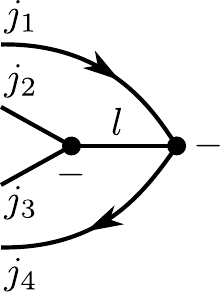}{0.6}
\ee
Therefore \Eq{thm4'} gives
\be
\sum_x d_x\sixj{j_1}{j_2}{x}{j_3}{j_4}{k}\sixj{j_1}{j_2}{x}{j_3}{j_4}{l} \quad = \quad \RealSymb{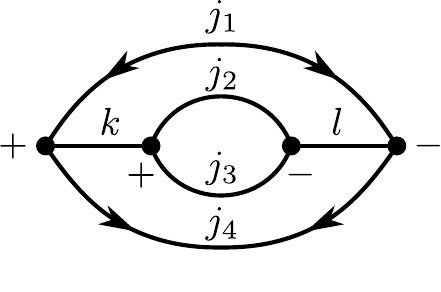}{0.6} \quad = \; \frac{1}{d_k}\delta_{kl},
\ee
where \Eqs{2arrows-opp}, \eqref{3j-orth-mm g} and \eqref{3j-theta g} were used to evaluate the resulting diagram. 

To conclude our overview of the graphical method, let us consider an example which illustrates the use of the machinery of graphical techniques in a practical calculation involving intertwiners. (The calculations in Chapter \ref{ch:elements} provide a large number of further examples.) The problem consists of expressing the six-valent intertwiner
\vspace{10pt}
\be\label{iota612 g}
\RealSymb{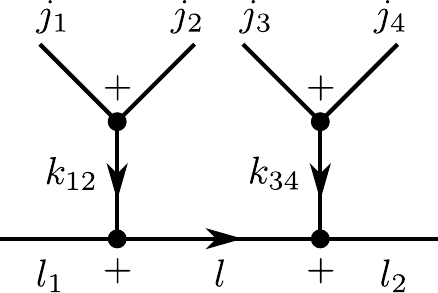}{0.6} \vspace{10pt}
\ee
in the basis formed by the intertwiners
\vspace{10pt}
\be\label{iota613 g}
\RealSymb{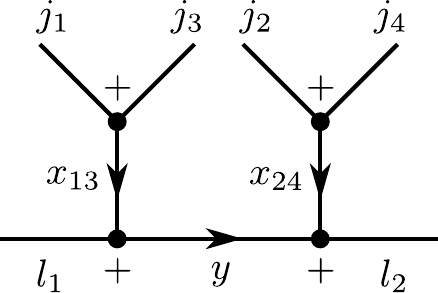}{0.6} \vspace{10pt}
\ee
The result of this exercise is used in section \ref{sec:CR-el} to compute the action of the curvature operator on a spin network node which contains an intertwiner of the form \eqref{iota612 g}.

\newpage

According to the generalization of \Eq{thm4} to six-valent invariant tensors, the intertwiner \eqref{iota612 g} can be expressed in terms of the intertwiners \eqref{iota613 g} as
\begin{align}
&\RealSymb{figA-iota612.pdf}{0.6} \notag \\[-12pt]
&\qquad\qquad = \quad \sum_{x_{13}x_{24}y} d_{x_{13}}d_{x_{24}}d_y\,K(x_{13},x_{24},y|k_{12},k_{34},l)\;\RealSymb{figA-iota613.pdf}{0.6}\label{iota612-start}
\end{align}
where the coefficient $K(x_{13},x_{24},y|k_{12},k_{34},l)$ is given by the contraction of the intertwiners \eqref{iota612 g} and \eqref{iota613 g}: \vspace{-12pt}
\be\label{iota6-K}
K(x_{13},x_{24},y|k_{12},k_{34},l) \quad = \quad \RealSymb{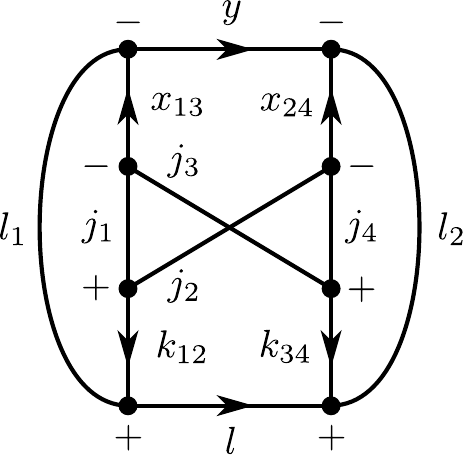}{0.6}
\ee
This diagram can be recognized as the so-called 12$j$-symbol of the first kind (see \eg \cite{Varshalovich} or \cite{YLV}), but we can use the fundamental theorem to break it down to the more familiar 6$j$- and 9$j$-symbols. The diagram clearly cannot be separated into two non-trivial pieces by cutting only two or three lines, so we will use \Eq{thm4'} to cut out the part consisting of the four nodes in the middle of the diagram, as indicated by the red dashed line in \Eq{iota6-cut}. For convenience, we make use of \Eqs{invarrow} and \eqref{3jarrows g} to reverse the direction of the arrows and interchange one pair of signs in \Eq{thm4'}, leading to
\be\label{iota6-cut}
\RealSymb{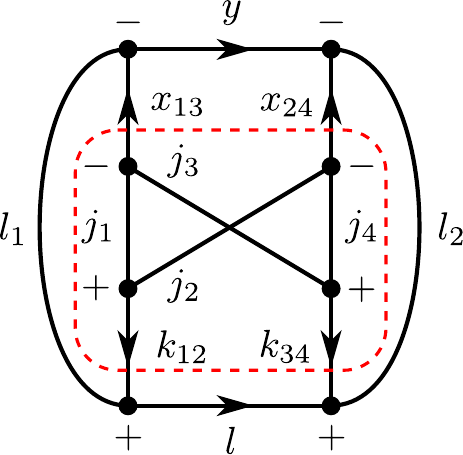}{0.6} \quad = \quad \sum_s d_s\quad\RealSymb{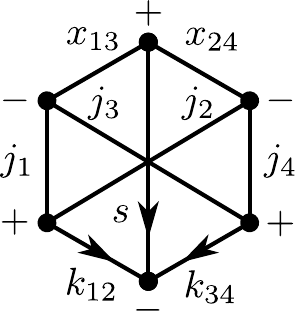}{0.6}\quad\RealSymb{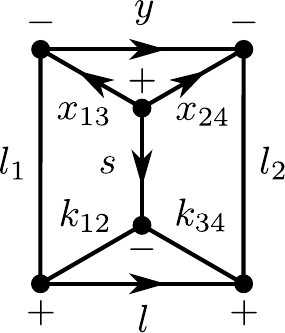}{0.6}
\ee
The first factor in the sum is simply the 9$j$-symbol of \Eq{9j g},
\be\label{iota6-cut1}
\RealSymb{figA-iota6-cut-part1.pdf}{0.6} \quad = \; \ninej{x_{24}}{j_4}{j_2}{x_{13}}{j_3}{j_1}{s}{k_{34}}{k_{12}} \; = \; \ninej{j_1}{j_2}{k_{12}}{j_3}{j_4}{k_{34}}{x_{13}}{x_{24}}{s},
\ee
since \Eq{3jarrows g} shows that the three arrows at the bottom node can be removed. In the second factor, we cancel the three arrows in the middle, and use \Eq{2arrows-opp} to introduce oppositely directed arrows on the lines carrying spins $l_1$ and $l_2$. After this we cut the diagram along the three vertical lines according to \Eq{thm3'}. Each of the resulting pieces is the 6$j$-symbol of \Eq{6j g}, up to powers of $(-1)$, which arise when \Eqs{invarrow} and \eqref{3jminus g} are used to adjust the arrows and the signs so that they agree with \Eq{6j g}. In this way we find
\be\label{iota6-cut2}
\RealSymb{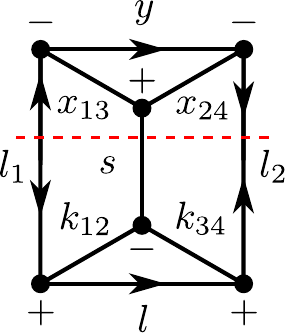}{0.6} \quad = \;\; (-1)^{k_{12}+k_{34}+s}\sixj{k_{12}}{k_{34}}{s}{l_2}{l_1}{l} (-1)^{x_{13}+x_{24}+s}\sixj{x_{13}}{x_{24}}{s}{l_2}{l_1}{y},
\ee
which completes the calculation. Inserting \Eqs{iota6-K}--\eqref{iota6-cut2} back into \Eq{iota612-start}, we conclude that the expansion of the intertwiner \eqref{iota612 g} in the basis of the intertwiners \eqref{iota613 g} is given by
\vspace{10pt}
\begin{align}
&\RealSymb{figA-iota612.pdf}{0.6} \quad = \quad \sum_{x_{13}x_{24}y} d_{x_{13}}d_{x_{24}}d_y(-1)^{k_{12}+k_{34}-x_{13}-x_{24}} \notag \\
&\times\sum_s d_s\sixj{k_{12}}{k_{34}}{s}{l_2}{l_1}{l}\sixj{x_{13}}{x_{24}}{s}{l_2}{l_1}{y}\ninej{j_1}{j_2}{k_{12}}{j_3}{j_4}{k_{34}}{x_{13}}{x_{24}}{s}\quad\RealSymb{figA-iota613.pdf}{0.6}\label{iota6change}
\end{align}

\newpage

\renewcommand{\refname}{{\LARGE Bibliography} \vspace{12pt}}

\end{document}